\documentclass[article]{emulateapj}
\usepackage{lscape}

\def\nhh     {\ifmmode{N({\rm H}_2)}\else{$N$(H$_2$)}\fi}

\newcommand{\hi}{H{\small{\sc I}}}

\newcommand{\msol}{$M_{\odot}$}

\newcommand{\kms}{km s$^{-1}$}
\newcommand{\mm}{$\mu$m}

\def\ltsim{\mathrel{\mathpalette\oversim<}} % less than or sim.
\def\oversim#1#2{\lower 2pt\vbox{\baselineskip 0pt \lineskip 1pt
    \ialign{$\nms#1\hfil##\hfil$\crcr#2\crcr\sim\crcr}}}
\def\nms{\mathsurround=0pt}

\usepackage{epsfig}
\usepackage{graphics}

\tighten
\shorttitle{THINGS: The \hi\ Nearby Galaxy Survey}
\shortauthors{Walter et al.}

\begin{document}

\title{THINGS: The \hi\ Nearby Galaxy Survey}

\author{Fabian Walter\altaffilmark{1},
Elias Brinks\altaffilmark{2}, 
W.J.G. de Blok\altaffilmark{3},
Frank Bigiel\altaffilmark{1},
Robert C. Kennicutt, Jr.\altaffilmark{4},
Michele D. Thornley\altaffilmark{5}
Adam K. Leroy\altaffilmark{1}
}

\altaffiltext{1}{Max-Planck-Institut f{\"u}r Astronomie,
  K{\"o}nigstuhl 17, D-69117, Heidelberg, Germany; walter@mpia.de} 

\altaffiltext{2}{Centre for Astrophysics Research, University of
  Hertfordshire, Hatfield AL10 9AB, U.K.}

\altaffiltext{3}{Department of Astronomy, University of Cape Town,
  Rondebosch 7700, South Africa}

\altaffiltext{4}{Institute of Astronomy, University of Cambridge,
  Madingley Road, Cambridge CB3 0HA, UK}

\altaffiltext{5}{Department of Physics and Astronomy, Bucknell
  University, Lewisburg, PA 17837, USA}

\begin{abstract}
  We present ``The \hi\ Nearby Galaxy Survey (THINGS)'', a high
  spectral ($\leq$5.2\,km\,s$^{-1}$) and spatial ($\sim6''$)
  resolution survey of \hi\ emission in 34 nearby galaxies obtained
  using the NRAO Very Large Array (VLA). The overarching scientific
  goal of THINGS is to investigate fundamental characteristics of the
  interstellar medium (ISM) related to galaxy morphology, star
  formation and mass distribution across the Hubble sequence. Unique
  characteristics of the THINGS database are the homogeneous
  sensitivity as well as spatial and velocity resolution of the \hi\
  data which is at the limit of what can be achieved with the VLA for
  a significant number of galaxies.  A sample of 34 objects at
  distances 2\,$\ltsim$\,$D$\,$\ltsim$\,15~Mpc (resulting in linear
  resolutions of $\sim$100 to 500\,pc) are targeted in THINGS,
  covering a wide range of star formation rates ($\sim10^{-3}$ to
  6\,$M_\odot$\,yr$^{-1}$), total \hi\ masses $M_{\rm HI}$ (0.01 to
  14$\times10^{9}$\,$M_{\odot}$), absolute luminosities $M_{\rm B}$
  (--11.5 to --21.7\,mag) and metallicities (7.5 to 9.2 in units of
  12+log[O/H]). We describe the setup of the VLA observations, the
  data reduction procedures and the creation of the final THINGS data
  products. We present an atlas of the integrated \hi\ maps, the
  velocity fields, the second moment (velocity dispersion) maps and
  individual channel maps of each THINGS galaxy. The THINGS data
  products are made publicly available through a dedicated webpage.
  Accompanying THINGS papers address issues such as the small--scale
  structure of the ISM, the (dark) matter distribution in THINGS
  galaxies, and the processes leading to star formation.
\end{abstract}

\keywords{surveys --- galaxies: structure --- galaxies: ISM --- ISM:
  general --- ISM: atoms --- radio lines: galaxies}

\section{Introduction}

For the past few decades studies of the atomic interstellar medium
(ISM), via observations of the 21--cm line of atomic hydrogen (HI),
have proven to be fundamental for our understanding of the processes
leading to star formation, the dynamics and structure of the ISM, and
the (dark) matter distribution, thereby touching on major issues
related to galaxy evolution. Since the detection of the \hi\ line
(whose forbidden hyperfine structure line had been predicted by van de
Hulst in 1945), this line has been used as the `workhorse' for studies
of the atomic gas in our own and other galaxies.  One of its benefits
is that, in contrast to optical or UV radiation, \hi\ emission does
not suffer from extinction by interstellar dust. In addition, its
Doppler shift provides information about the velocity of the emitting
gas.  This provides important information on the physical properties
of the interstellar gas and the associated kinematics of the ISM.
Furthermore, 21\,cm emission is (under most circumstances) optically
thin; this means that the total amount of \hi\ seen along a particular
line of sight (\hi\ column density) can be converted directly into an
\hi\ mass density and, integrated over an entire galaxy, a total \hi\
mass.

Early studies, using single dish telecopes (yielding resolutions of
$\sim10'$), naturally concentrated on performing detailed studies of
the Galaxy and obtaining some global measurements of other nearby
systems. Only after radio interferometers came into operation, did it
become feasible to obtain detailed, spatially resolved \hi\ images of
external galaxies.  However, given the intrinsically low surface
brightness of the \hi\ emission, large collecting areas are needed, in
particular if high--resolution ($<$1$'$) imaging is desired. 

Ever since the first pioneering observations with large
interferometers in the late 1970's, many galaxies have been mapped
with ever increasing resolution and sensitivity (see, e.g., Bosma
1981a/b, Brinks \& Bajaja 1986, Begeman 1987, Kamphuis et al.\ 1991,
Puche et al.\ 1992, Braun et al.\ 1995, Staveley--Smith et al.\ 1997,
Walter \& Brinks 1999, Kim et al.\ 1999, de~Blok \& Walter 2000,
Verheijen \& Sancisi 2001, Walter \& Brinks 2001, Ott et al.\ 2001,
Walter et al.\ 2002, Swaters et al.\ 2002).  Although remarkable
progress has been achieved in these studies, the lack of
high--resolution \hi\ observations in a significant sample of nearby
galaxies precludes a systematic study of the physical characteristics
and dynamics of the atomic ISM.

This paper describes `The \hi\ Nearby Galaxy Survey' (THINGS) which
was obtained at the Very Large Array of the National Radio Astronomy
Observatory\footnote{The National Radio Astronomy Observatory is a
  facility of the National Science Foundation operated under
  cooperative agreement by Associated Universities, Inc.}. The goal of
THINGS was to obtain high quality observations of the atomic ISM of a
substantial sample of nearby galaxies, covering a wide range of Hubble
types, star formation rates, absolute luminosities, and metallicities
to address key science questions regarding the ISM in galaxies.  A key
characteristic of the THINGS database is the homogeneous sensitivity,
as well as spatial and velocity resolution that is at the limit of
what can be achieved in studies of extragalactic \hi\ with the VLA.
Most of the galaxies in THINGS were drawn from the ``{\em Spitzer}
Infrared Nearby Galaxies Survey (SINGS)'' (Kennicutt et al.\ 2003), a
multi--wavelength project designed to study the properties of the
dusty ISM in nearby galaxies, to ensure that multi--wavelength
observations for each galaxy are available for further analysis.

Science capitalizing on the unique properties of the THINGS data
products will be presented in accompanying papers: de~Blok et al.\
(2008) present a detailed analysis of high--resolution rotation curves
and mass models (including new constraints for the stellar
mass--to--light ratio) for the majority of THINGS galaxies.
Trachternach et al.\ (2008) present a detailed analysis of the centres
of the THINGS targets and perform a harmonic decomposition of the
velocity fields to constrain the non--circular motions in these
objects.  Oh et al.\ (2008) present a new method to remove these
non--circular motions from the velocity fields to derive accurate
rotation curves unaffected by such motions.  The star formation
properties are addressed in papers by Leroy et al.\ (2008) and Bigiel
et al.\ (2008). Leroy et al.\ test various star formation and
threshold recipes that have been proposed to explain the observed star
formation in galaxies. They also describe a new method for creating
star formation rate maps of individual galaxies.  Bigiel et al.\
perform a pixel--by--pixel comparison of the SFR and measured gas
densities for a sample of THINGS galaxies to constrain the `star
formation law' on small scales. Tamburro et al.\ (2008) use the offset
seen between the \hi\ and 24 micron emission to derive an average
timescale for star formation to commence in spiral arm environments.
The wealth of small--scale structure in the ISM revealed in the THINGS
data products is analyzed by Bagetakos et al.\ (in prep.) and Usero et
al.\ (in prep.). Bagetakos et al.\ concentrate on structures generally
referred to as \hi\ holes whereas Usero et al.\ study the \hi\ peak
brightness temperature distribution. Brinks et al.\ (in prep.) present
a study of the \hi\ edges of a sample of THINGS galaxies. Zwaan et
al.\ (2008) use the THINGS data to compare the \hi\ properties in
nearby galaxies to the HI absorption properties found in Damped
Lyman--$\alpha$ Absorbers systems at high redshift. Finally, the
THINGS dataset is of sufficiently high quality that alternative data
imaging techniques can be explored. This is described in Rich et al.\
(2008) who use the THINGS data to investigate the applicability of the
Multi--Scale CLEAN algorithm.

This paper is organized as follows: in Sec.~2 we give a description of
the sample selection and the setup of the observations.  The
observations and data reduction details are described in Sec.~3.
Sec.~4 presents the resulting THINGS data products (cubes and moment
maps). Finally, a summary is presented in Sec.~5.

\section{THINGS Survey Design}

\subsection{Sample Selection}

To ensure multi--wavelength coverage of the sample galaxies, THINGS
targets were mostly drawn from SINGS (Kennicutt et al.\ 2003). Some of
the THINGS targets are not part of SINGS (as SINGS did not include a
few key targets covered in {\em Spitzer Space Telescope} guaranteed
time observations).  Galaxies were chosen to cover the full range of
physical properties, from low--mass, metal--poor, quiescent dwarf
galaxies to massive spiral galaxies. The THINGS galaxies and their
properties are listed in Table~1. Some of our sample galaxies form
part of the WHISP survey obtained at the Westerbork radio telescope
(e.g., Swaters et al.\ 2002) and we have labeled these targets in
Table~1.

Early type (E/S0) galaxies were excluded from the THINGS target list
as their ISM is dominated by hot gas with little, if any, in a neutral
atomic phase. Edge--on systems (e.g., M\,82) were also excluded as the
radial structure of the ISM cannot be disentangled from projection
effects in these systems. The sample was furthermore limited to
galaxies at distances $D<$\,15\,Mpc to ensure that structures smaller
than 500\,pc can be resolved (this is limited by the maximum \hi\
resolution of 6$''$ that can realistically be achieved at the VLA, as
discussed below).  Local Group galaxies were also excluded because of
their large size on the sky. The 34 galaxies in THINGS are listed in
Table~1; the galaxy parameters listed in this table are adopted for all
THINGS papers listed in the introduction, unless explicitly stated
otherwise. Throughout this paper, galaxies are sorted by Right
Ascension (RA).

\subsection{Spatial Resolution}

The main scientific goals for THINGS require an angular resolution of
a few hundred parsecs in order to resolve neutral atomic gas
complexes, to trace the fine structure of the \hi\ and to resolve
spiral arms. This resolution is also required to accurately derive the
central shape and slope of the rotation curve of a galaxy. Given the
distances to our sample galaxies, this necessitates an angular
resolution of $\sim$6$''$, which can be achieved using the VLA in its
B~array configuration (baselines: 210\,m to 11.4\,km). B~array
observations, however, are only sensitive to small--scale structure.
Therefore, additional D~array (35\,m to 1.03\,km) and C~array (35\,m
to 3.4\,km) data are needed to recover extended emission in the
objects.  Objects at declinations lower than about $-20$ degrees are
observed with the extended north arm (BnA array) to ensure a
homogeneous $uv$ coverage.  Reaching even higher spatial resolution at
the VLA is in principle possible (using the A configuration) but would
require integration times per object of $\sim 50$\,hours which rules
out a large survey. We note that our final resolution of 6$''$ is well
matched to that typically obtained for the other data products from
the {\em Spitzer Space Telescope} (e.g., resolution at 24\mm:
$\sim$6$''$) and the {\em Galaxy Evolution Explorer GALEX} (resolution
in Near--Ultraviolet: $\sim$5$''$).

\subsection{Velocity Resolution}

Many of the THINGS science goals require high {\em velocity}
resolution: the $1\,\sigma$ velocity dispersion of the warm neutral
medium is about $6-7$\,km\,s$^{-1}$ (full width half maximum, FWHM:
$\sim16$\,km\,s$^{-1}$); therefore all observations are done with at
least 5\,km\,s$^{-1}$ velocity resolution in order to Nyquist sample
the \hi\ line. The latter constraint implied that many observations in
the VLA archive could not be incorporated in our analysis (as typical
velocity resolutions chosen for those observations were poorer).
Taking into account the velocity range over which HI was detected in
earlier studies (and the limited configurations of the current VLA
correlator), two galaxies have been observed at 1.3 km\,s$^{-1}$
resolution, 15 at 2.6 km\,s$^{-1}$ and 17 at 5.2 km\,s$^{-1}$ (see
Table~2 for a detailed summary). Two targets (NGC\,2841 and NGC\,3521)
had to be observed twice over two adjacent velocity ranges to cover
the large \hi\ line--width.  Most observations are done using two
orthogonal polarizations (see Table~2 for details).  Depending on the
width of the \hi\ profile, either correlator mode 2AD or mode~4 is
used.  The latter setup results in 2~IFs, each IF using 2
polarizations.  The IFs can be tuned independently and are set at
velocities such that the last channels of one IF overlap with the
first few channels of the second IF, allowing roughly twice the
bandwidth of a single IF to be covered in a single observation.

\subsection{Sensitivity Considerations}

High sensitivity is also a key requirement for many THINGS science
goals.  In B--array, the gain is of order 15\,K per mJy for a $6''$
beam.  Observations of on average 7 hours on source, observing 2
polarizations, reach a $1 \sigma$ rms noise of typically
0.4\,mJy\,beam$^{-1}$ (or 6\,K) which corresponds to a column density
limit of $\sim 3.2\times 10^{20}$\,cm$^{-2}$ (a detection in 2
channels of 5 km\,s$^{-1}$ width at the $3\,\sigma$ level).  Higher
surface brightness sensitivities can be reached by convolving the
data. THINGS observations have typical column density sensitivities of
$4\times 10^{19}$\,cm$^{-2}$ at 30$''$ resolution which is adequate to
trace the outer regions of the galaxies. Typical observing times were:
1.5, 2.5, and 7 hours for the D, C and B array configurations
(observations are longer in the extended array to account for the
reduced surface brightness sensitivity). The amount of observing times
required at the VLA and the different array configurations needed are
summarized in Table~2. We have included as much archival data in our
analysis as possible and the respective project IDs are also given in
Table~2. These archival data were observed in various setups, but with
at least the minimum equivalent observing time, and spatial/velocity
resolution as required for the new observations. The total observing
time for this project (including the archival data) is
$\sim500$\,hours.

Given the substantial amount of observing time required at the VLA for
THINGS, data for this project were taken over two years, from Summer
2003 to Summer 2005. Based on the requirements outlined in Sec.~2, the
key observational data for each galaxy are summarized in Table~2:
\begin{description}
\item[column 1:] galaxy name 
\item[column 2:] VLA array used 
\item[column 3:] program code (AW605: THINGS; others: archival data)
\item[column 4:] observing date (local time: MST/MDT)
\item[column 5/6:] start/end of observations (IAT) 
\item[column 7:] duration of observations (hours/minutes)
\item[column 8/9:] name of phase calibrator and its flux density 
\item[column 10/11:] coordinates of pointing centre of the array (equinox given in column 16) 
\item[column 12:] correlator mode
\item[column 13:] total bandwidth in MHz 
\item[column 14:] number of channels per IF
\item[column 15:] channel width in km\,s$^{-1}$ 
\item[column 16:] equinox
\item[column 17:] Barycentric (heliocentric) central velocities in
  km\,s$^{-1}$. The observations were done assuming the optical
  definition for the Doppler shift. In case two independent IFs were
  used, the central velocity of the second IF is also listed.

\end{description}

\section{Observations and Data Reduction}

\subsection{Amplitude/Phase calibration}

The calibration and data reduction is performed using standard
routines in the {\sc AIPS} package\footnote{The Astronomical Image
  Processing System ({\sc AIPS}) has been developed by the NRAO.}.
The absolute flux scale for the data is determined by observing one of
the standard primary calibrators used at the VLA (3C286=J1331+305;
3C48=J0137+331 or 3C147=J0542+498, using the flux scale of Baars et
al.\ 1977, task {\sc setjy}). The primary calibrator is used for the
bandpass corrections as well (task {\sc bpass}). The time variable
phase and amplitude calibration is done using nearby, secondary
calibrators which are unresolved for the arrays used ({\sc calib}).
The names and fluxes of these secondary calibrators are given in
Table~2 for each observing run.  Note that in some cases different
array configurations required the choice of different calibrators for
an individual galaxy. The uncertainty of the flux calibration
($\sim5$\%) dominates all measurements of HI flux densities derived in
the remainder of this paper.

In those cases where the target galaxy is close in velocity to
\ion{H}{1} emission from our Galaxy, each calibrator observation
(primary and secondary) is split into two parts to form a pair of
observations, one part having the velocity shifted by typically $+300$
\kms\ and the other by $-300$ \kms\ to avoid contamination by Galactic
emission (the exact velocity offset depends on correlator
configuration and systemic velocity of the object). This information
is also listed in Table~2 (last column).

\subsection{Flagging of bad visibilities} 

The $uv$--data are inspected by eye for each configuration and bad
data points due to either interference or cross--talk between antennae
are removed (tasks {\sc tvflg, uvflg}), after which the data are
calibrated ({\sc calib}, {\sc clcal}).  In some cases, to ensure
proper calibration, solar interference on the shortest baselines had
to be removed from the calibrator observations by deleting the
affected baselines (in these cases short spacing information was still
available from observations in other array configurations). We note
though that most of the D--array observations were deliberately
obtained during nighttime and were thus not affected by solar
interference.  After final editing, all data are combined to form a
single dataset which was subsequently used for mapping (task {\sc
  dbcon}).

\subsection{Continuum subtraction} 

In order to separate continuum from line emission we first determine
the line--free channels in our observation and subsequently subtract
the continuum emission in the $uv$--plane by fitting a first order
polynomial to the line--free channels (task {\sc uvlin}). In a few
cases, given the limitations of the current VLA correlator, only very
few channels ($\sim$5) of line--free emission were available (but even
in those cases the continuum subtraction worked satisfactorily).  The
line free channels were also used to create continuum maps of our
galaxies.  The quality of these continuum maps varies significantly
from galaxy to galaxy (depending on the number of line--free
channels). In general these maps have limited sensitivity and are
therefore not presented here. For high sensitivity radio continuum
images of the SINGS galaxies the reader is referred to Braun et al.\
(2007).

\subsection{Mapping/Data cubes} 

After calibration and editing, datacubes are produced using the task
{\sc imagr}. For each galaxy, one data cube is calculated with natural
weighting to obtain maps at the highest surface brightness sensitivity
(hereafter referred to `NA cubes/maps'). A second cube is produced
using the {\sc robust} weighting scheme (Briggs 1995) for every galaxy
(`RO cubes/maps').  With this scheme a data cube can be created with
the sensitivity approaching that of natural weighting, but with a
well--behaved synthesized beam at a resolution closer to that of
uniform weighting ($\sim$6$''$). After experimenting with various {\sc
  robust} parameters we used a common parameter ({\sc robust}=0.5) for
all galaxies. We typically cleaned down to a flux level of 2.5 times
the noise level (no clean boxes were used).

Table 3 summarizes the final mapping parameters of the data cubes:
\begin{description}
\item[column 1:] galaxy name 
\item[column 2:] weighting scheme used. NA: Natural, RO: Robust=0.5 weighting
\item[column 3/4:] major and minor axis of synthesized beam in arcseconds 
\item[column 5:] position angle of synthesized beam in degrees (measured north to east) 
\item[column 6:] noise in one channel map in mJy\,beam$^{-1}$
\item[column 7:] image size in pixels 
\item[column 8:] pixel size in arcseconds 
\item[column 9:] number of channels in final data cube 
\item[column 10:] channel width in km\,s$^{-1}$.
\end{description}

Throughout the paper we give the noise in mJy\,beam$^{-1}$. The noise
in Kelvin can be calculated using the following equation:
\begin{equation}
{T}_{\rm B}[K] = 6.07\times10^5 \times {S\rm{[Jy\,beam}^{-1}]} / ({\rm
  FWHM}_{\rm maj}[\arcsec] \times {\rm FWHM}_{\rm min}[\arcsec]).
\end{equation}

where FWHM$_{\rm maj}$ and FWHM$_{\rm min}$ are the major and minor
axis of the beam in arcsec.

\subsection{Flux densities}

Obtaining the correct \ion{H}{1} flux of each channel and hence the
global \ion{H}{1}\ profiles of our targets is not straightforward. As
first pointed out by J\"ors\"ater \& van Moorsel
(1995), fully cleaned maps do not exist and any
cleaned map consists of the sum of two maps: one containing the
restored clean components and the other the residual map. In the
former, the unit is Jansky per clean beam area, and in the latter
Jansky per dirty beam area. Usually, fluxes are determined on the
combined map, assuming that the clean beam is the correct one for the
entire map. In reality, the flux is calculated correctly only for the
cleaned fraction of the map; for the residual map the flux is
overestimated by a factor $\epsilon$=$\Omega_{\rm dirty}/\Omega_{\rm
  clean}$ where $\Omega_{\rm dirty}$ and $\Omega_{\rm clean}$ are the
dirty and clean beam sizes, respectively.  This becomes an issue for
extended objects, especially in the case of combined array data, like
our THINGS datasets. For a full discussion on this topic see
J\"ors\"ater \& van Moorsel (1995) and Walter \&
Brinks (1999). Following their prescriptions, the real flux of a
channel map is given by:
\begin{equation}
G=\frac{D \times C}{D-R}
\end{equation}
where $C$ is the cleaned flux, and $R$ and $D$ are the 'erroneous'
residual flux and the 'erroneous' dirty flux over the same area of a
channel map, respectively. The AIPS task {\sc imagr} allows one to
correctly scale the residual image (i.e., the division by $\epsilon$)
so that the correct flux is obtained in the fully restored map. We
have carefully compared the {\sc imagr} results (using the paramaters
{\sc imagrprm(5)=1, imagrprm(6)=50, imagrprm(7)=50}, which determine
the half--width of the box used to determine the dirty beam area) to
the analytical expression given above and conclude that {\sc imagr}
correctly scales the residual maps.

We stress that this flux correction works only for those areas that
contain real emission (defined by the master cube discussed below).
It is important to note that the noise outside the regions containing
emission is artificially decreased in the rescaled cubes as the fluxes
in the residual map (which represent only noise in the areas that do
not contain emission) have been scaled down using this technique.

In summary, there are a two different sets of cubes that can be used
for future analysis:

\noindent $\bullet$ `standard' cube: this is the standard output from
{\sc imagr} with no flux correction which has uniform noise
properties.  This cube is used for all analysis that requires data
selection that is based on the noise in the cubes.

\noindent $\bullet$ blanked, flux `rescaled' cube: this cube contains
only the regions with emission (the blanking procedure is described in
the next sub--section), i.e. it does not contain regions with no
emission (noise). The \hi\ fluxes in these channel maps are correct
and these cubes have been used in the creation of the moment maps
described in the following.

\subsection{Blanking the data cubes}

To separate real emission from noise, we consider only those regions
for further analysis which show emission in three consecutive channels
above a set level ($2\sigma$) in standard cubes that have been
convolved to 30$''$ resolution. These areas of genuine emission are
stored as masks in a master cube. Using this cube as reference when
calculating maps of the total \hi\ surface brightness, velocity, and
velocity dispersion has the advantage that the same regions are
included when inspecting cubes at different resolutions and with
different signal--to--noise ratios.

In the next step, we blank the areas that contain noise in our
residual--rescaled data cubes (Sec.~3.5) using these master cubes
(using the task {\sc blank}).  This way we make sure that only regions
of genuine emission are considered. The moment maps were calculated
using the task {\sc xmom}, by using all pixels in the blanked data
cubes. As a last step, we correct the integrated \hi\ map for primary
beam attenuation using the task {\sc pbcor} (note that primary beam
correction has no effect on the first and second moment maps).

\subsection{\hi\ Spectra and \hi\ Masses}

To create global \hi\ profiles (spectra) for each individual galaxy we
calculated the \hi\ flux in each channel map in the residual flux
scaled NA data cube (after blanking and correction for primary beam
attenuation) for each galaxy and plot the resulting spectra in Fig.~1.
The \hi\ spectra show a wealth of properties: the dwarfs show narrow
and faint emission (given their low dynamical masses and low HI
contents); their spectra can be approximated by Gaussian functions.
The spirals on the other hand show broad and bright line emission.
Some profiles show the typical double--horned profile indicative of
flat rotation curves. The individual \hi\ spectra of the THINGS sample
are given in Fig.~1 and Table~4 (only one example galaxy is printed
here, the tables for all other galaxies are available in the online
version of the Journal). These spectra were also used to determine the
central velocity of the individual galaxies that are listed in
Table~5. This central value was obtained by taking the average
velocity of those velocities corresponding to the 20\% peak level on
each side of the spectrum.  Table 5 also lists the velocity widths at
50\% and 20\% of the peak level. A typical uncertainty for the both
central velocities and the velocity widths is the width of one channel
(as summarized in Tab.~2).

The total \hi\ fluxes in units of Jy\,km\,s$^{-1}$ and $M_\odot$ are
listed in Table~5 and their errors are dominated by uncertainties in
the flux calibration (see Sec.~3.1). The \hi\ masses have been
calculated from the total fluxes using:

\begin{equation}M_{HI}[M_\odot] = 2.36\times10^5 D^2 \times \sum_i S_i \, \Delta v\end{equation}

where $\sum_i(S_i \Delta v)$ is the summation over the total emission
in each channel in units of Jy\,km\,s$^{-1}$ and $D$ is the distance
to the source in Mpc.  This assumes that the \hi\ is optically thin.
If there were significant self--absorption, this would cause the
derived \hi\ masses to be underestimated. As Zwaan et al.\ (1997)
argue that the flux correction due to self absorption is around 10\%
(in a statistical sample) and given the fact that most of our objects
are seen at low inclinations, we did not attempt to correct for any
(unknown) optical depth effects.

Although our data have not been corrected for potentially missing
short spacings we note that our D--array observations are sensitive to
emission up to spatial scales of 15$'$.  For comparison, we also
summarize single dish fluxes (compiled from various measurements
available in the literature by Paturel et al.\ 2003) in Table~5
(column 4). We also compare our fluxes to the study by Fisher \& Tully
1981 as there is a significant overlap between our sample and their
study. A comparison shows that our fluxes are broadly in agreement
with the published ones. As the single dish fluxes from Paturel et
al.\ 2003 have been compiled from a variety of sources, we consider
this agreement satisfactory. The only cases where our fluxes are
considerably lower (by a factor 2 or more) than the single dish
measurements are those galaxies which are more extended than our
primary beam (i.e. the area mapped by us).  In Tab.~5 we also compare
our central velocities and velocity widths to those obtained by Fisher
\& Tully (1981) and find excellent agreement between the two studies.

\section{THINGS Data Products}

In the following we present the \hi\ data cubes (channel maps), the
integrated \hi\ maps (moment 0), the velocity fields (moment 1) and
the dispersion maps (moment 2) for each of the THINGS galaxies.  All
these THINGS data products (natural and robust weighting) are made
publicly available at http://www.mpia.de/THINGS.

\subsection{Channel maps}

Every even--numbered figure (Figs 2, 4, ...68) shows the channel maps
for each of the THINGS targets.  To save space, not every single
channel map is shown (see figure caption for details). In order to
show the noise structure in the individual channels, the non--blanked
(and non--rescaled) cubes are shown.  To emphasize extended structure,
the channel maps derived from the NA data cubes are presented; the
greyscale represents the \hi\ intensity on a linear scale (given in
the caption).  The central velocities for each channel map are given
in the top right corner and the channel width is given in the caption.
In general, a lot of complex fine--scale structure is visible in the
individual channel maps; most of this information is lost when adding
the channel maps together to create the integrated \hi\ maps (moment
0).  The following page shows the moment maps created from the data
cubes (as discussed in the following).

\notetoeditor{Please make sure that figures 2,3 and figures 4,5 etc...
  68, 69 are facing each other in the printed version of the journal
  as they refer to the same galaxy. Thank you.}

\subsection{Integrated \hi\ maps}

The integrated \hi\ maps (based on the NA data) are shown in the top
left panel of every odd--numbered figure (Figs.~3, 5, ...69). For each
galaxy we show the same area as in the corresponding channel maps. The
integrated \hi\ maps have been created from the blanked data cubes by
integrating:

\begin{equation}
I_{HI}=\displaystyle\sum_{i} S_i \times \Delta v,
\end{equation}

where $i$ denotes the i--th channel, $S_i$ is the emission in one
channel map in Jy\,beam$^{-1}$ and $\Delta v$ is the velocity width of
one channel.  The resulting map has subsequently been corrected for
primary beam attenuation. The units of the integrated \hi\ map are
Jy\,beam$^{-1}$\,km\,s$^{-1}$ which can be converted to column
densities $N_{\rm HI}$ (in units of cm$^{-2}$) using:

\begin{equation}
  N_{\rm HI} = 1.823\times 10^{18} \times \displaystyle\sum_{i} T_{\rm B,i} \,\Delta  v,
\end{equation}

where $T_{\rm B,i} \,\Delta v$ is the velocity integrated surface
brightness temperature in units of K\,km\,s$^{-1}$ (see Eq.~1).

The \hi\ is shown on a linear scale, and the range of \hi\ intensities
shown is given in the respective figure caption for reference. The
beamsize is indicated in the lower left corner (see Table~3). Even
though a lot of the fine structure seen in the channel maps is lost,
these images still reveal detailed structure, mostly in the form of
spiral arms and HI holes.  In the top right panel of each figure we
show an optical image of the same area for comparison (taken from the
Digitized Sky Survey, plotted on a logarithmic scale).

\subsection{Velocity fields}

The bottom left corner of the odd--numbered figures (Fig.~3, 5, ...69)
shows the velocity field of each galaxy. The velocity field (first
moment) was calculated from the blanked NA data cube by evaluating:

\begin{equation}
{\rm <v>}= \frac{\sum_{i} S_i \times v}{\sum_{i} S_i},
\end{equation}

i.e. the intensity--weighted average of the velocities. In the adopted
colour scheme dark colour represents redshifted (receding) emission
whereas lighter colour indicates blue--shifted (approaching) emission
(both with respect to the systemic velocity).  The thick black contour
indicates the systemic velocity (given in the figure caption). The
black and white contours thus show the approaching and receding
iso--velocity contours at a fixed velocity interval (the interval is
given in the figure captions). We note that the velocity fields
presented here are created by applying equation~6 to all of the
information present in the (blanked) data cubes (i.e., including
material that may not be directly associated with the bulk of the
galaxy's disk). Therefore, they can only give some first indication of
the overall kinematics in the individual systems and should be treated
with caution when attempting a detailed kinematical analysis.  de~Blok
et al.\ (2008), Oh et al.\ (2008) and Trachternach et al.\ (2008)
discuss in detail the various methods that need to be employed for a
proper kinematical analysis of the THINGS galaxies.

\subsection{Velocity Dispersion maps}

Finally, in the bottom right corner of the odd--numbered figures (Fig.~3, 5,
...69) we present the velocity dispersion (second moment) map for each
galaxy calculated from:

\begin{equation} 
{\rm \sigma}=\sqrt{\frac{\sum_{i} S_i
  \times (v - \langle v \rangle)^2}{\sum_{i} S_i}},
\end{equation}

again based on the NA weighted data.  Contours indicate
iso--dispersion values (numbers are given in the respective figure
captions) for the individual galaxies. Again we stress that all
emission in the (blanked) data cubes was used to create the maps - in
the case of tidal material in the field (which is usually located at
velocities that are quite different from those in the galaxy's disk)
this will lead to high apparent velocity dispersion (second moment)
values in the map which have no immediate physical meaning. For
quiescent regions, the velocity dispersion is a measure of the
turbulence of the ISM; however a more sophisticated analysis is needed
to constrain the true velocity dispersions of the different phases
along individual lines of sight.

\section{Summary}

In this paper we have described the observations and data products of
the THINGS project -- ``The \hi\ Nearby Galaxy Survey''. THINGS is a
large observational project executed at the NRAO Very Large Array
(VLA) to obtain 21--cm \hi\ observations of the highest quality
(spatial resolution $\sim6''$, velocity resolution
$\leq$5.2\,km\,s$^{-1}$) of nearby galaxies.  A key characteristic of
the THINGS dataset is the homogeneous sensitivity as well as spatial
and velocity resolution of the \hi\ data which is at the limit of what
can be achieved with the VLA for a significant sample of galaxies.
The sample includes 34 objects at distances
2\,$\ltsim$\,$D$\,$\ltsim$\,15~Mpc (resulting in linear resolutions of
$\sim$100 to 500\,pc). The objects cover a wide range of star
formation rates ($\sim10^{-3}$ to 6\,$M_\odot$\,yr$^{-1}$), total \hi\
masses $M_{\rm HI}$ (0.01 to 14$\times10^{9}$\,$M_{\odot}$), absolute
luminosities $M_{\rm B}$ (--11.5 to --21.7\,mag) and metallicities
(7.5 to 9.2 in units of 12+log(O/H)).

We have presented a detailed description of the sample selection, the
observational requirements, and the setup of the individual
observations. We have described the data reduction procedures in
detail and presented an atlas of the THINGS data products. This atlas
contains channel maps for each galaxy as well as the integrated \hi\
(moment 0), velocity (moment 1) and velocity dispersion maps (moment
2).  These THINGS data products enable investigations of key
characteristics of the interstellar medium related to galaxy
morphology, star formation and mass distribution across the Hubble
sequence.  The THINGS data products (cubes and moment maps) are
available at http://www.mpia.de/THINGS. Accompanying THINGS papers
address issues such as the small--scale structure of the ISM, the
(dark) matter distribution in THINGS galaxies, and the processes
leading to star formation.

E.B. gratefully acknowledges financial support through an EU Marie
Curie International Reintegration Grant (Contract No.
MIRG-CT-6-2005-013556). The work of W.J.G.d.B. is based upon research
supported by the South African Research Chairs Initiative of the
Department of Science and Technology and National Research Foundation.
F.B. acknowledges support from the Deutsche Forschungsgemeinschaft
(DFG) Priority Program 1177.  The National Radio Astronomy Observatory
is a facility of the National Science Foundation operated under
cooperative agreement by Associated Universities, Inc.  We have made
use of the Extragalactic Database (NED), which is operated by the Jet
Propulsion Laboratory, California Institute of Technology, under
contract with the National Aeronautics and Space Administration. This
research has also made use of NASA's Astrophysics Data System (ADS).
We acknowledge the usage of the HyperLeda database
(http://leda.univ-lyon1.fr).

\newpage

\clearpage
\begin{landscape}
\begin{deluxetable*}{llllrccccccccc}
\tabletypesize{\scriptsize}
%\rotate
\tablewidth{23cm}
\tablecaption{Properties of THINGS sample galaxies.}
\tablehead{
\colhead{1}&\colhead{2}&\colhead{3}&\colhead{4}&\colhead{5}&\colhead{6}&\colhead{7}&\colhead{8}&\colhead{9}&\colhead{10}&\colhead{11}&\colhead{12}&\colhead{13}&\colhead{14}\\
\colhead{Galaxy}	& alt. name  &\colhead{RA(2000.0)}  	&\colhead{Dec(2000.0)}	&\colhead{$D$}	  &\colhead{log($D_{25}$)}&\colhead{$m_B$} &\colhead{$M_B$}  &\colhead{Incl}	 &\colhead{P.A.}	       &\colhead{Metal.}     &\colhead{SFR}                  &\colhead{Type}  & \colhead{Notes}\\
 \colhead{}            &{}   &\colhead{hh mm ss.s}     &\colhead{dd mm ss.s} 	&\colhead{Mpc}    &\colhead{log(0.1$'$)}  &\colhead{mag}	&\colhead{mag} &\colhead{$^{\circ}$} &\colhead{$^{\circ}$}   &\colhead{12+log(O/H)}&\colhead{$M_\odot$\,yr$^{-1}$} &\colhead{---} & \colhead{---} \\
}
\startdata
NGC\,628	& M\,74, UGC\,1149 &01 36 41.8		&+15 47 00	&7.3  [K04]	&1.99	&9.35	&-19.97	&7      &20 [T08]	&8.33 [M08]	&1.21 [L08]	&5  & \\
NGC\,925	& UGC\,1913     &02 27 16.5  [T08]	&+33 34 44	&9.2  [F01]	&2.03	&9.77	&-20.04	&66	&287 [dB08]	&8.24 [M08]	&1.09 [L08]	&7  & \\
NGC\,1569	& UGC\,3056     &04 30 49.0		&+64 50 53	&2.0 [K04]	&1.60	&8.33	&-18.12	&63	&112 [M05]	&8.16 [S89]	&0.06 [K98]	&10 & \\
NGC\,2366	& UGC\,2366     &07 28 53.4  [O08]	&+69 12 51	&3.4 [H01]	&1.64	&10.51	&-17.17	&64	&40 [O08]	&7.96 [S89]	&---	        &10 & B\\
NGC\,2403	& UGC\,3918     &07 36 51.1  [T08]	&+65 36 03	&3.2 [F01]	&2.20	&8.11	&-19.43	&63	&124 [dB08]	&8.31 [M08]	&0.85 [L08]	&6  & B\\
Ho\,II          & UGC\,4305     &08 19 05.0		&+70 43 12	&3.4 [K04]      &1.82	&10.78	&-16.87	&41	&177 [P92]	&7.68 [M08]	&0.07 [L08]	&10 & W\\
M81 DwA		&               &08 23 56.0		&+71 01 45	&3.6 [K04]	&1.1*	&16.26  &-11.49       &23     &49 [B06]	&---	        &---            &10 & \\
DDO\,53		& UGC\,4459     &08 34 07.2  		&+66 10 54	&3.6 [K04]      &0.89	&14.31	&-13.45	&31	&132 [B06]	&7.77 [M08]	&0.008 [L08]	&10 & \\
\enddata
\tablecomments{Comments on columns:\\
  {\em column 1:} galaxy name\\
  {\em column 2:} alternative names (UGC and Messier catalogs)\\
  {\em column 3/4:} Coordinates in J2000.0. No entry: coordinates taked from NED; T08: Trachternach et al.\ 2008\\
  {\em column 5:} distances in Mpc. vflow: distances calculated from
  NED using Hubble flow distances (corrected for Virgo infall); dB08:
  de Blok et al.\ 2008, O08: Oh et al.\ 2008, K02: Karachentsev et
  al.\ 2002, K04: Karachentsev et al.\ 2004, H01: Hunter et al.\ 2001,
  F01: Freedman et al.\ 2001, L02: Leonard et al.\ 2002, M01: Macri et
  al.\ 2001, D00: Drozdovsky \& Karachentsev 2000\\
  {\em column 6:} optical size (taken from LEDA, {\sc leda.univ-lyon.fr}); K04: Karachentsev et al.\ 2004; Note: Holmberg diameter is given for M\,81\,dwA\\
  {\em column 7:} apparent blue magnitude, corrected for Galactic foreground extinction and extinction internal to the galaxy from LEDA; B06: Begum et al.\ 2006\\
  {\em column 8}: absolute B magnitude derived from columns 7 and 4\\
  {\em column 9/10:} Inclination and position angle; No entry: taken from
  LEDA, dB08: de Blok et al.\ 2008, B06: Begum et al.\ 2006, O01: Ott
  et al.\ 2001, P92: Puche et al.\ 1992, O08: Oh et al.\ 2008, M05:
  M\"uhle et al.\ 2005, T08: Tamburro et al.\ 2008, H98: Hunter et
  al.\ 1998, T93: Tilanus \& Allen 1993, B81: Bosma et al.\ 1981 \\
  {\em column 11:} metallicities. M08: Moustakas et al.\ 2008, S89:
  Skillman et al.\ 1989, Z94: Zaritzky et al.\ 1994, S94: Storchi et al.\ 1994\\
  {\em column 12:} Star formation rates. Derived from fluxes published in
  the literature and corrected for our adopted distances: L08: Lee et
  al.\ 2008, K98: Kennicutt 1998, K03: Kennicutt et al.\ 2003\\
  {\em column 13:} Morphological type from LEDA\\
  {\em column 14:} Notes: W: WHISP target (e.g. Swaters et al.\ 2002), B: target studied by Braun (1995).\\
}
\end{deluxetable*}
\clearpage
\end{landscape}

\newpage

\clearpage
\begin{landscape}
\begin{deluxetable*}{lcclrrrclrrccccccc}
\tabletypesize{\scriptsize}
%\rotate
\tablewidth{25cm}
\tablecaption{Summary of THINGS observing setups.}
%\tablenum{}
%\tablecolumns{}
\tablehead{
\colhead{1}&\colhead{2}&\colhead{3}&\colhead{4}&\colhead{5}&\colhead{6}&\colhead{7}&\colhead{8}&\colhead{9}&\colhead{10}&\colhead{11}&\colhead{12}&\colhead{13}&\colhead{14}&\colhead{15}&\colhead{16}&\colhead{17}&\colhead{18}\\
\colhead{Name} & \colhead{Conf.} &\colhead{Project}&\colhead{Date}       &\colhead{Start}  &\colhead{End}    &\colhead{Dur.}  &\colhead{cal.}      &\colhead{Flux}  &\colhead{RA}           &\colhead{DEC}        &\colhead{mode}  &\colhead{BW}    &\colhead{chan} &\colhead{$\Delta$V}   &\colhead{equ.} &\colhead{Vhel1} &\colhead{Vhel2}\\
  \colhead{} & \colhead{} &\colhead{} &\colhead{yy-mm-dd} &
  \colhead{} & \colhead{} & \colhead{} & \colhead{} & \colhead{Jy} &
  \colhead{hh mm ss} & \colhead{dd mm ss} &\colhead{} & \colhead{MHz}
  &\colhead{\#}  &\colhead{km\,s$^{-1}$} & \colhead{} 
  &\colhead{km\,s$^{-1}$} &\colhead{km\,s$^{-1}$} \\} 
\startdata
NGC\, 628 & B	& AW605	&2005-04-02   &16 45 &00 15  & 7 00 &0202+149  & 3.578$\pm$0.003 &01 34 01.00  & 15 31 55.0  &2AD &1.56   &128  & 2.6 &1950    & 655.353 & \nodata  \\
NGC\, 628 & C	& AW605	&2004-02-22   &20 45 &23 00  & 2 00 &0202+149  & 3.745$\pm$0.003 &01 34 01.00  & 15 31 55.0  &2AD &1.56   &128  & 2.6 &1950    & 655.353 & \nodata  \\
NGC\, 628 & D	& AW605	&2004-07-05   &16 30 &18 15  & 1 15 &0202+149  & 3.717$\pm$0.006 &01 34 01.00  & 15 31 55.0  &2AD &1.56   &128  & 2.6 &1950    & 655.353 & \nodata  \\
\tableline
NGC\,925 & B	& AW605	&2004-01-08   &22 45 &05 45  & 6 00 &0234+285  & 2.075$\pm$0.008 &02 24 16.90  & 33 21 19.0  &2AD &1.56   &128  & 2.6 &1950    & 565.0   & \nodata  \\
NGC\,925 & C	& AW430	&1996-04-27   &22 30 &01 30  & 2 15 &0234+285  & 2.288$\pm$0.007 &02 24 16.90  & 33 21 19.0  &2AD &1.56   &128  & 2.6 &1950    & 565.0   & \nodata  \\
NGC\,925 & C	& AW430	&1996-04-28   &23 30 &01 15  & 2 15 &0234+285  & 2.301$\pm$0.006 &02 24 16.90  & 33 21 19.0  &2AD &1.56   &128  & 2.6 &1950    & 565.0   & \nodata  \\
NGC\,925 & D	& AW436	&1996-08-30   &14 45 &15 30  & 0 30 &0234+285  & 2.266$\pm$0.005 &02 24 16.90  & 33 21 19.0  &2AD &1.56   &128  & 2.6 &1950    & 565.0   & \nodata  \\
\tableline
NGC\,1569 & B	& AW605 &2003-11-10   &05 00 &12 30  & 6 00 &0404+768  & 5.507$\pm$0.003 &04 26 06.00  & 64 44 18.0  &2AD &1.56   &128  & 2.6 &1950    & -90.0   & \nodata  \\
\enddata
\end{deluxetable*}

\newpage
\clearpage
%\begin{landscape}
\begin{deluxetable*}{lcclrrrclrrccccccc}
\tablenum{2}
\tabletypesize{\scriptsize}
%\rotate
\tablewidth{25cm}
\tablecaption{Summary of THINGS observing setups.}
%\tablenum{}
%\tablecolumns{}
\tablehead{
\colhead{1}&\colhead{2}&\colhead{3}&\colhead{4}&\colhead{5}&\colhead{6}&\colhead{7}&\colhead{8}&\colhead{9}&\colhead{10}&\colhead{11}&\colhead{12}&\colhead{13}&\colhead{14}&\colhead{15}&\colhead{16}&\colhead{17}&\colhead{18}\\
\colhead{Name} & \colhead{Conf.} &\colhead{Project}&\colhead{Date}       &\colhead{Start}  &\colhead{End}    &\colhead{Dur.}  &\colhead{cal.}      &\colhead{Flux}  &\colhead{RA}           &\colhead{DEC}        &\colhead{mode}  &\colhead{BW}    &\colhead{chan} &\colhead{$\Delta$V}   &\colhead{equ.} &\colhead{Vhel1} &\colhead{Vhel2}\\
  \colhead{} & \colhead{} &\colhead{} &\colhead{yy-mm-dd} &
  \colhead{} & \colhead{} & \colhead{} & \colhead{} & \colhead{Jy} &
  \colhead{hh mm ss} & \colhead{dd mm ss} &\colhead{} & \colhead{MHz}
  &\colhead{\#}  &\colhead{km\,s$^{-1}$} & \colhead{} 
  &\colhead{km\,s$^{-1}$} &\colhead{km\,s$^{-1}$} \\} 
\startdata
\newpage
NGC\,3031 & B	& AW343 &1993-02-28   &02 30 &10 30  & 2 45 &0917+624  & 1.106$\pm$0.002 &09 50 49.00  & 69 24 20.0  &2AD &1.56   &128  & 2.6 &1950    &  40.0   & \nodata  \\
NGC\,3031 & B	& AW343 &1993-02-28   &02 30 &10 30  & 2 45 &0917+624  & 1.097$\pm$0.002 &09 52 07.00  & 69 12 00.0  &2AD &1.56   &128  & 2.6 &1950    &-125.0   & \nodata  \\
NGC\,3031 & B	& AW343 &1993-03-05   &03 30 &09 00  & 2 00 &0917+624  & 1.052$\pm$0.007 &09 50 49.00  & 69 24 20.0  &2AD &1.56   &128  & 2.6 &1950    &  40.0   & \nodata  \\
NGC\,3031 & B	& AW343 &1993-03-05   &03 30 &09 00  & 2 00 &0917+624  & 1.050$\pm$0.007 &09 52 07.00  & 69 12 00.0  &2AD &1.56   &128  & 2.6 &1950    &-125.0   & \nodata  \\
NGC\,3031 & B	& AW343 &1993-03-06   &01 30 &10 30  & 3 15 &0917+624  & 1.041$\pm$0.002 &09 50 49.00  & 69 24 20.0  &2AD &1.56   &128  & 2.6 &1950    &  40.0   & \nodata  \\
NGC\,3031 & B	& AW343 &1993-03-06   &01 30 &10 30  & 3 15 &0917+624  & 1.035$\pm$0.002 &09 52 07.00  & 69 12 00.0  &2AD &1.56   &128  & 2.6 &1950    &-125.0   & \nodata  \\
NGC\,3031 & C	& AW326 &1993-07-03   &16 45 &05 00  & 3 45 &0836+710  & 3.585$\pm$0.004 &09 50 49.00  & 69 24 20.0  &2AD &1.56   &128  & 2.6 &1950    &  40.0   & \nodata  \\
NGC\,3031 & C	& AW326 &1993-07-03   &16 45 &05 00  & 3 45 &0836+710  & 3.568$\pm$0.004 &09 52 07.00  & 69 12 00.0  &2AD &1.56   &128  & 2.6 &1950    &-125.0   & \nodata  \\
\enddata
\end{deluxetable*}

\newpage
\clearpage
%\begin{landscape}
\begin{deluxetable*}{lcclrrrclrrccccccc}
\tablenum{2}
\tabletypesize{\scriptsize}
%\rotate
\tablewidth{25cm}
\tablecaption{Summary of THINGS observing setups.}
%\tablenum{}
%\tablecolumns{}
\tablehead{
\colhead{1}&\colhead{2}&\colhead{3}&\colhead{4}&\colhead{5}&\colhead{6}&\colhead{7}&\colhead{8}&\colhead{9}&\colhead{10}&\colhead{11}&\colhead{12}&\colhead{13}&\colhead{14}&\colhead{15}&\colhead{16}&\colhead{17}&\colhead{18}\\
\colhead{Name} & \colhead{Conf.} &\colhead{Project}&\colhead{Date}       &\colhead{Start}  &\colhead{End}    &\colhead{Dur.}  &\colhead{cal.}      &\colhead{Flux}  &\colhead{RA}           &\colhead{DEC}        &\colhead{mode}  &\colhead{BW}    &\colhead{chan} &\colhead{$\Delta$V}   &\colhead{equ.} &\colhead{Vhel1} &\colhead{Vhel2}\\
  \colhead{} & \colhead{} &\colhead{} &\colhead{yy-mm-dd} &
  \colhead{} & \colhead{} & \colhead{} & \colhead{} & \colhead{Jy} &
  \colhead{hh mm ss} & \colhead{dd mm ss} &\colhead{} & \colhead{MHz}
  &\colhead{\#}  &\colhead{km\,s$^{-1}$} & \colhead{} 
  &\colhead{km\,s$^{-1}$} &\colhead{km\,s$^{-1}$} \\} 
\startdata
%\newpage
\tableline
NGC\,3627 & B	& AW605 &2005-05-28   &22 00 &04 30  & 6 15 &1120+143  & 2.393$\pm$0.009 &11 20 15.02  & 12 59 29.5  &  4 &1.56   & 64  & 5.2 &2000    & 849.257 &  600.723 \\
NGC\,3627 & C	& AW605 &2004-03-11   &05 15 &06 30  & 1 00 &1120+143  & 2.385$\pm$0.010 &11 20 15.02  & 12 59 29.5  &  4 &1.56   & 64  & 5.2 &2000    & 849.257 &  600.723 \\
NGC\,3627 & C	& AW605 &2004-03-11   &06 30 &09 30  & 0 45 &1120+143  & 2.394$\pm$0.006 &11 20 15.02  & 12 59 29.5  &  4 &1.56   & 64  & 5.2 &2000    & 849.257 &  600.723 \\
NGC\,3627 & DnC	& AW605 &2004-06-12   &23 00 &01 00  & 1 45 &1120+143  & 2.386$\pm$0.005 &11 20 15.02  & 12 59 29.5  &  4 &1.56   & 64  & 5.2 &2000    & 849.257 &  600.723 \\
NGC\,3627 & D	& AS750 &2003-03-22   &04 15 &08 15  & 2 00 &1120+143  & 2.370$\pm$0.002 &11 20 15.02  & 12 59 29.5  &  4 &1.56   & 64  & 5.2 &2000    & 352.600 &  600.723 \\
NGC\,3627 & D	& AS750 &2003-03-22   &04 15 &08 15  & 2 00 &1120+143  & 2.370$\pm$0.002 &11 20 15.02  & 12 59 29.5  &  4 &1.56   & 64  & 5.2 &2000    & 849.257 & 1098.203 \\
\tableline
NGC\,4214 & B	& AM418	&1994-06-30   &23 00 &06 00  & 5 45 &1227+365  & 2.053$\pm$0.003 &12 15 39.50  & 36 19 39.0  &2AD &0.78   &128  & 1.3 &2000    & 291.0   & \nodata  \\
NGC\,4214 & C	& AM418	&1993-08-09   &20 15 &00 45  & 4 00 &1227+365  & 2.069$\pm$0.006 &12 15 39.50  & 36 19 39.0  &2AD &0.78   &128  & 1.3 &2000    & 291.0   & \nodata  \\
\enddata
\end{deluxetable*}
\newpage
\clearpage
%\begin{landscape}
\begin{deluxetable*}{lcclrrrclrrccccccc}
\tablenum{2}
\tabletypesize{\scriptsize}
%\rotate
\tablewidth{25cm}
\tablecaption{Summary of THINGS observing setups.}
%\tablenum{}
%\tablecolumns{}
\tablehead{
\colhead{1}&\colhead{2}&\colhead{3}&\colhead{4}&\colhead{5}&\colhead{6}&\colhead{7}&\colhead{8}&\colhead{9}&\colhead{10}&\colhead{11}&\colhead{12}&\colhead{13}&\colhead{14}&\colhead{15}&\colhead{16}&\colhead{17}&\colhead{18}\\
\colhead{Name} & \colhead{Conf.} &\colhead{Project}&\colhead{Date}       &\colhead{Start}  &\colhead{End}    &\colhead{Dur.}  &\colhead{cal.}      &\colhead{Flux}  &\colhead{RA}           &\colhead{DEC}        &\colhead{mode}  &\colhead{BW}    &\colhead{chan} &\colhead{$\Delta$V}   &\colhead{equ.} &\colhead{Vhel1} &\colhead{Vhel2}\\
  \colhead{} & \colhead{} &\colhead{} &\colhead{yy-mm-dd} &
  \colhead{} & \colhead{} & \colhead{} & \colhead{} & \colhead{Jy} &
  \colhead{hh mm ss} & \colhead{dd mm ss} &\colhead{} & \colhead{MHz}
  &\colhead{\#}  &\colhead{km\,s$^{-1}$} & \colhead{} 
  &\colhead{km\,s$^{-1}$} &\colhead{km\,s$^{-1}$} \\} 
\startdata
\newpage
\tableline
NGC\,7793 & BnA	& AW605	&2003-09-23   &03 45 &10 15  & 5 45 &0025-260  & 8.550$\pm$0.044 &23 57 49.00  &-32 35 27.0  &2AD &1.56   &128  & 2.6 &2000    & 230.0   & \nodata  \\
NGC\,7793 & CnB	& AW605	&2004-02-03   &21 00 &23 30  & 2 00 &0025-260  & 8.535$\pm$0.010 &23 57 49.00  &-32 35 27.0  &2AD &1.56   &128  & 2.6 &2000    & 230.0   & \nodata  \\
NGC\,7793 & DnC	& AW605	&2004-06-05   &13 30 &15 45  & 1 45 &0025-260  & 8.527$\pm$0.012 &23 57 49.00  &-32 35 27.0  &2AD &1.56   &128  & 2.6 &2000    & 230.0   & \nodata  \\
\enddata
\end{deluxetable*}
\clearpage
\end{landscape}

\newpage
\clearpage
\begin{deluxetable*}{llrrrrccrc}
\tablenum{2}
\tabletypesize{\small}
%\rotate
\tablewidth{16cm}
%\tablenum{}
%\tablecolumns{}
\tablecaption{Mapping/Noise parameters of THINGS galaxies}
\tablehead{
\colhead{1}&\colhead{2}&\colhead{3}&\colhead{4}&\colhead{5}&\colhead{6}&\colhead{7}&\colhead{8}&\colhead{9}&\colhead{10}\\
\colhead{Name}      & \colhead{Weighting} & \colhead{$B_{\rm maj}$} & \colhead{$B_{\rm min}$} & \colhead{BPA} & \colhead{noise} & \colhead{size} & \colhead{pixel} & \colhead{channels} & \colhead{width} \\
\colhead{}          & \colhead{}          & \colhead{$''$}          & \colhead{$''$}          & \colhead{$^{\circ}$} & \colhead{mJy\,beam$^{-1}$}& \colhead{pixels} & \colhead{$''$} & \colhead{}& \colhead{km\,s$^{-1}$} \\}
\startdata
NGC\,628      &  NA  & 11.88 &  9.30 &  -70.3  & 0.60   &   1024  &  1.5 &   58  &  2.6 \\
          &  RO  &  6.8  &  5.57 &  -61.4  & 0.66   &         &      &       & \\
NGC\,925     &  NA  &  5.94 &  5.71 &   30.6  & 0.57   &   1024  &  1.5 &  100  &  2.6 \\
          &  RO  &  4.85 &  4.65 &   36.9  & 0.68   &         &      &       & \\
NGC\,1569     &  NA  &  7.71 &  7.04 &   16.5  & 0.81   &   1024  &  1.5 &   75  &  2.6 \\
          &  RO  &  5.76 &  5.04 &   26.7  & 0.76   &         &      &       & \\
NGC\,2366     &  NA  & 13.10 & 11.85 &   -1.6  & 0.52   &   1024  &  1.5 &   66  &  2.6 \\
          &  RO  &  6.96 &  5.94 &   20.2  & 0.61   &         &      &       & \\
\enddata
\tablecomments{Notes on columns:\\
{\em column 1:} galaxy name\\
{\em column 2:} weighting scheme used. NA: Natural, RO: Robust=0.5 weighting\\
{\em column 3/4:} major and minor axis of synthesized beam in arcsec\\
{\em column 5:} position angle of synthesized beam in degrees (measured north to east)\\
{\em column 6:} noise in one channel map in mJy\,beam$^{-1}$\\
{\em column 7:} image size in pixels\\
{\em column 8:} pixel size in arcsec\\
{\em column 9:} number of channels in final data cube\\
{\em column 10:} channel width in km\,s$^{-1}$.
}
\end{deluxetable*}

\begin{deluxetable*}{rc}
\tabletypesize{\small}
%\rotate
\tablewidth{5cm} \tablecaption{\hi\ spectrum of M81\,dwA based on the
  blanked Natural weighted cubes (see the online version of the
  Journal for the tables of the other THINGS targets).}
%\tablenum{}
%\tablecolumns{}
\tablehead{\colhead{velocity}       &  \colhead{\hi\ flux density} \\
 \colhead{km\,s$^{-1}$}             &  \colhead{Jy} \\}
\startdata
93.7   &  0.023  \\ 
95.0   &  0.036  \\ 
96.2   &  0.037  \\ 
97.5   &  0.054  \\ 
98.8   &  0.068  \\ 
100.1  &  0.076  \\ 
101.4  &  0.079  \\ 
102.7  &  0.095  \\ 
104.0  &  0.107  \\ 
105.3  &  0.198  \\ 
106.6  &  0.131  \\ 
107.8  &  0.149  \\ 
109.1  &  0.161  \\ 
110.4  &  0.182  \\ 
111.7  &  0.188  \\ 
113.0  &  0.205  \\ 
114.3  &  0.198  \\ 
115.6  &  0.194  \\ 
116.9  &  0.180  \\ 
118.2  &  0.169  \\ 
119.4  &  0.147  \\ 
120.7  &  0.123  \\ 
122.0  &  0.098  \\ 
123.3  &  0.087  \\ 
124.6  &  0.076  \\ 
125.9  &  0.068  \\ 
127.2  &  0.050  \\ 
128.5  &  0.036  \\ 
129.8  &  0.032  \\ 
131.0  &  0.019  \\ 
\enddata
\end{deluxetable*}

\clearpage

\begin{landscape}
\begin{deluxetable*}{lrrrrrrrrr}
\tabletypesize{\small}
%\rotate
\tablewidth{20cm}
\tablecaption{THINGS \hi\ fluxes, velocities, line--widths and masses}
%\tablenum{}
%\tablecolumns{}
\tablehead{\colhead{1}&\colhead{2}&\colhead{3}&\colhead{4}&\colhead{5}&\colhead{6}&\colhead{7}&\colhead{8}&\colhead{9}&\colhead{10}\\
\colhead{name}       &  \colhead{$S_{\rm HI}$}     & \colhead{$M_{\rm HI}$}    &  \colhead{$S_{\rm HI}^{\rm Paturel}$} & \colhead{$S_{\rm HI}^{\rm Fisher\&Tully}$} &\colhead{$v_{\rm cen}$}  &\colhead{$v_{\rm cen}^{\rm Fisher\&Tully}$}  & \colhead{$W_{50}$} & \colhead{$W_{20}$} & \colhead{$W_{20}^{\rm Fisher\&Tully}$} \\
 \colhead{}          &  \colhead{Jy\,km\,s$^{-1}$} & \colhead{10$^{8}$\,\msol} &  \colhead{Jy\,km\,s$^{-1}$}&  \colhead{Jy\,km\,s$^{-1}$} & \colhead{km\,s$^{-1}$} & \colhead{km\,s$^{-1}$} & \colhead{km\,s$^{-1}$} & \colhead{km\,s$^{-1}$}& \colhead{km\,s$^{-1}$} \\}
\startdata
NGC\,628       &      302        &  38.0          & 252    & 381$\pm$17   &659.1 & 659    & 56.4    & 75.4   & 74$\pm$5    \\
NGC\,925       &      232        &  45.8          & 197    & 301$\pm$14   &552.5 & 554    & 207.1   & 222.6  & 224$\pm$8   \\
NGC\,1569      &       84.0      &   0.75         & 49.4   & ---          &-85.6 & ---    & 100.8   & 123.8  & ---         \\
NGC\,2366      &      233        &   6.49         & 188    & 274$\pm$14   &100.1 & 102    & 101.2   & 115.8  & 116$\pm$5   \\
NGC\,2403      &     1054        &  25.8          & 925    & 1458$\pm$47  &133.1 & 132    & 240.6   & 254.7  & 247$\pm$20  \\
Ho\,II         &      219        &   5.95         &  197   & 388$\pm$15   &157.1 & 158    & 57.4    & 70.8   & 76$\pm$5    \\
M81\,dwA       &        4.1      &   0.12         & ---    & ---          &112.0 & ---    & 24.7    & 36.6   & ---         \\
DDDO\,53       &       20.0      &   0.60         & ---    & ---          &17.7  & ---    & 28.3    & 46.0   & ---         \\
\enddata
\tablecomments{Notes on columns:\\Derived HI fluxes, masses and
  velocity widths.\\
  {\em Column 1:} Galaxy name,\\
  {\em Column 2:} THINGS \hi\ flux. The total uncertainty is dominated by the accuracy of the flux calibration (of order 5\%).\\
  {\em Column 3:} THINGS \hi\ mass,\\
  {\em Column 4:} Single dish \hi\ flux compiled from the literature
  by Paturel et al.\ (2003).  Values with $^+$ indicate that galaxies
  are more extended than the
  area (primary beam) mapped by us.\\
  {\em Column 5:} Single dish \hi flux from Fisher \& Tully 1981,\\
  {\em Column 6:} Velocity of the HI spectra, derived from the central velocity of the 20\% level of the HI profile (column 9). A typical uncertainty is the width of one velocity channel (Tab.~2). \\
  {\em Column 7:} Heliocentric velocities from Fisher \& Tully 1981,\\
  {\em Column 8:} Profile width at 50\% of the peak intensity. A typical uncertainty is the width of one velocity channel (Tab.~2),\\
  {\em Column 9:} Profile width at 20\% of the peak intensity. A typical uncertainty is the width of one velocity channel (Tab.~2),\\
  {\em Column 10:} Profile width at 20\% of the peak intensity from
  Fisher \& Tully 1981.  }
\end{deluxetable*}
\clearpage
\end{landscape}

%
%
% HI SPECTRA -- 6 galaxies per page, i.e. a total of 6 pages.
%
%

\notetoeditor{Figure 1 will run over 6 pages}

\begin{figure*}
      \begin{center}
\begin{minipage}[t]{70mm}{
\resizebox{70mm}{!}{
\includegraphics[angle=0]{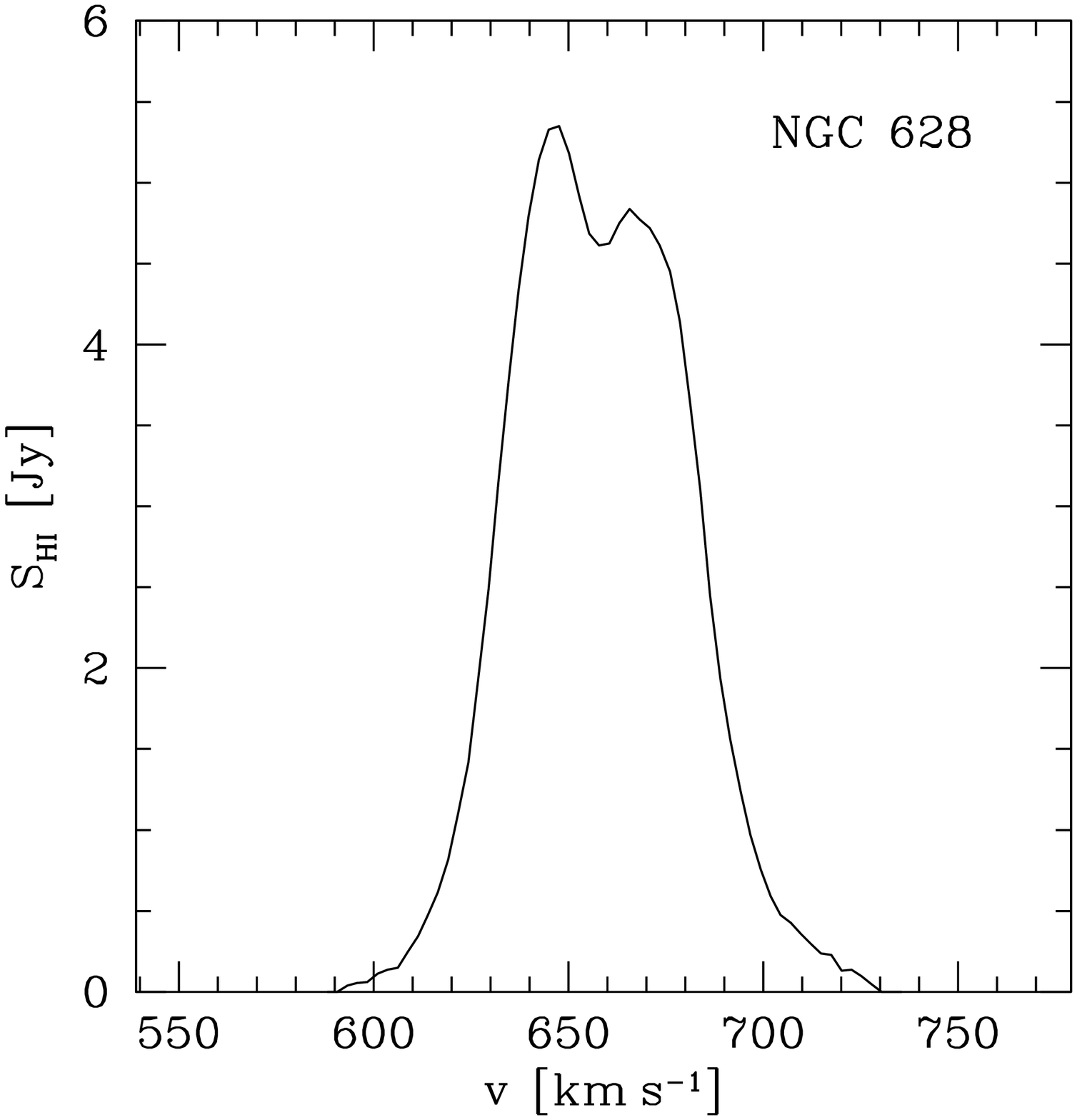}}
}
\end{minipage}
\begin{minipage}[t]{70mm}{
\resizebox{70mm}{!}{
\includegraphics[angle=0]{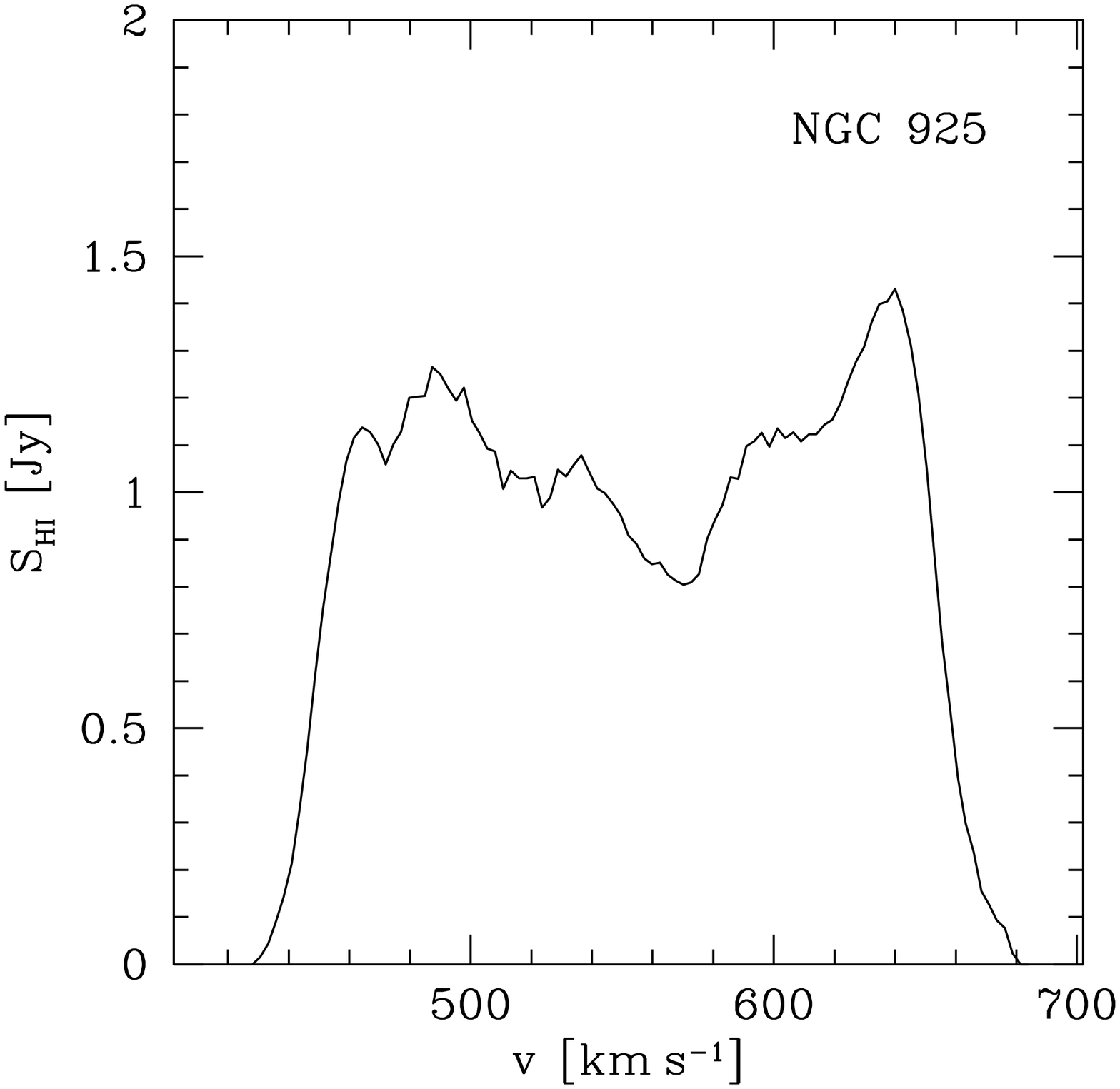}}
}
\end{minipage}
\begin{minipage}[t]{70mm}{
\resizebox{70mm}{!}{
\includegraphics[angle=0]{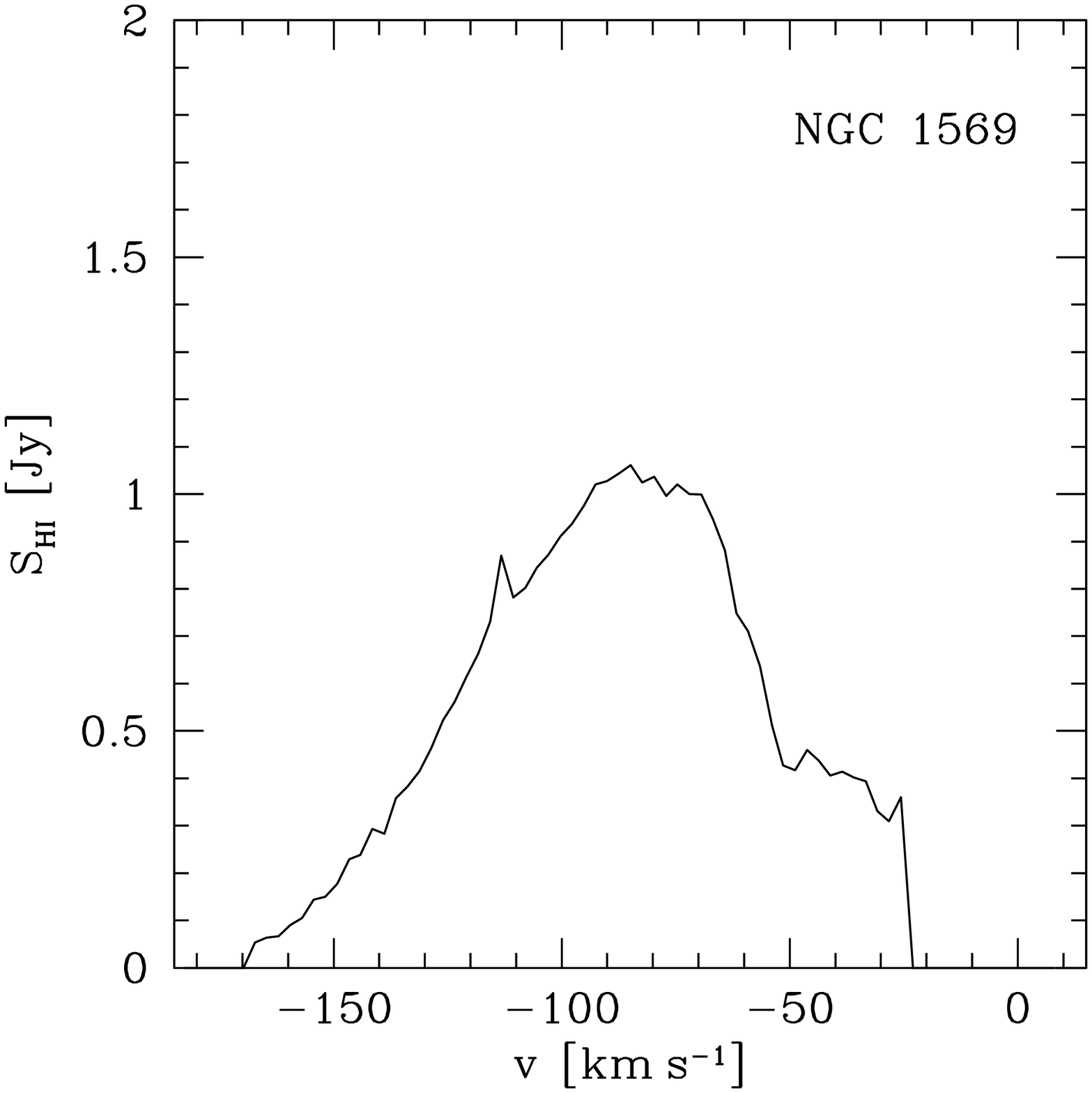}}
}
\end{minipage}
\begin{minipage}[t]{70mm}{
\resizebox{70mm}{!}{
\includegraphics[angle=0]{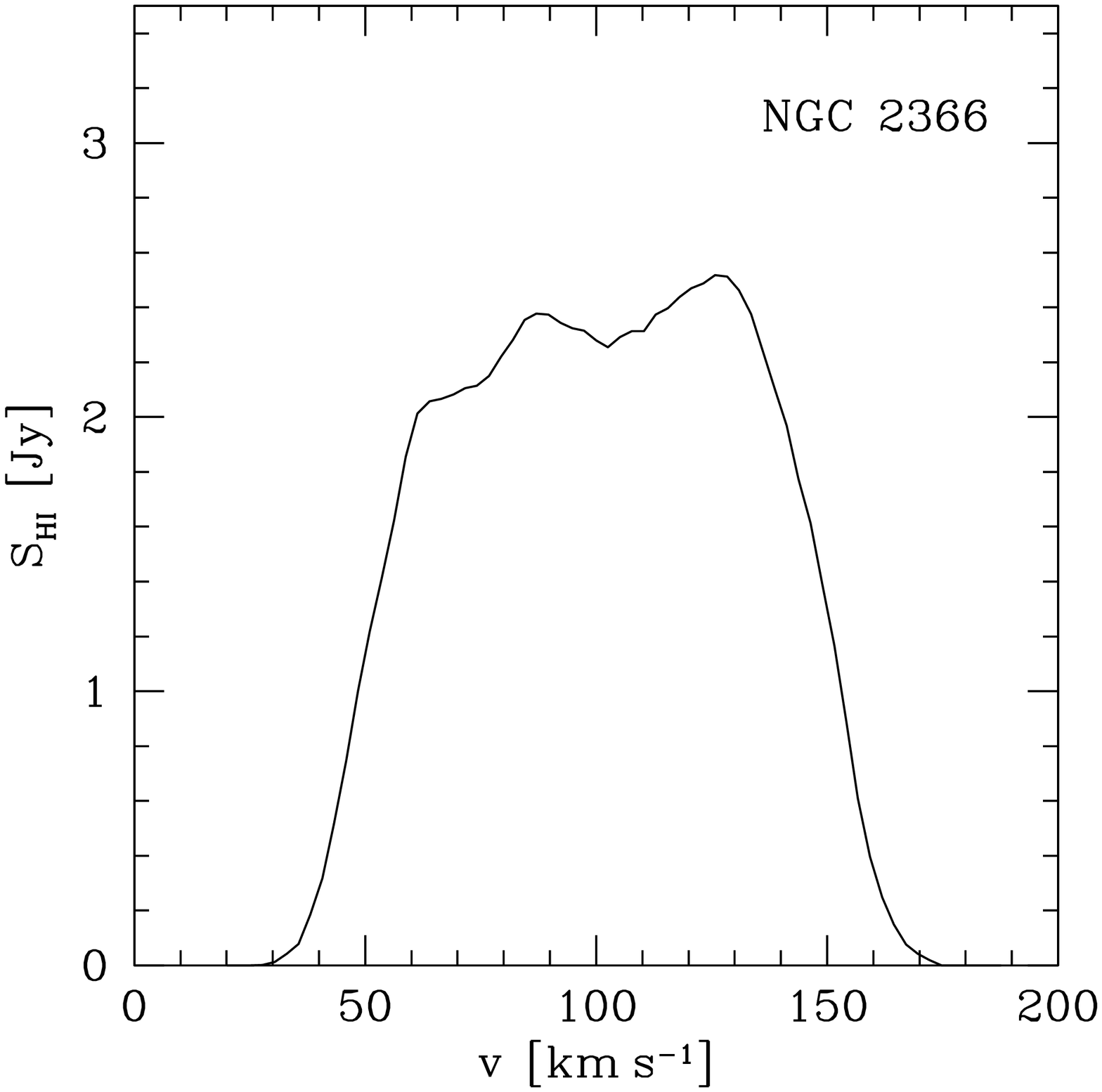}}
}
\end{minipage}
\begin{minipage}[t]{70mm}{
\resizebox{70mm}{!}{
\includegraphics[angle=0]{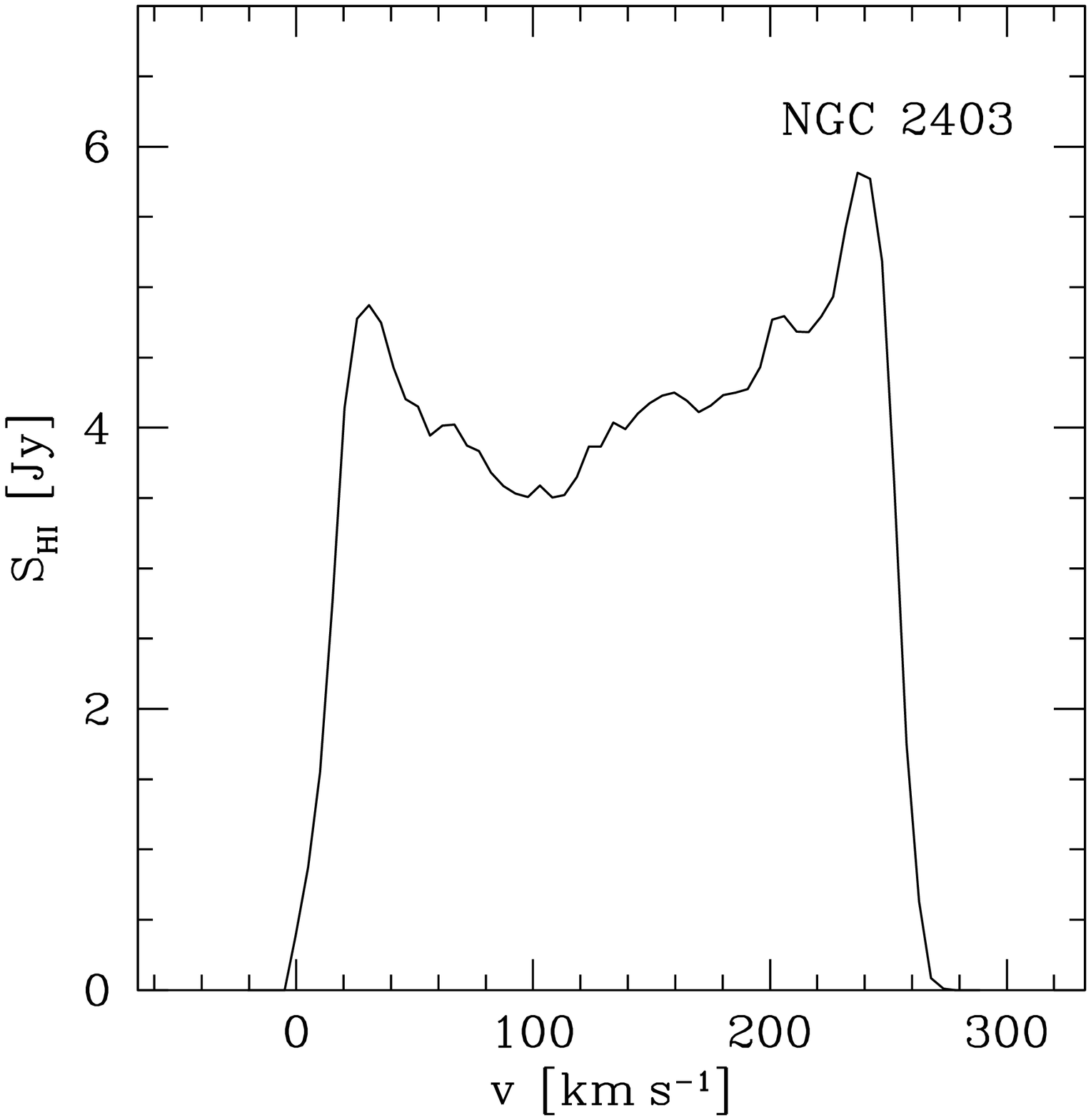}}
}
\end{minipage}
\begin{minipage}[t]{70mm}{
\resizebox{70mm}{!}{
\includegraphics[angle=0]{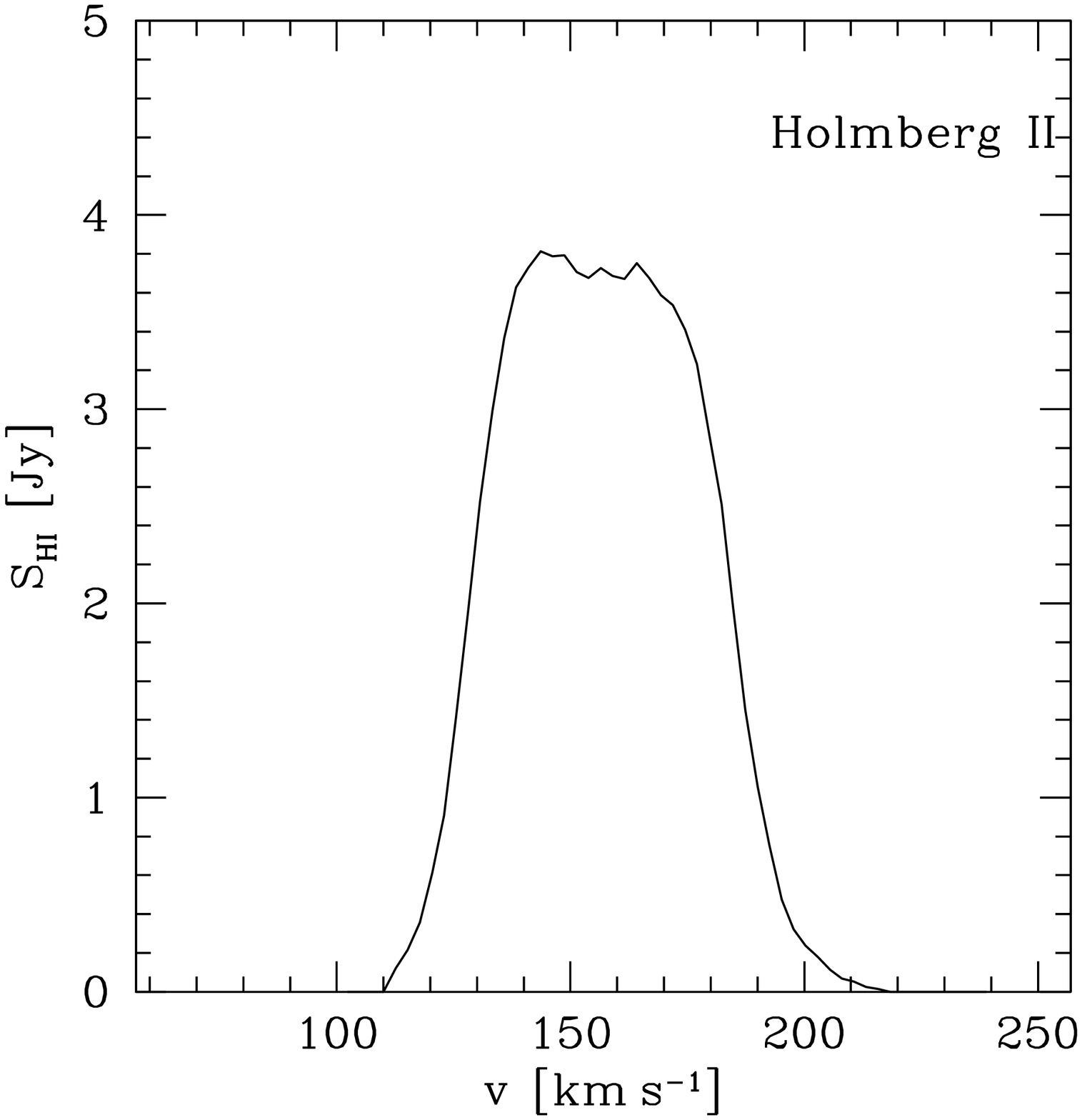}}
}
\end{minipage}
%\vspace*{-7mm}

\caption{Global \hi\ profiles for all THINGS galaxies.}
\end{center}
\end{figure*}

\begin{figure*}
\figurenum{1}
      \begin{center}
\begin{minipage}[t]{70mm}{
\resizebox{70mm}{!}{
\includegraphics[angle=0]{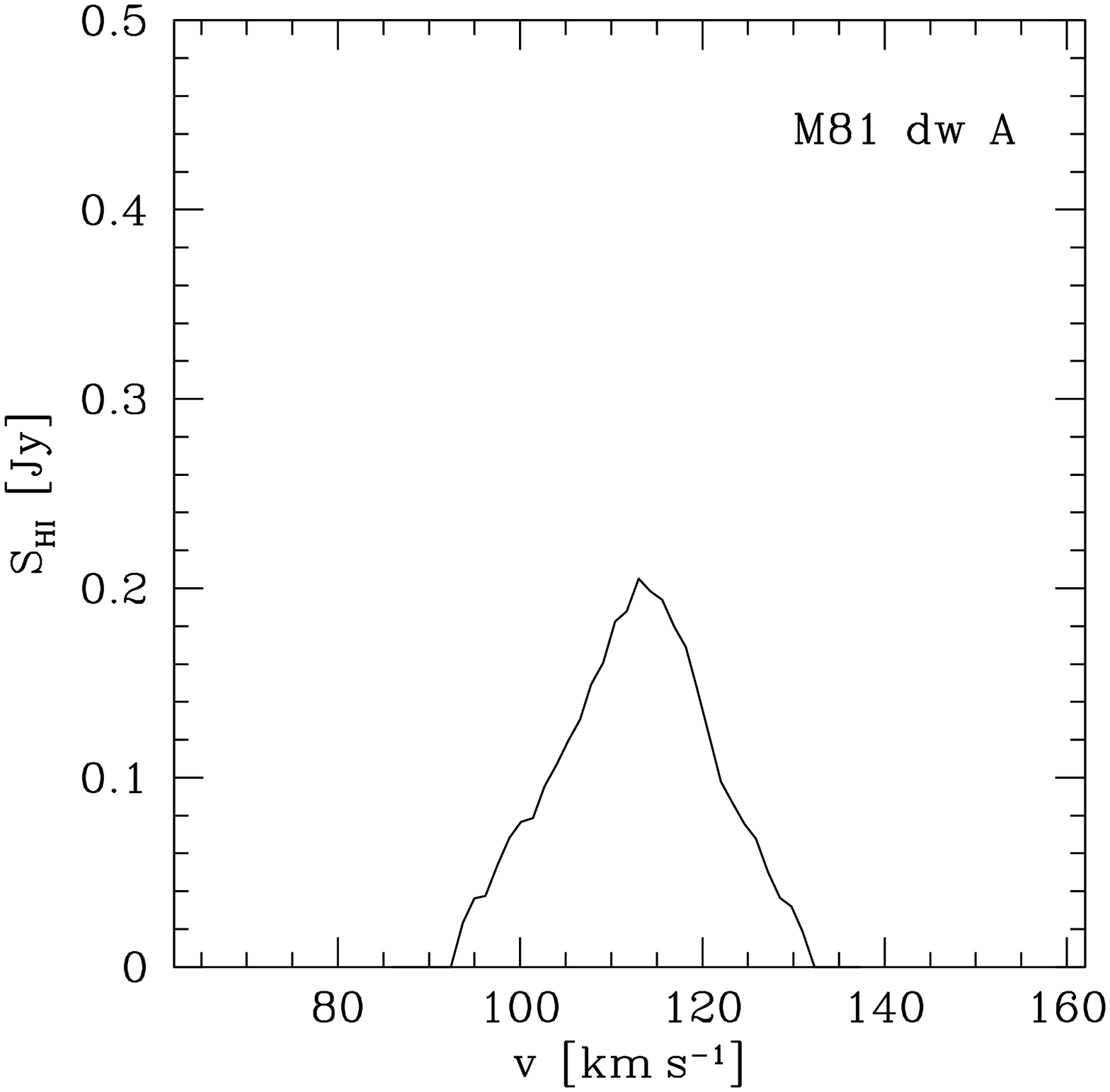}}
}
\end{minipage}
\begin{minipage}[t]{70mm}{
\resizebox{70mm}{!}{
\includegraphics[angle=0]{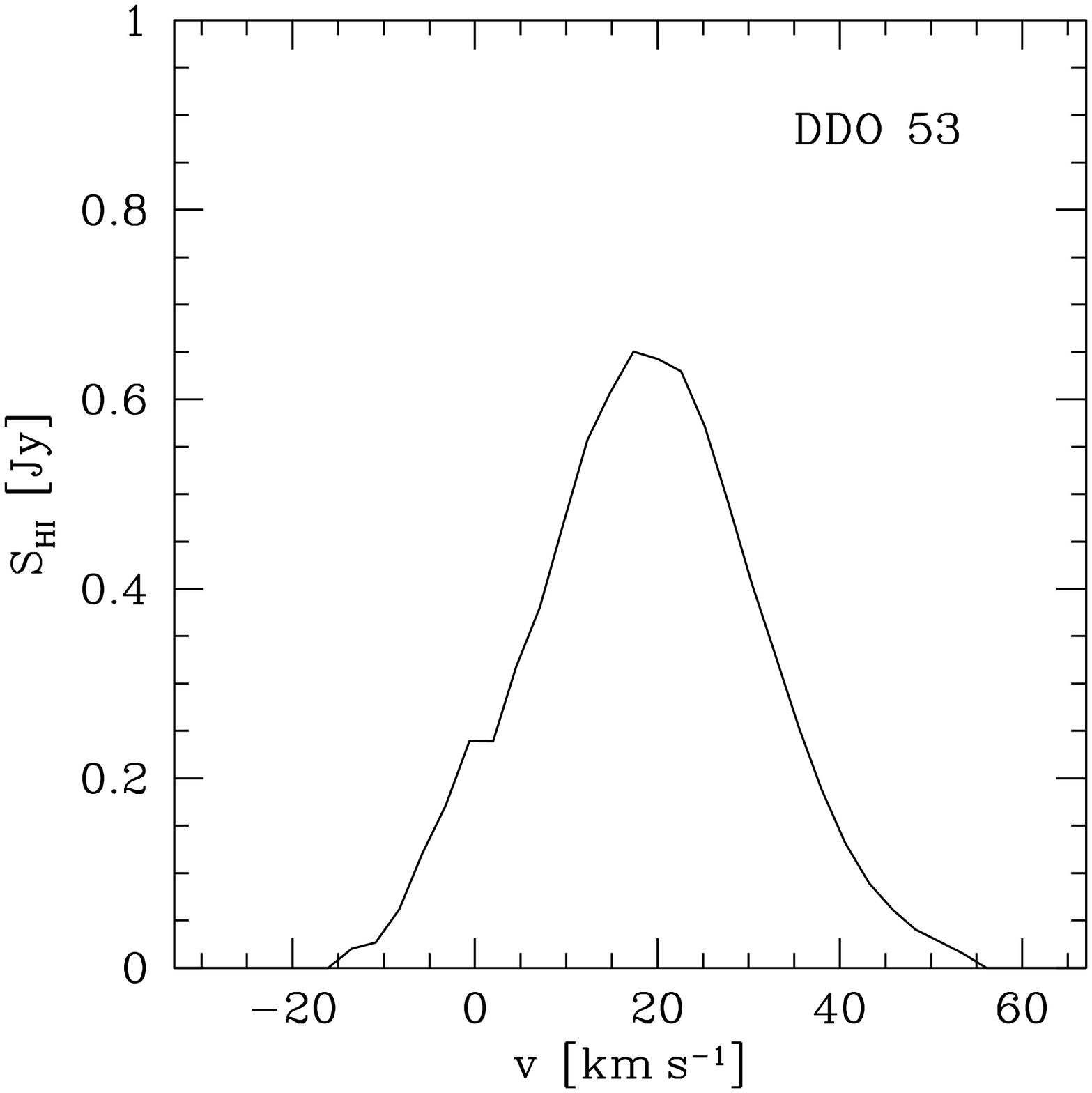}}
}
\end{minipage}
\begin{minipage}[t]{70mm}{
\resizebox{70mm}{!}{
\includegraphics[angle=0]{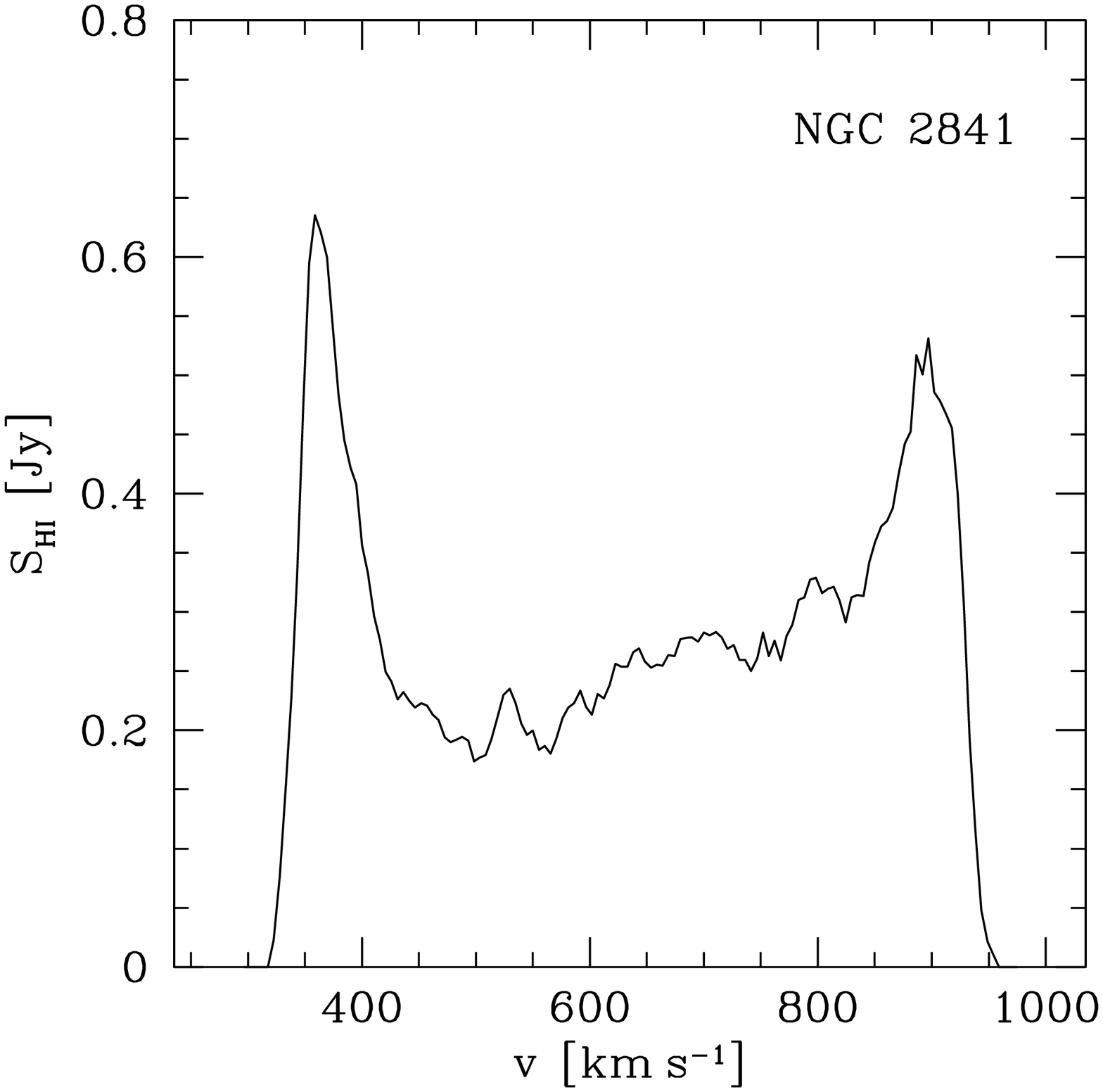}}
}
\end{minipage}
\begin{minipage}[t]{70mm}{
\resizebox{70mm}{!}{
\includegraphics[angle=0]{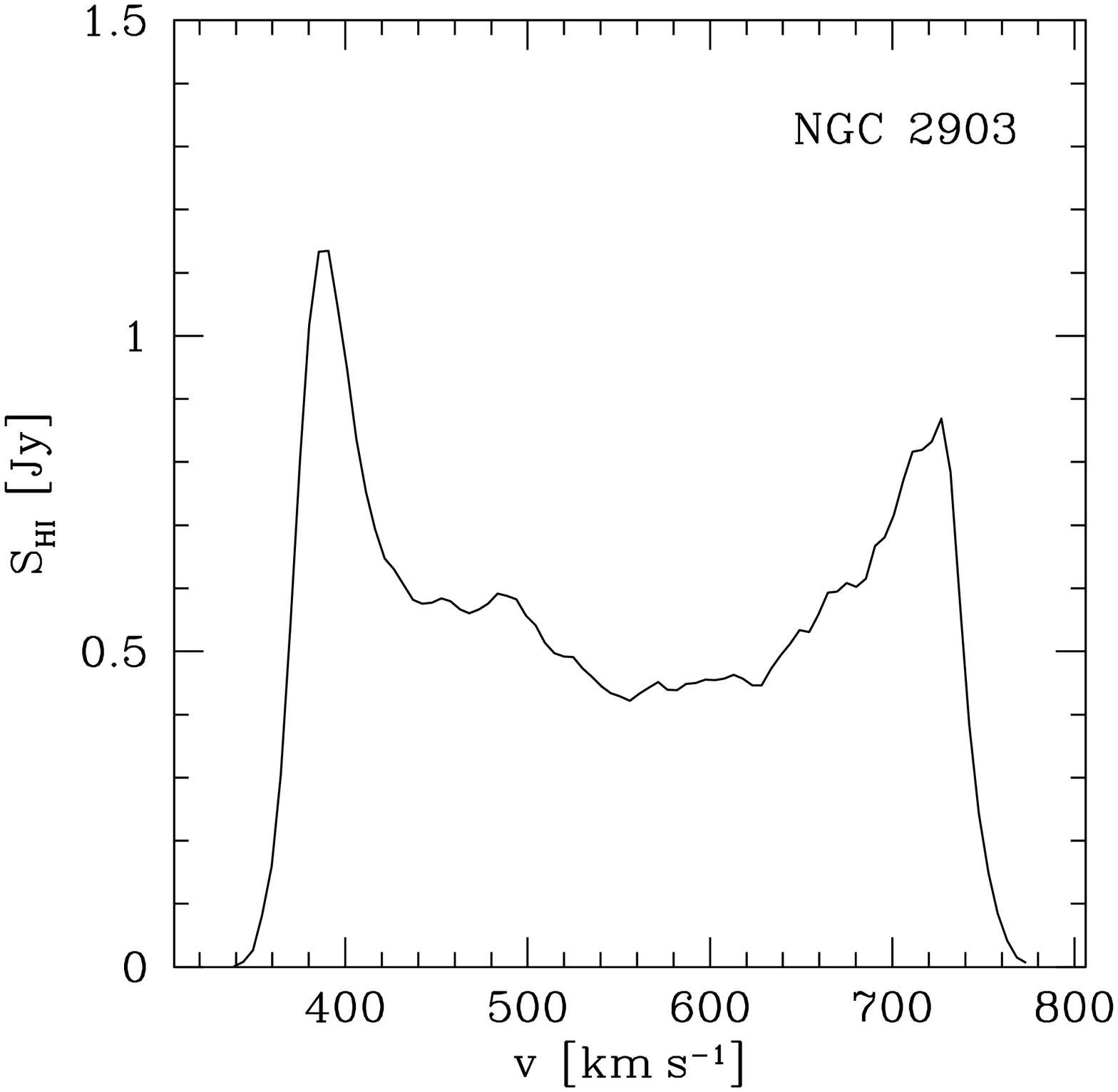}}
}
\end{minipage}
\begin{minipage}[t]{70mm}{
\resizebox{70mm}{!}{
\includegraphics[angle=0]{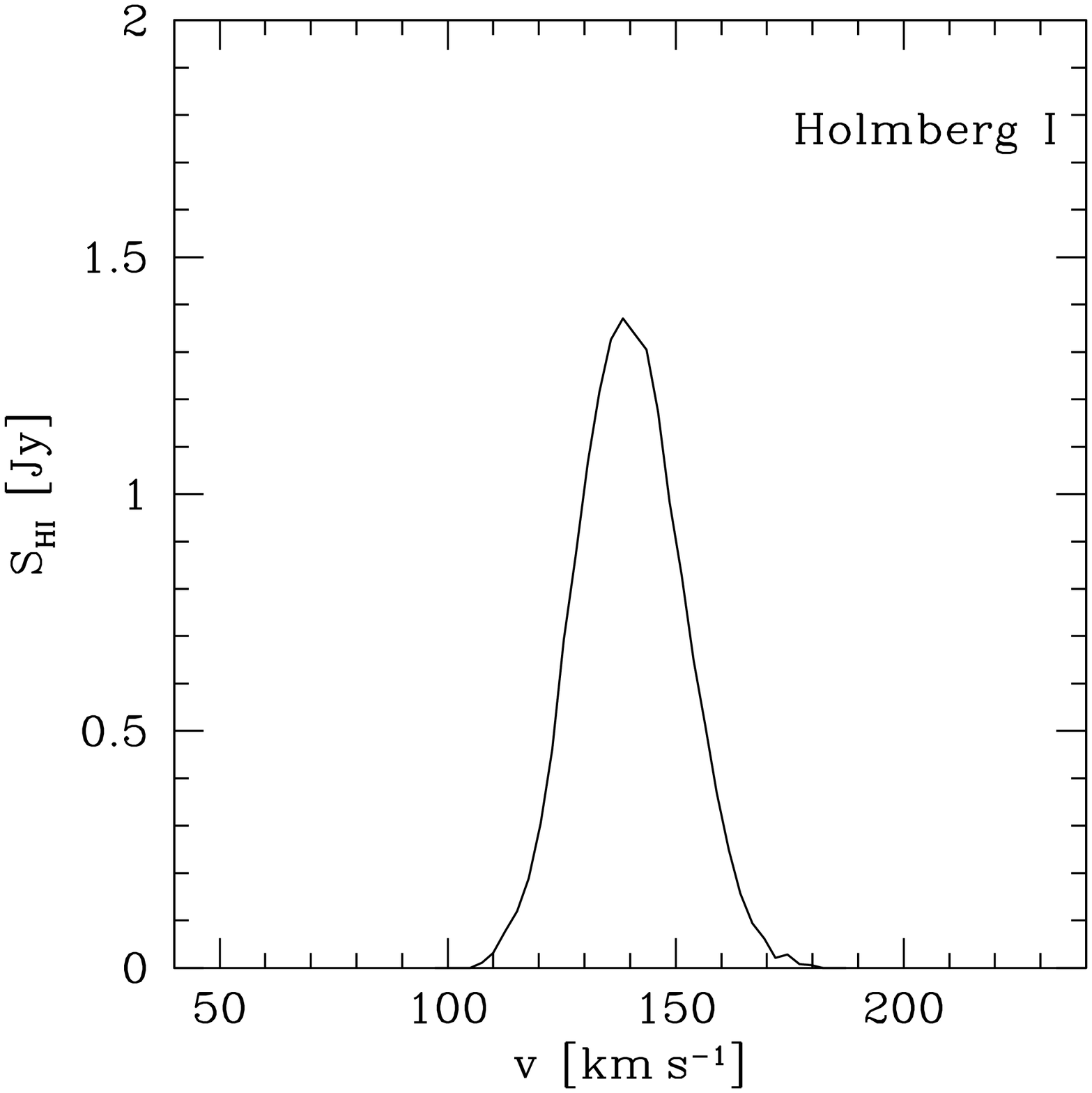}}
}
\end{minipage}
\begin{minipage}[t]{70mm}{
\resizebox{70mm}{!}{
\includegraphics[angle=0]{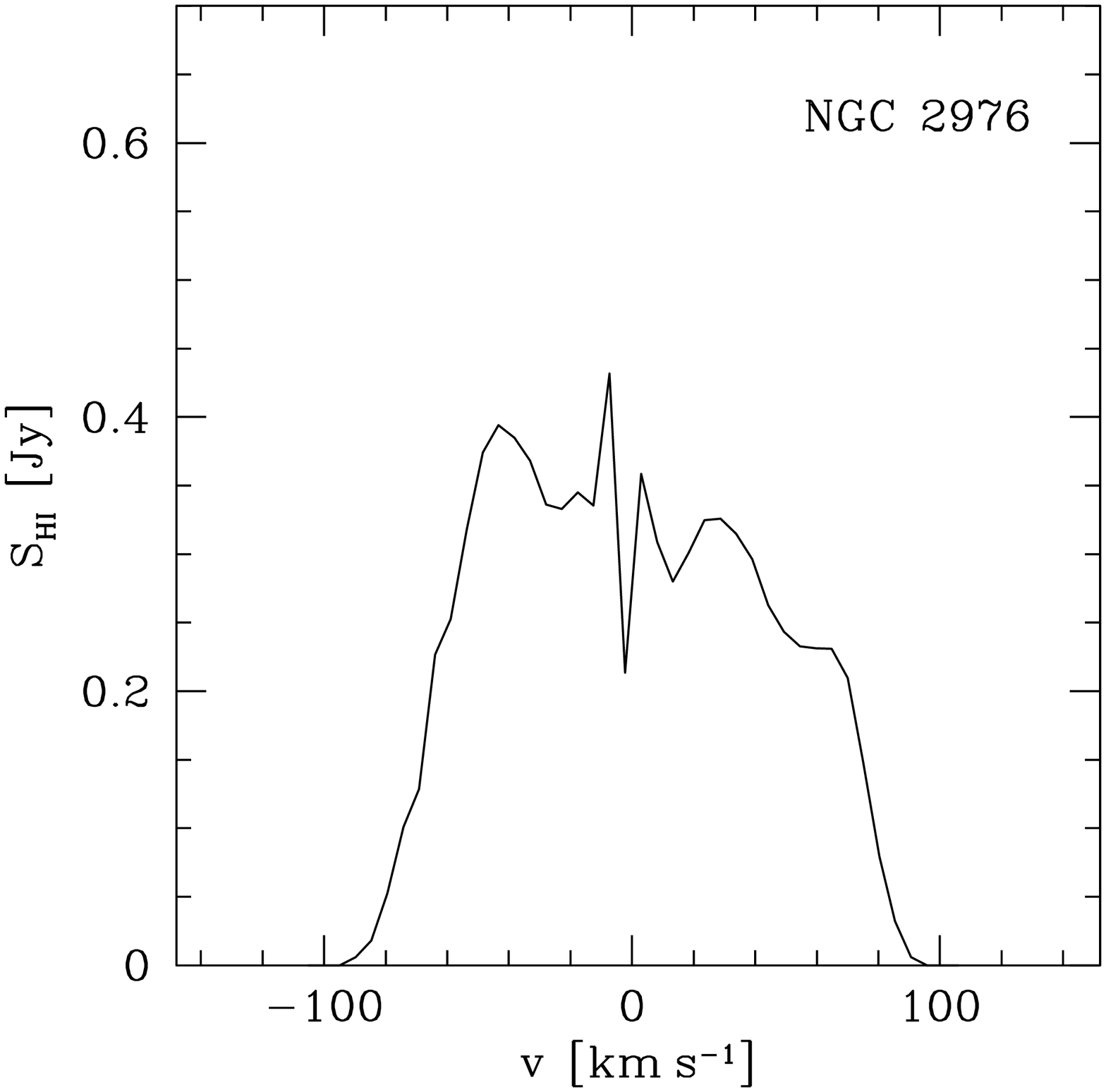}}
}
\end{minipage}
%\vspace*{-7mm}

\caption{Global \hi\ profiles for all THINGS galaxies (continued).}
\end{center}
\end{figure*}

\begin{figure*}
\figurenum{1}
      \begin{center}
\begin{minipage}[t]{70mm}{
\resizebox{70mm}{!}{
\includegraphics[angle=0]{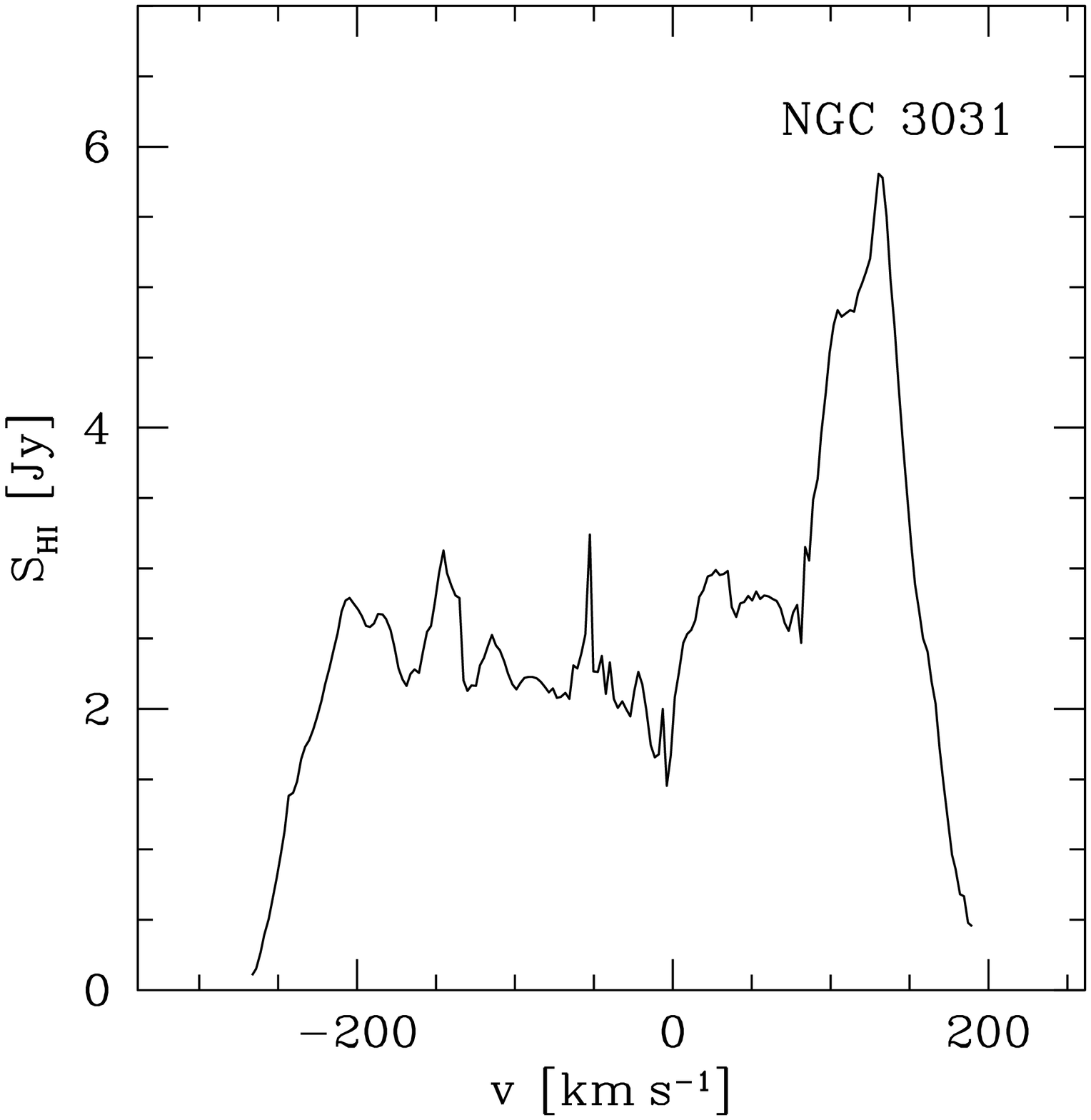}}
}
\end{minipage}
\begin{minipage}[t]{70mm}{
\resizebox{70mm}{!}{
\includegraphics[angle=0]{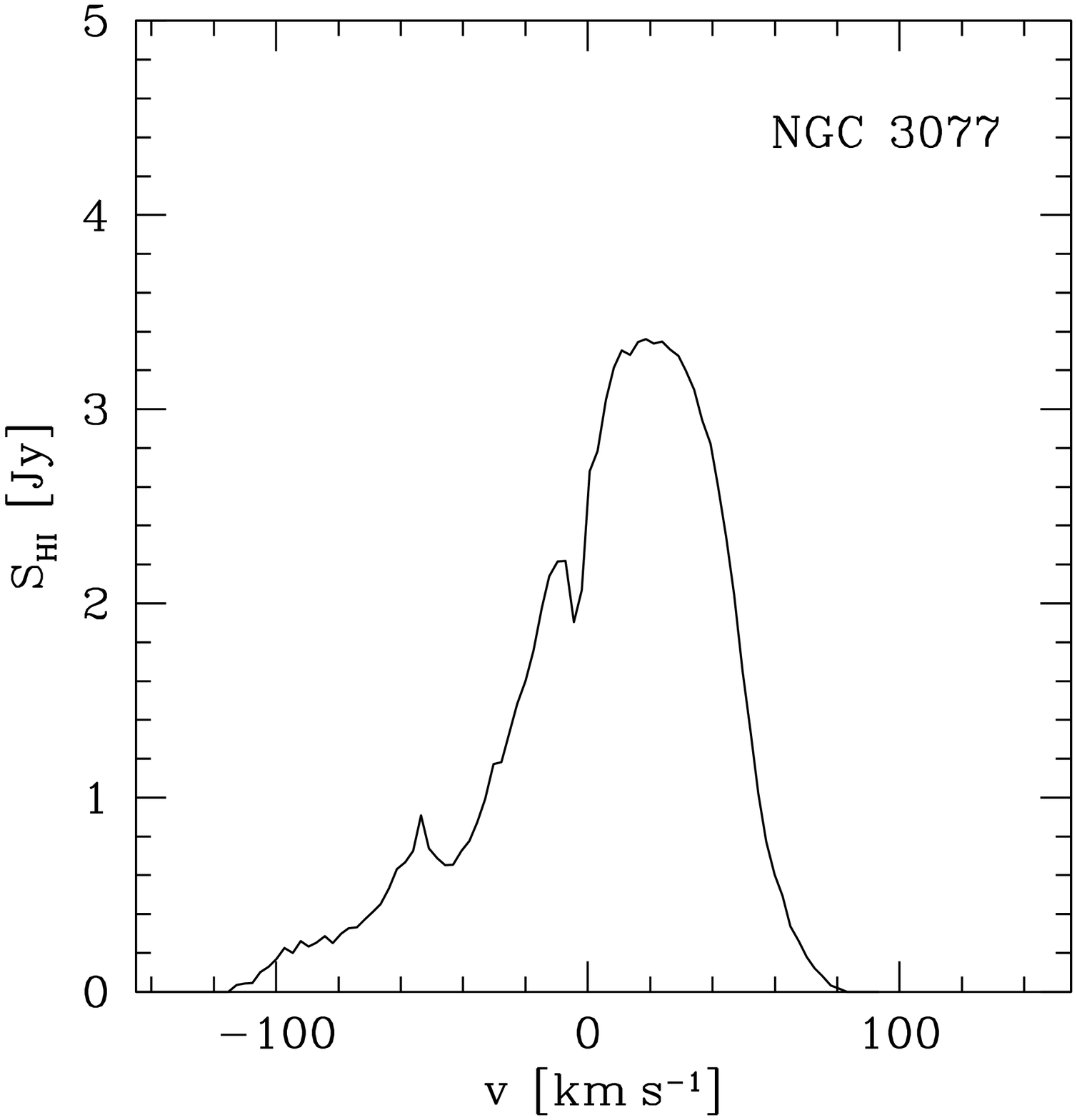}}
}
\end{minipage}
\begin{minipage}[t]{70mm}{
\resizebox{70mm}{!}{
\includegraphics[angle=0]{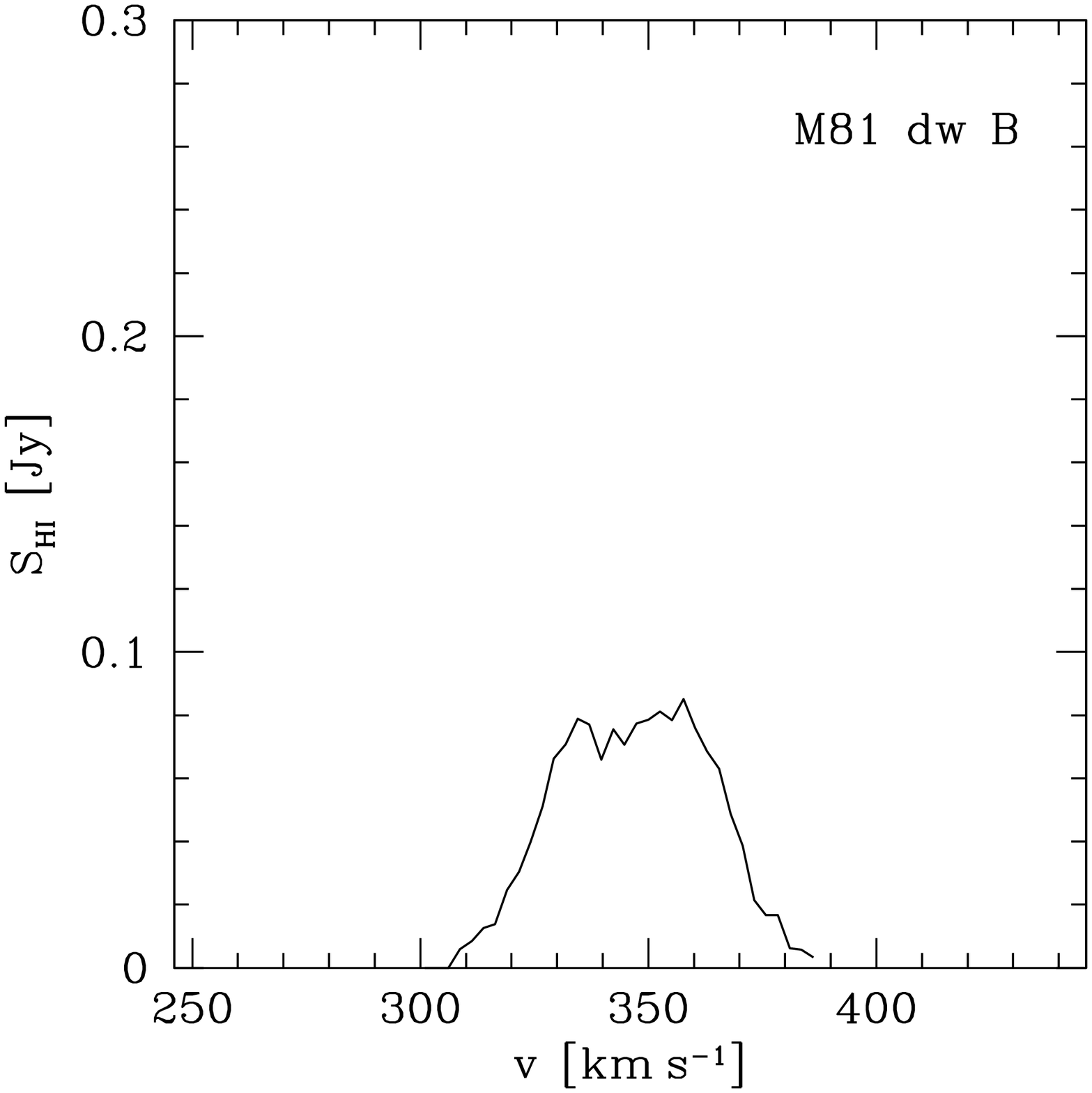}}
}
\end{minipage}
\begin{minipage}[t]{70mm}{
\resizebox{70mm}{!}{
\includegraphics[angle=0]{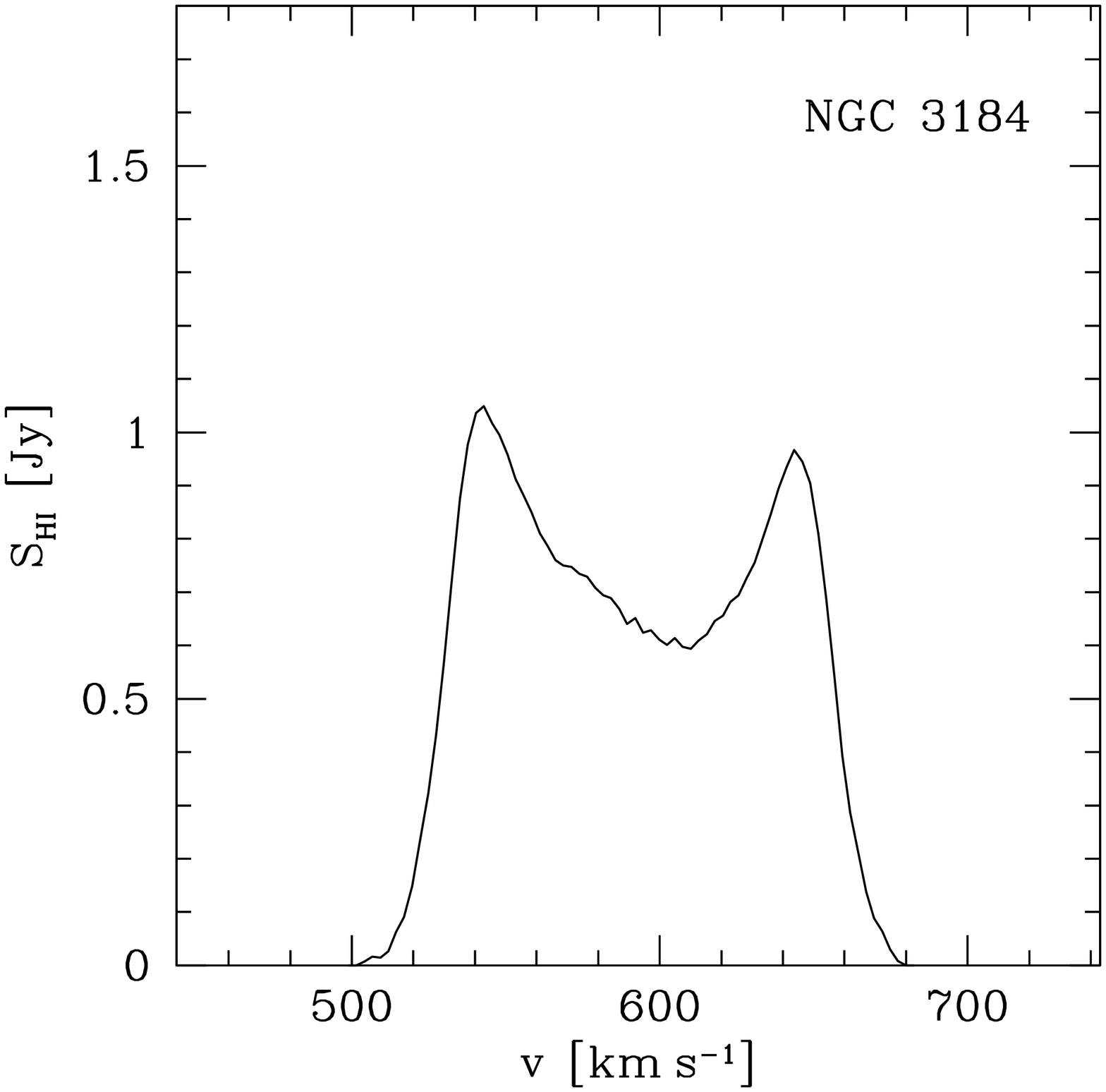}}
}
\end{minipage}
\begin{minipage}[t]{70mm}{
\resizebox{70mm}{!}{
\includegraphics[angle=0]{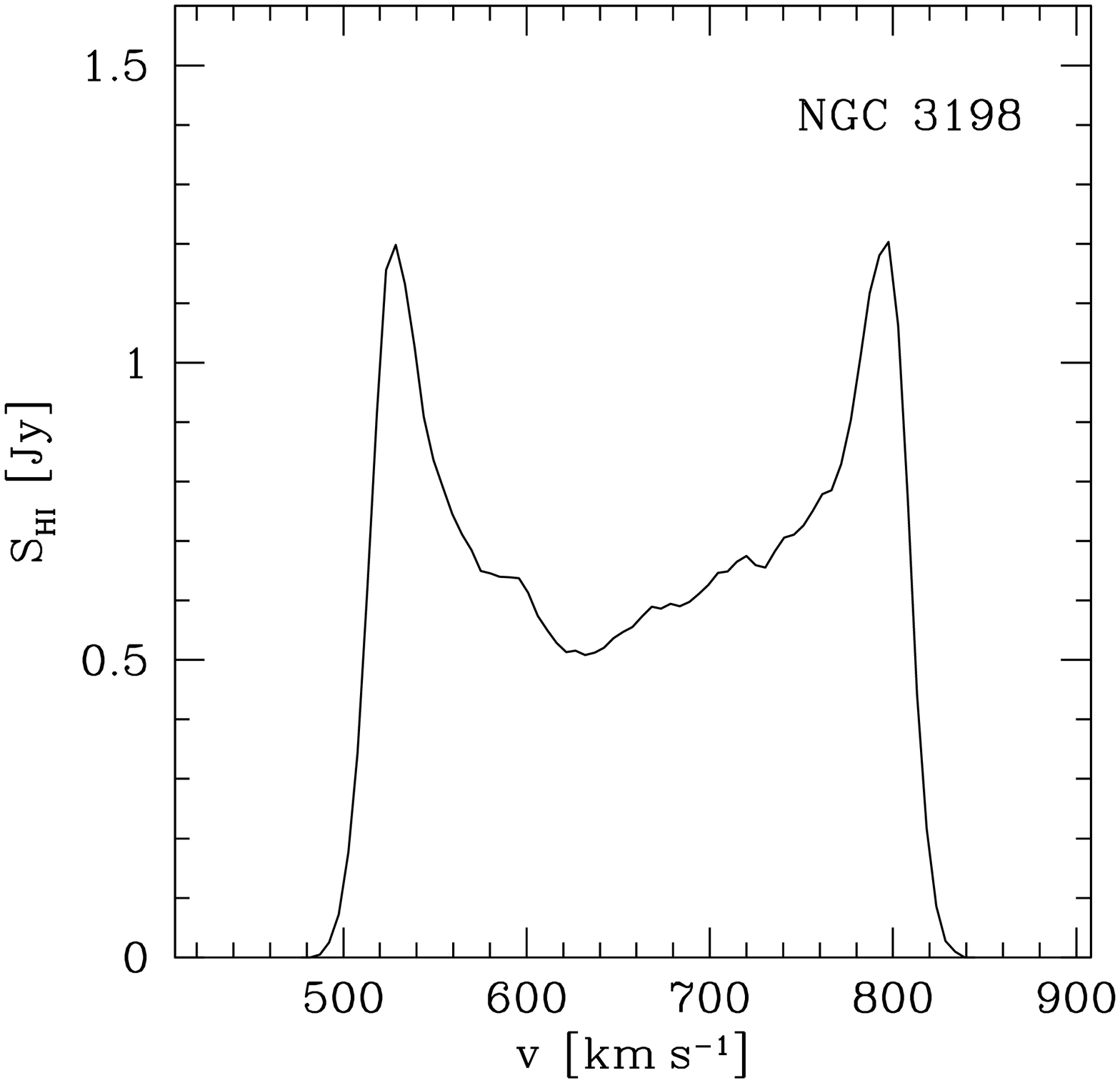}}
}
\end{minipage}
\begin{minipage}[t]{70mm}{
\resizebox{70mm}{!}{
\includegraphics[angle=0]{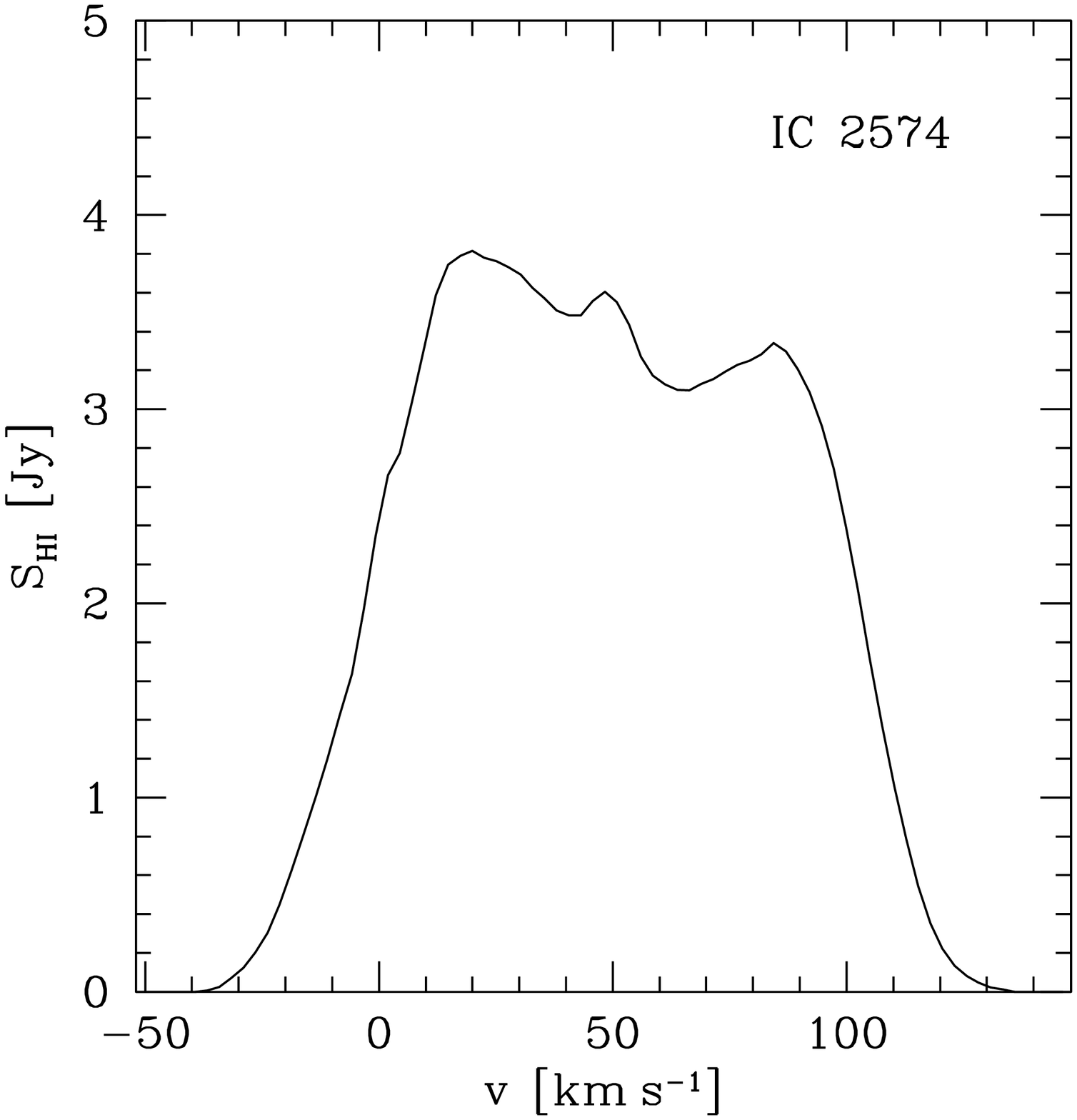}}
}
\end{minipage}
%\vspace*{-7mm}
\caption{Global \hi\ profiles for all THINGS galaxies (continued).}
\end{center}
\end{figure*}

\begin{figure*}
\figurenum{1}
      \begin{center}
\begin{minipage}[t]{70mm}{
\resizebox{70mm}{!}{
\includegraphics[angle=0]{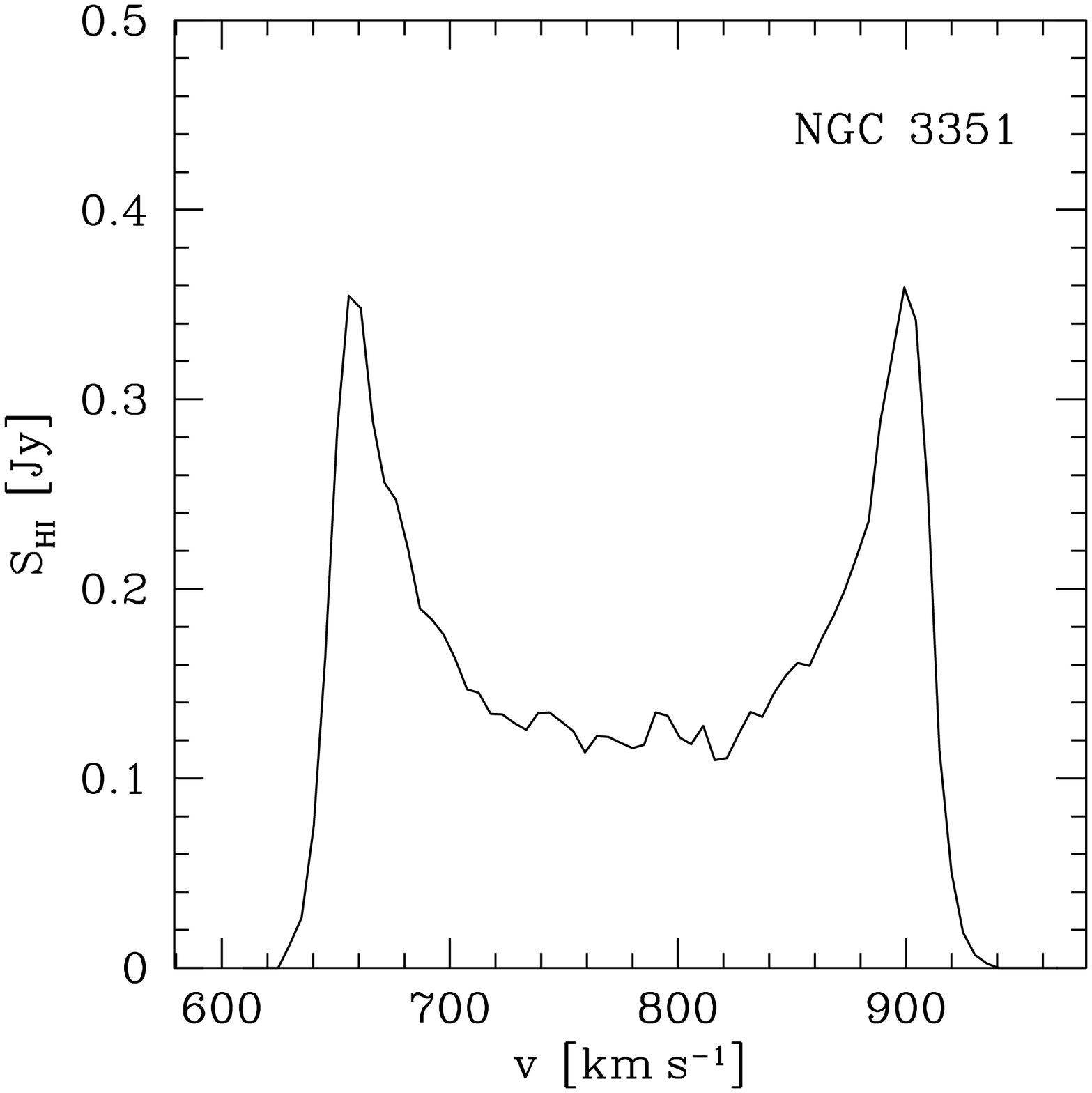}}
}
\end{minipage}
\begin{minipage}[t]{70mm}{
\resizebox{70mm}{!}{
\includegraphics[angle=0]{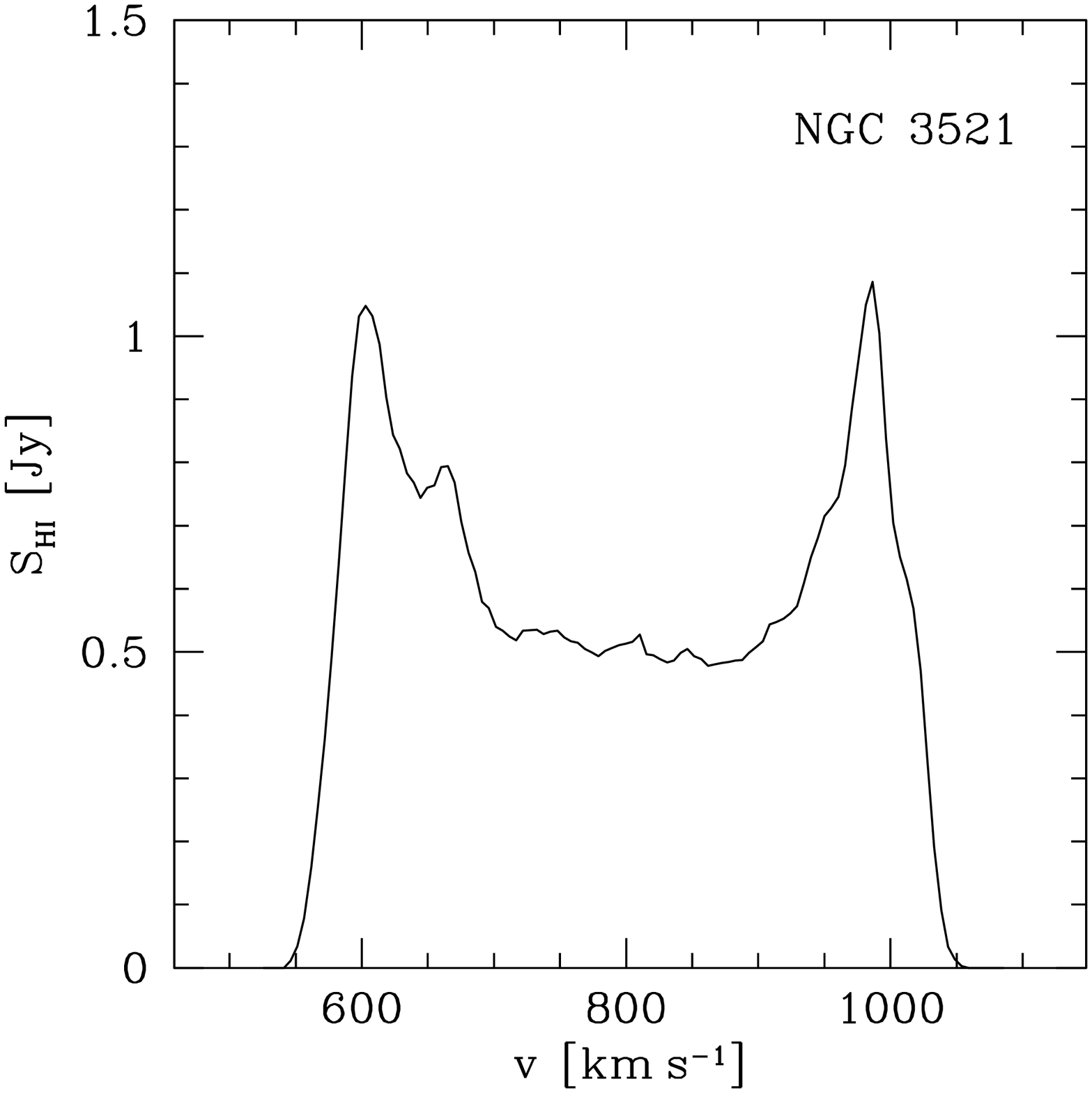}}
}
\end{minipage}
\begin{minipage}[t]{70mm}{
\resizebox{70mm}{!}{
\includegraphics[angle=0]{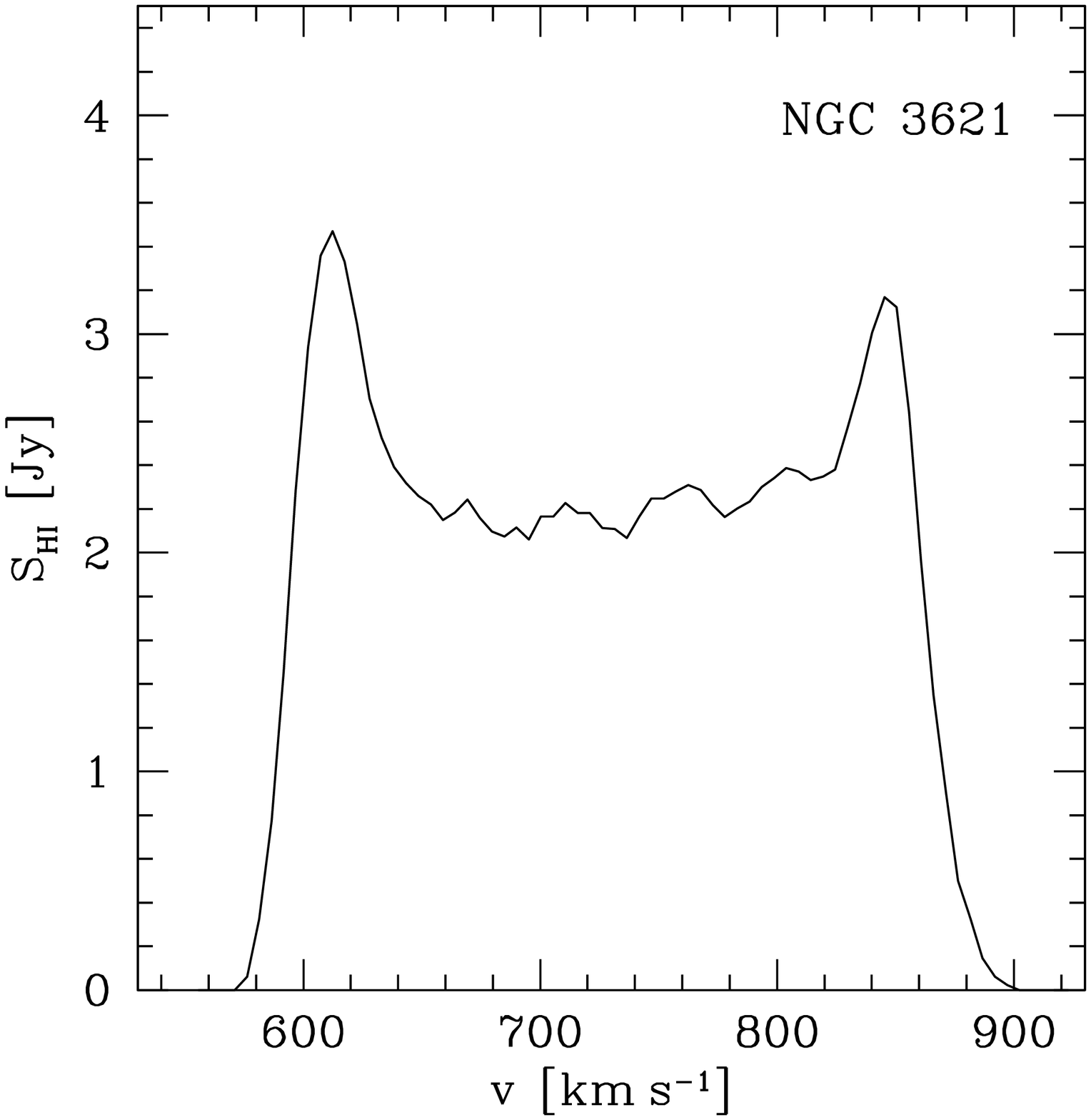}}
}
\end{minipage}
\begin{minipage}[t]{70mm}{
\resizebox{70mm}{!}{
\includegraphics[angle=0]{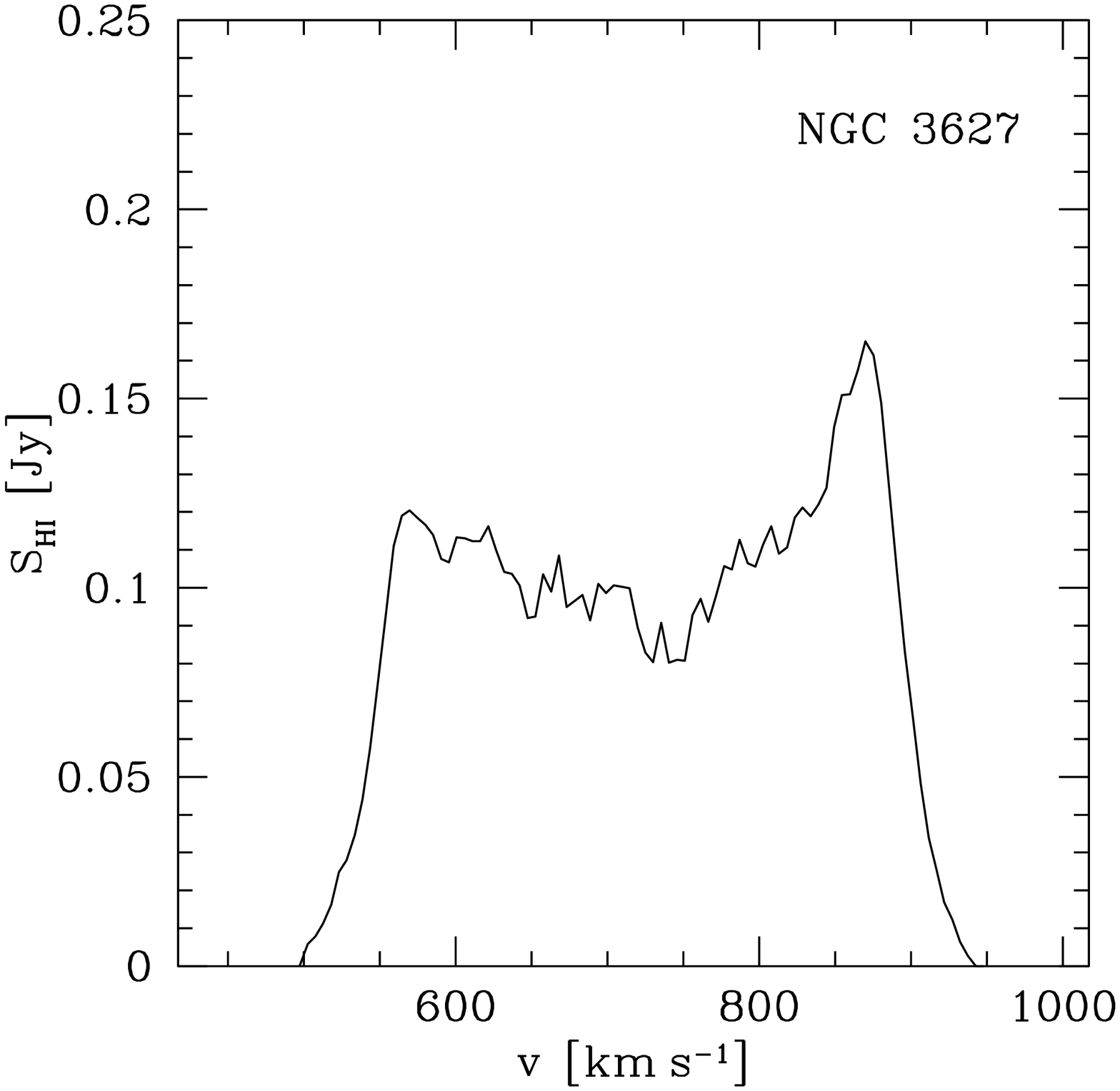}}
}
\end{minipage}
\begin{minipage}[t]{70mm}{
\resizebox{70mm}{!}{
\includegraphics[angle=0]{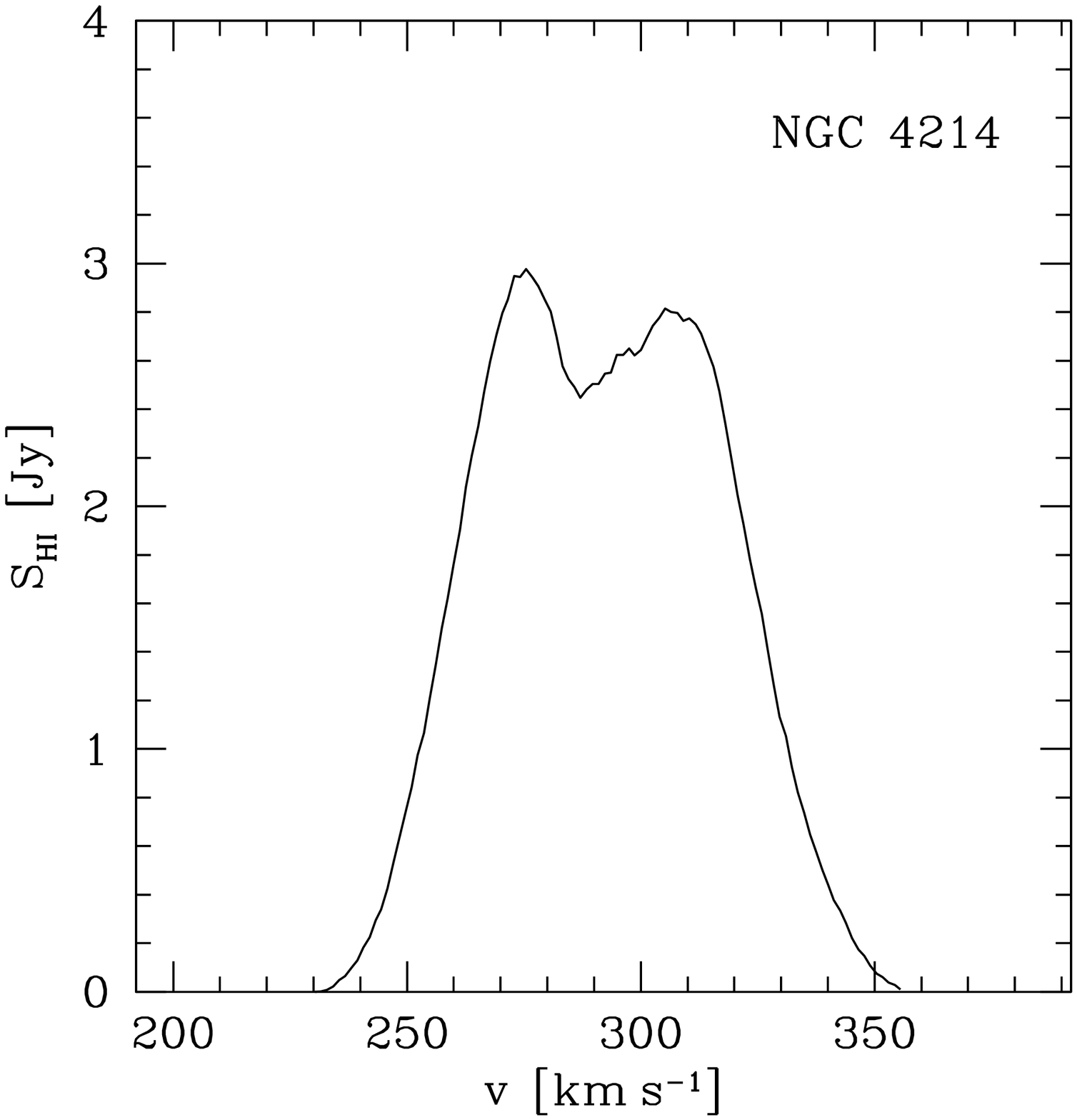}}
}
\end{minipage}
\begin{minipage}[t]{70mm}{
\resizebox{70mm}{!}{
\includegraphics[angle=0]{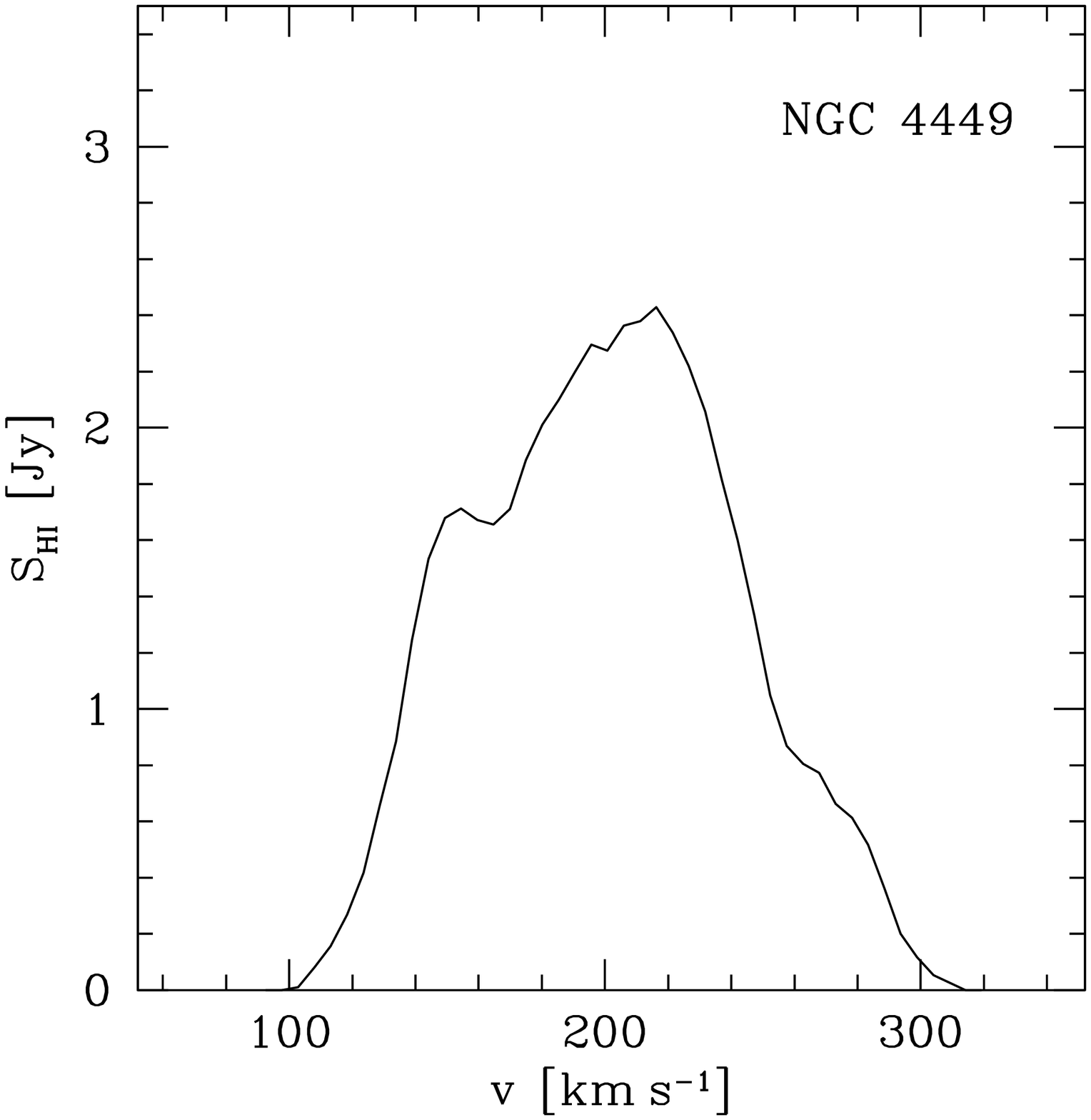}}
}
\end{minipage}
%\vspace*{-7mm}

\caption{Global \hi\ profiles for all THINGS galaxies (continued).}
\end{center}
\end{figure*}

\begin{figure*}
\figurenum{1}
      \begin{center}
\begin{minipage}[t]{70mm}{
\resizebox{70mm}{!}{
\includegraphics[angle=0]{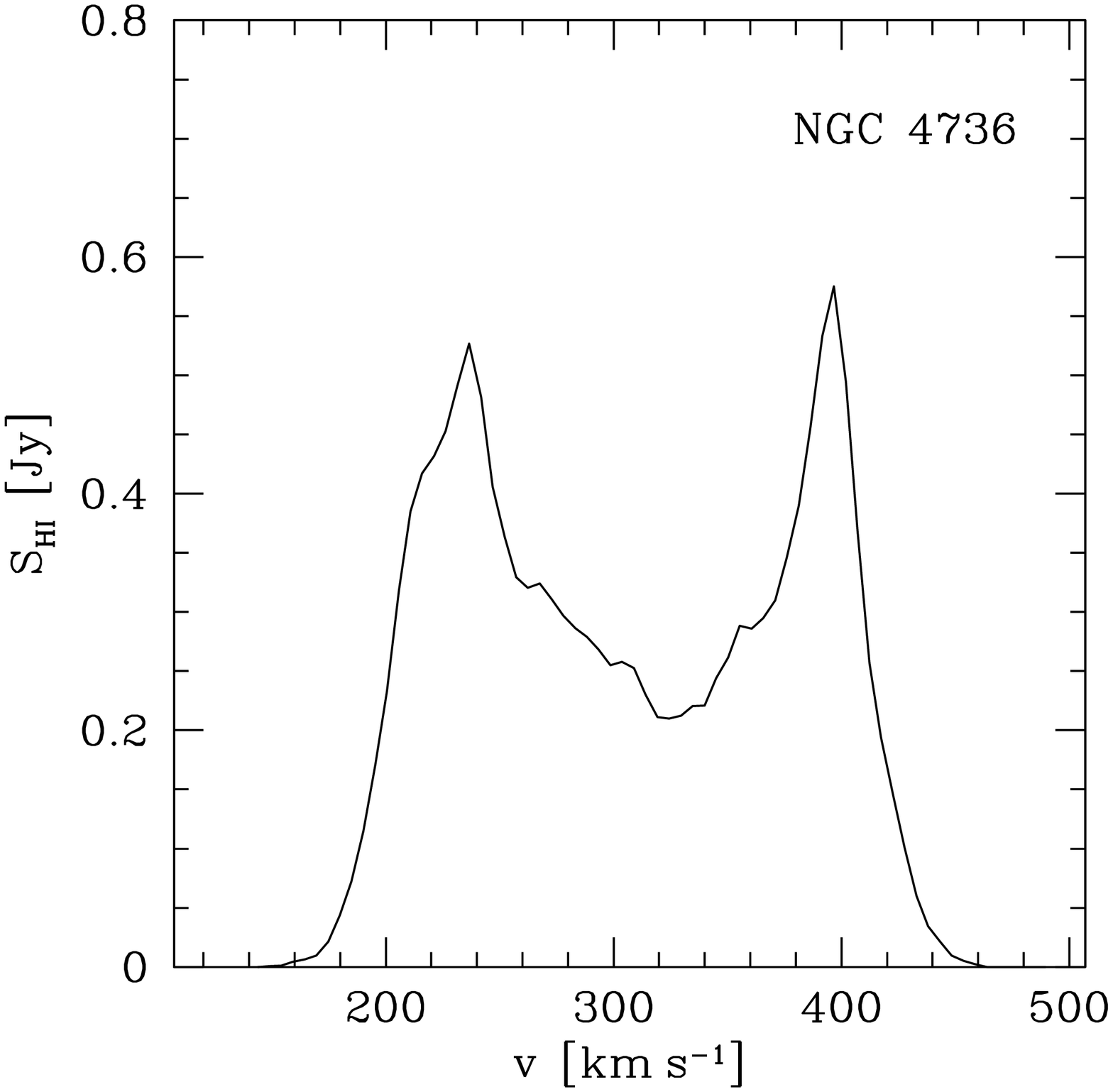}}
}
\end{minipage}
\begin{minipage}[t]{70mm}{
\resizebox{70mm}{!}{
\includegraphics[angle=0]{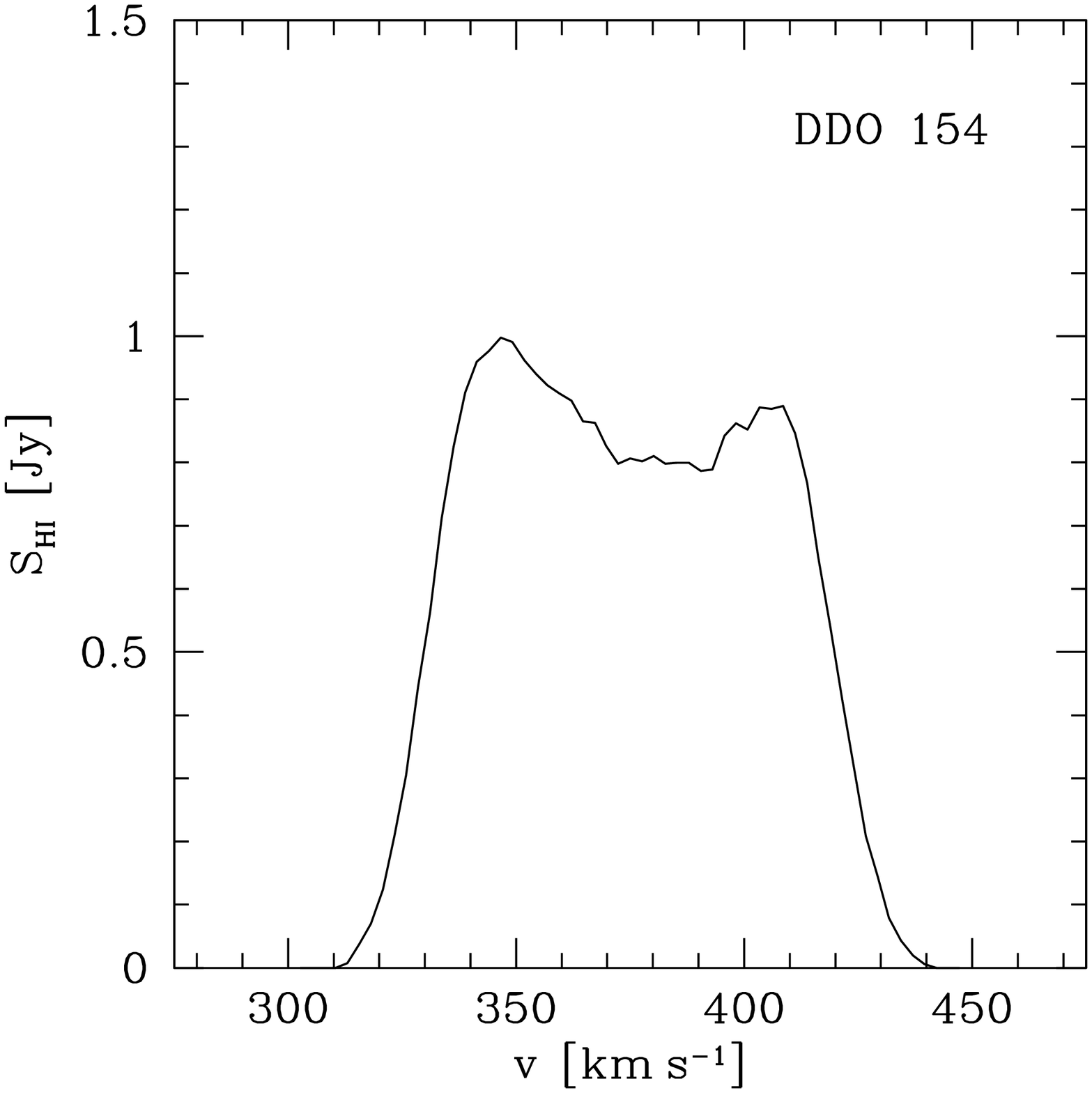}}
}
\end{minipage}
\begin{minipage}[t]{70mm}{
\resizebox{70mm}{!}{
\includegraphics[angle=0]{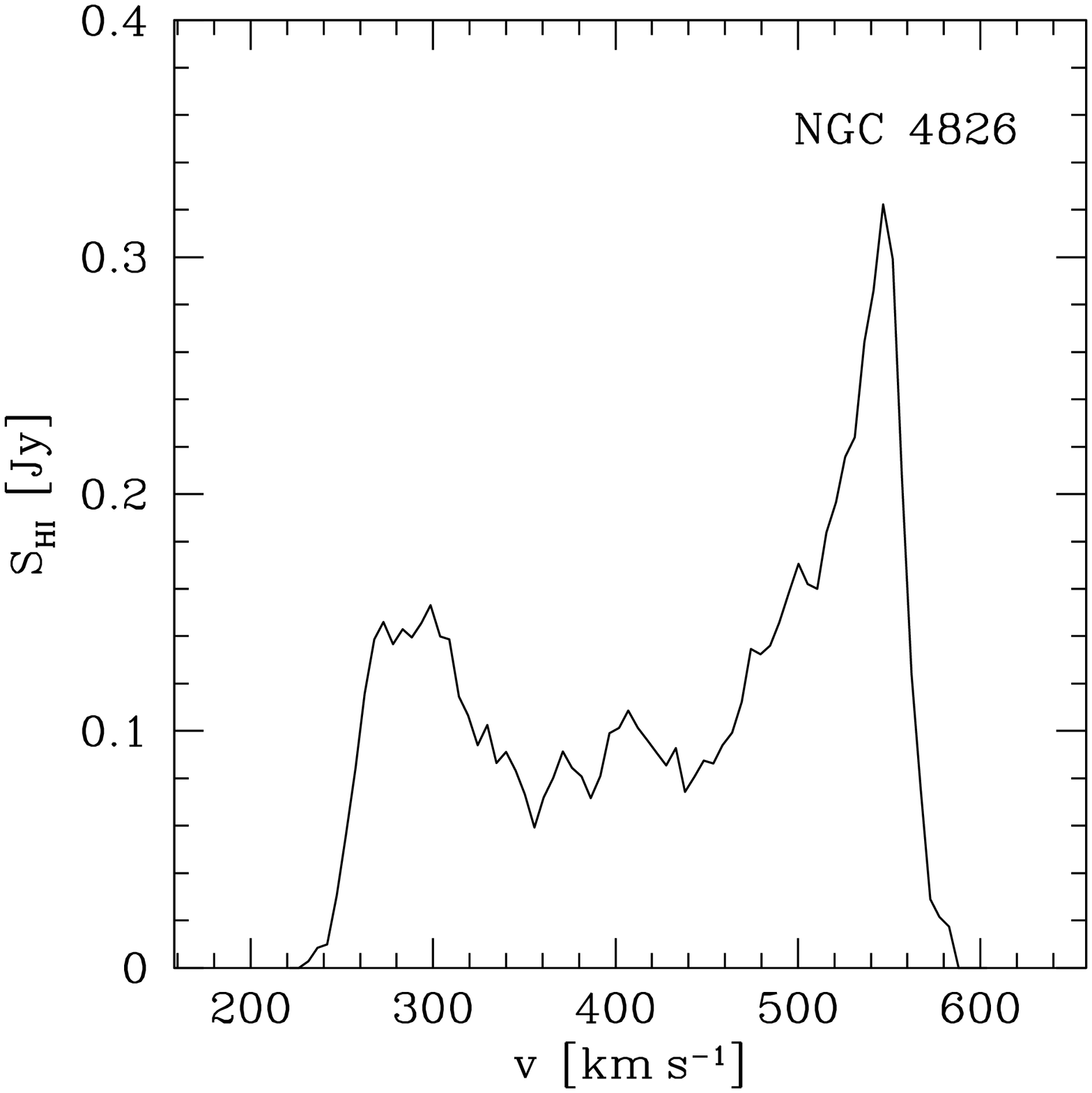}}
}
\end{minipage}
\begin{minipage}[t]{70mm}{
\resizebox{70mm}{!}{
\includegraphics[angle=0]{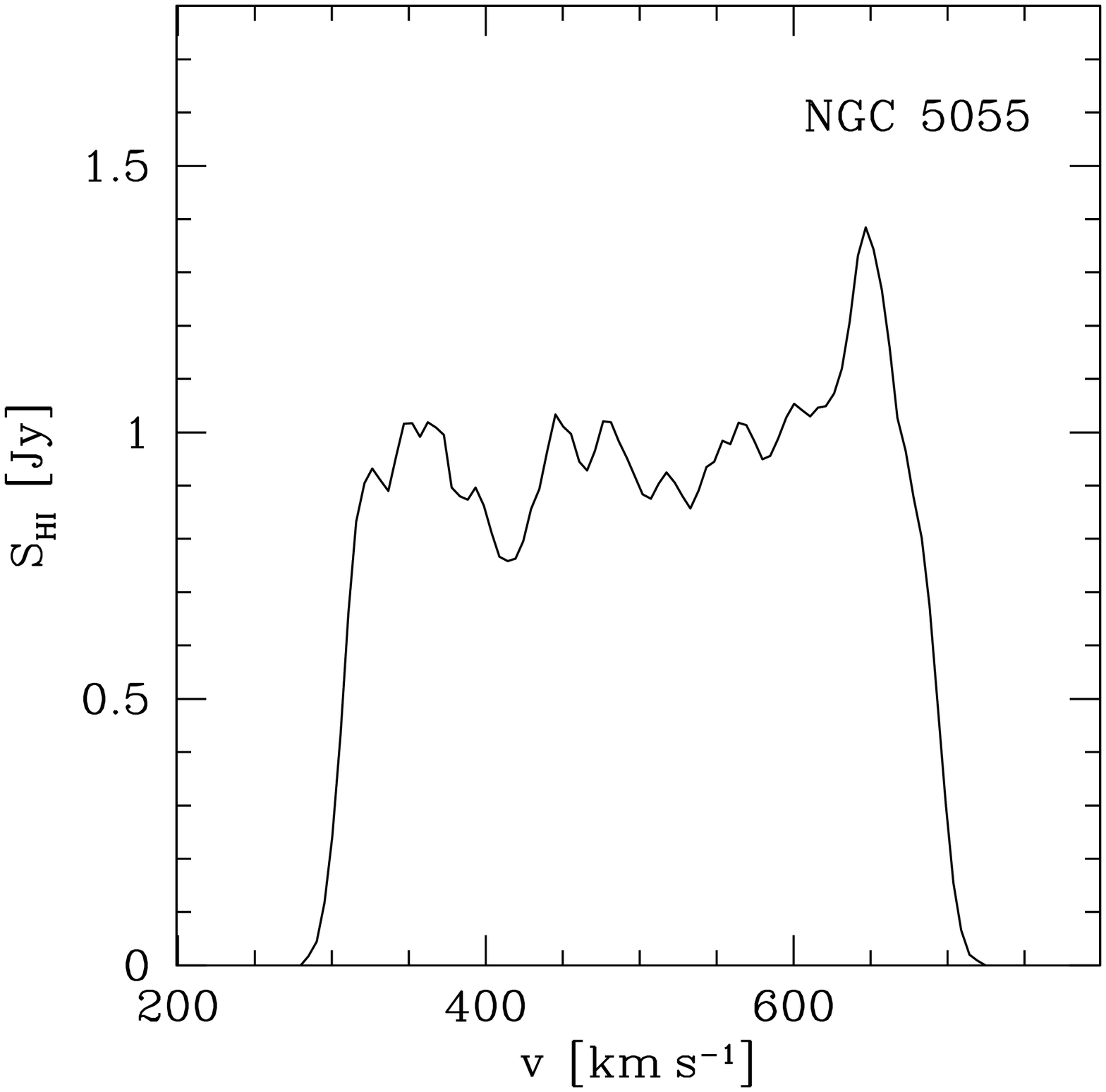}}
}
\end{minipage}
\begin{minipage}[t]{70mm}{
\resizebox{70mm}{!}{
\includegraphics[angle=0]{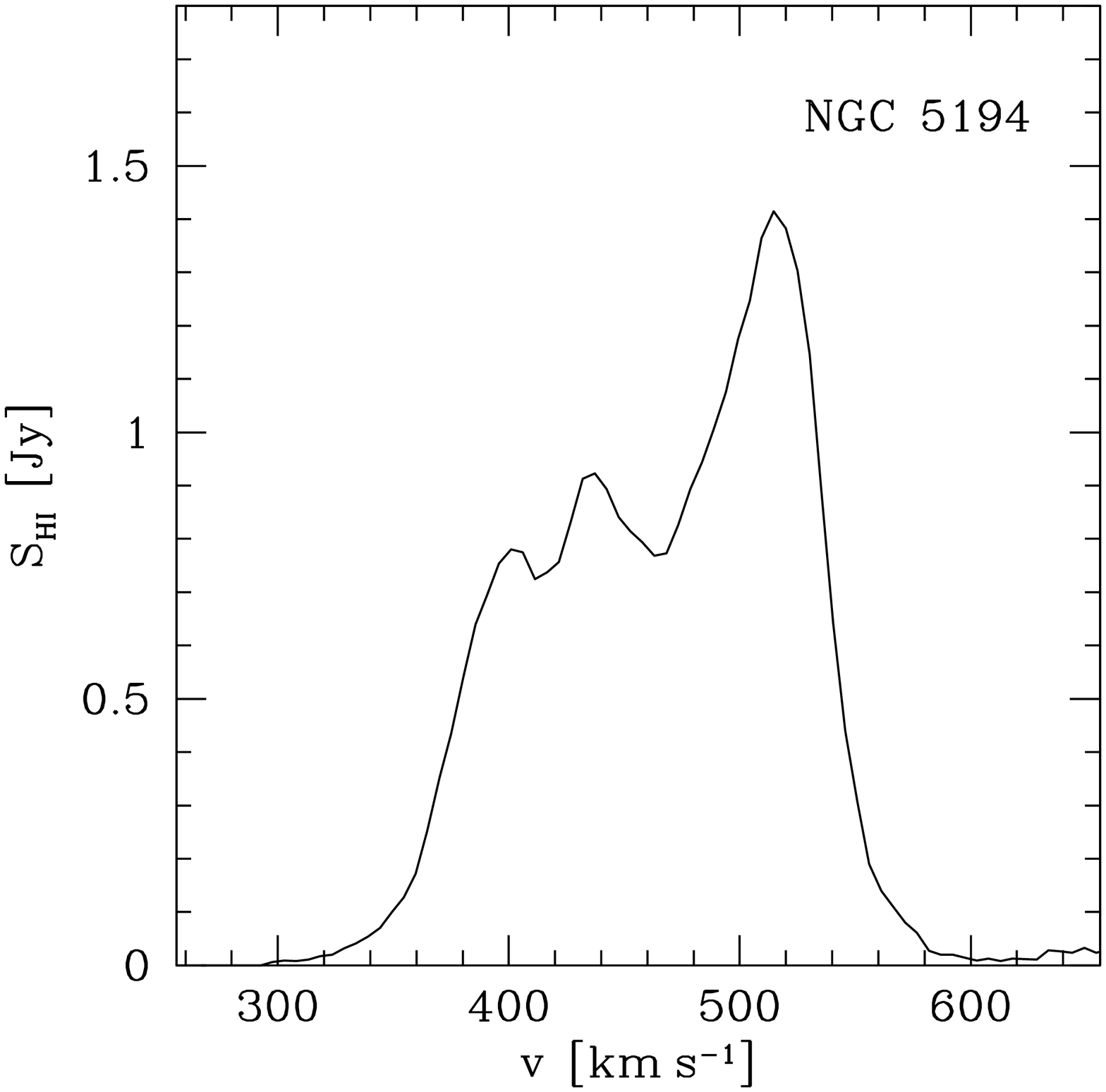}}
}
\end{minipage}
\begin{minipage}[t]{70mm}{
\resizebox{70mm}{!}{
\includegraphics[angle=0]{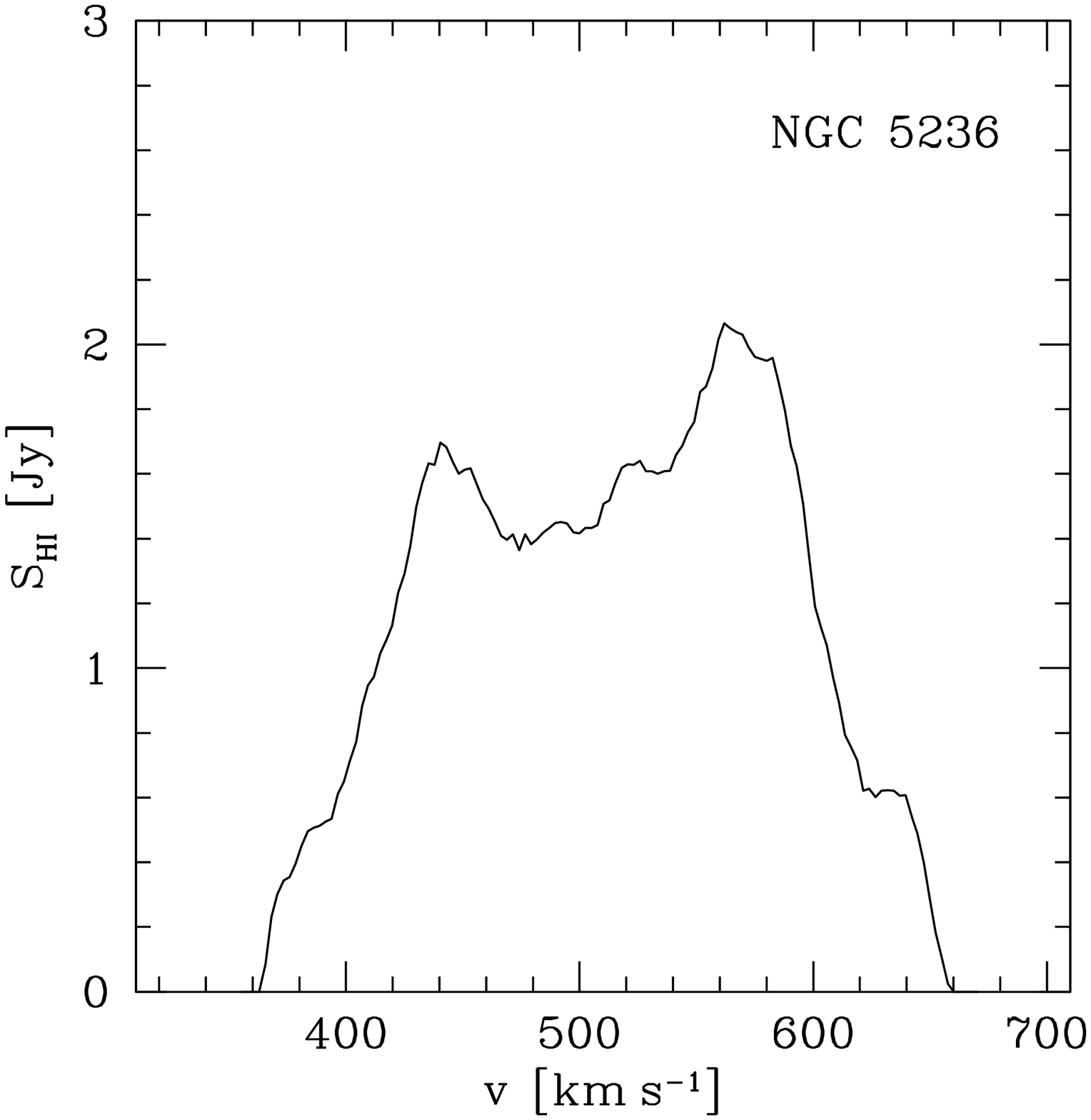}}
}
\end{minipage}
%\vspace*{-7mm}

\caption{Global \hi\ profiles for all THINGS galaxies (continued).}
\end{center}
\end{figure*}

\begin{figure*}
\figurenum{1}
      \begin{center}
\begin{minipage}[t]{70mm}{
\resizebox{70mm}{!}{
\includegraphics[angle=0]{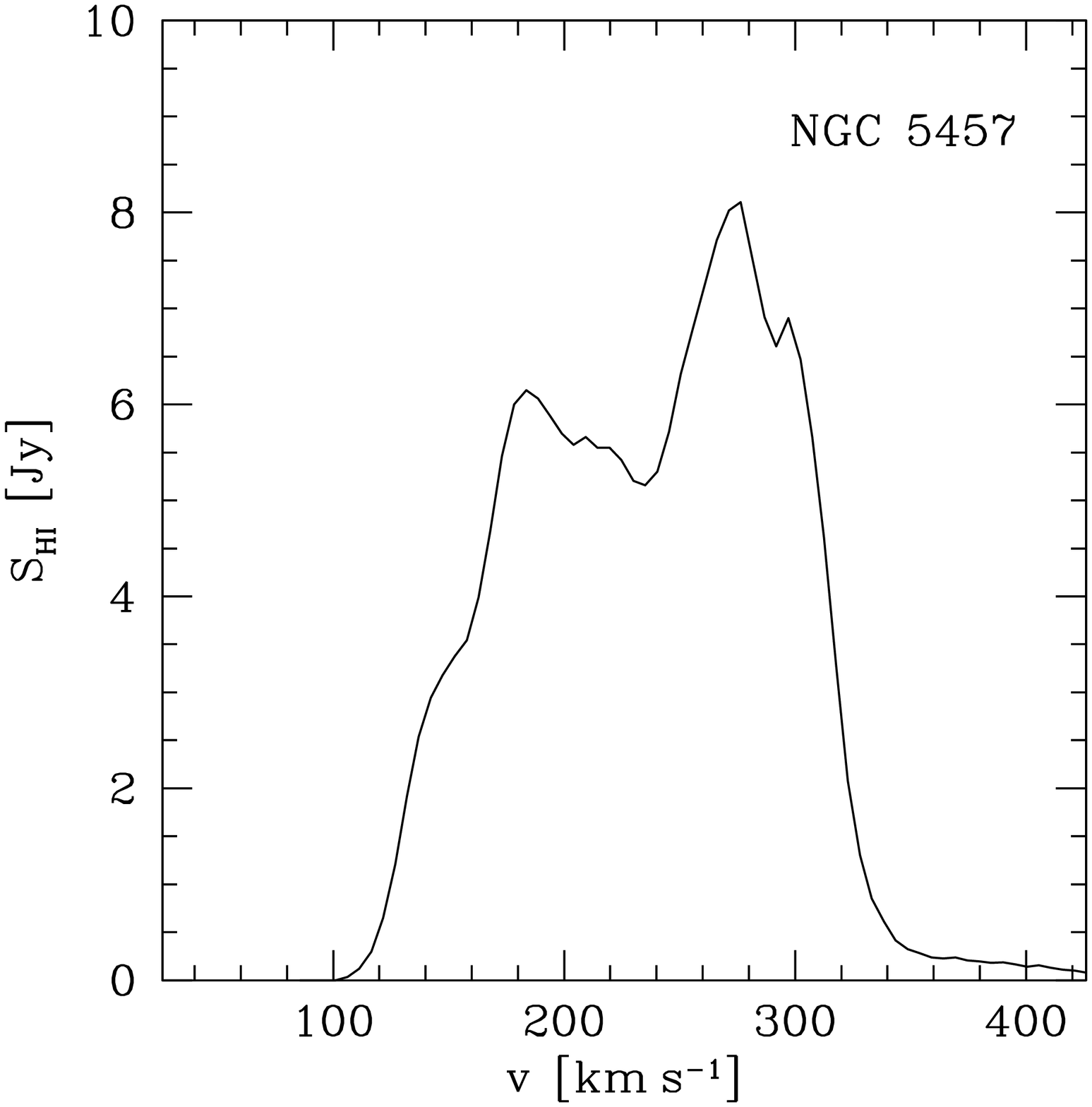}}
}
\end{minipage}
\begin{minipage}[t]{70mm}{
\resizebox{70mm}{!}{
\includegraphics[angle=0]{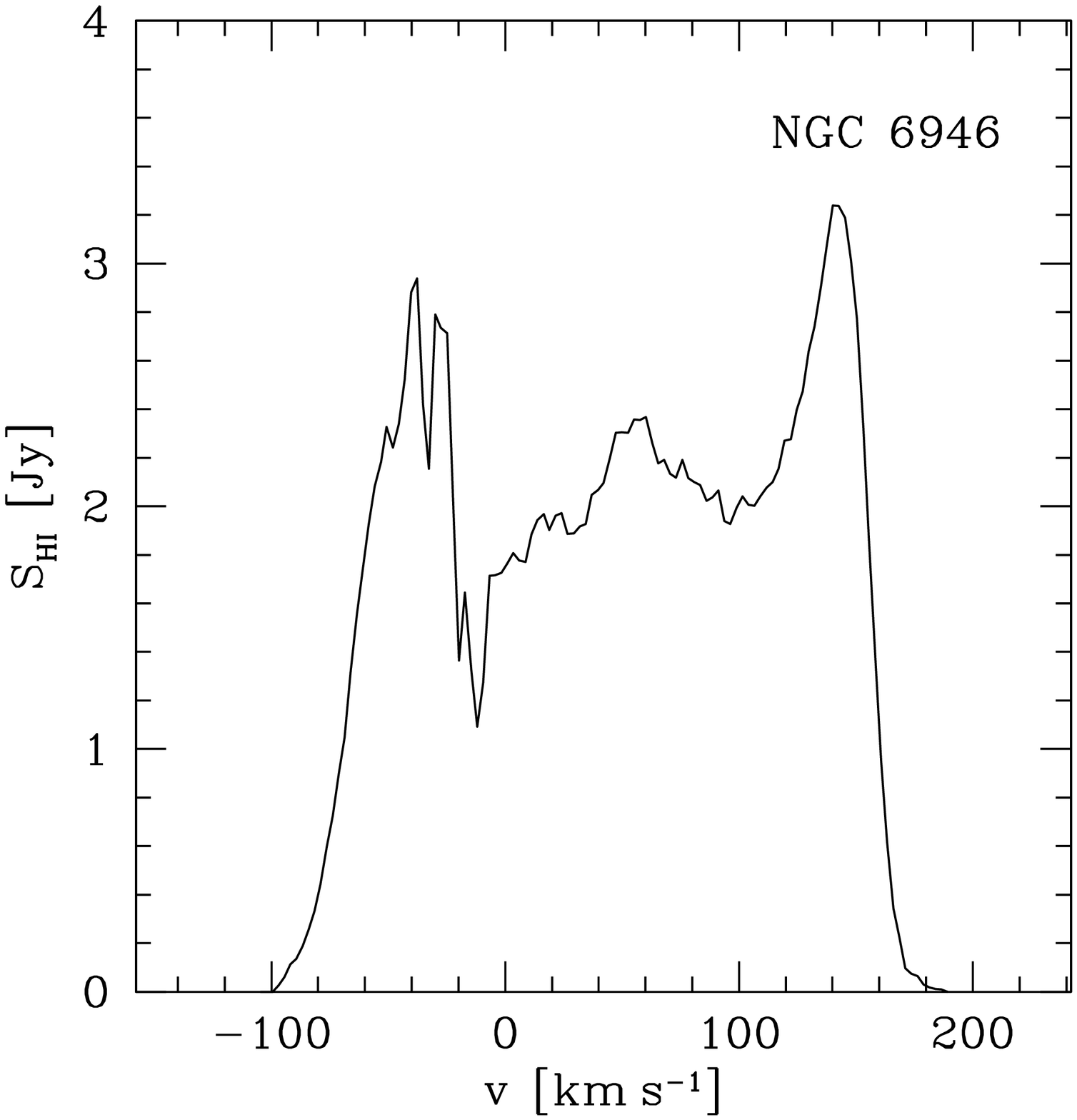}}
}
\end{minipage}
\begin{minipage}[t]{70mm}{
\resizebox{70mm}{!}{
\includegraphics[angle=0]{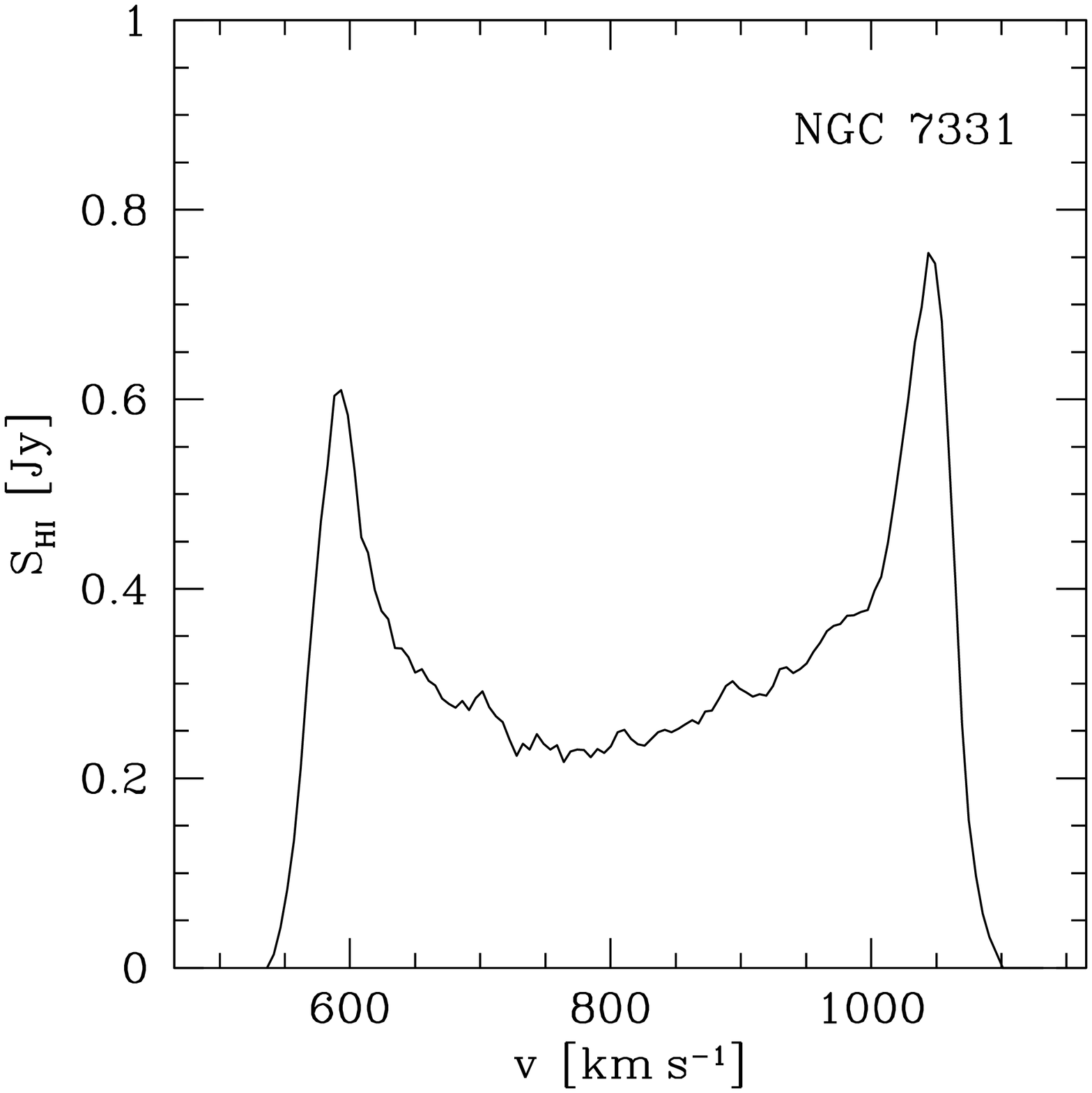}}
}
\end{minipage}
\begin{minipage}[t]{70mm}{
\resizebox{70mm}{!}{
\includegraphics[angle=0]{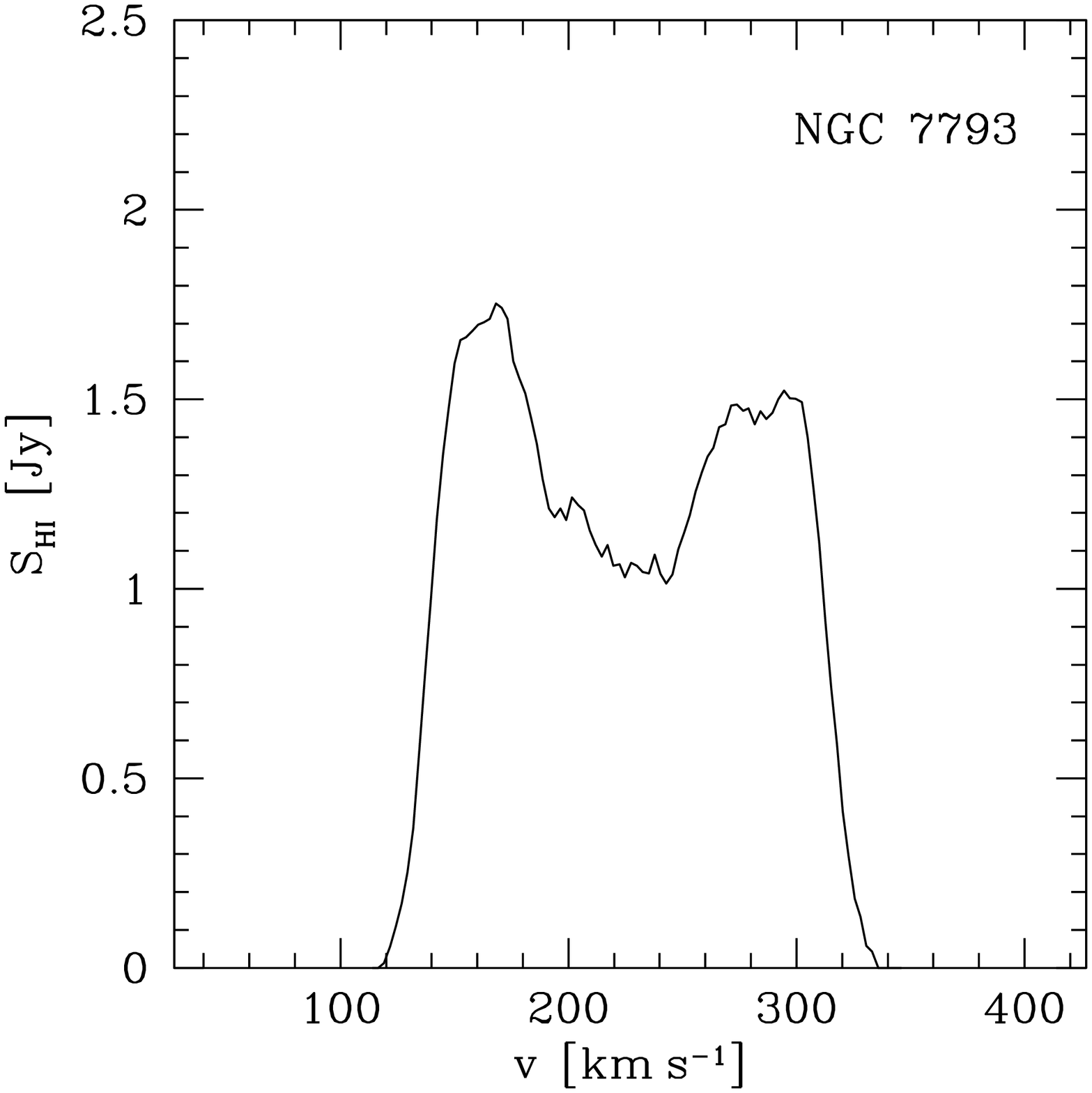}}
}
\end{minipage}
%
%\vspace*{-7mm}

\caption{Global \hi\ profiles for all THINGS galaxies (continued).}
\end{center}
\end{figure*}

%
%
%   GALAXIES  -  CHANNEL MAPS AND MOMENT MAPS
%
%

%\end{document}

%
% NGC\,628
%

\clearpage
\begin{figure}
\epsscale{1.0}
\plotone{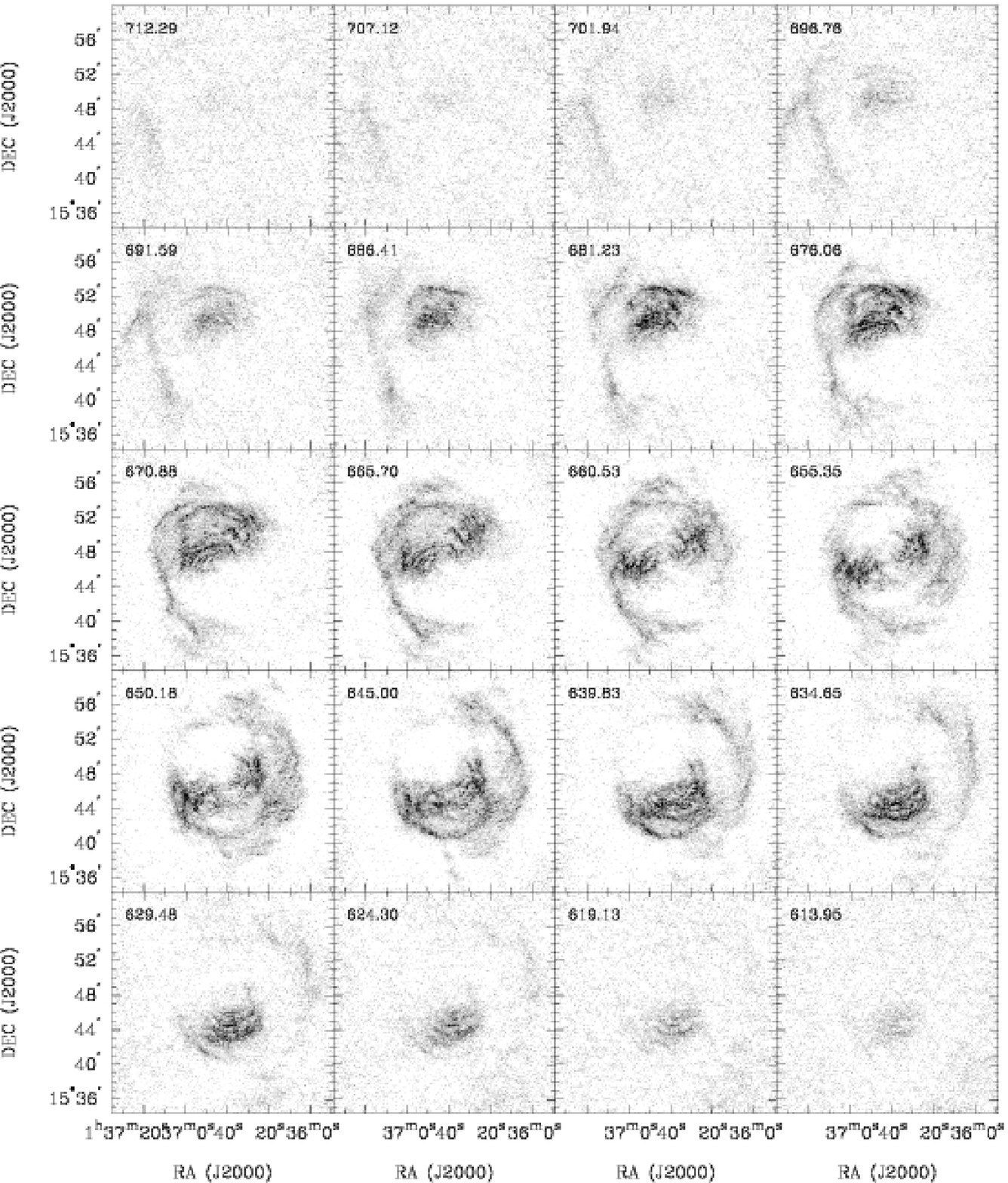}
\caption{{\bf NGC~628:} Channel maps based on the NA cube (greyscale
  range: --0.02 to 8 mJy\,beam$^{-1}$).  Every second channel is shown
  (channel width: 2.6\,km\,s$^{-1}$). The area shown in each panel is
  identical to the area shown on the next figure}
\end{figure}

\clearpage
\begin{figure}
\vspace{0cm}  \epsscale{1.1}
\plotone{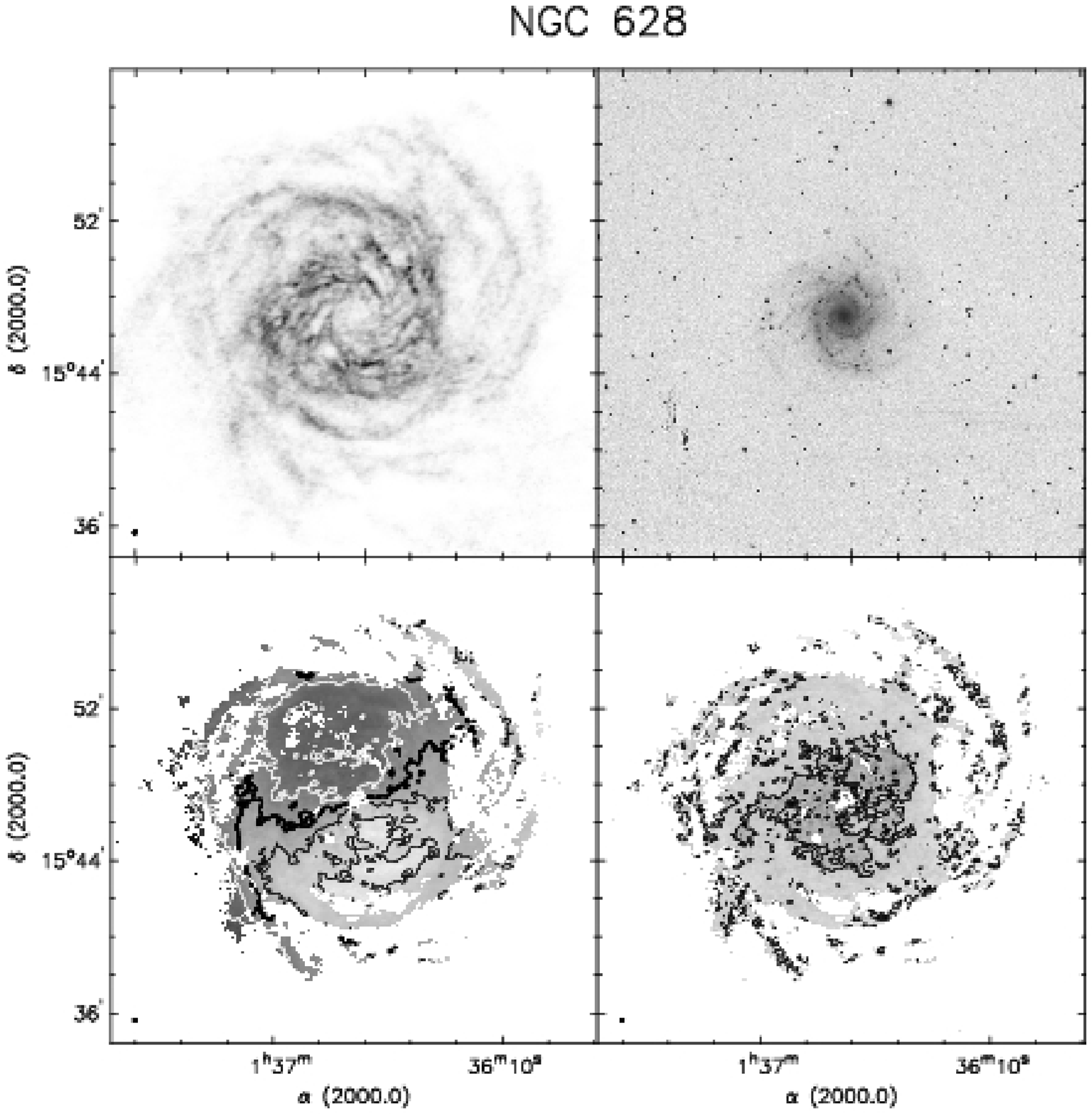}
\vspace{-2.5cm}
\caption{{\bf NGC~628}. {\em Top left:} integrated \hi\ map (moment 0).
  Greyscale range from 0--218 Jy\,km\,s$^{-1}$. {\em Top right:}
  Optical image from the digitized sky survey (DSS). {\em Bottom
    left:} Velocity field (moment 1). Black contours (lighter
  greyscale) indicate approaching emission, white contours (darker
  greyscale) receding emission. The thick black contour is the
  systemic velocity ($v_{\rm sys}$=659.1 \,km\,s$^{-1}$), the
  iso--velocity contours are spaced by $\Delta\,v$=12.5\,km\,s$^{-1}$.
  {\em Bottom right:} Velocity dispersion map (moment 2). Contours are plotted
  at 5, 10 and 20\,km\,s$^{-1}$.}
\end{figure}

%
% NGC\,925
%

\clearpage
\begin{figure}
\epsscale{1.0}
\plotone{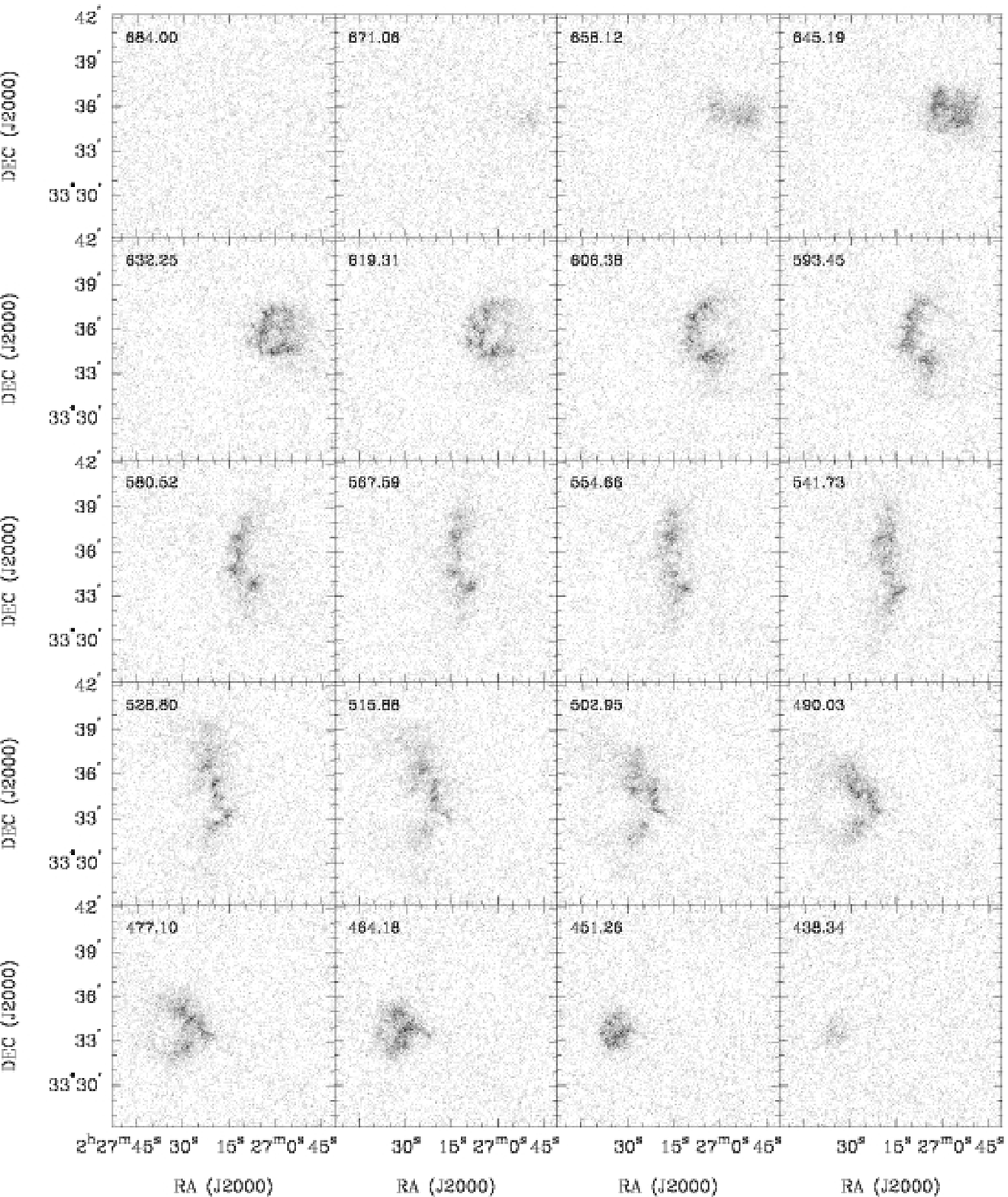}
\caption{{\bf NGC~925:} Channel maps based on the NA cube (greyscale
  range: --0.02 to 8 mJy\,beam$^{-1}$).  Every fifth channel is shown
  (channel width: 2.6\,km\,s$^{-1}$). The area shown in each panel is
  identical to the area shown on the next figure}
\end{figure}

\clearpage
\begin{figure}
\vspace{0cm}  \epsscale{1.1}
\plotone{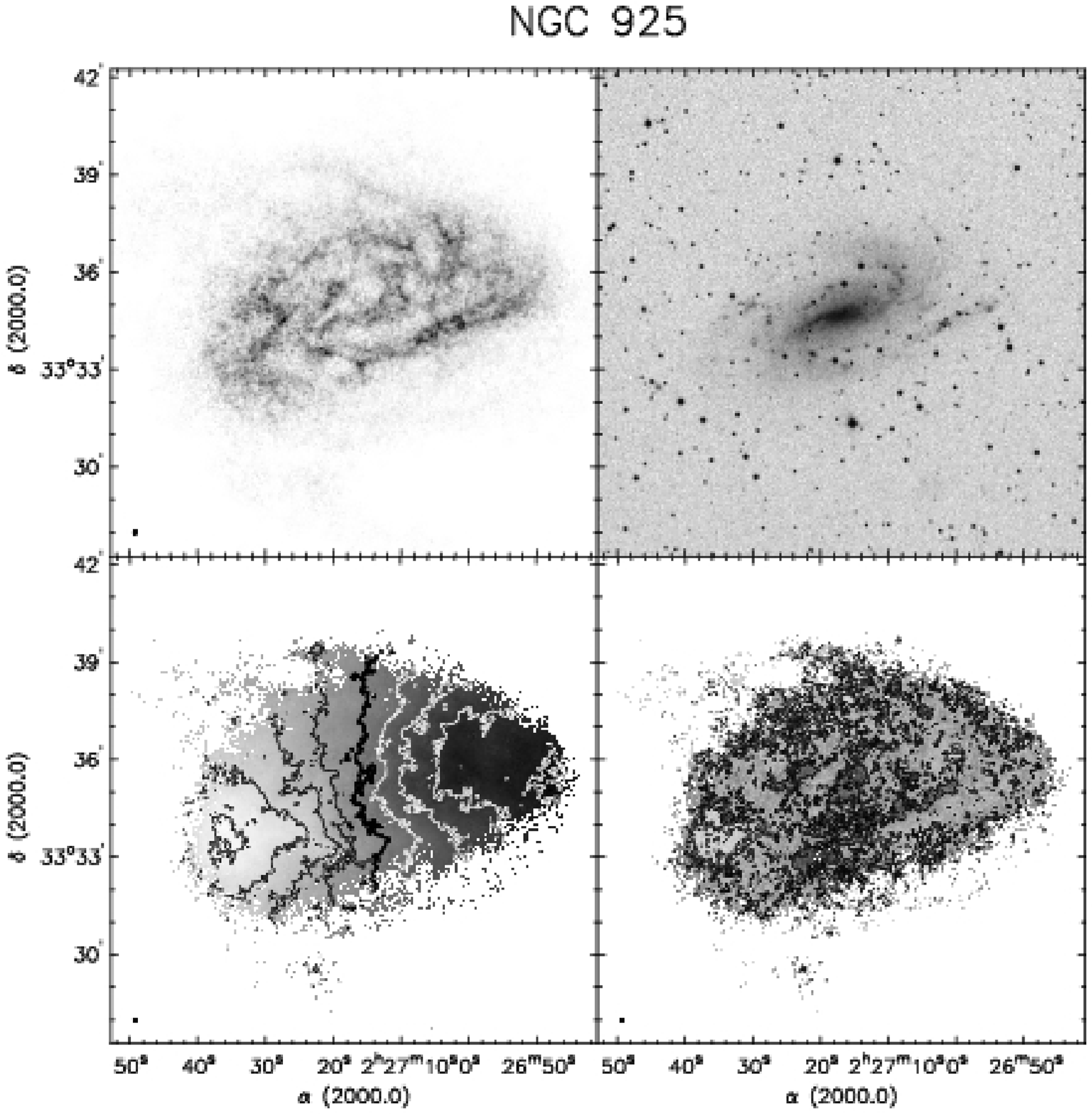}
\vspace{-2.5cm}
\caption{{\bf NGC~925}. {\em Top left:} integrated \hi\ map (moment 0).
  Greyscale range from 0--157 Jy\,km\,s$^{-1}$. {\em Top right:}
  Optical image from the digitized sky survey (DSS). {\em Bottom
    left:} Velocity field (moment 1). Black contours (lighter
  greyscale) indicate approaching emission, white contours (darker
  greyscale) receding emission. The thick black contour is the
  systemic velocity ($v_{\rm sys}$=552.5 \,km\,s$^{-1}$), the
  iso--velocity contours are spaced by $\Delta\,v$=25\,km\,s$^{-1}$.
  {\em Bottom right:} Velocity dispersion map (moment 2). Contours are plotted
  at 5, 10 and 20\,km\,s$^{-1}$.}
\end{figure}

%
% NGC\,1569
%

\clearpage
\begin{figure}
\epsscale{1.0}
\plotone{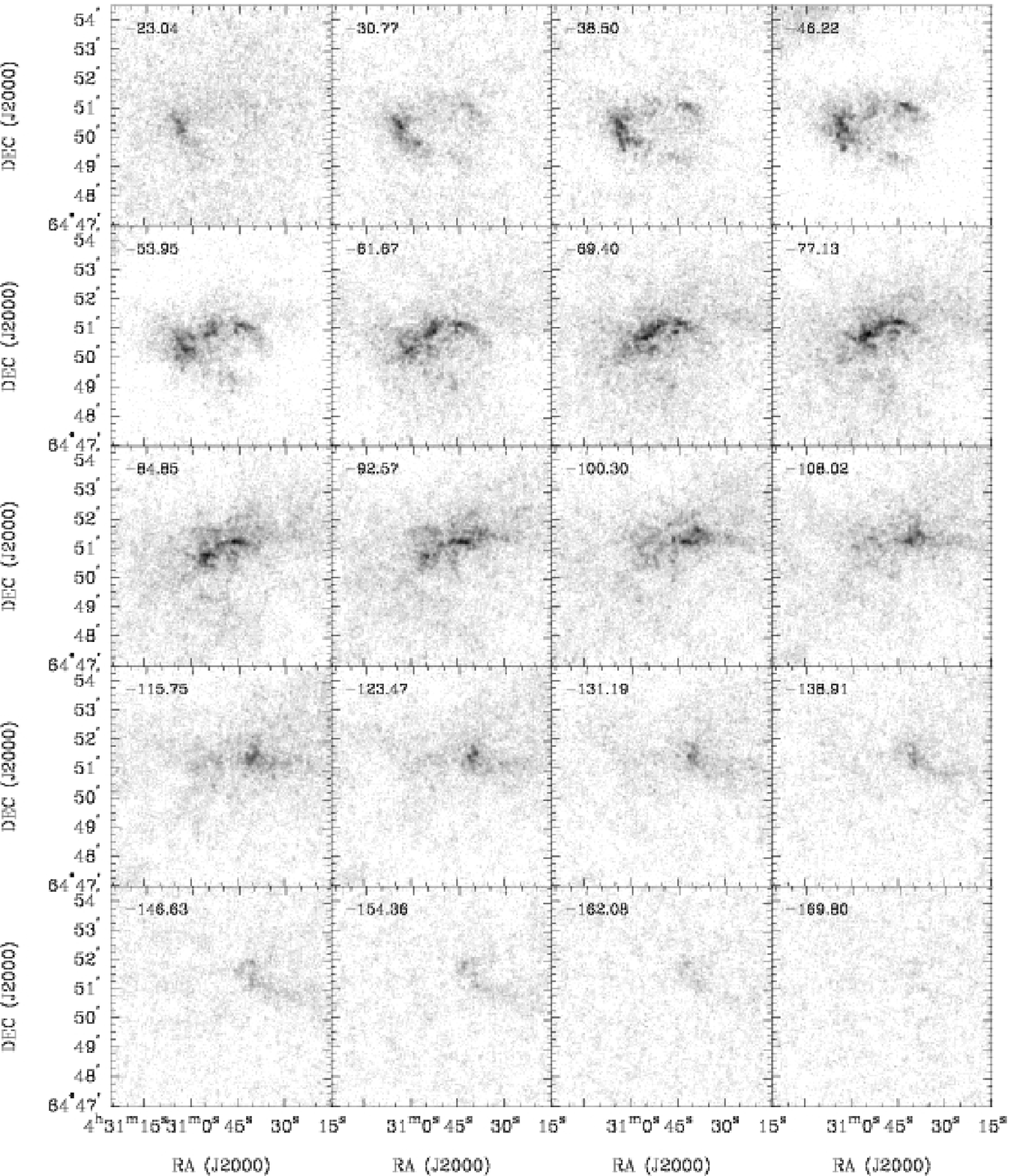}
\caption{{\bf NGC~1569:} Channel maps based on the NA cube (greyscale
  range: --0.02 to 10 mJy\,beam$^{-1}$).  Every third channel is shown
  (channel width: 2.6\,km\,s$^{-1}$). The area shown in each panel is
  identical to the area shown on the next figure}
\end{figure}

\clearpage
\begin{figure}
\vspace{0cm}  \epsscale{1.1}
\plotone{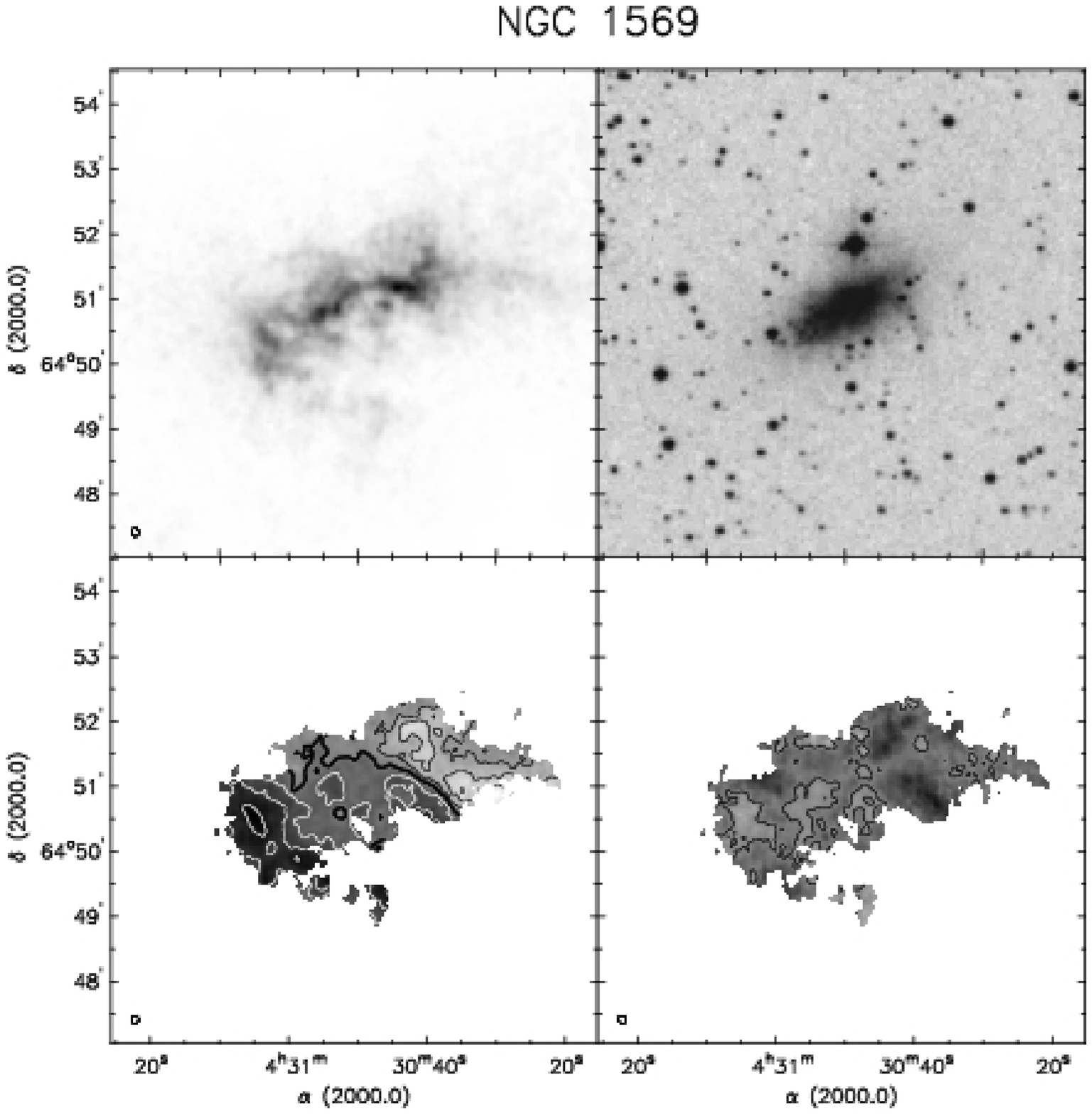}
\vspace{-2.5cm}
\caption{{\bf NGC~1569}. {\em Top left:} integrated \hi\ map (moment 0).
  Greyscale range from 0--426 Jy\,km\,s$^{-1}$. {\em Top right:}
  Optical image from the digitized sky survey (DSS). {\em Bottom
    left:} Velocity field (moment 1). Black contours (lighter
  greyscale) indicate approaching emission, white contours (darker
  greyscale) receding emission. The thick black contour is the
  systemic velocity ($v_{\rm sys}$=--85.57 \,km\,s$^{-1}$), the
  iso--velocity contours are spaced by $\Delta\,v$=12.5\,km\,s$^{-1}$.
  {\em Bottom right:} Velocity dispersion map (moment 2). Contours are plotted
  at 5\,km\,s$^{-1}$.}
\end{figure}

%
% NGC\,2366
%

\clearpage
\begin{figure}
\epsscale{1.0}
\plotone{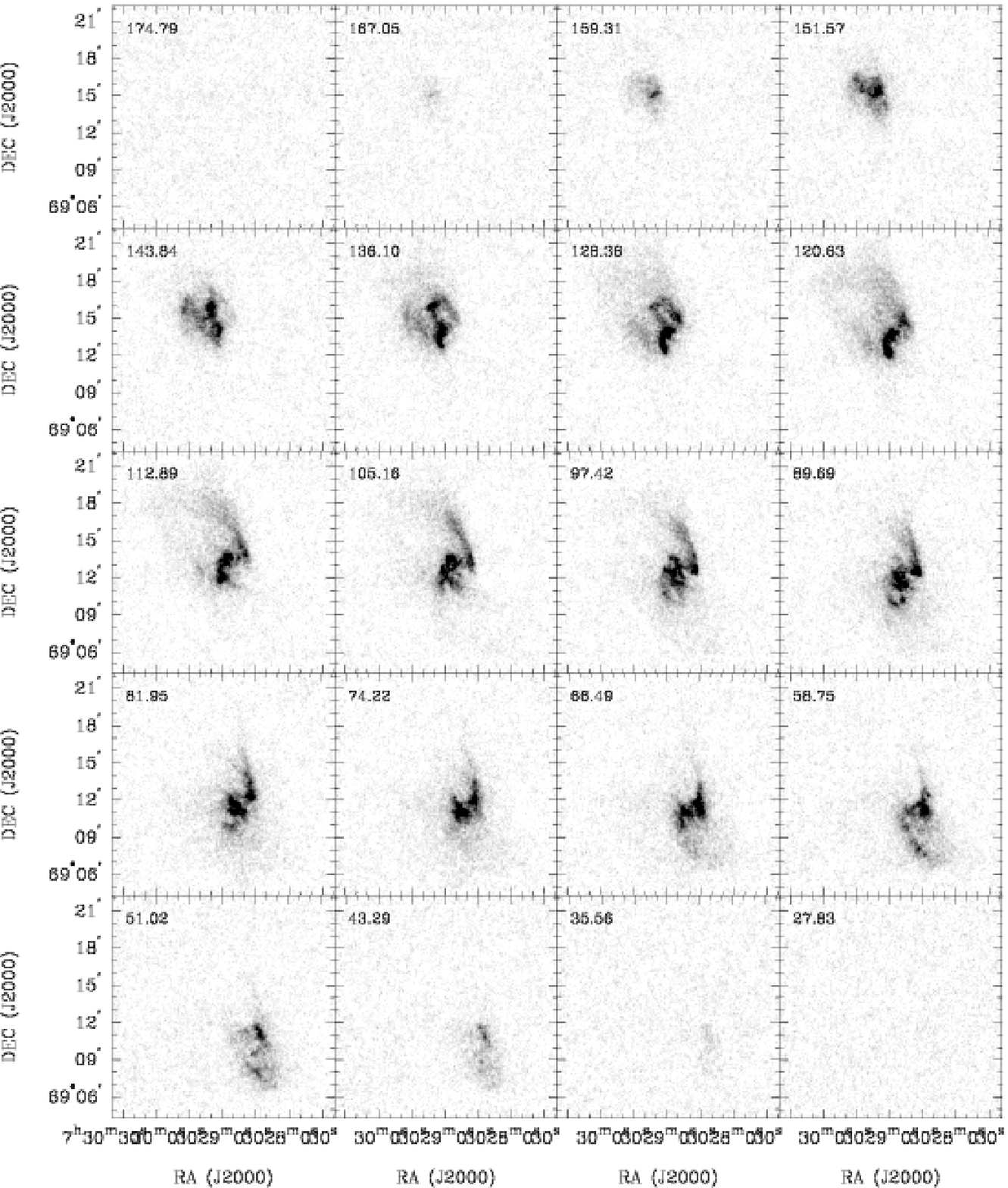}
\caption{{\bf NGC~2366:} Channel maps based on the NA cube (greyscale
  range: --0.02 to 13 mJy\,beam$^{-1}$).  Every third channel is shown
  (channel width: 2.6\,km\,s$^{-1}$). The area shown in each panel is
  identical to the area shown on the next figure}
\end{figure}

\clearpage
\begin{figure}
\vspace{0cm}  \epsscale{1.1}
\plotone{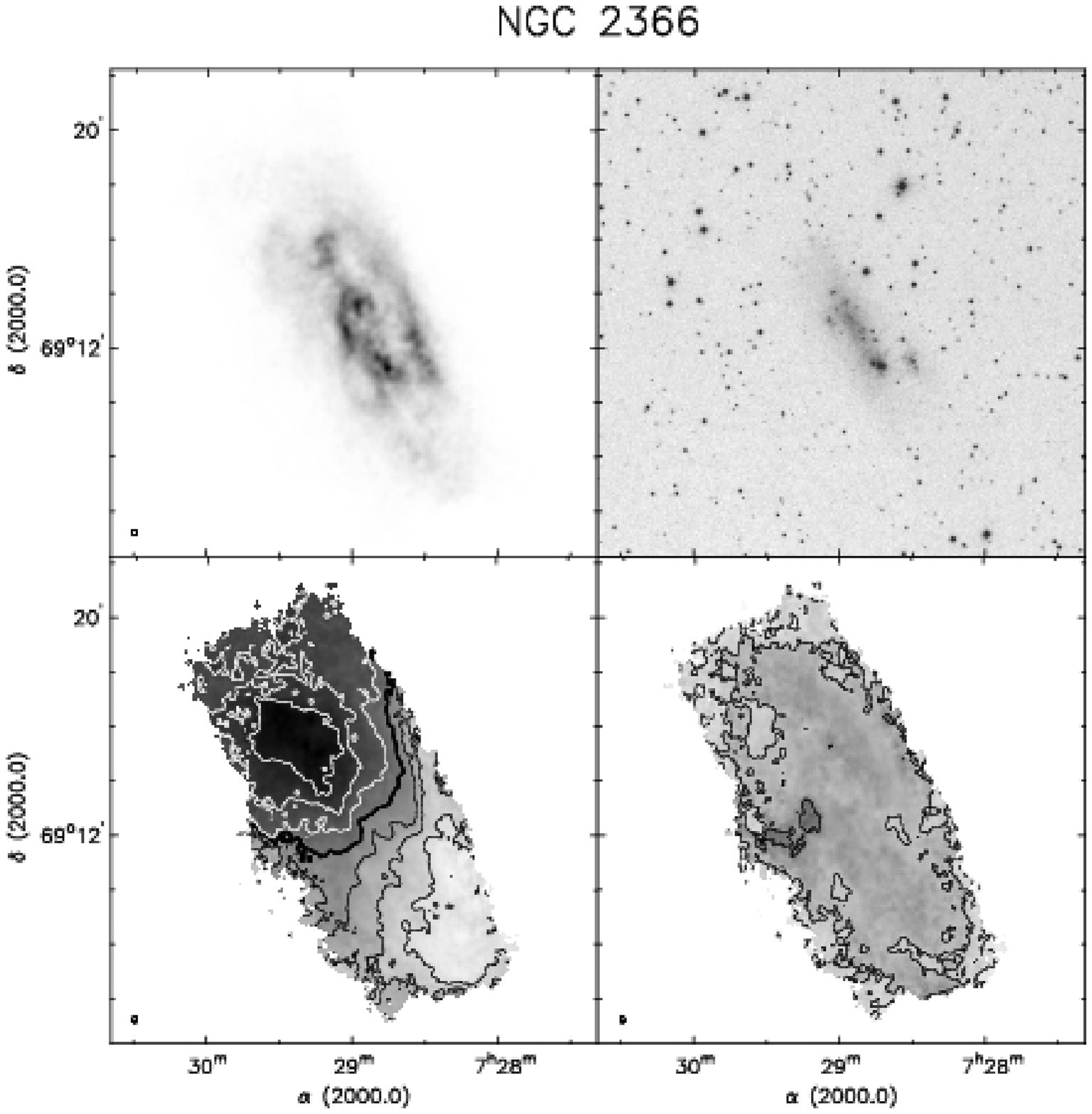}
\vspace{-2.5cm}
\caption{{\bf NGC~2366}. {\em Top left:} integrated \hi\ map (moment 0).
  Greyscale range from 0--1020 Jy\,km\,s$^{-1}$. {\em Top right:}
  Optical image from the digitized sky survey (DSS). {\em Bottom
    left:} Velocity field (moment 1). Black contours (lighter
  greyscale) indicate approaching emission, white contours (darker
  greyscale) receding emission. The thick black contour is the
  systemic velocity ($v_{\rm sys}$=100.1 \,km\,s$^{-1}$), the
  iso--velocity contours are spaced by $\Delta\,v$=12.5\,km\,s$^{-1}$.
  {\em Bottom right:} Velocity dispersion map (moment 2). Contours are plotted
  at 5, 10 and 20\,km\,s$^{-1}$.}
\end{figure}

%
% NGC\,2403
%

\clearpage
\begin{figure}
\epsscale{1.0}
\plotone{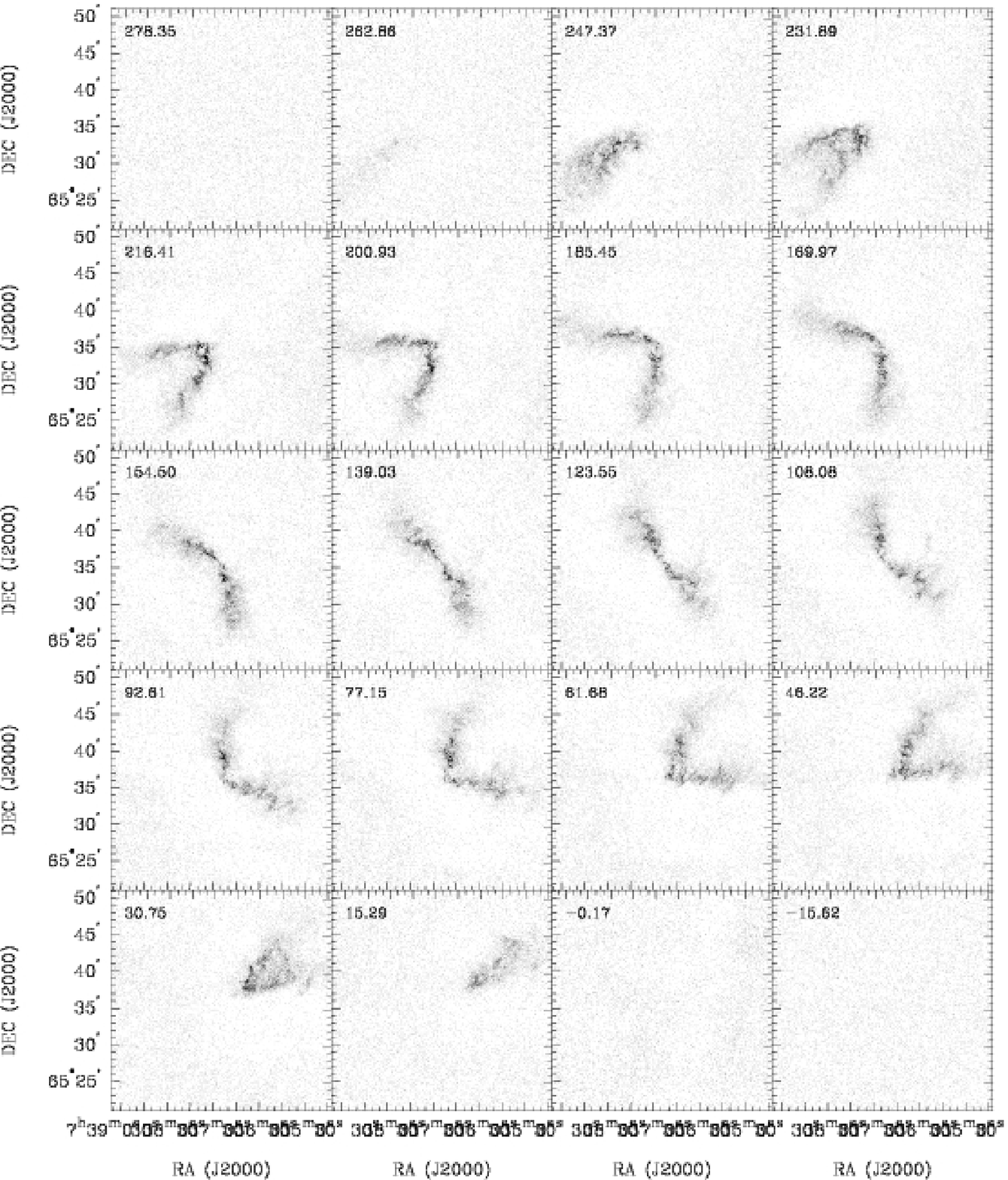}
\caption{{\bf NGC~2403:} Channel maps based on the NA cube (greyscale
  range: --0.02 to 10 mJy\,beam$^{-1}$).  Every third channel is shown
  (channel width: 5.2\,km\,s$^{-1}$). The area shown in each panel is
  identical to the area shown on the next figure}
\end{figure}

\clearpage
\begin{figure}
\vspace{0cm}  \epsscale{1.1}
\plotone{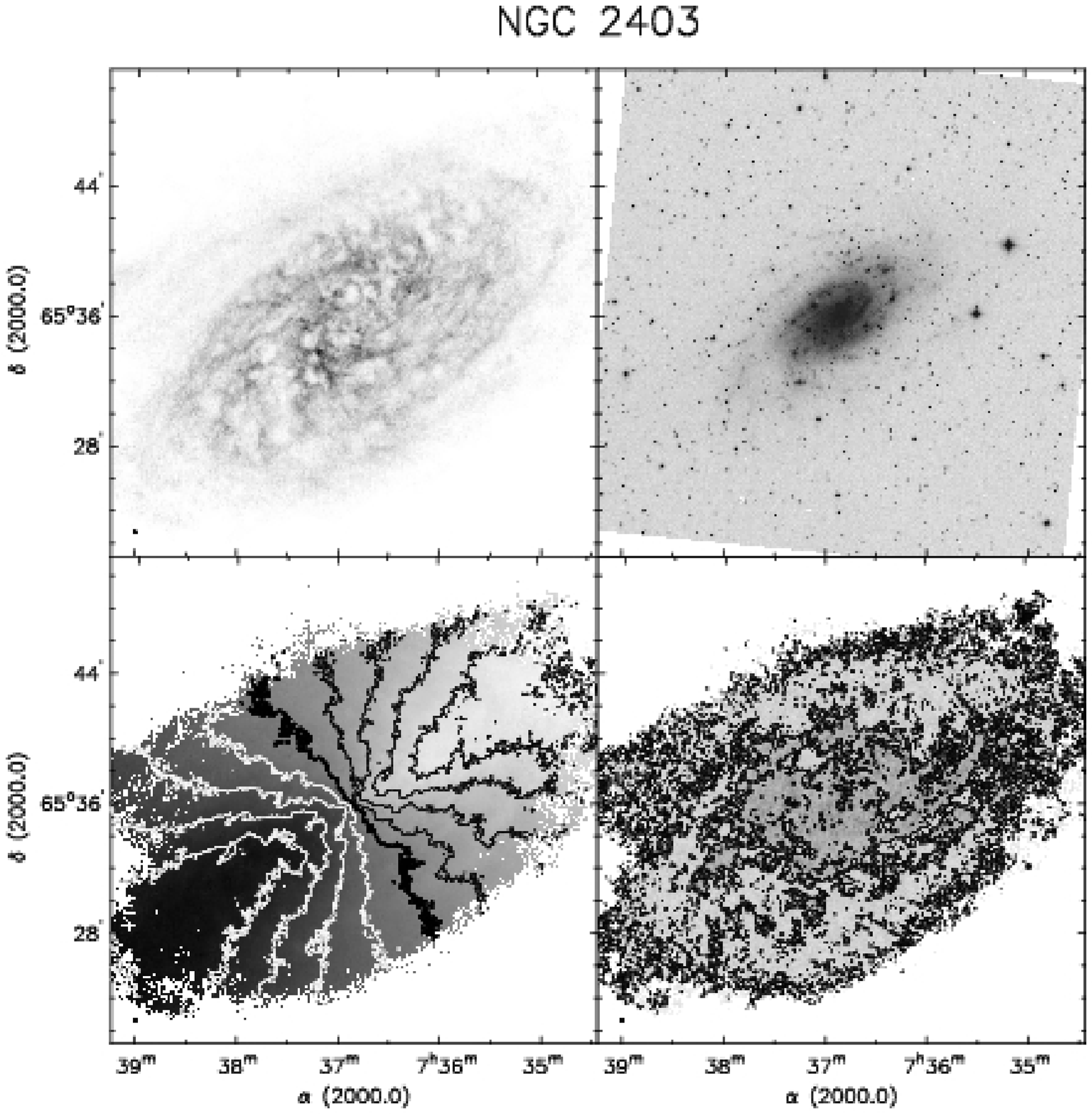}
\vspace{-2.5cm}
\caption{{\bf NGC~2403}. {\em Top left:} integrated \hi\ map (moment 0).
  Greyscale range from 0--357 Jy\,km\,s$^{-1}$. {\em Top right:}
  Optical image from the digitized sky survey (DSS). {\em Bottom
    left:} Velocity field (moment 1). Black contours (lighter
  greyscale) indicate approaching emission, white contours (darker
  greyscale) receding emission. The thick black contour is the
  systemic velocity ($v_{\rm sys}$=133.1\,km\,s$^{-1}$), the
  iso--velocity contours are spaced by $\Delta\,v$=25\,km\,s$^{-1}$.
  {\em Bottom right:} Velocity dispersion map (moment 2). Contours are plotted
  at 5, 10 and 20\,km\,s$^{-1}$.}
\end{figure}

%
% HOLMBERG~II
%

\clearpage
\begin{figure}
\epsscale{1.0}
\plotone{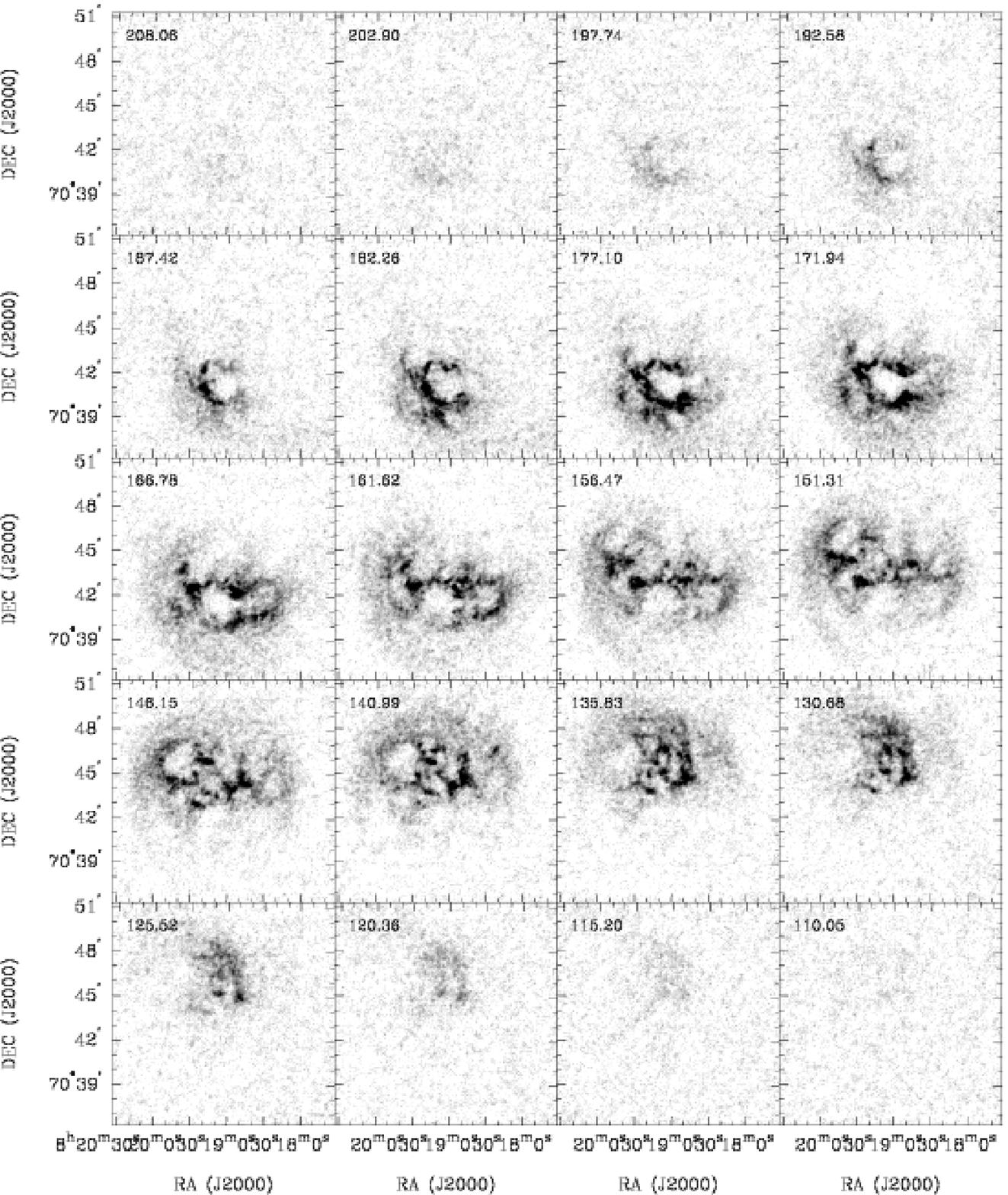}
\caption{{\bf Holmberg~II:} Channel maps based on the NA cube
  (greyscale range: --0.02 to 13 mJy\,beam$^{-1}$).  Every second channel
  is shown (channel width: 2.6\,km\,s$^{-1}$). The area shown in each
  panel is identical to the area shown on the next figure}
\end{figure}

\clearpage
\begin{figure}
\vspace{0cm}  \epsscale{1.1}
\plotone{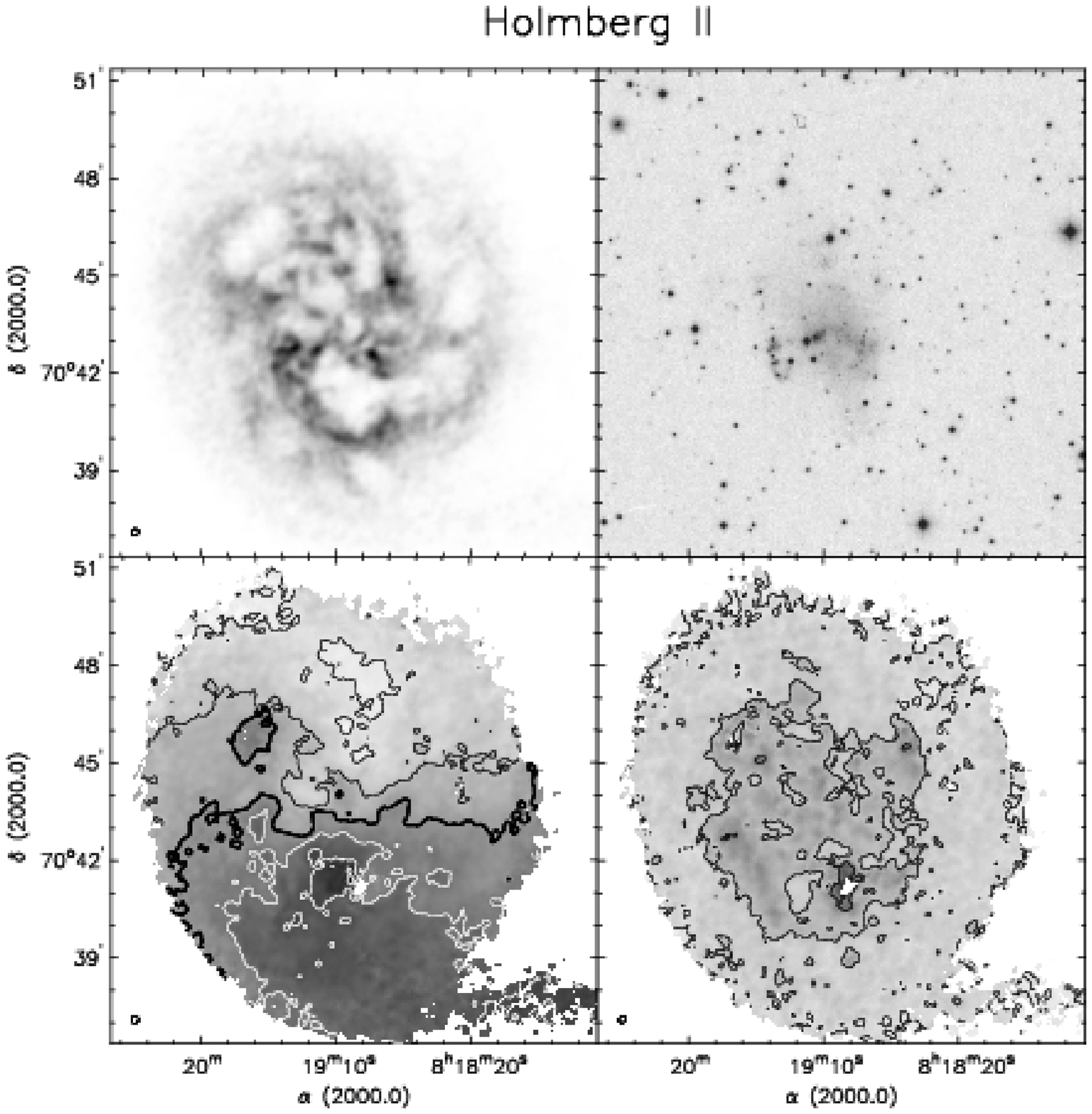}
\vspace{-2.5cm}
\caption{{\bf Holmberg~II}. {\em Top left:} integrated \hi\ map (moment
  0). Greyscale range from 0--599 Jy\,km\,s$^{-1}$. {\em Top right:}
  Optical image from the digitized sky survey (DSS). {\em Bottom
    left:} Velocity field (moment 1). Black contours (lighter
  greyscale) indicate approaching emission, white contours (darker
  greyscale) receding emission. The thick black contour is the
  systemic velocity ($v_{\rm sys}$=157.1 \,km\,s$^{-1}$), the
  iso--velocity contours are spaced by $\Delta\,v$=12.5\,km\,s$^{-1}$.
  {\em Bottom right:} Velocity dispersion map (moment 2). Contours are plotted
  at 5, 10 and 20\,km\,s$^{-1}$.}
\end{figure}

%
%  M81 DWARF A
%

\clearpage
\begin{figure}
\epsscale{1.0}
\plotone{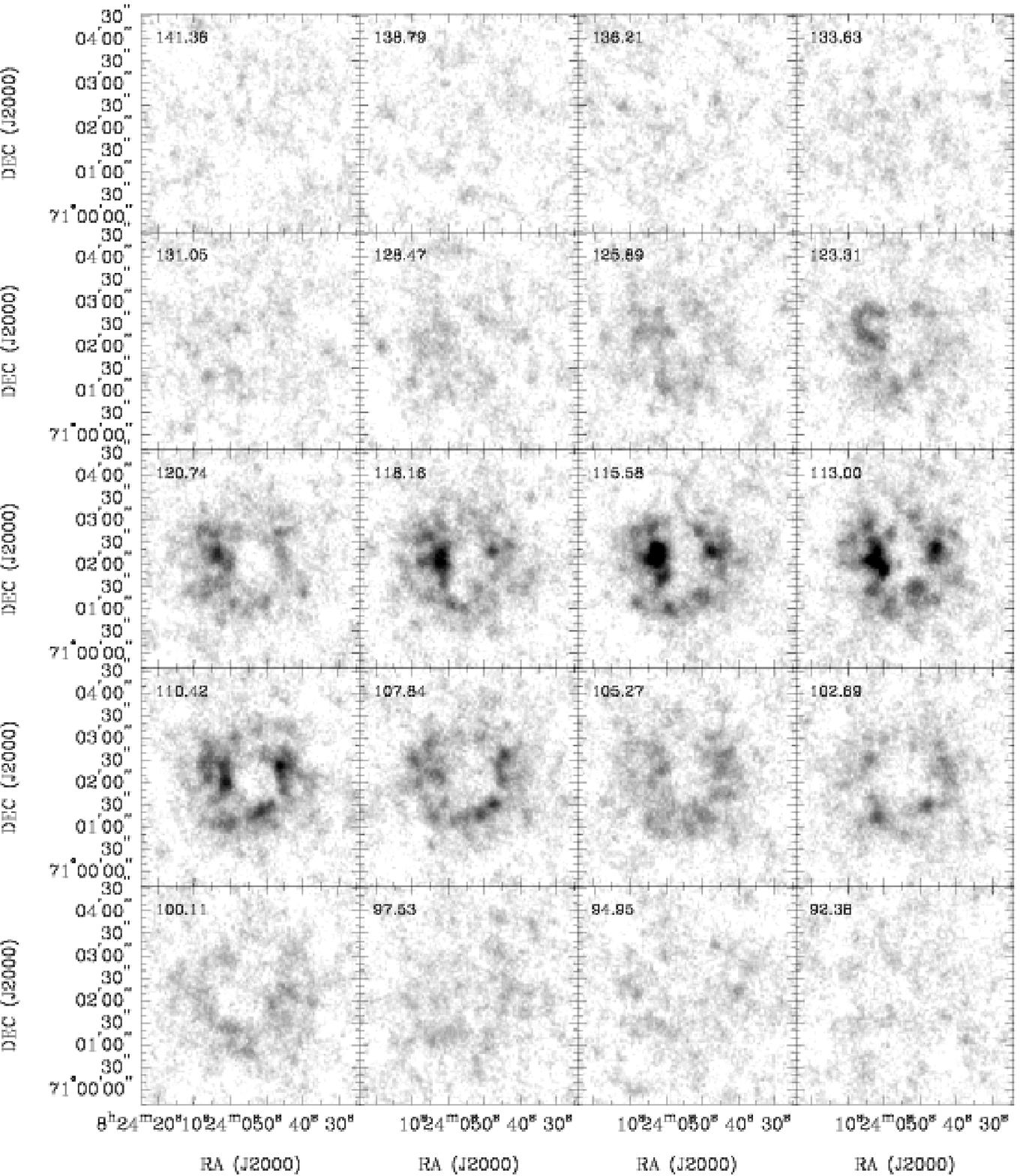}
\caption{{\bf M\,81 dwarf A:} Channel maps based on the NA cube
  (greyscale range: --0.2 to 8 mJy\,beam$^{-1}$).  Every second channel
  is shown (channel width:~1.3\,km\,s$^{-1}$). The area shown in each
  panel is identical to the area shown on the next figure.}
\end{figure}

\clearpage
\begin{figure}
\vspace{0cm}
\epsscale{1.1}
\plotone{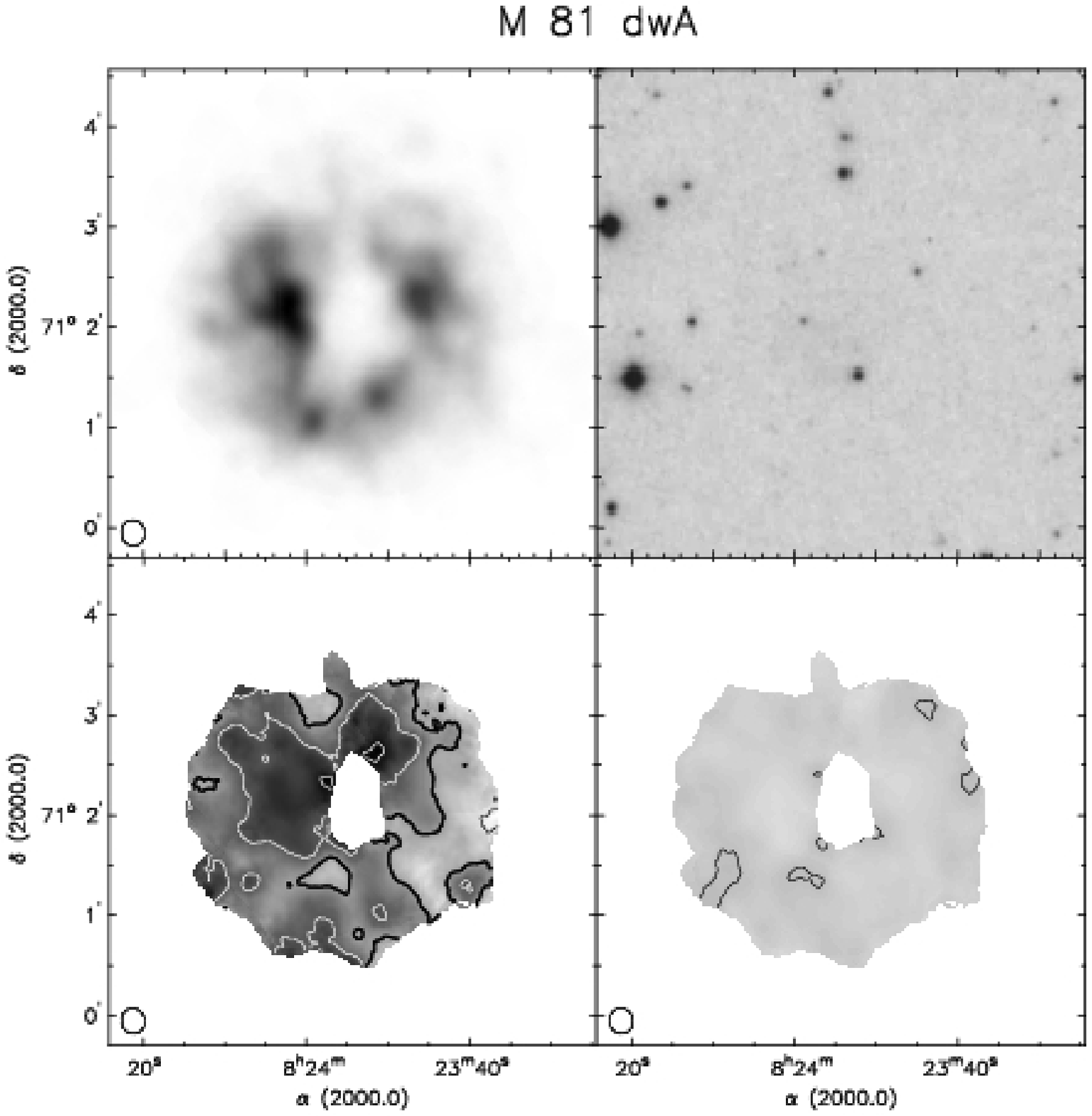}
\vspace{-2.5cm}
\caption{{\bf M\,81 dwarf A}. {\em Top left:} integrated \hi\ map
  (moment 0). Greyscale range from 0--125 Jy\,km\,s$^{-1}$. {\em Top
    right:} Optical image from the digitized sky survey (DSS). {\em
    Bottom left:} Velocity field (moment 1). Black contours (lighter
  greyscale) indicate approaching emission, white contours (darker
  greyscale) receding emission. The thick black contour is the
  systemic velocity ($v_{\rm sys}$=110.76 \,km\,s$^{-1}$), the
  iso--velocity contours are spaced by $\Delta\,v$=3.1\,km\,s$^{-1}$.
  {\em Bottom right:} Velocity dispersion map (moment 2). A contour is plotted
  at 5.0\,km\,s$^{-1}$.}
\end{figure}

%
% DDO\,53
%

\clearpage
\begin{figure}
\epsscale{1.0}
\plotone{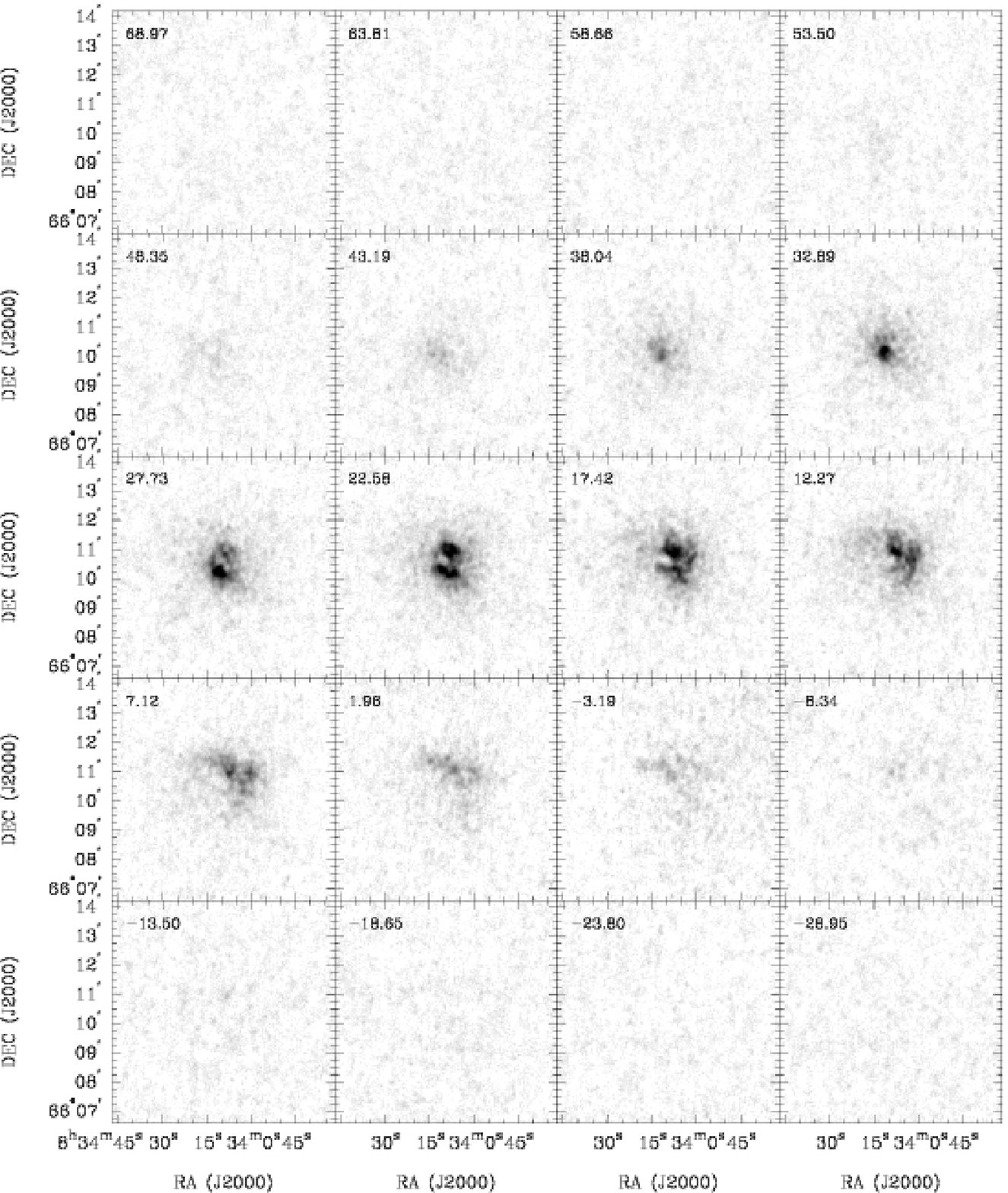}
\caption{{\bf DDO~53:} Channel maps based on the NA cube (greyscale
  range: --0.02 to 12 mJy\,beam$^{-1}$).  Every second channel is
  shown (channel width: 2.6\,km\,s$^{-1}$). The area shown in each
  panel is identical to the area shown on the next figure}
\end{figure}

\clearpage
\begin{figure}
\vspace{0cm}  \epsscale{1.1}
\plotone{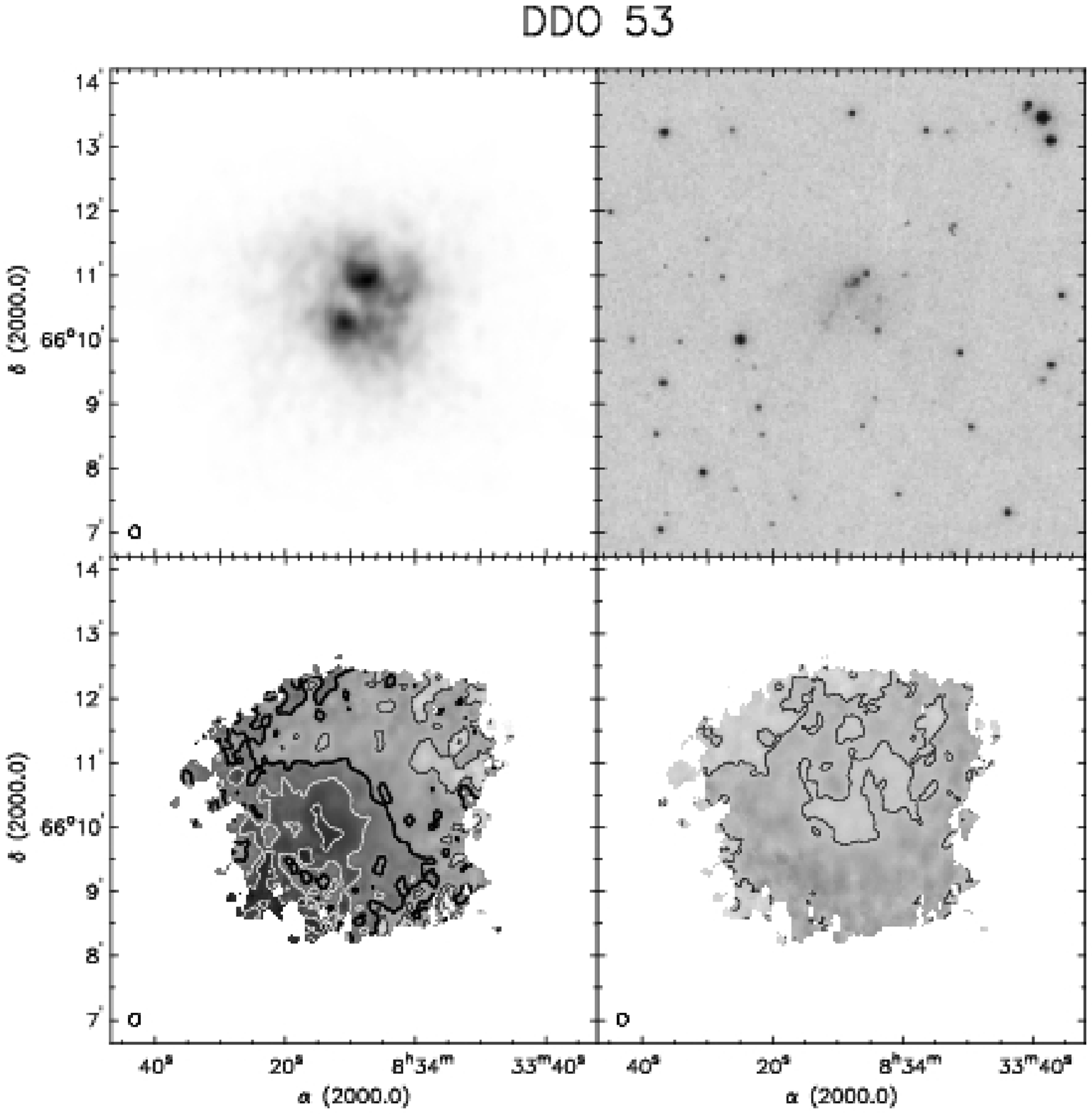}
\vspace{-2.5cm}
\caption{{\bf DDO~53}. {\em Top left:} integrated \hi\ map (moment 0).
  Greyscale range from 0--338 Jy\,km\,s$^{-1}$. {\em Top right:}
  Optical image from the digitized sky survey (DSS). {\em Bottom
    left:} Velocity field (moment 1). Black contours (lighter
  greyscale) indicate approaching emission, white contours (darker
  greyscale) receding emission. The thick black contour is the
  systemic velocity ($v_{\rm sys}$=17.71 \,km\,s$^{-1}$), the
  iso--velocity contours are spaced by $\Delta\,v$=6.25\,km\,s$^{-1}$.
  {\em Bottom right:} Velocity dispersion map (moment 2). A contour is plotted
  at 5.0\,km\,s$^{-1}$.}
\end{figure}

%
% NGC\,2841
%

\clearpage
\begin{figure}
\epsscale{1.0}
\plotone{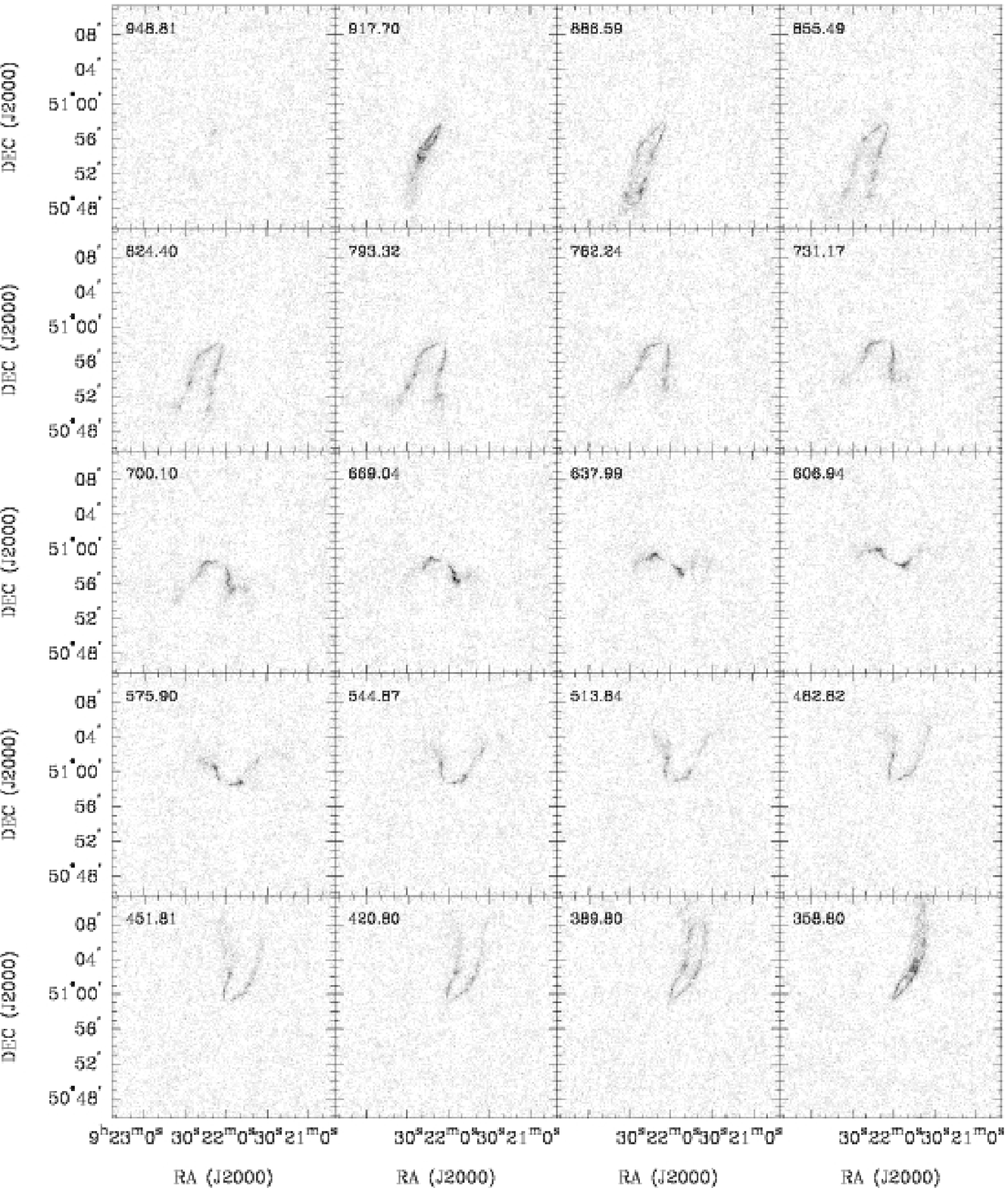}
\caption{{\bf NGC~2841:} Channel maps based on the NA cube (greyscale
  range: --0.02 to 7 mJy\,beam$^{-1}$).  Every sixth channel is shown
  (channel width: 5.2\,km\,s$^{-1}$). The area shown in each panel is
  identical to the area shown on the next figure}
\end{figure}

\clearpage
\begin{figure}
\vspace{0cm}  \epsscale{1.1}
\plotone{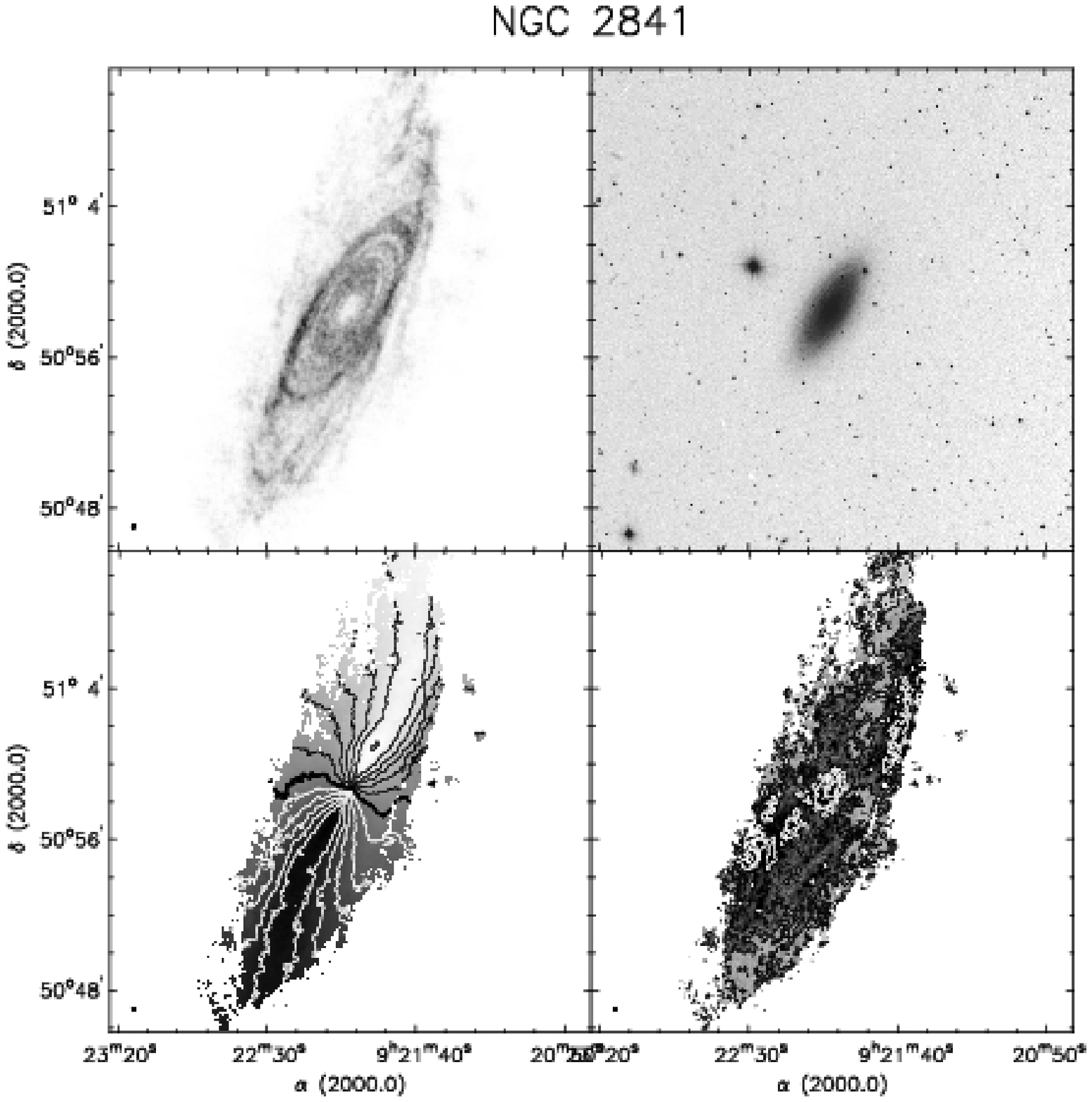}
\vspace{-2.5cm}
\caption{{\bf NGC~2841}. {\em Top left:} integrated \hi\ map (moment 0).
  Greyscale range from 0--203 Jy\,km\,s$^{-1}$. {\em Top right:}
  Optical image from the digitized sky survey (DSS). {\em Bottom
    left:} Velocity field (moment 1). Black contours (lighter
  greyscale) indicate approaching emission, white contours (darker
  greyscale) receding emission. The thick black contour is the
  systemic velocity ($v_{\rm sys}$=635.2 \,km\,s$^{-1}$), the
  iso--velocity contours are spaced by $\Delta\,v$=50\,km\,s$^{-1}$.
  {\em Bottom right:} Velocity dispersion map (moment 2). Contours are plotted
  at 5, 10 and 20\,km\,s$^{-1}$ (white contours: 50 and
  100\,km\,s$^{-1}$).}
\end{figure}

%
% NGC\,2903
%

\clearpage
\begin{figure}
\epsscale{1.0}
\plotone{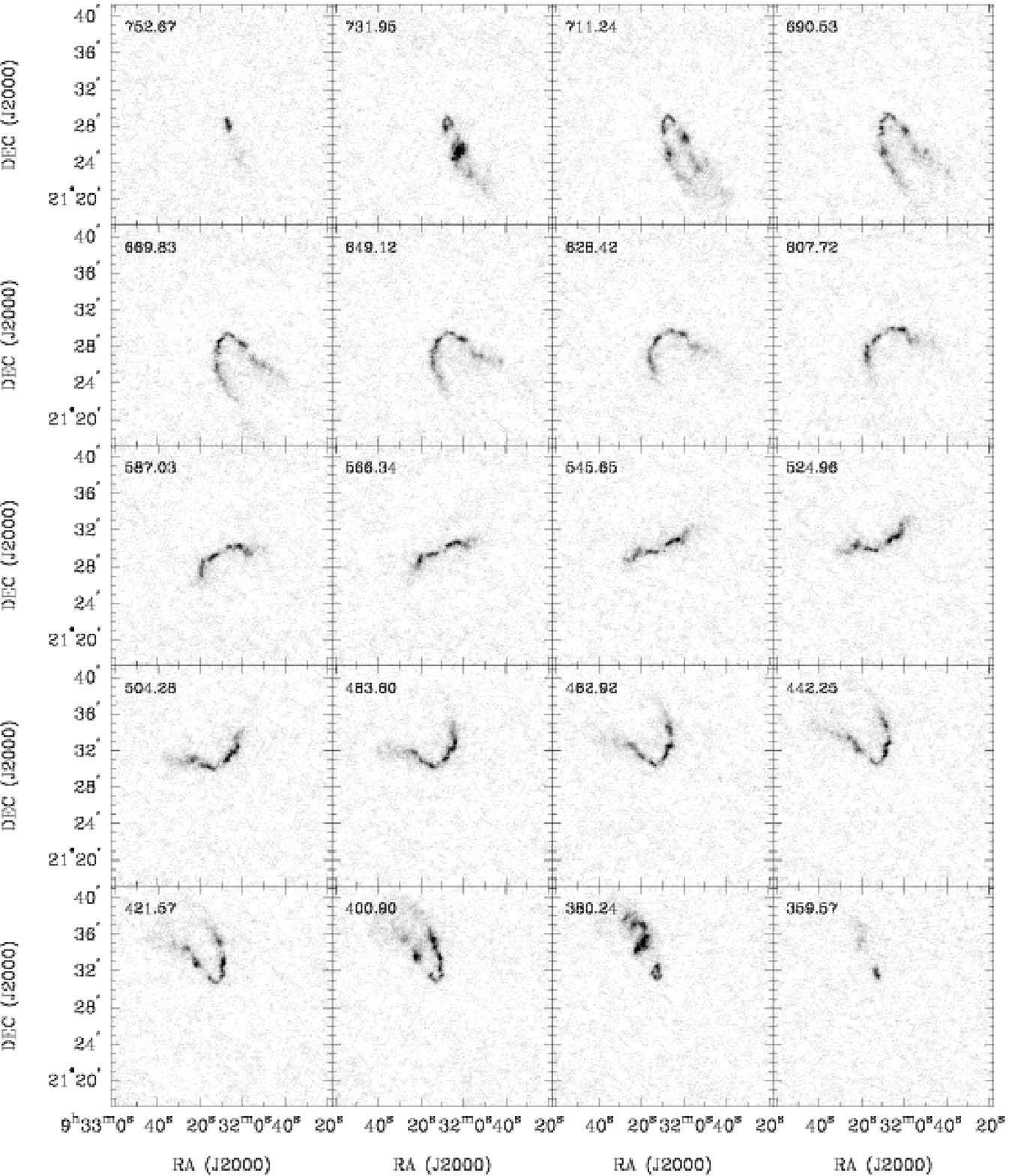}
\caption{{\bf NGC~2903:} Channel maps based on the NA cube (greyscale
  range: --0.02 to 10 mJy\,beam$^{-1}$).  Every fourth channel is shown
  (channel width: 5.2\,km\,s$^{-1}$). The area shown in each panel is
  identical to the area shown on the next figure}
\end{figure}

\clearpage
\begin{figure}
\vspace{0cm}  \epsscale{1.1}
\plotone{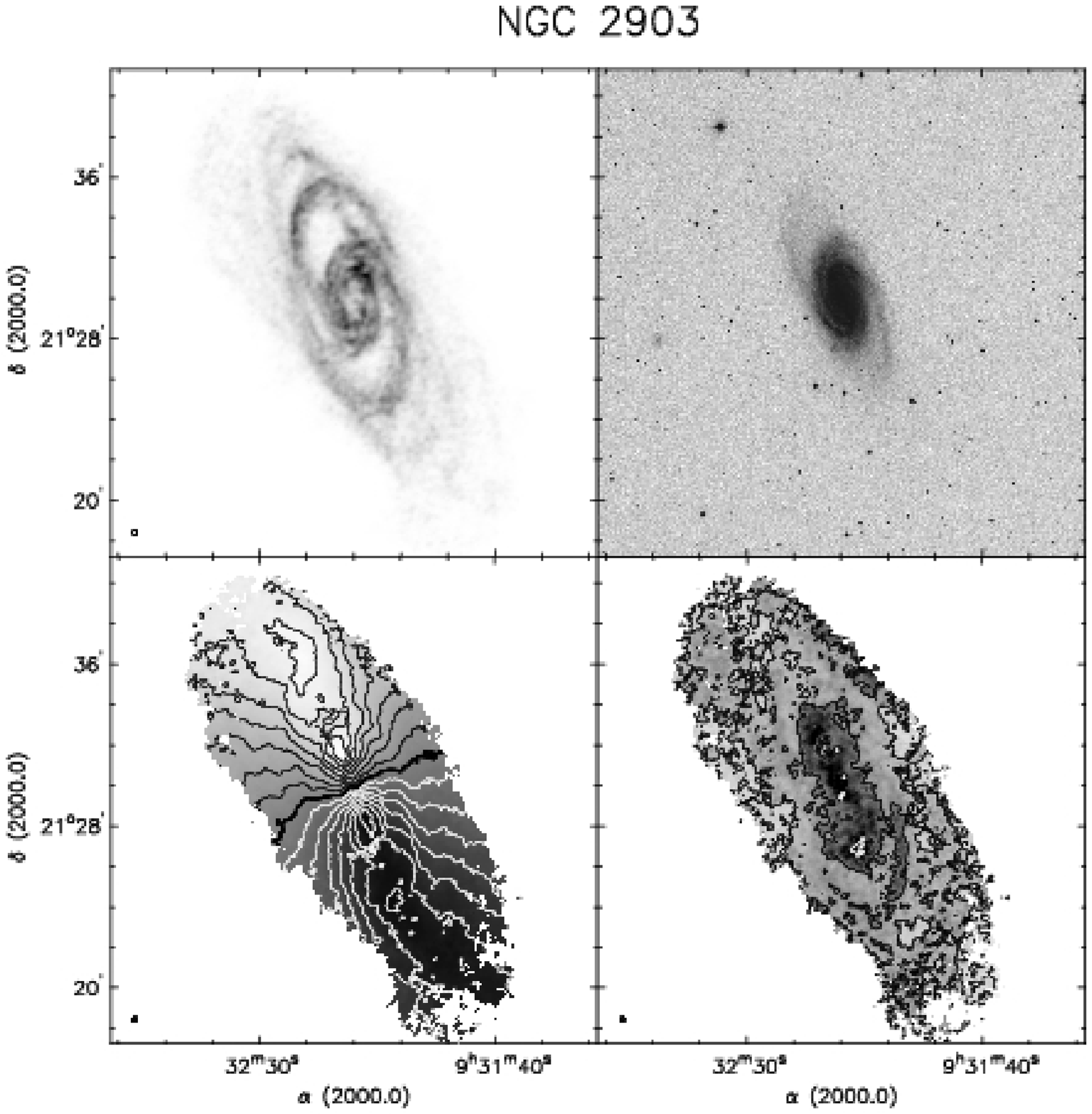}
\vspace{-2.5cm}
\caption{{\bf NGC~2903}. {\em Top left:} integrated \hi\ map (moment 0).
  Greyscale range from 0--577 Jy\,km\,s$^{-1}$. {\em Top right:}
  Optical image from the digitized sky survey (DSS). {\em Bottom
    left:} Velocity field (moment 1). Black contours (lighter
  greyscale) indicate approaching emission, white contours (darker
  greyscale) receding emission. The thick black contour is the
  systemic velocity ($v_{\rm sys}$=2.6 \,km\,s$^{-1}$), the
  iso--velocity contours are spaced by $\Delta\,v$=25\,km\,s$^{-1}$.
  {\em Bottom right:} Velocity dispersion map (moment 2). Contours are plotted
  at 5, 10 and 20\,km\,s$^{-1}$ (white contour: 50\,km\,s$^{-1}$).}
\end{figure}

%
% Holmberg I
%

\clearpage
\begin{figure}
  \epsscale{1.0} \plotone{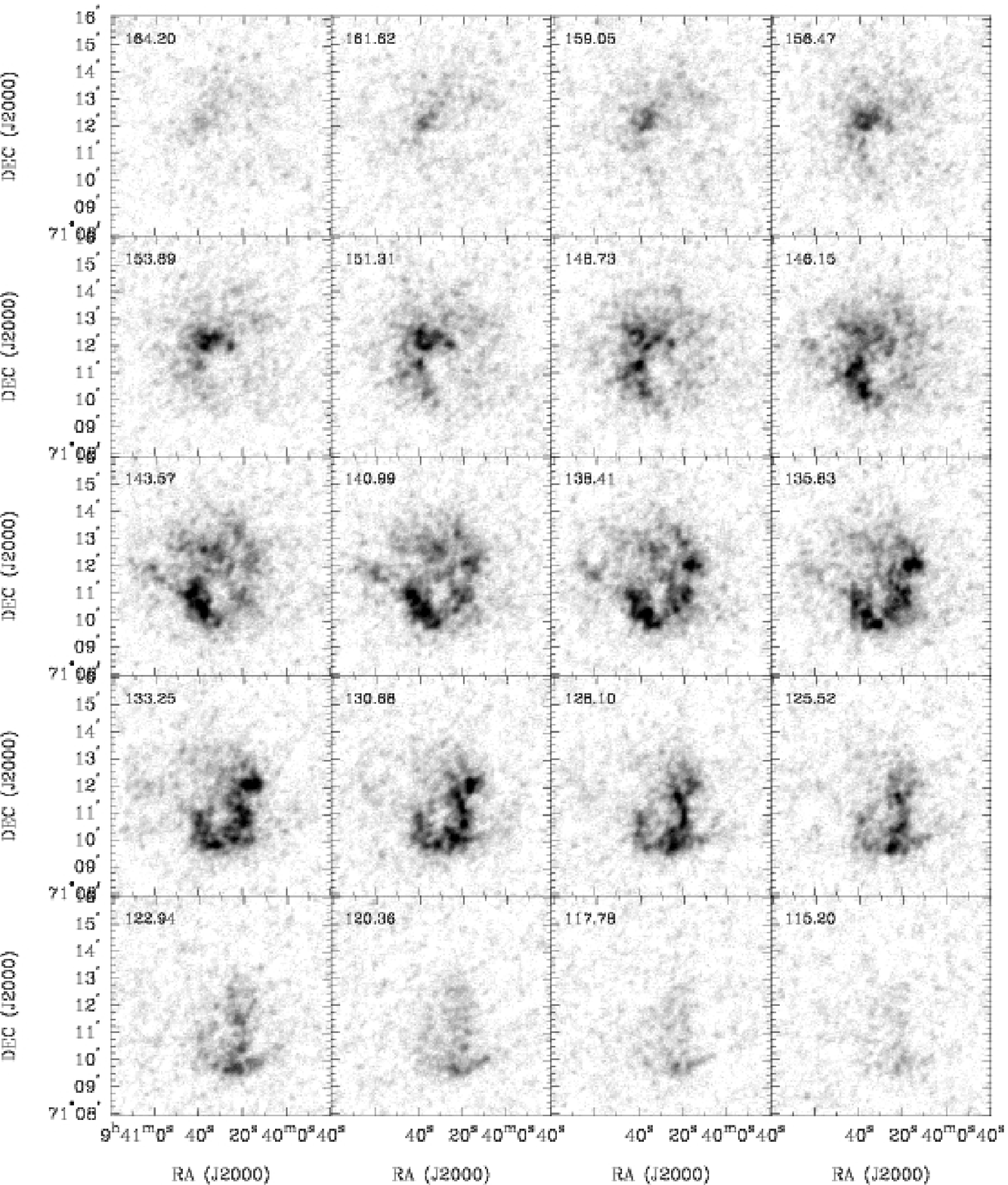}
  \caption{{\bf Holmberg~I:} Channel maps based on the NA cube
    (greyscale range: --0.02 to 15 mJy\,beam$^{-1}$).  Every channel is
    shown (channel width: 2.6\,km\,s$^{-1}$). The area shown in each
    panel is identical to the area shown on the next figure}
\end{figure}

\clearpage
\begin{figure}
\vspace{0cm}  \epsscale{1.1}
\plotone{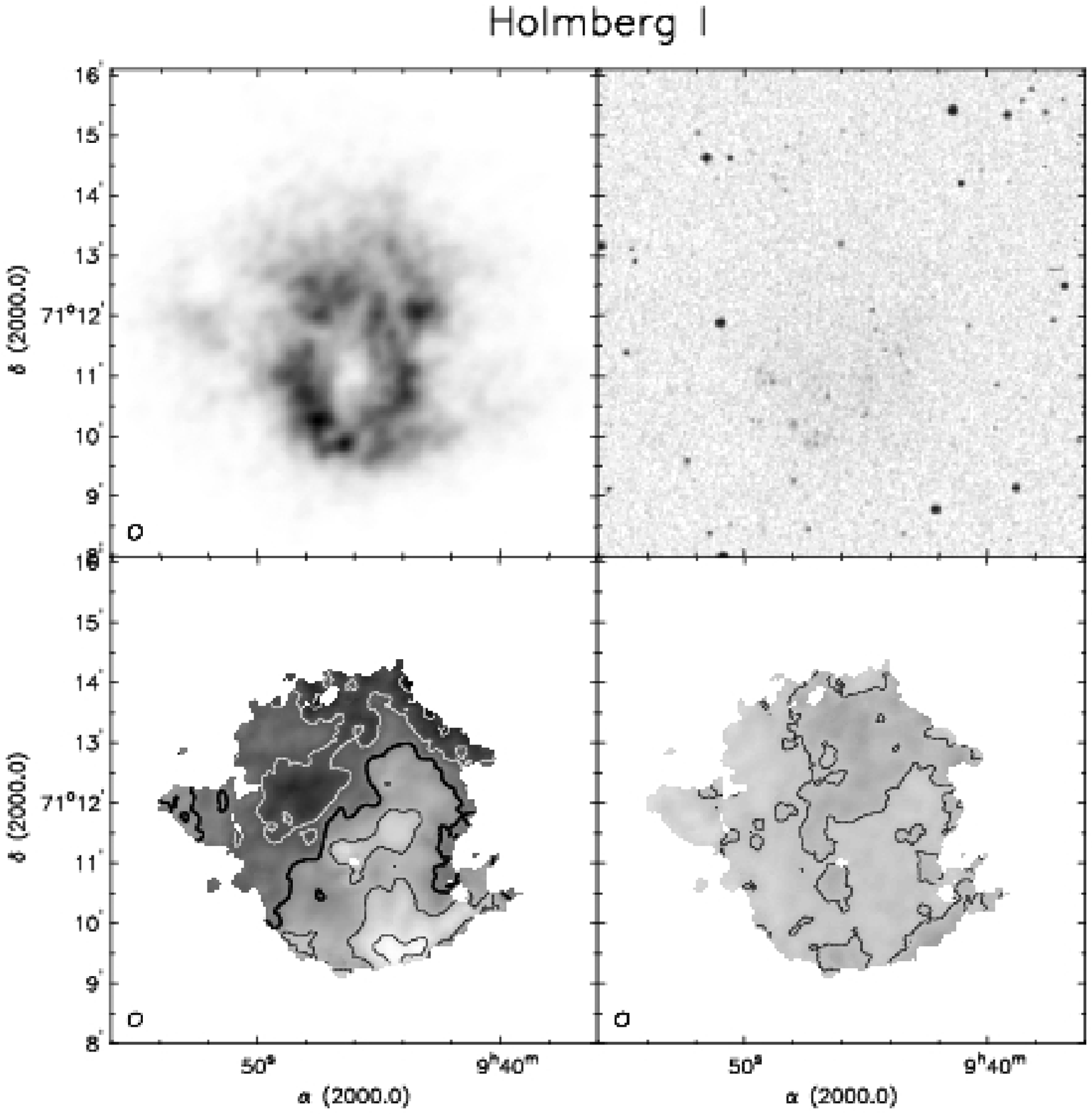}
\vspace{-2.5cm}
\caption{{\bf Holmberg~I}. {\em Top left:} integrated \hi\ map (moment
  0). Greyscale range from 0--382 Jy\,km\,s$^{-1}$. {\em Top right:}
  Optical image from the digitized sky survey (DSS). {\em Bottom
    left:} Velocity field (moment 1). Black contours (lighter
  greyscale) indicate approaching emission, white contours (darker
  greyscale) receding emission. The thick black contour is the
  systemic velocity ($v_{\rm sys}$=140.39 \,km\,s$^{-1}$), the
  iso--velocity contours are spaced by $\Delta\,v$=6.25\,km\,s$^{-1}$.
  {\em Bottom right:} Velocity dispersion map (moment 2). A contour is plotted
  at 5.0\,km\,s$^{-1}$.}
\end{figure}

%
% NGC\,2976
%

\clearpage
\begin{figure}
\epsscale{1.0}
\plotone{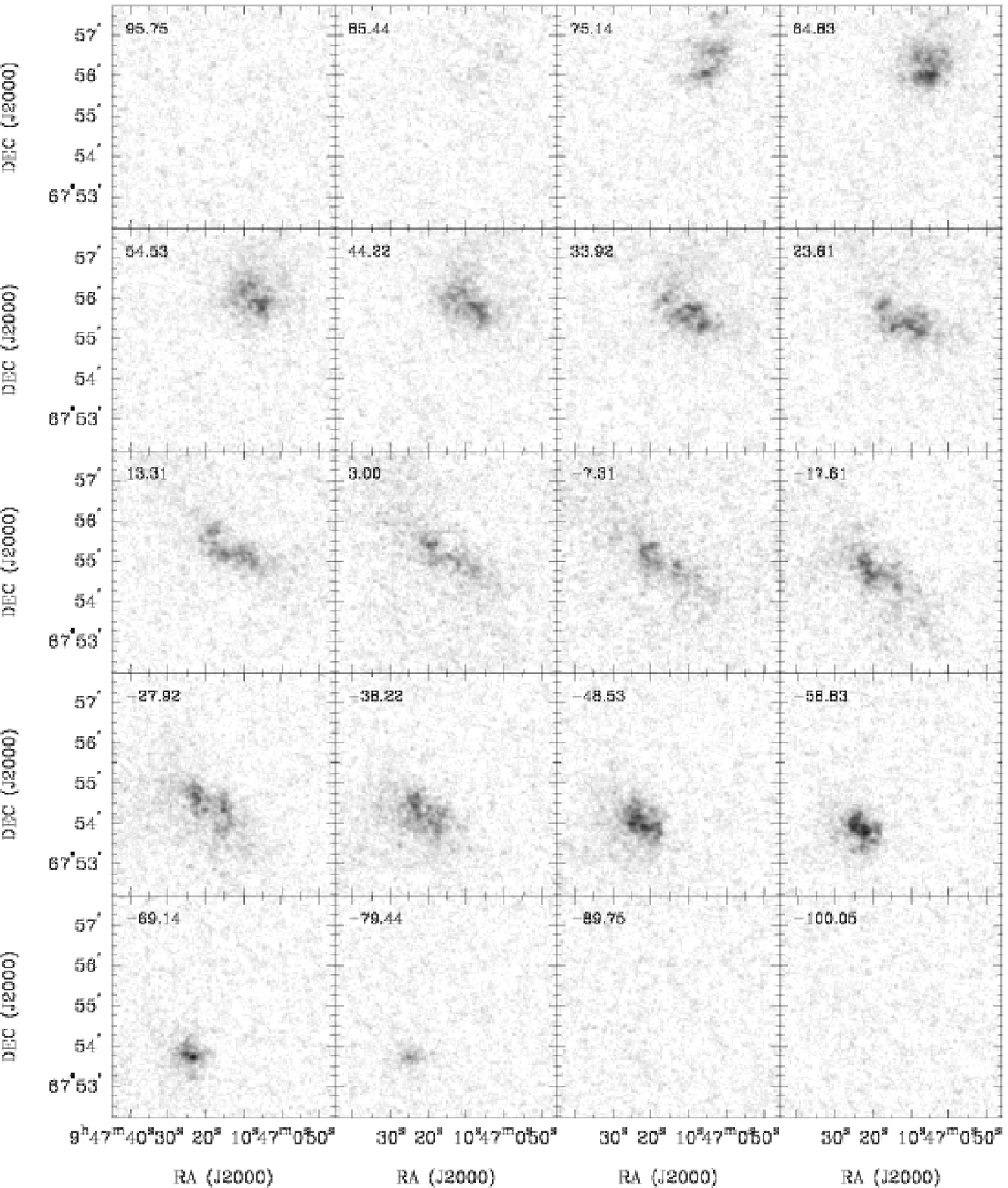}
\caption{{\bf NGC~2976:} Channel maps based on the NA cube (greyscale
  range: --0.02 to 8 mJy\,beam$^{-1}$).  Every second channel is
  shown (channel width: 5.2\,km\,s$^{-1}$). The area shown in each
  panel is identical to the area shown on the next figure}
\end{figure}

\clearpage
\begin{figure}
\vspace{0cm}  \epsscale{1.1}
\plotone{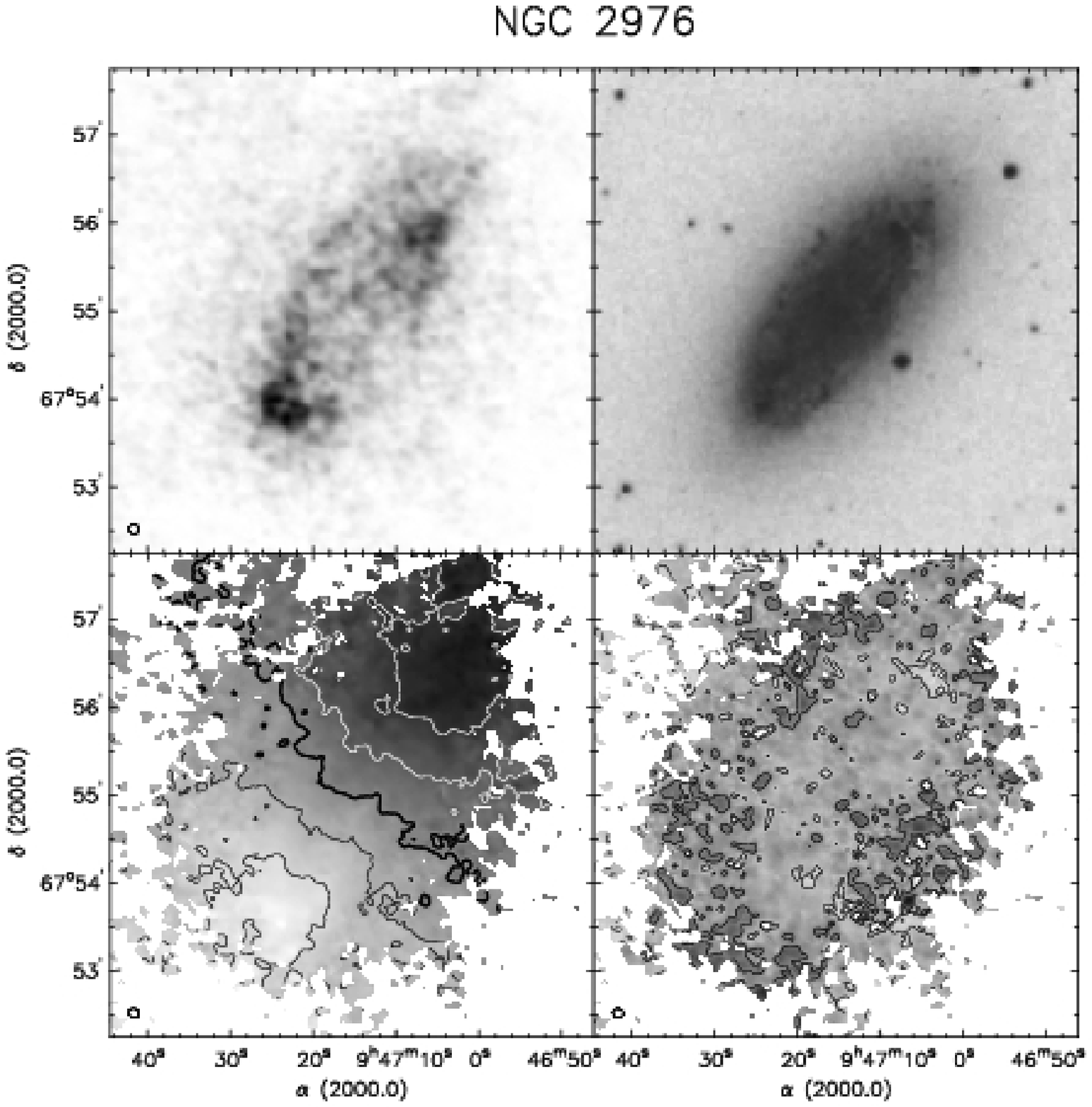}
\vspace{-2.5cm}
\caption{{\bf NGC~2976}. {\em Top left:} integrated \hi\ map (moment 0).
  Greyscale range from 0--192 Jy\,km\,s$^{-1}$. {\em Top right:}
  Optical image from the digitized sky survey (DSS). {\em Bottom
    left:} Velocity field (moment 1). Black contours (lighter
  greyscale) indicate approaching emission, white contours (darker
  greyscale) receding emission. The thick black contour is the
  systemic velocity ($v_{\rm sys}$=2.60 \,km\,s$^{-1}$), the
  iso--velocity contours are spaced by $\Delta\,v$=25\,km\,s$^{-1}$.
  {\em Bottom right:} Velocity dispersion map (moment 2). Contours are plotted
  at 5, 10\,km\,s$^{-1}$.}
\end{figure}

%
% NGC\,3031
%

\clearpage
\begin{figure}
\epsscale{1.0}
\plotone{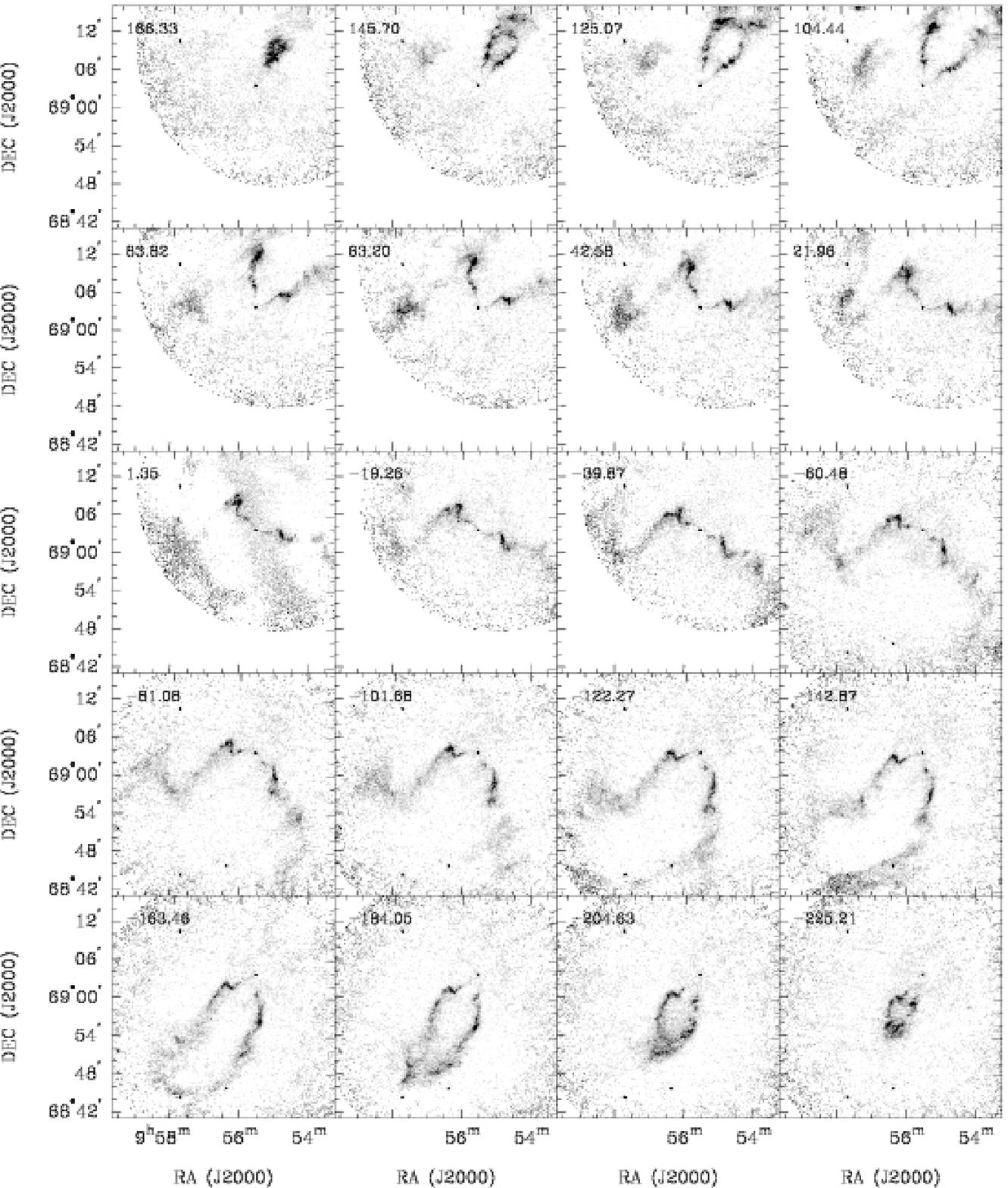}
\caption{{\bf NGC~3031:} Channel maps based on the NA cube (greyscale
  range: --0.02 to 10 mJy\,beam$^{-1}$).  Every eigth channel is shown
  (channel width: 2.6\,km\,s$^{-1}$). The area shown in each panel is
  identical to the area shown on the next figure}
\end{figure}

\clearpage
\begin{figure}
\vspace{0cm}  \epsscale{1.1}
\plotone{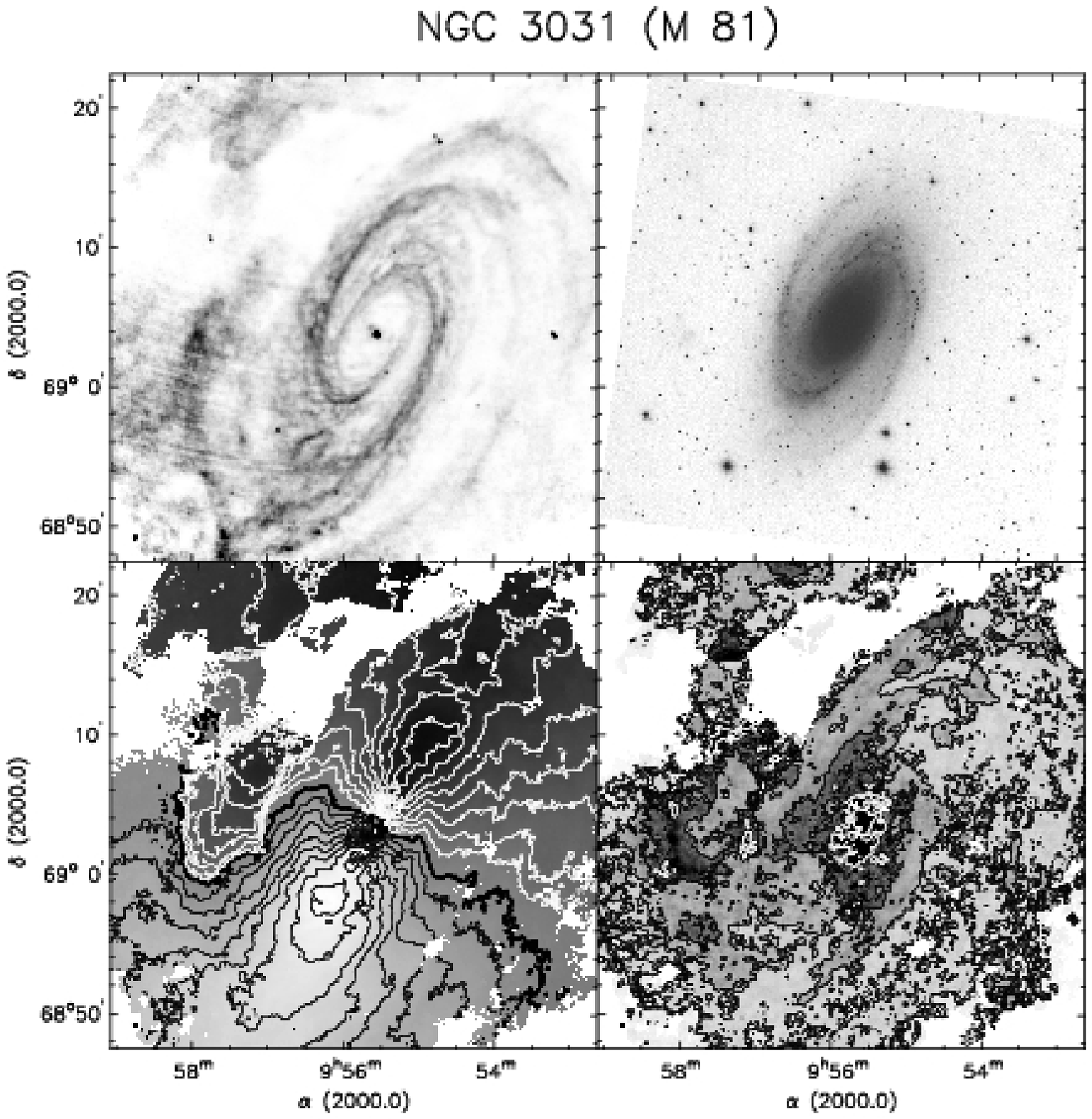}
\vspace{-2.5cm}
\caption{{\bf NGC~3031}. {\em Top left:} integrated \hi\ map (moment 0).
  Greyscale range from 0--700 Jy\,km\,s$^{-1}$. Note that the
  continuum (point sources) have not been subracted. {\em Top right:}
  Optical image from the digitized sky survey (DSS). {\em Bottom
    left:} Velocity field (moment 1). Black contours (lighter
  greyscale) indicate approaching emission, white contours (darker
  greyscale) receding emission. The thick black contour is the
  systemic velocity ($v_{\rm sys}$=--39.4 \,km\,s$^{-1}$), the
  iso--velocity contours are spaced by $\Delta\,v$=25\,km\,s$^{-1}$.
  {\em Bottom right:} Velocity dispersion map (moment 2). Contours are plotted
  at 5, 10 and 20 \,km\,s$^{-1}$ (white contours: 50 and 100
  \,km\,s$^{-1}$).}
\end{figure}

%
% NGC\,3077
%

\clearpage
\begin{figure}
\epsscale{1.0}
\plotone{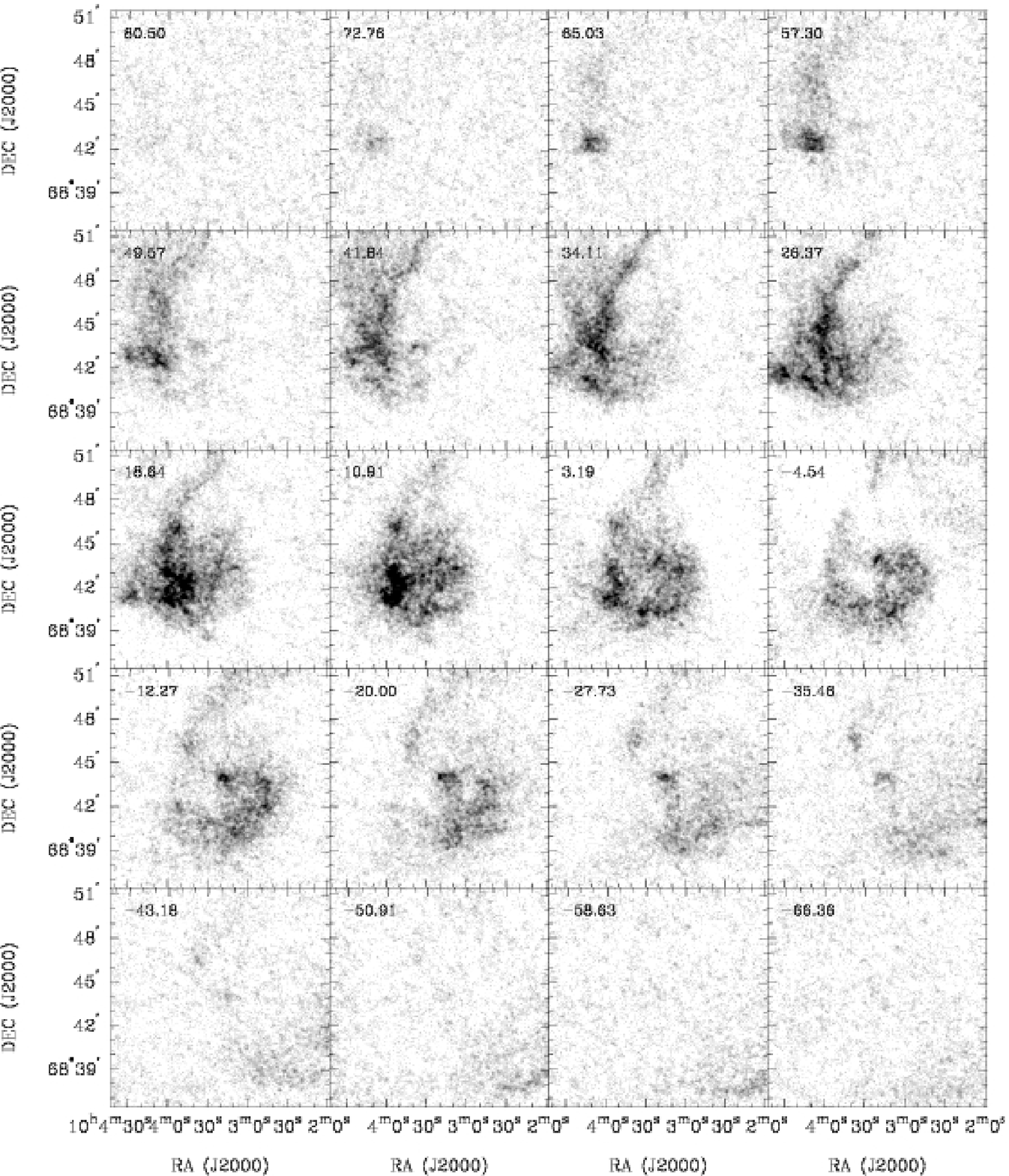}
\caption{{\bf NGC~3077:} Channel maps based on the NA cube (greyscale
  range: --0.02 to 10 mJy\,beam$^{-1}$).  Every third channel is shown
  (channel width: 2.6\,km\,s$^{-1}$). The area shown in each panel is
  identical to the area shown on the next figure}
\end{figure}

\clearpage
\begin{figure}
\vspace{0cm}  \epsscale{1.1}
\plotone{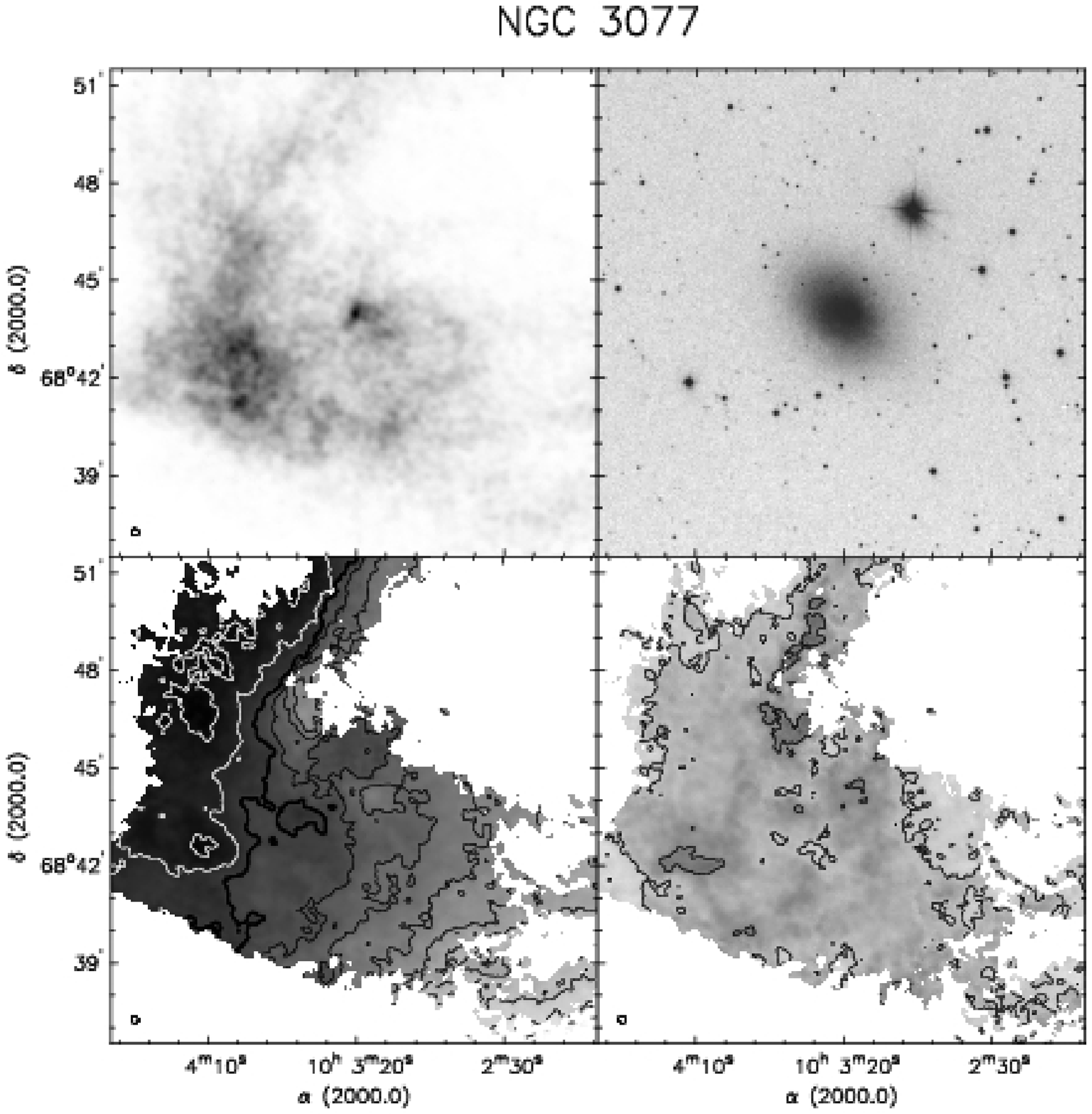}
\vspace{-2.5cm}
\caption{{\bf NGC~3077}. {\em Top left:} integrated \hi\ map (moment 0).
  Greyscale range from 0--529 Jy\,km\,s$^{-1}$. {\em Top right:}
  Optical image from the digitized sky survey (DSS). {\em Bottom
    left:} Velocity field (moment 1). Black contours (lighter
  greyscale) indicate approaching emission, white contours (darker
  greyscale) receding emission. The thick black contour is the
  systemic velocity ($v_{\rm sys}$=19.8 \,km\,s$^{-1}$), the
  iso--velocity contours are spaced by $\Delta\,v$=12.5\,km\,s$^{-1}$.
  {\em Bottom right:} Velocity dispersion map (moment 2). Contours are plotted
  at 5 and 10\,km\,s$^{-1}$.}
\end{figure}

%
% M81 DWARF B
%
\clearpage
\begin{figure}
\epsscale{1.0}
\plotone{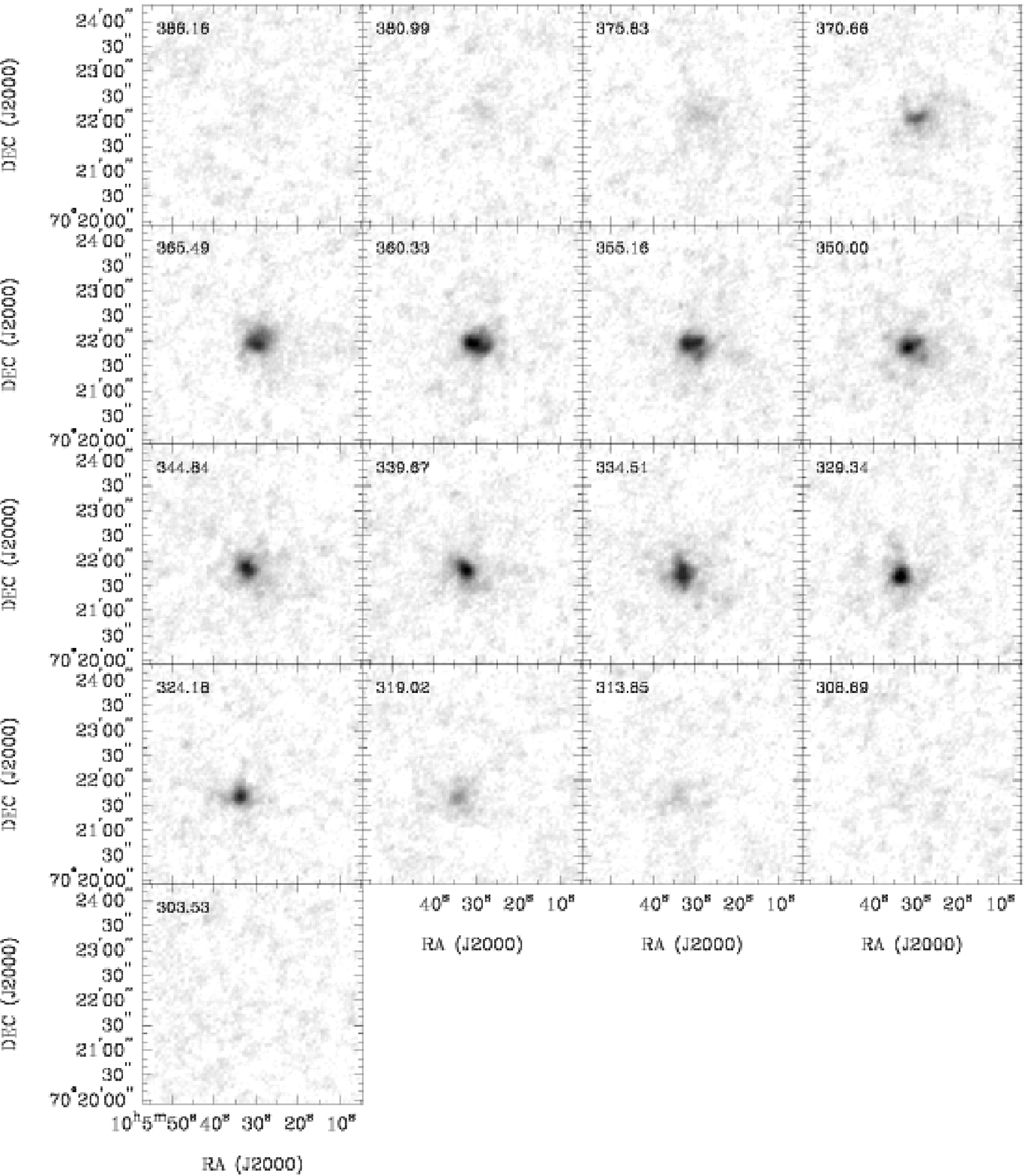}
\caption{{\bf M\,81 dwarf B:} Channel maps based on the NA cube
  (greyscale range: --0.02 to 12 mJy\,beam$^{-1}$).  Every second
  channel is shown (channel width: 2.6\,km\,s$^{-1}$). The area shown
  in each panel is identical to the area shown on the next figure.}
\end{figure}

\clearpage
\begin{figure}
\vspace{0cm}  \epsscale{1.1}
\plotone{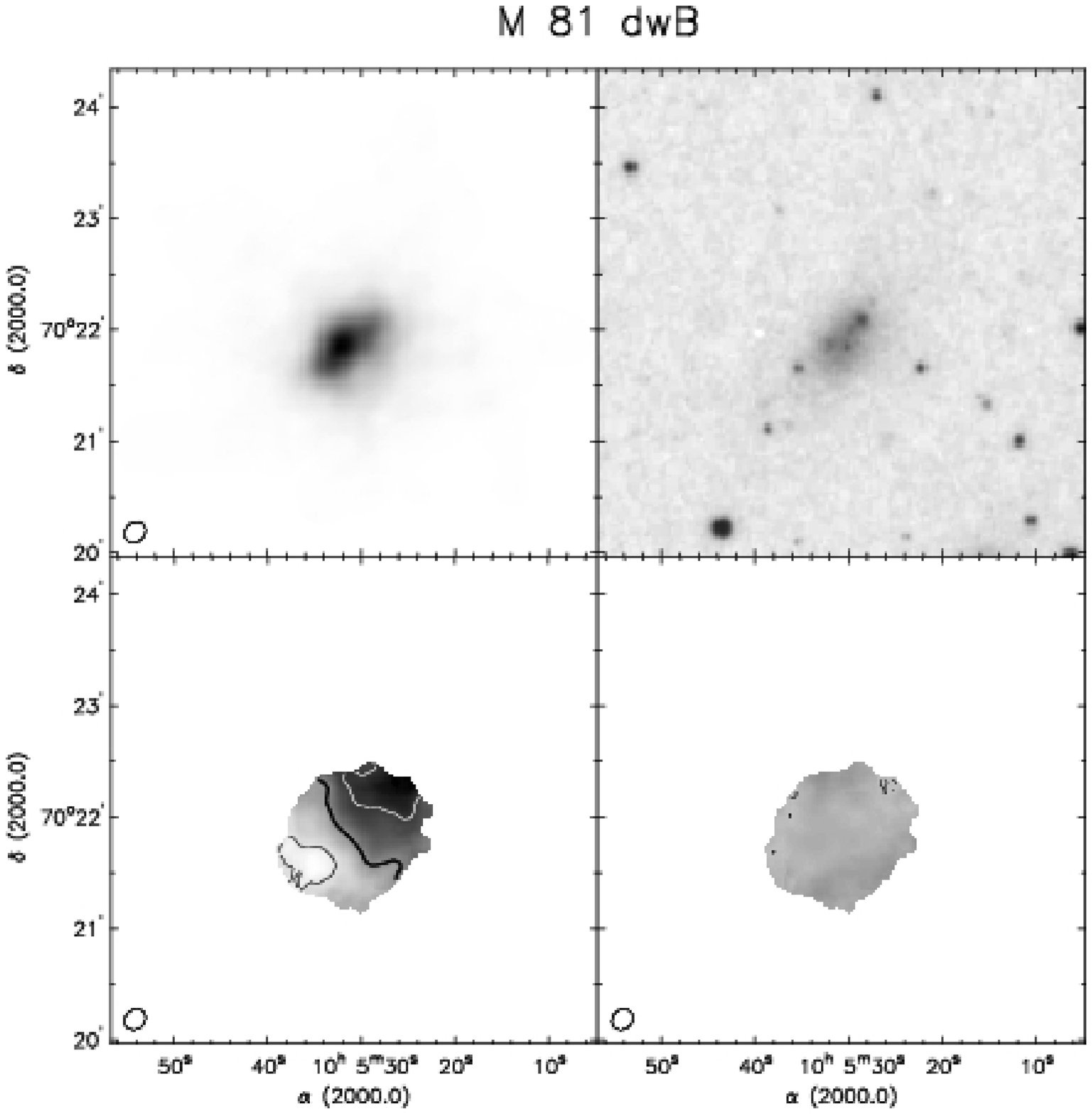}
\vspace{-2.5cm}
\caption{{\bf M\,81 dwarf B}. {\em Top left:} integrated \hi\ map
  (moment 0). Greyscale range from 0--364 Jy\,km\,s$^{-1}$. {\em Top
    right:} Optical image from the digitized sky survey (DSS). {\em
    Bottom left:} Velocity field (moment 1). Black contours (lighter
  greyscale) indicate approaching emission, white contours (darker
  greyscale) receding emission. The thick black contour is the
  systemic velocity ($v_{\rm sys}$=346.41 \,km\,s$^{-1}$), the
  iso--velocity contours are spaced by
  $\Delta\,v$=12.50\,km\,s$^{-1}$. {\em Bottom right:} Velocity dispersion map
  (moment 2). A contours is plotted at 5.0\,km\,s$^{-1}$.}
\end{figure}

%
% NGC\,3184
%

\clearpage
\begin{figure}
\epsscale{1.0}
\plotone{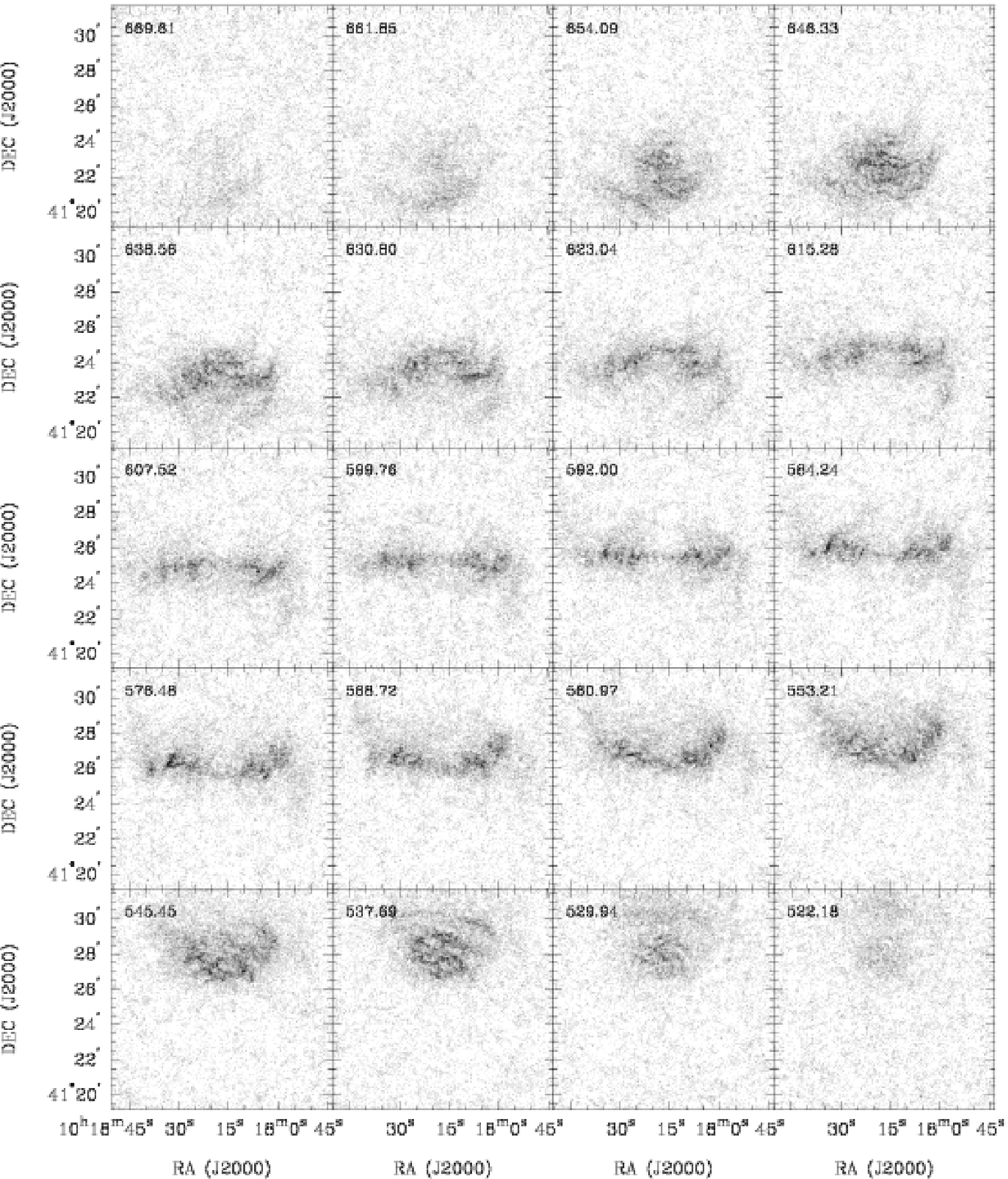}
\caption{{\bf NGC~3184:} Channel maps based on the NA cube (greyscale
  range: --0.02 to 5 mJy\,beam$^{-1}$).  Every third channel is shown
  (channel width: 5.2\,km\,s$^{-1}$). The area shown in each panel is
  identical to the area shown on the next figure}
\end{figure}

\clearpage
\begin{figure}
\vspace{0cm}  \epsscale{1.1}
\plotone{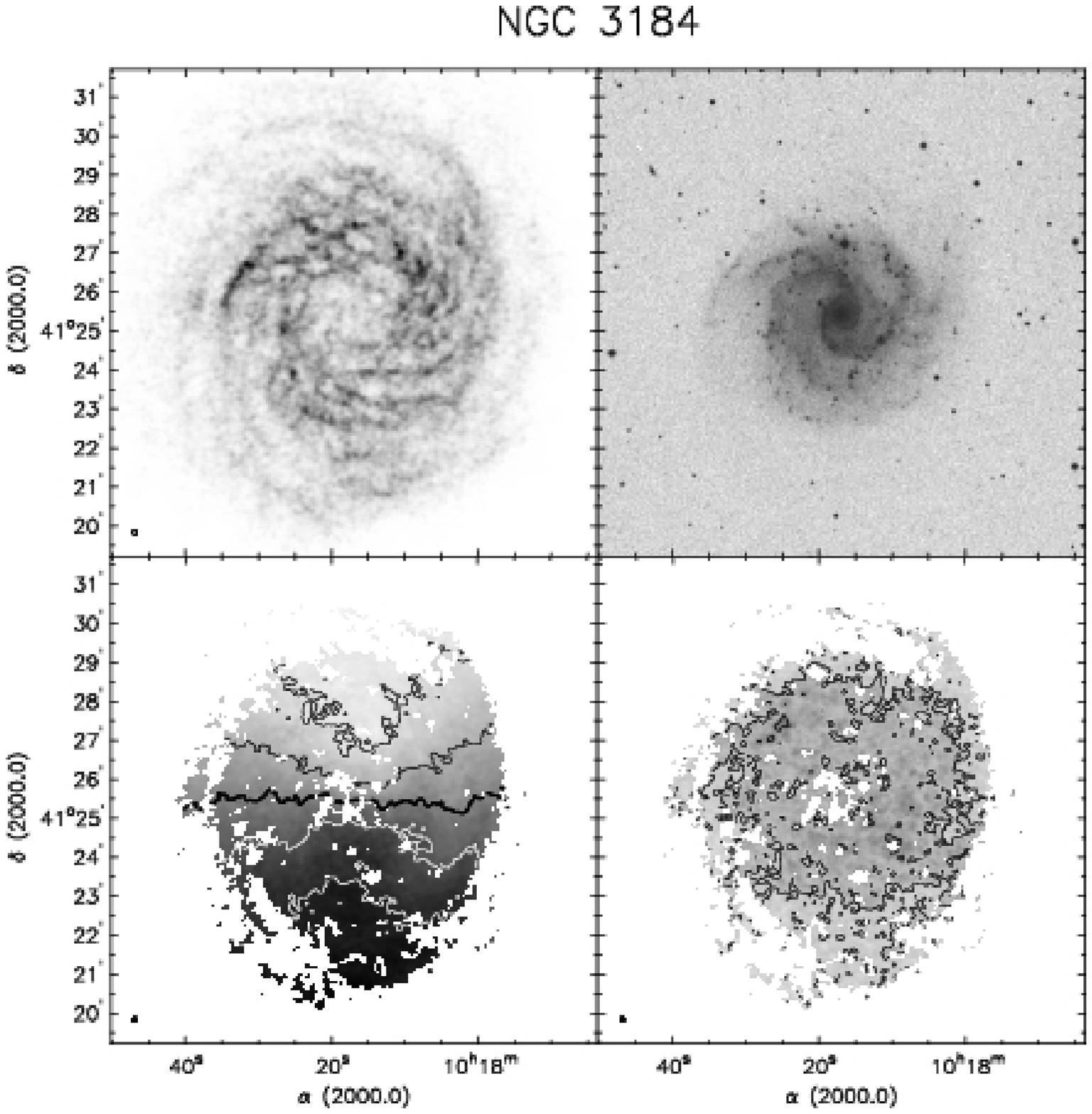}
\vspace{-2.5cm}
\caption{{\bf NGC~3184}. {\em Top left:} integrated \hi\ map (moment 0).
  Greyscale range from 0--95 Jy\,km\,s$^{-1}$. {\em Top right:}
  Optical image from the digitized sky survey (DSS). {\em Bottom
    left:} Velocity field (moment 1). Black contours (lighter
  greyscale) indicate approaching emission, white contours (darker
  greyscale) receding emission. The thick black contour is the
  systemic velocity ($v_{\rm sys}$=593.3 \,km\,s$^{-1}$), the
  iso--velocity contours are spaced by $\Delta\,v$=25\,km\,s$^{-1}$.
  {\em Bottom right:} Velocity dispersion map (moment 2). Contours are plotted
  at 5 and 10\,km\,s$^{-1}$.}
\end{figure}

% NGC\,3198
%

\clearpage
\begin{figure}
\epsscale{1.0}
\plotone{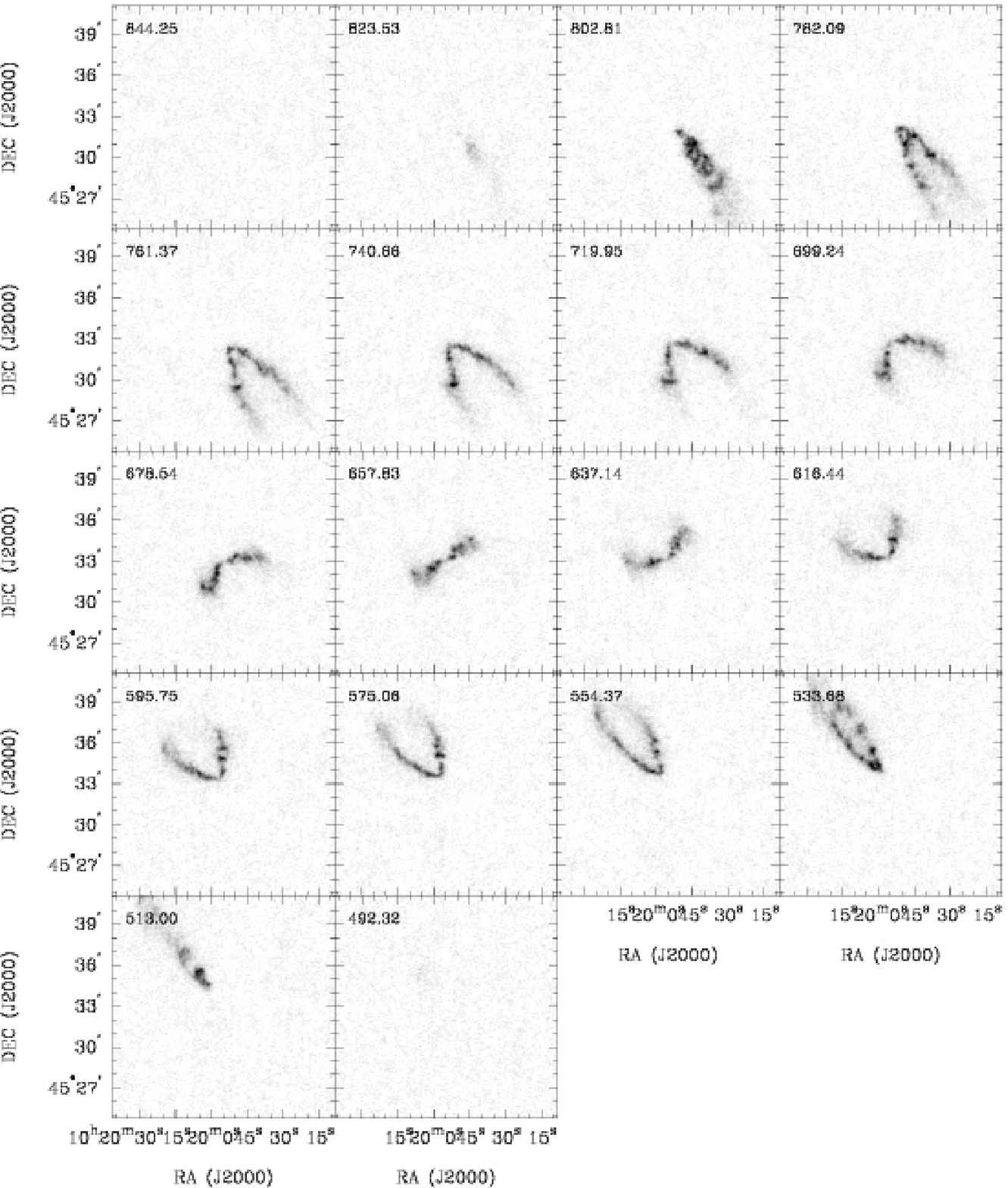}
\caption{{\bf NGC~3198:} Channel maps based on the NA cube (greyscale
  range: --0.02 to 12 mJy\,beam$^{-1}$).  Every fourth channel is
  shown (channel width: 5.2\,km\,s$^{-1}$). The area shown in each
  panel is identical to the area shown on the next figure}
\end{figure}

\clearpage
\begin{figure}
\vspace{0cm}  \epsscale{1.1}
\plotone{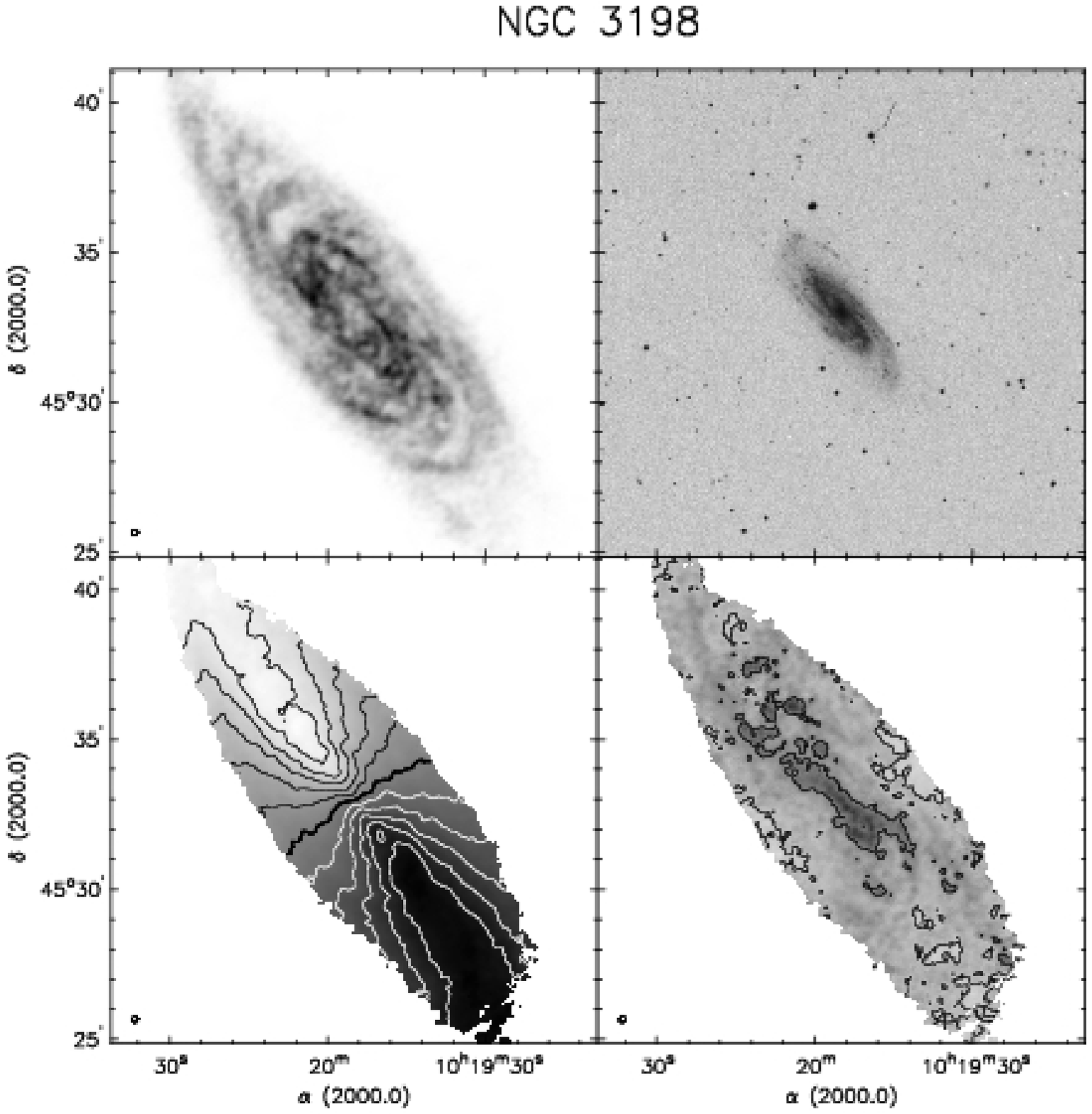}
\vspace{-2.5cm}
\caption{{\bf NGC~3198}. {\em Top left:} integrated \hi\ map (moment 0).
  Greyscale range from 0--421 Jy\,km\,s$^{-1}$. {\em Top right:}
  Optical image from the digitized sky survey (DSS). {\em Bottom
    left:} Velocity field (moment 1). Black contours (lighter
  greyscale) indicate approaching emission, white contours (darker
  greyscale) receding emission. The thick black contour is the
  systemic velocity ($v_{\rm sys}$=661.2 \,km\,s$^{-1}$), the
  iso--velocity contours are spaced by $\Delta\,v$=25\,km\,s$^{-1}$.
  {\em Bottom right:} Velocity dispersion map (moment 2). Contours are plotted
  at 5, 10 and 20\,km\,s$^{-1}$.}
\end{figure}

%
% IC~2574
%

\clearpage
\begin{figure}
\epsscale{1.0}
\plotone{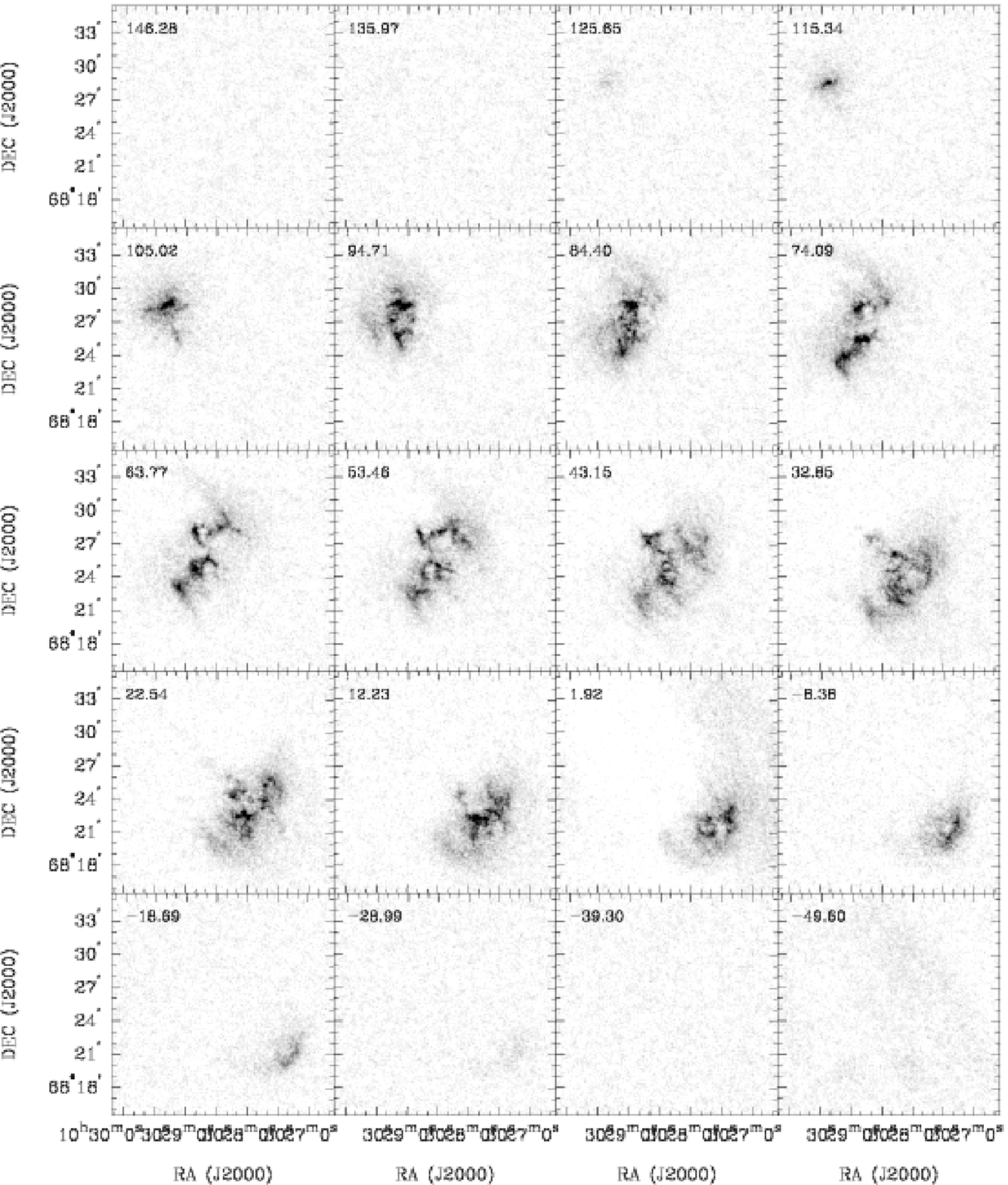}
\caption{{\bf IC~2574:} Channel maps based on the NA cube (greyscale
  range: --0.02 to 15 mJy\,beam$^{-1}$).  Every fourth channel is
  shown (channel width: 2.6\,km\,s$^{-1}$). The area shown in each
  panel is identical to the area shown on the next figure}
\end{figure}

\clearpage
\begin{figure}
\vspace{0cm}  \epsscale{1.1}
\plotone{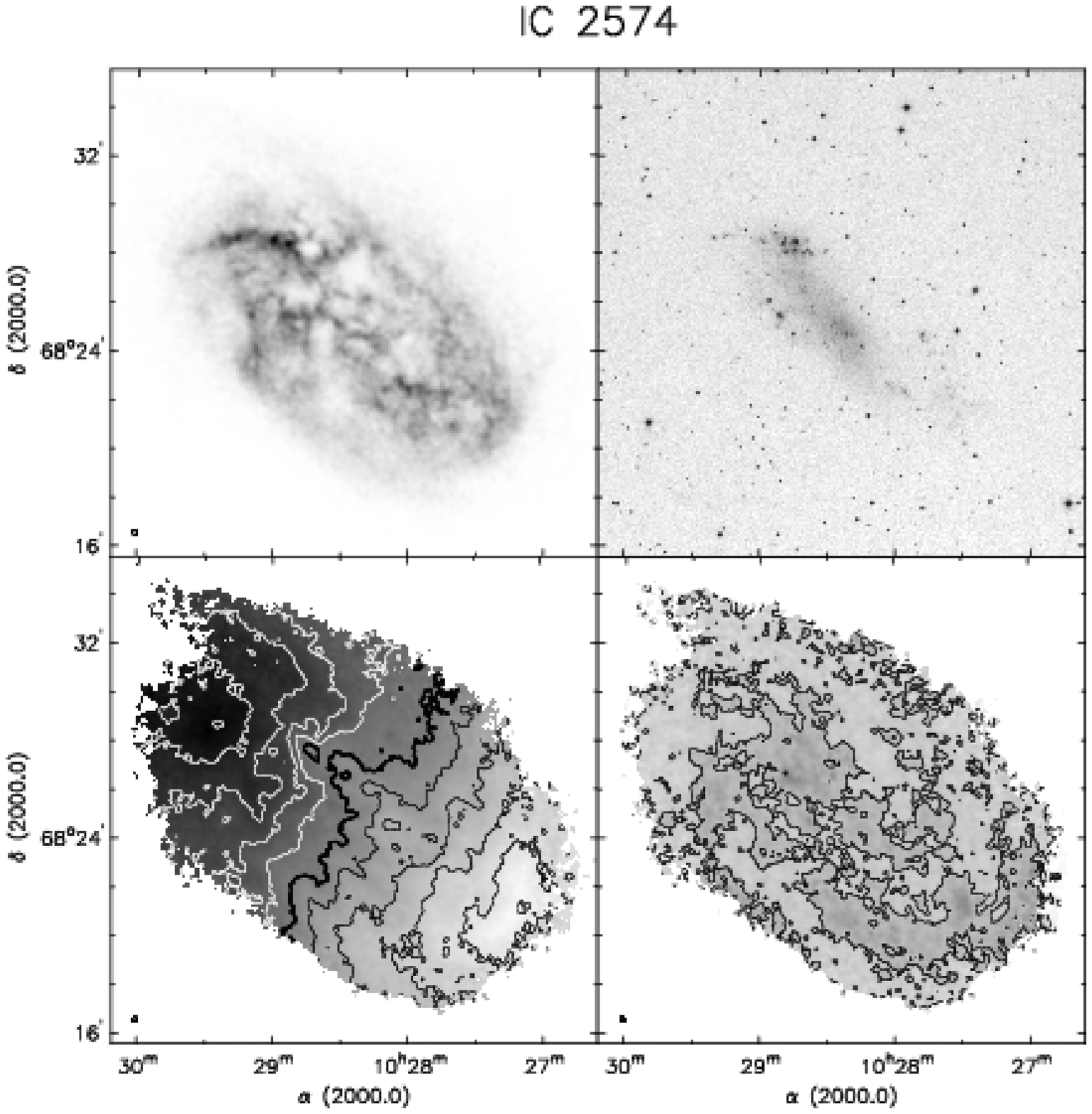}
\vspace{-2.5cm}
\caption{{\bf IC~2574}. {\em Top left:} integrated \hi\ map (moment 0).
  Greyscale range from 0--639 Jy\,km\,s$^{-1}$. {\em Top right:}
  Optical image from the digitized sky survey (DSS). {\em Bottom
    left:} Velocity field (moment 1). Black contours (lighter
  greyscale) indicate approaching emission, white contours (darker
  greyscale) receding emission. The thick black contour is the
  systemic velocity ($v_{\rm sys}$=48.6 \,km\,s$^{-1}$), the
  iso--velocity contours are spaced by $\Delta\,v$=12.5\,km\,s$^{-1}$.
  {\em Bottom right:} Velocity dispersion map (moment 2). Contours are plotted
  at 5 and 10\,km\,s$^{-1}$.}
\end{figure}

%
% NGC\,3351
%

\clearpage
\begin{figure}
\epsscale{1.0}
\plotone{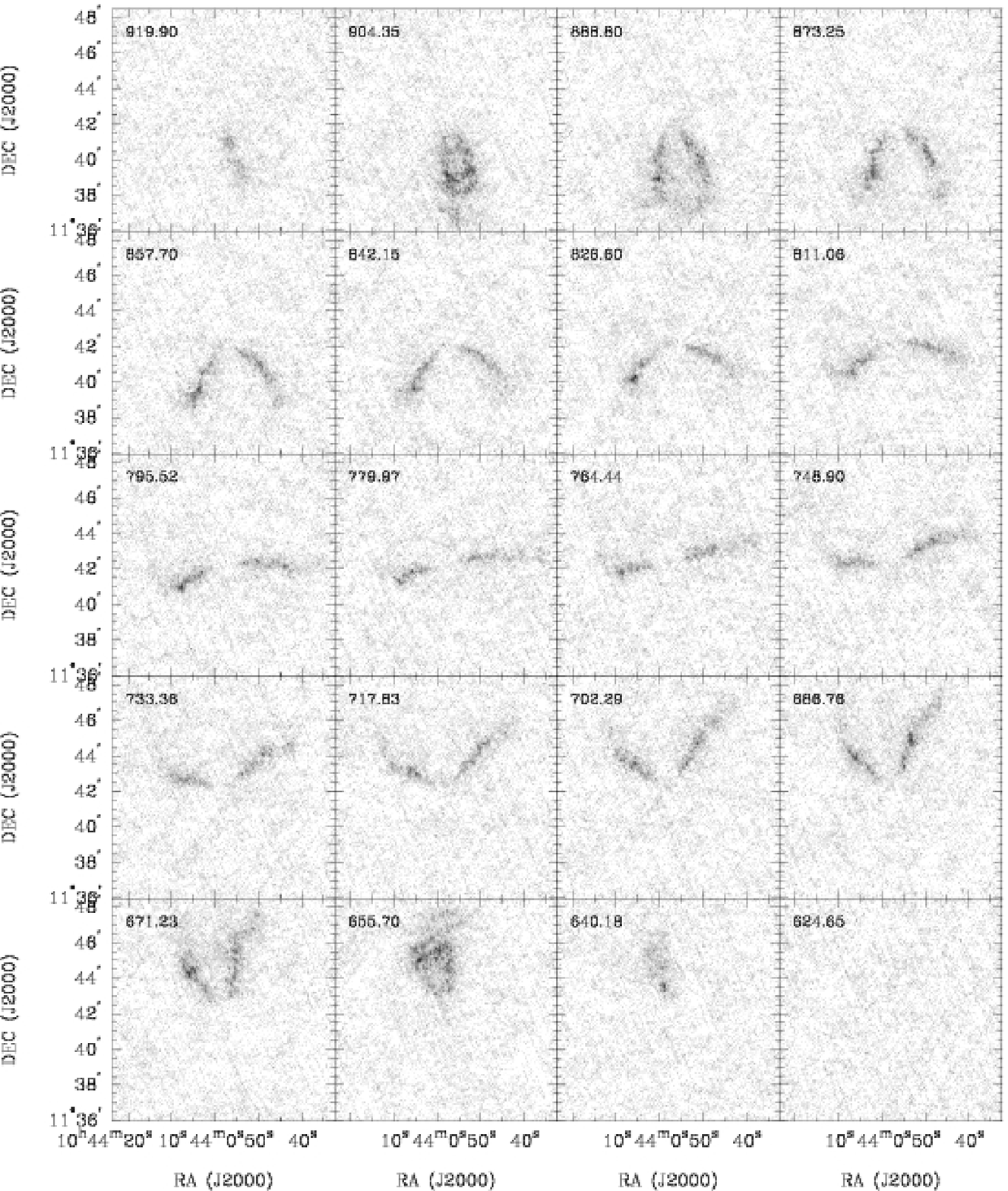}
\caption{{\bf NGC~3351:} Channel maps based on the NA cube (greyscale
  range: --0.02 to 5 mJy\,beam$^{-1}$).  Every third channel is shown
  (channel width: 5.2\,km\,s$^{-1}$). The area shown in each panel is
  identical to the area shown on the next figure}
\end{figure}

\clearpage
\begin{figure}
\vspace{0cm}  \epsscale{1.1}
\plotone{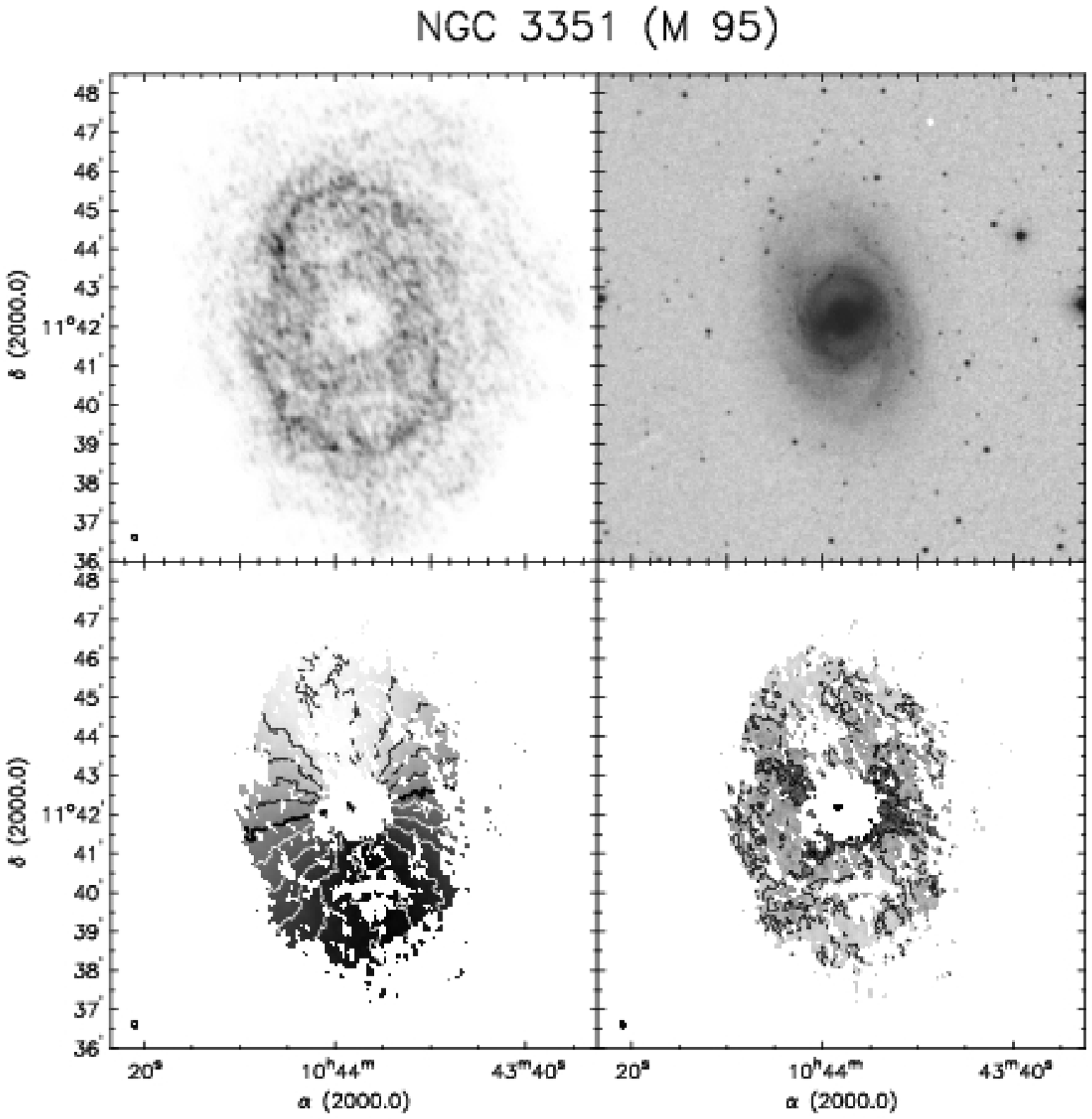}
\vspace{-2.5cm}
\caption{{\bf NGC~3351}. {\em Top left:} integrated \hi\ map (moment 0).
  Greyscale range from 0--83 Jy\,km\,s$^{-1}$. {\em Top right:}
  Optical image from the digitized sky survey (DSS). {\em Bottom
    left:} Velocity field (moment 1). Black contours (lighter
  greyscale) indicate approaching emission, white contours (darker
  greyscale) receding emission. The thick black contour is the
  systemic velocity ($v_{\rm sys}$=779.0\,km\,s$^{-1}$), the
  iso--velocity contours are spaced by $\Delta\,v$=25.0\,km\,s$^{-1}$.
  {\em Bottom right:} Velocity dispersion map (moment 2). Contours are plotted
  at 5, 10 and 20\,km\,s$^{-1}$.}
\end{figure}

%
% NGC\,3521
%

\clearpage
\begin{figure}
\epsscale{1.0}
\plotone{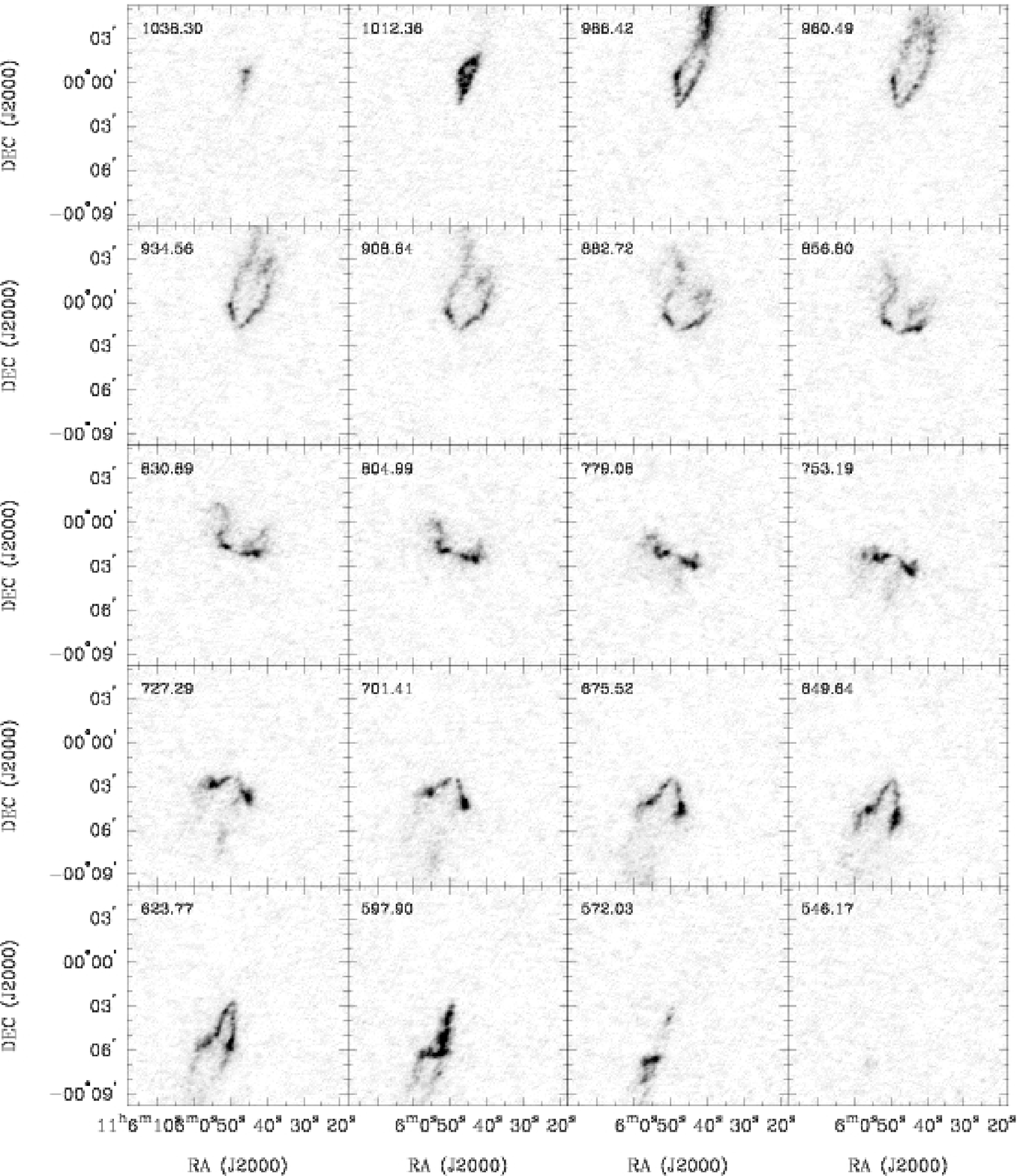}
\caption{{\bf NGC~3521:} Channel maps based on the NA cube (greyscale
  range: --0.02 to 12 mJy\,beam$^{-1}$).  Every fifth channel is shown
  (channel width: 5.2\,km\,s$^{-1}$). The area shown in each panel is
  identical to the area shown on the next figure}
\end{figure}

\clearpage
\begin{figure}
\vspace{0cm}  \epsscale{1.1}
\plotone{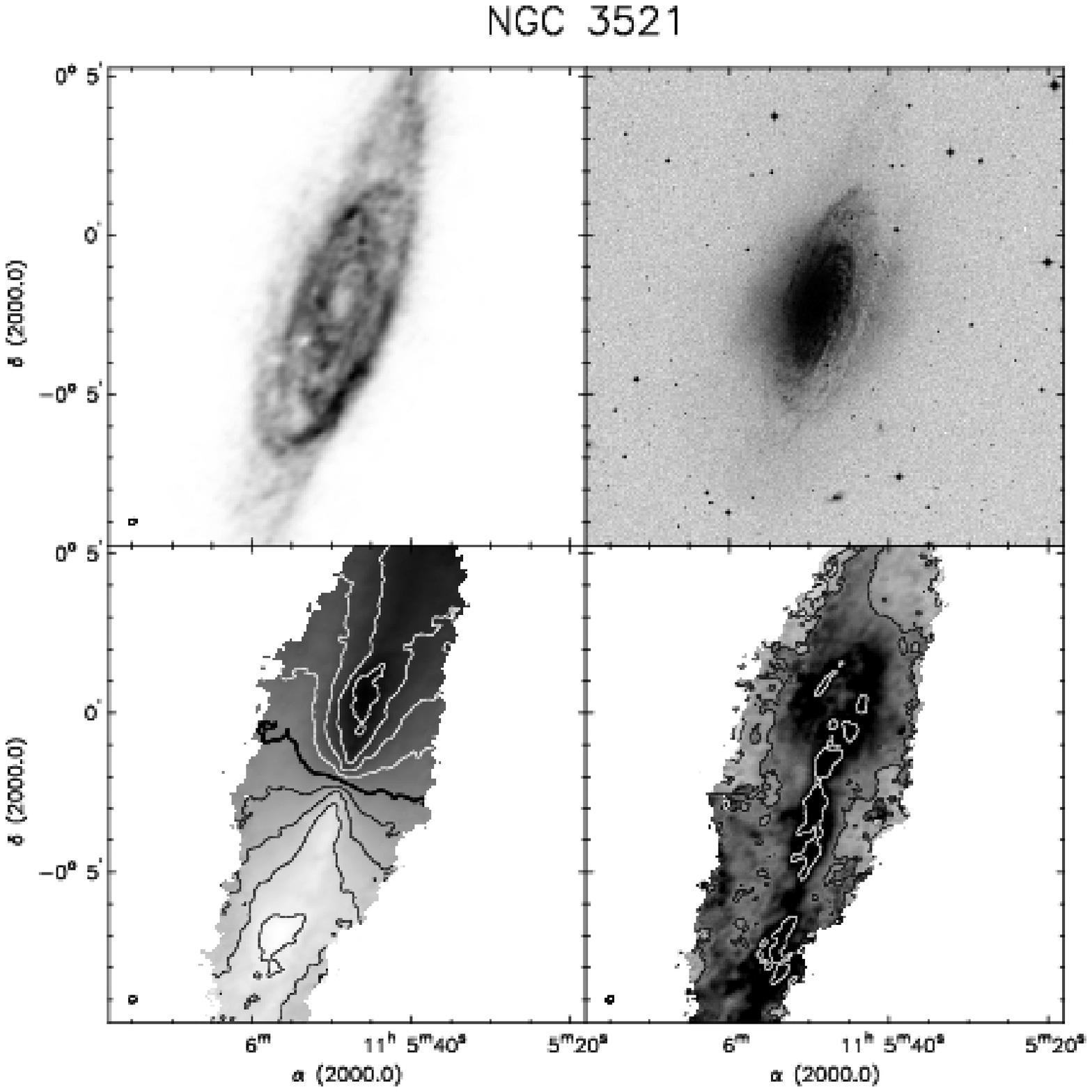}
\vspace{-2.5cm}
\caption{{\bf NGC~3521}. {\em Top left:} integrated \hi\ map (moment 0).
  Greyscale range from 0--740 Jy\,km\,s$^{-1}$. {\em Top right:}
  Optical image from the digitized sky survey (DSS). {\em Bottom
    left:} Velocity field (moment 1). Black contours (lighter
  greyscale) indicate approaching emission, white contours (darker
  greyscale) receding emission. The thick black contour is the
  systemic velocity ($v_{\rm sys}$=798.2 \,km\,s$^{-1}$), the
  iso--velocity contours are spaced by $\Delta\,v$=50\,km\,s$^{-1}$.
  {\em Bottom right:} Velocity dispersion map (moment 2). Contours are plotted
  at 5, 10 and 20\,km\,s$^{-1}$ (white contour: 50\,km\,s$^{-1}$).}
\end{figure}

%
% NGC\,3621
%

\clearpage
\begin{figure}
\epsscale{1.0}
\plotone{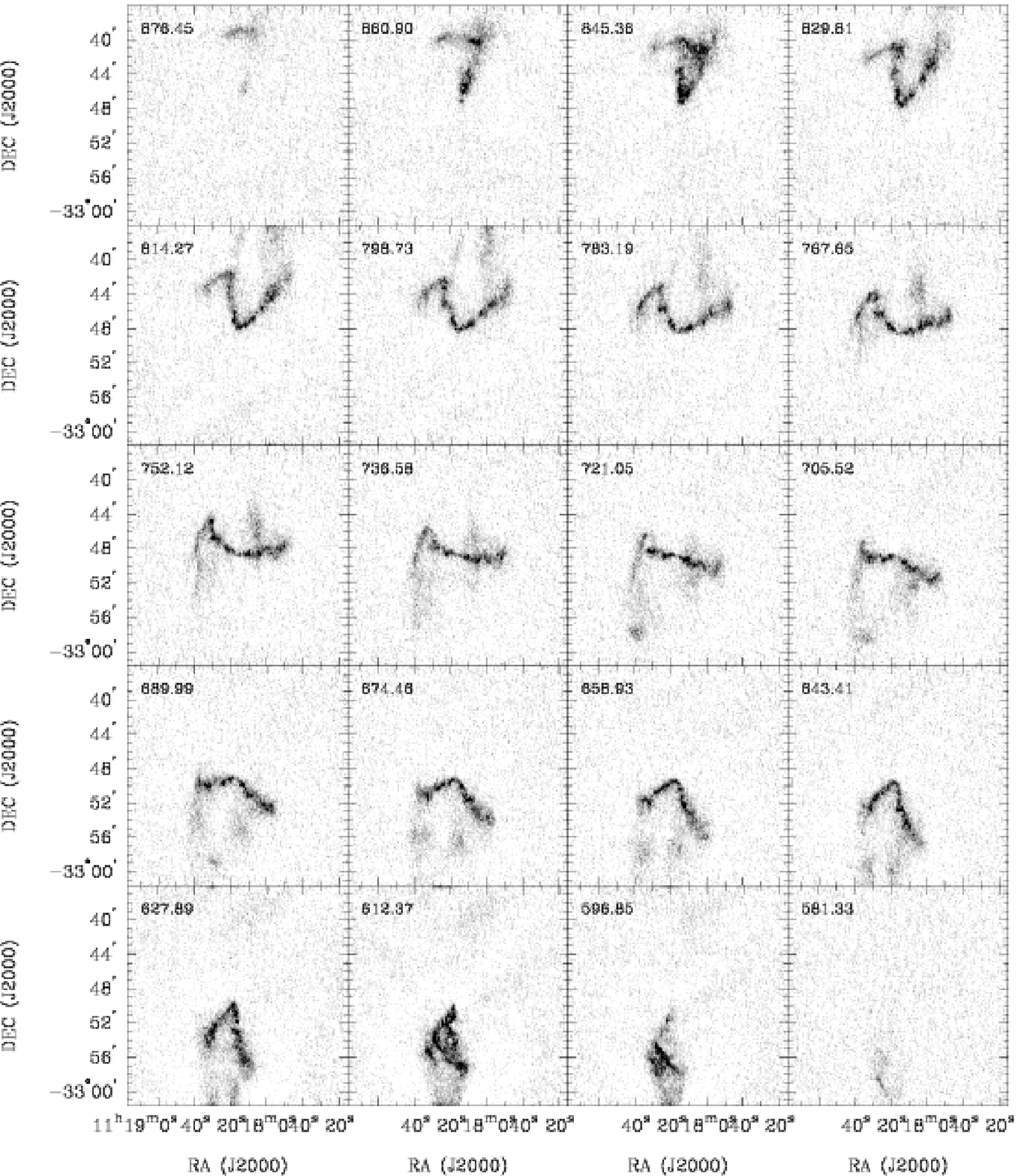}
\caption{{\bf NGC~3621:} Channel maps based on the NA cube (greyscale
  range: --0.02 to 10 mJy\,beam$^{-1}$).  Every third channel is shown
  (channel width: 5.2\,km\,s$^{-1}$). The area shown in each panel is
  identical to the area shown on the next figure}
\end{figure}

\clearpage
\begin{figure}
\vspace{0cm}  \epsscale{1.1}
\plotone{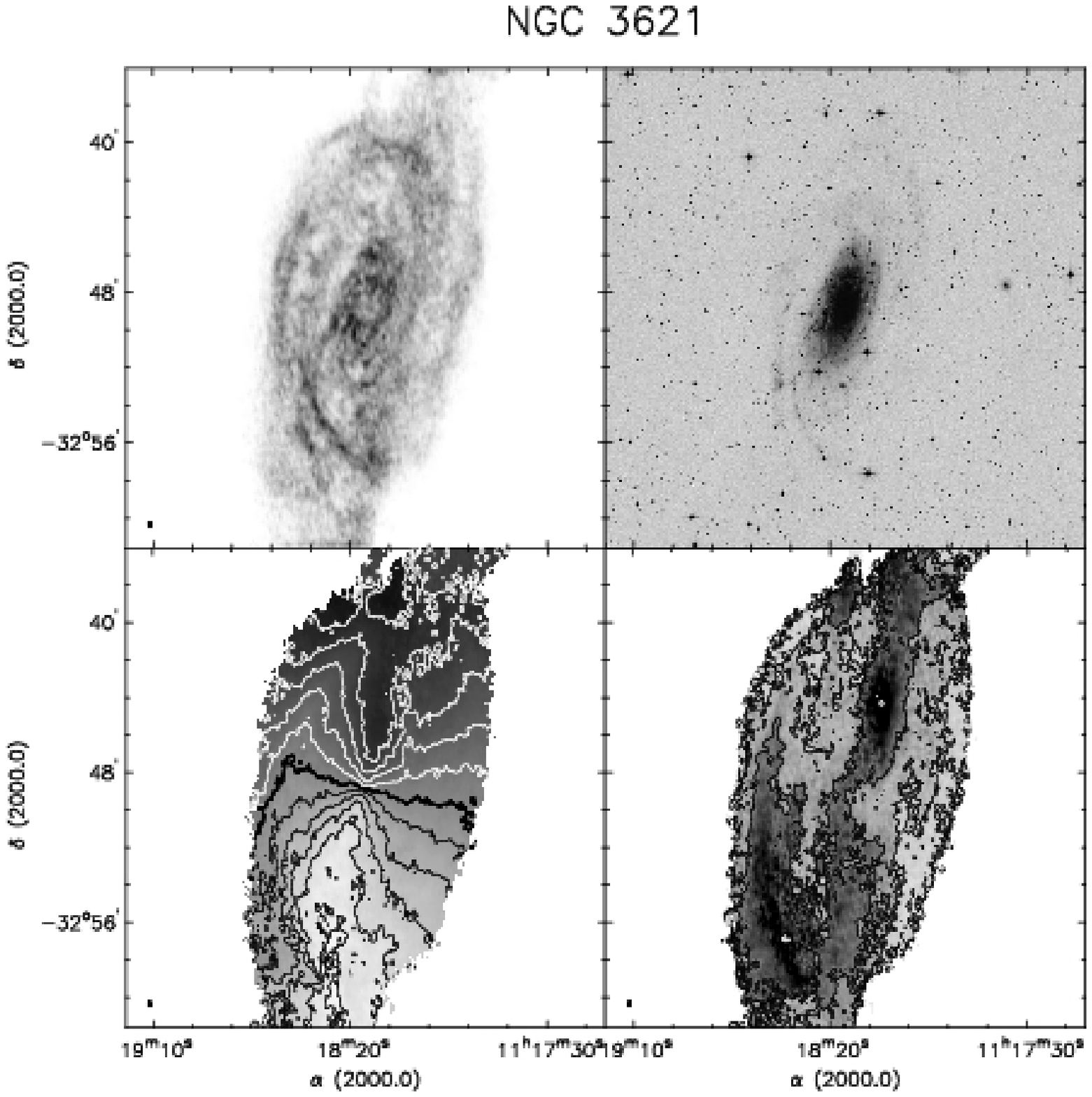}
\vspace{-2.5cm}
\caption{{\bf NGC~3621}. {\em Top left:} integrated \hi\ map (moment 0).
  Greyscale range from 0--525 Jy\,km\,s$^{-1}$. {\em Top right:}
  Optical image from the digitized sky survey (DSS). {\em Bottom
    left:} Velocity field (moment 1). Black contours (lighter
  greyscale) indicate approaching emission, white contours (darker
  greyscale) receding emission. The thick black contour is the
  systemic velocity ($v_{\rm sys}$=730.1 \,km\,s$^{-1}$), the
  iso--velocity contours are spaced by $\Delta\,v$=25\,km\,s$^{-1}$.
  {\em Bottom right:} Velocity dispersion map (moment 2). Contours are plotted
  at 5, 10 and 20\,km\,s$^{-1}$ (white contour: 50\,km\,s$^{-1}$).}
\end{figure}

%
% NGC\,3627
%

\clearpage
\begin{figure}
\epsscale{1.0}
\plotone{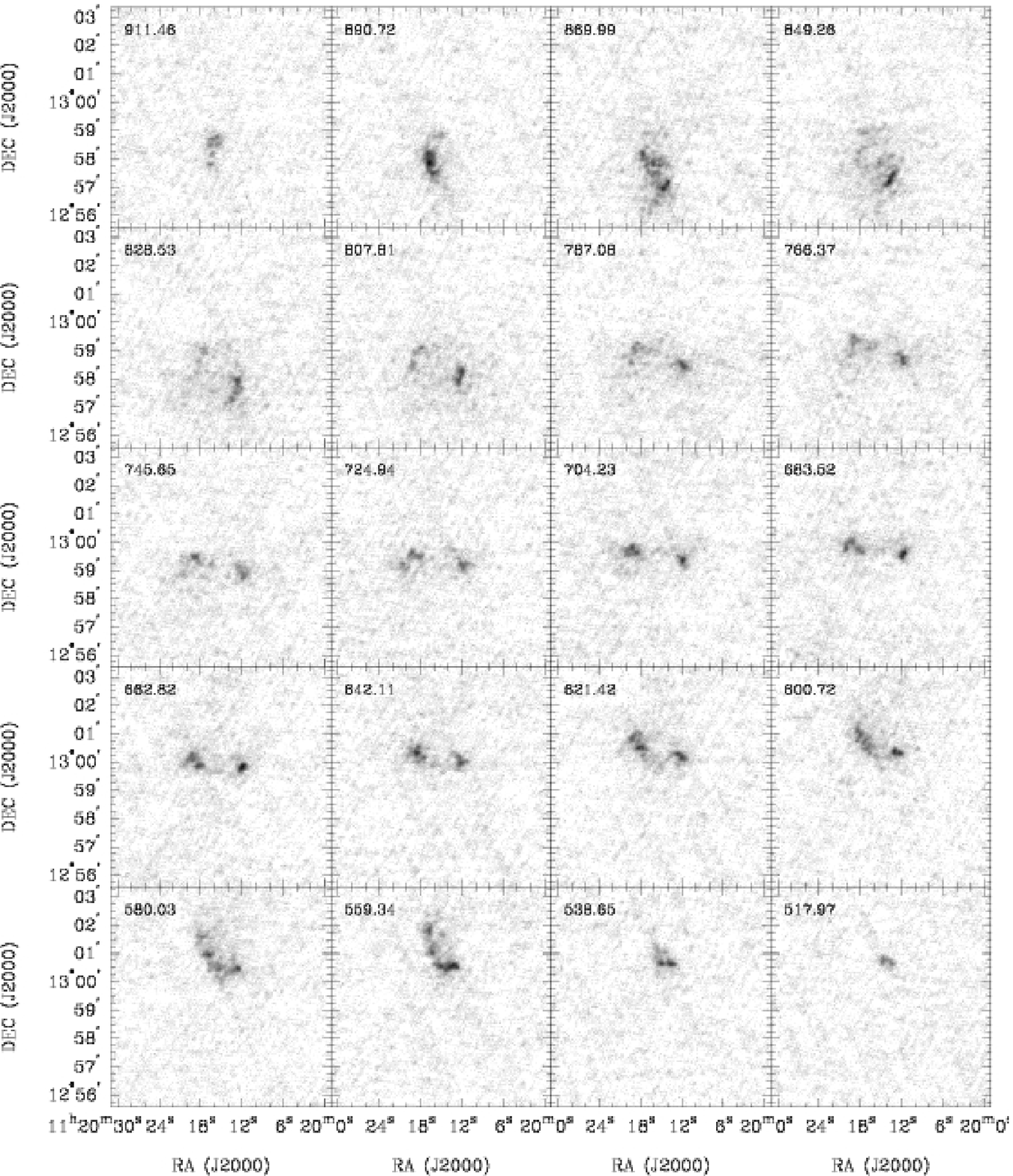}
\caption{{\bf NGC~3627:} Channel maps based on the NA cube (greyscale
  range: --0.02 to 7 mJy\,beam$^{-1}$).  Every fourth channel is shown
  (channel width: 5.2\,km\,s$^{-1}$). The area shown in each panel is
  identical to the area shown on the next figure}
\end{figure}

\clearpage
\begin{figure}
\vspace{0cm}  \epsscale{1.1}
\plotone{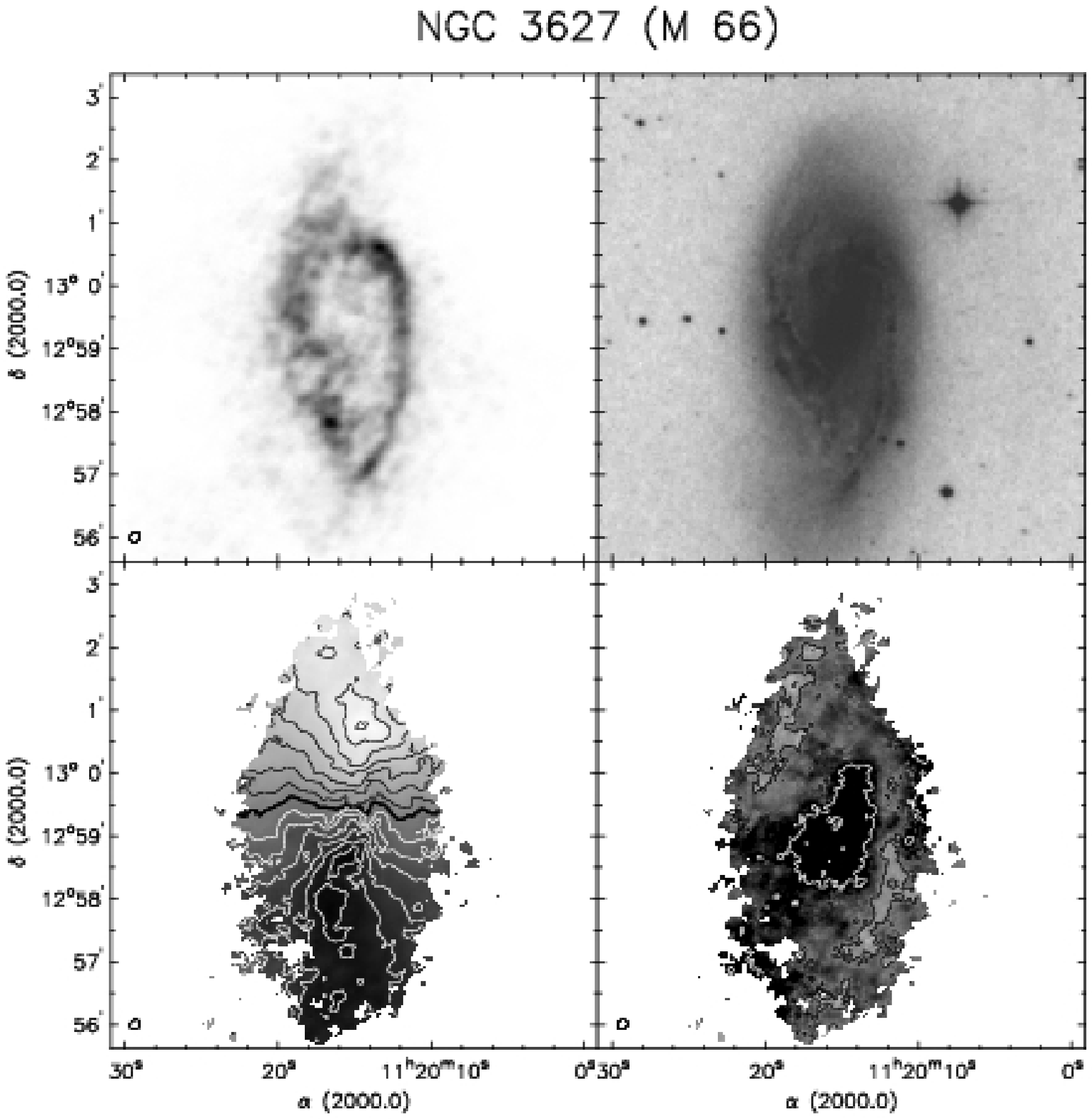}
\vspace{-2.5cm}
\caption{{\bf NGC~3627}. {\em Top left:} integrated \hi\ map (moment 0).
  Greyscale range from 0--293 Jy\,km\,s$^{-1}$. {\em Top right:}
  Optical image from the digitized sky survey (DSS). {\em Bottom
    left:} Velocity field (moment 1). Black contours (lighter
  greyscale) indicate approaching emission, white contours (darker
  greyscale) receding emission. The thick black contour is the
  systemic velocity ($v_{\rm sys}$=717.3 \,km\,s$^{-1}$), the
  iso--velocity contours are spaced by $\Delta\,v$=25\,km\,s$^{-1}$.
  {\em Bottom right:} Velocity dispersion map (moment 2). Contours are plotted
  at 5, 10 and 20\,km\,s$^{-1}$ (white contours: 50 and
  100\,km\,s$^{-1}$).}
\end{figure}

%
% NGC\,4214
%

\clearpage
\begin{figure}
\epsscale{1.0}
\plotone{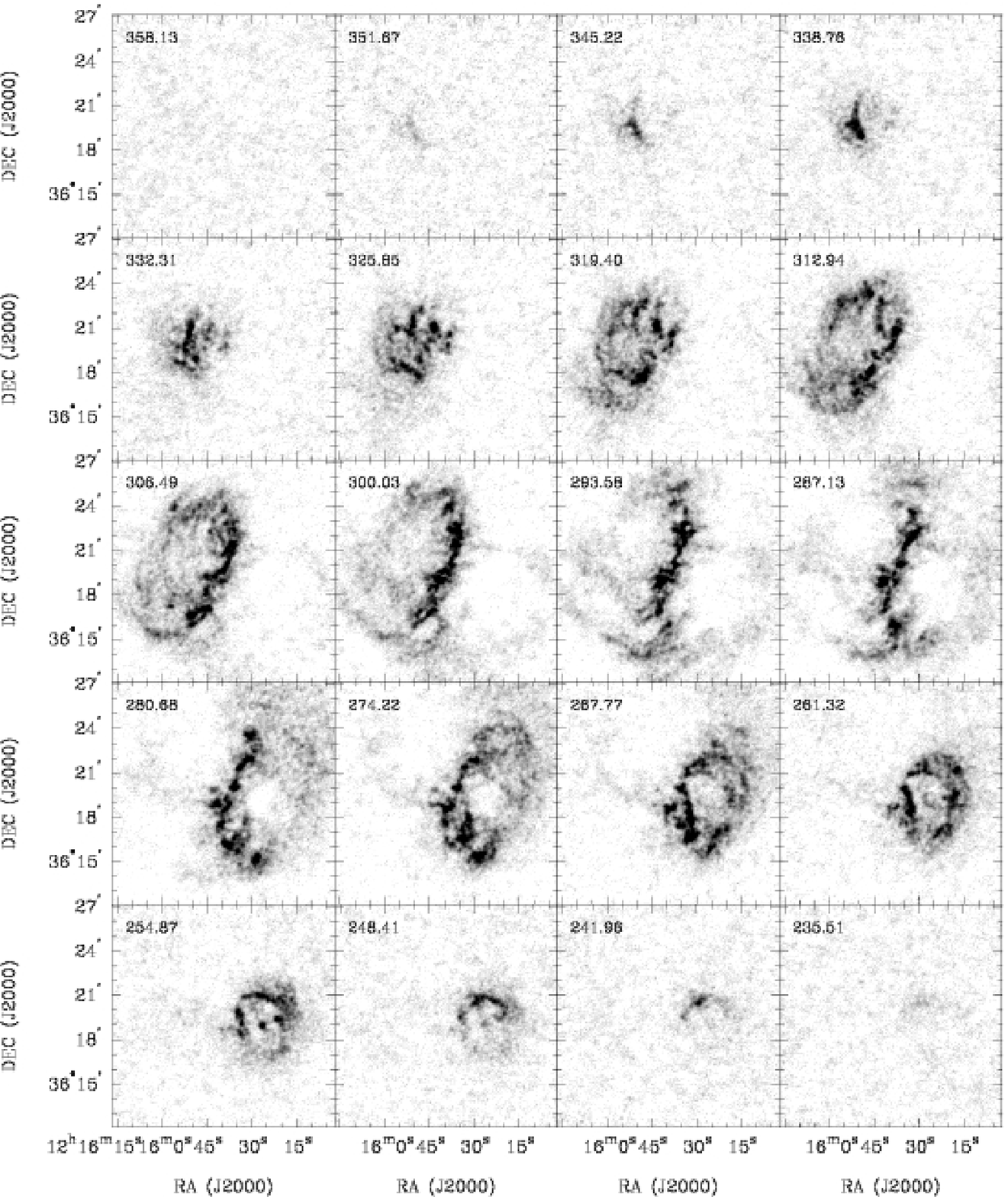}
\caption{{\bf NGC~4214:} Channel maps based on the NA cube (greyscale
  range: --0.02 to 10 mJy\,beam$^{-1}$).  Every fifth channel is shown
  (channel width: 1.3\,km\,s$^{-1}$). The area shown in each panel is
  identical to the area shown on the next figure}
\end{figure}

\clearpage
\begin{figure}
\vspace{0cm}  \epsscale{1.1}
\plotone{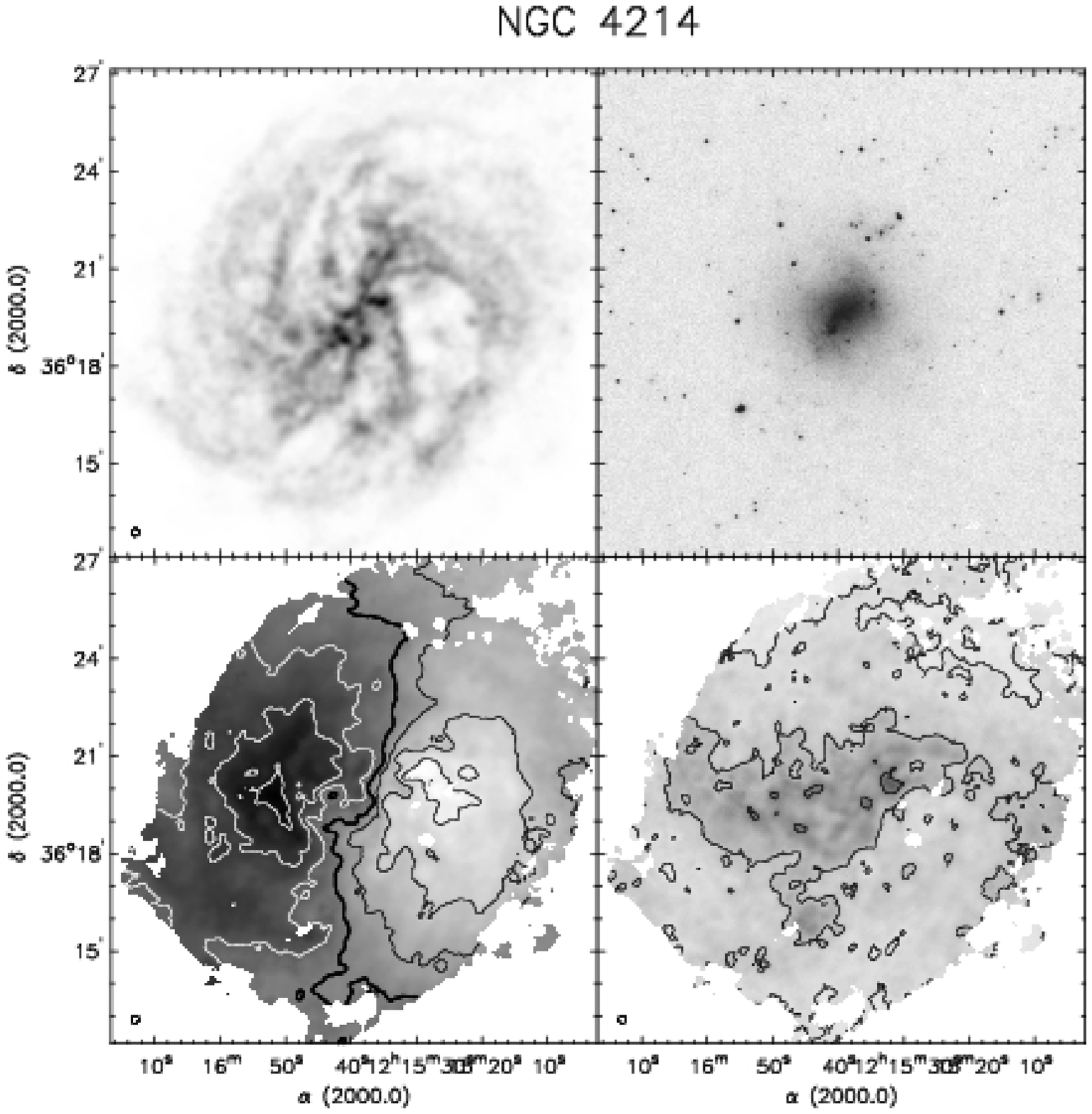}
\vspace{-2.5cm}
\caption{{\bf NGC~4214}. {\em Top left:} integrated \hi\ map (moment 0).
  Greyscale range from 0--610Jy\,km\,s$^{-1}$. {\em Top right:}
  Optical image from the digitized sky survey (DSS). {\em Bottom
    left:} Velocity field (moment 1). Black contours (lighter
  greyscale) indicate approaching emission, white contours (darker
  greyscale) receding emission. The thick black contour is the
  systemic velocity ($v_{\rm sys}$=292.8 \,km\,s$^{-1}$), the
  iso--velocity contours are spaced by $\Delta\,v$=12.5\,km\,s$^{-1}$.
  {\em Bottom right:} Velocity dispersion map (moment 2). Contours are plotted
  at 5, 10 and 20\,km\,s$^{-1}$.}
\end{figure}

%
% NGC\,4449
%

\clearpage
\begin{figure}
\epsscale{1.0}
\plotone{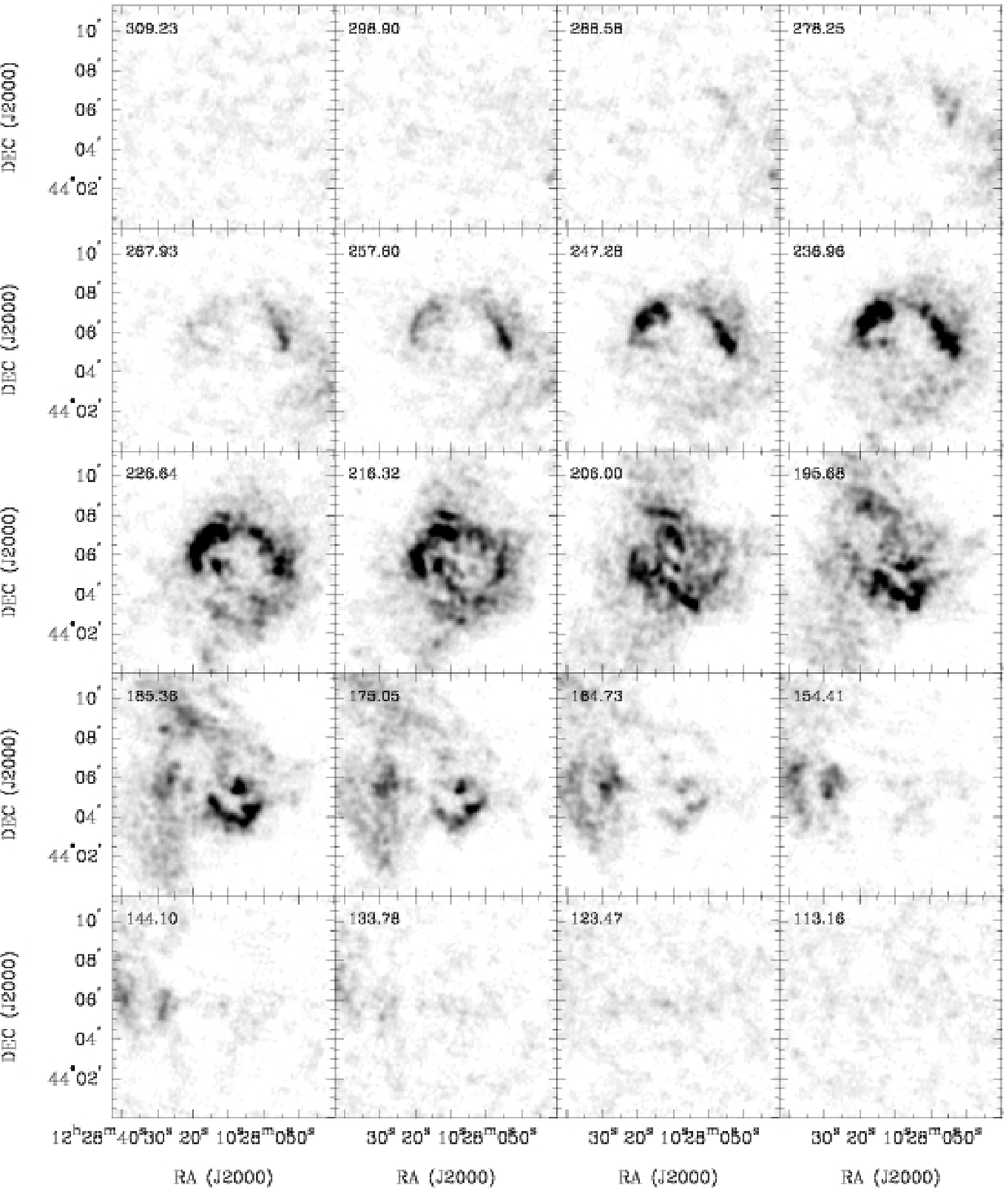}
\caption{{\bf NGC~4449:} Channel maps based on the NA cube (greyscale
  range: --0.02 to 20 mJy\,beam$^{-1}$).  Every second channel is
  shown (channel width: 5.2\,km\,s$^{-1}$). The area shown in each
  panel is identical to the area shown on the next figure}
\end{figure}

\clearpage
\begin{figure}
\vspace{0cm}  \epsscale{1.1}
\plotone{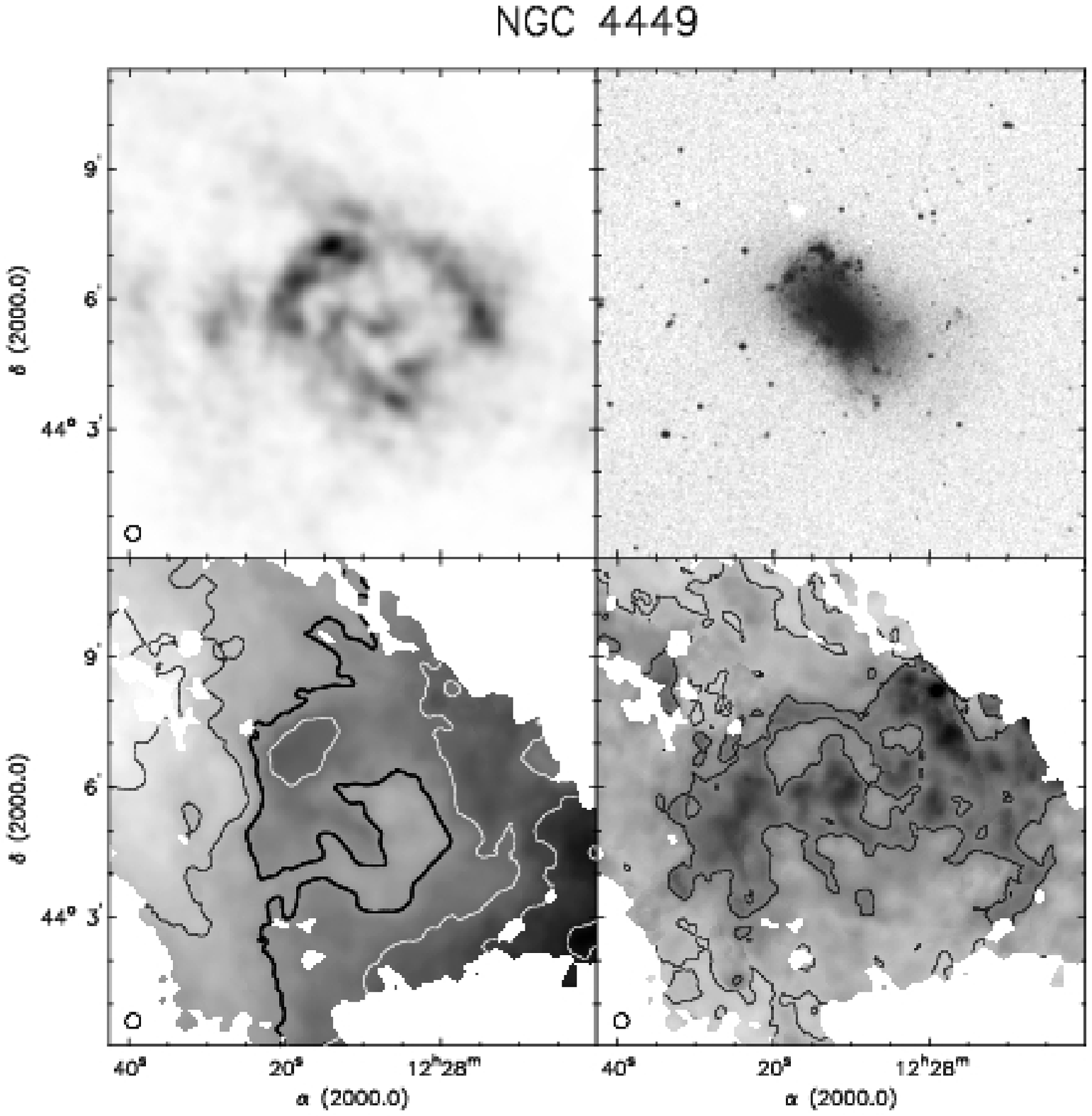}
\vspace{-2.5cm}
\caption{{\bf NGC~4449}. {\em Top left:} integrated \hi\ map (moment 0).
  Greyscale range from 0--1830 Jy\,km\,s$^{-1}$. {\em Top right:}
  Optical image from the digitized sky survey (DSS). {\em Bottom
    left:} Velocity field (moment 1). Black contours (lighter
  greyscale) indicate approaching emission, white contours (darker
  greyscale) receding emission. The thick black contour is the
  systemic velocity ($v_{\rm sys}$=202.7 \,km\,s$^{-1}$), the
  iso--velocity contours are spaced by $\Delta\,v$=25.0\,km\,s$^{-1}$.
  {\em Bottom right:} Velocity dispersion map (moment 2). Contours are plotted
  at 5, 10 and 20\,km\,s$^{-1}$.}
\end{figure}

%
% NGC\,4736
%

\clearpage
\begin{figure}
\epsscale{1.0}
\plotone{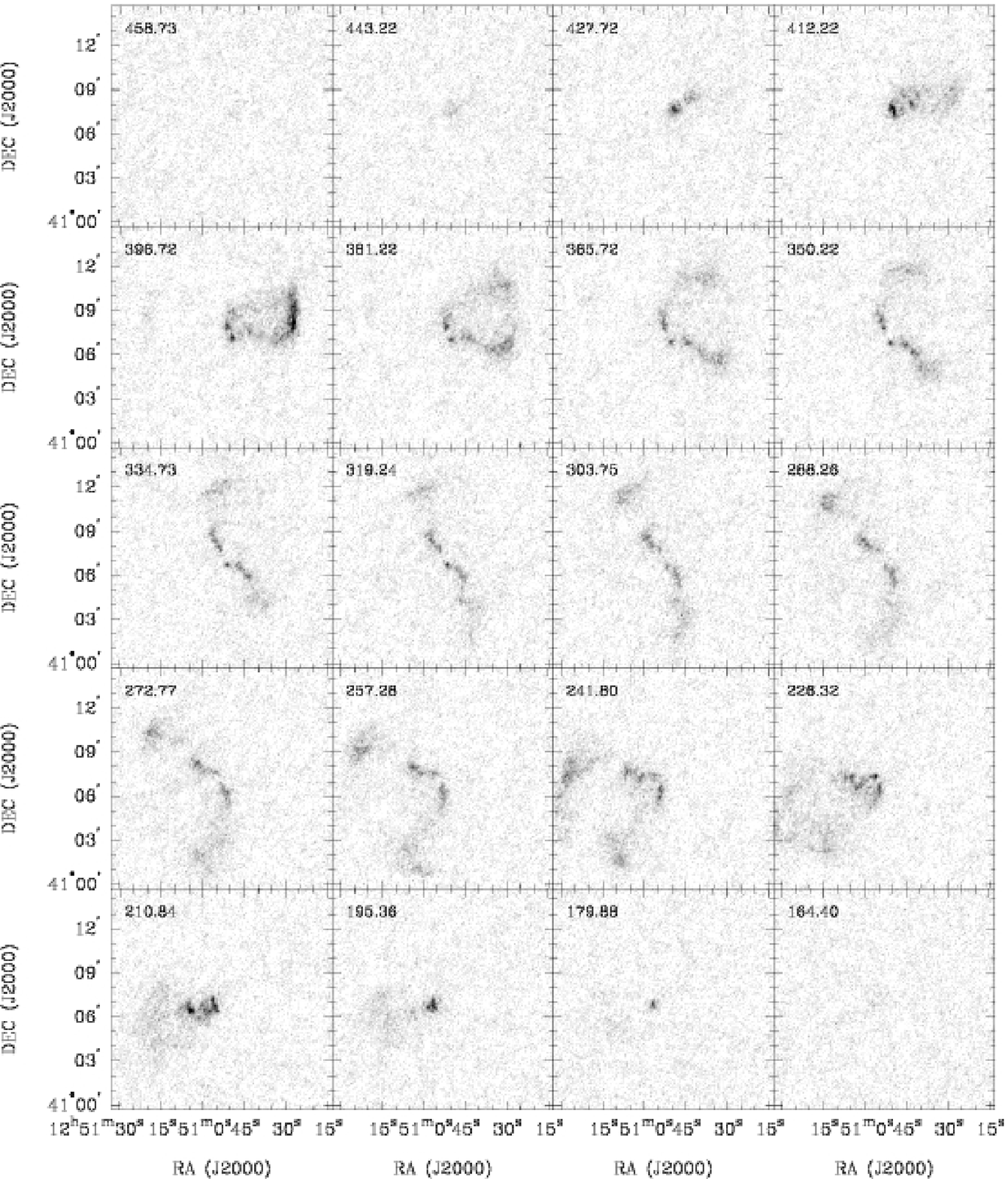}
\caption{{\bf NGC~4736:} Channel maps based on the NA cube (greyscale
  range: --0.02 to 6 mJy\,beam$^{-1}$).  Every third channel is shown
  (channel width: 5.2\,km\,s$^{-1}$). The area shown in each panel is
  identical to the area shown on the next figure}
\end{figure}

\clearpage
\begin{figure}
\vspace{0cm}  \epsscale{1.1}
\plotone{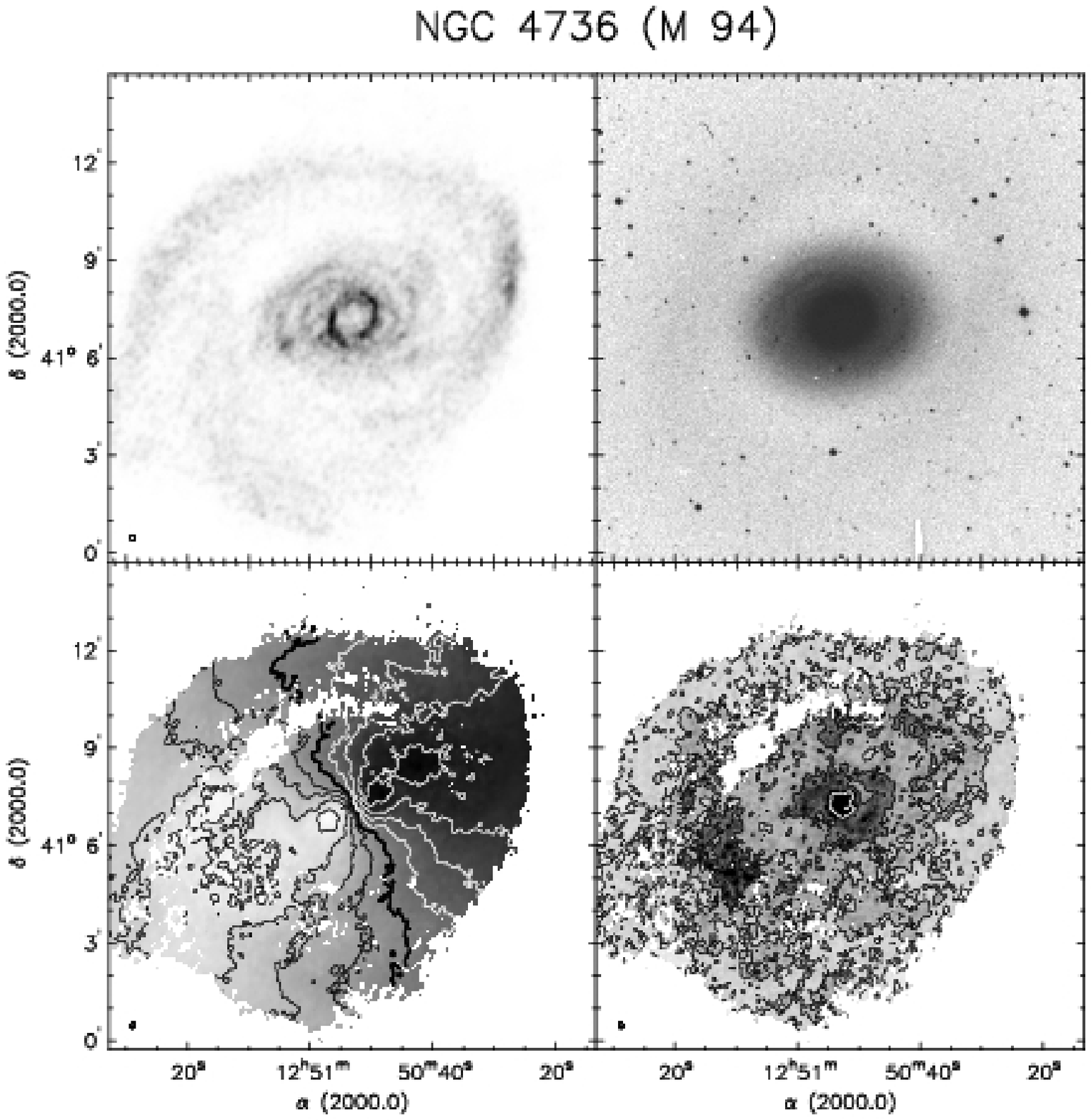}
\vspace{-2.5cm}
\caption{{\bf NGC~4736}. {\em Top left:} integrated \hi\ map (moment 0).
  Greyscale range from 0--209 Jy\,km\,s$^{-1}$. {\em Top right:}
  Optical image from the digitized sky survey (DSS). {\em Bottom
    left:} Velocity field (moment 1). Black contours (lighter
  greyscale) indicate approaching emission, white contours (darker
  greyscale) receding emission. The thick black contour is the
  systemic velocity ($v_{\rm sys}$=307.6 \,km\,s$^{-1}$), the
  iso--velocity contours are spaced by $\Delta\,v$=25.0\,km\,s$^{-1}$.
  {\em Bottom right:} Velocity dispersion map (moment 2). Contours are plotted
  at 5, 10 and 20 \,km\,s$^{-1}$ (white contour: 50\,km\,s$^{-1}$).}
\end{figure}

%
% DDO~154
%

\clearpage
\begin{figure}
\epsscale{1.0}
\plotone{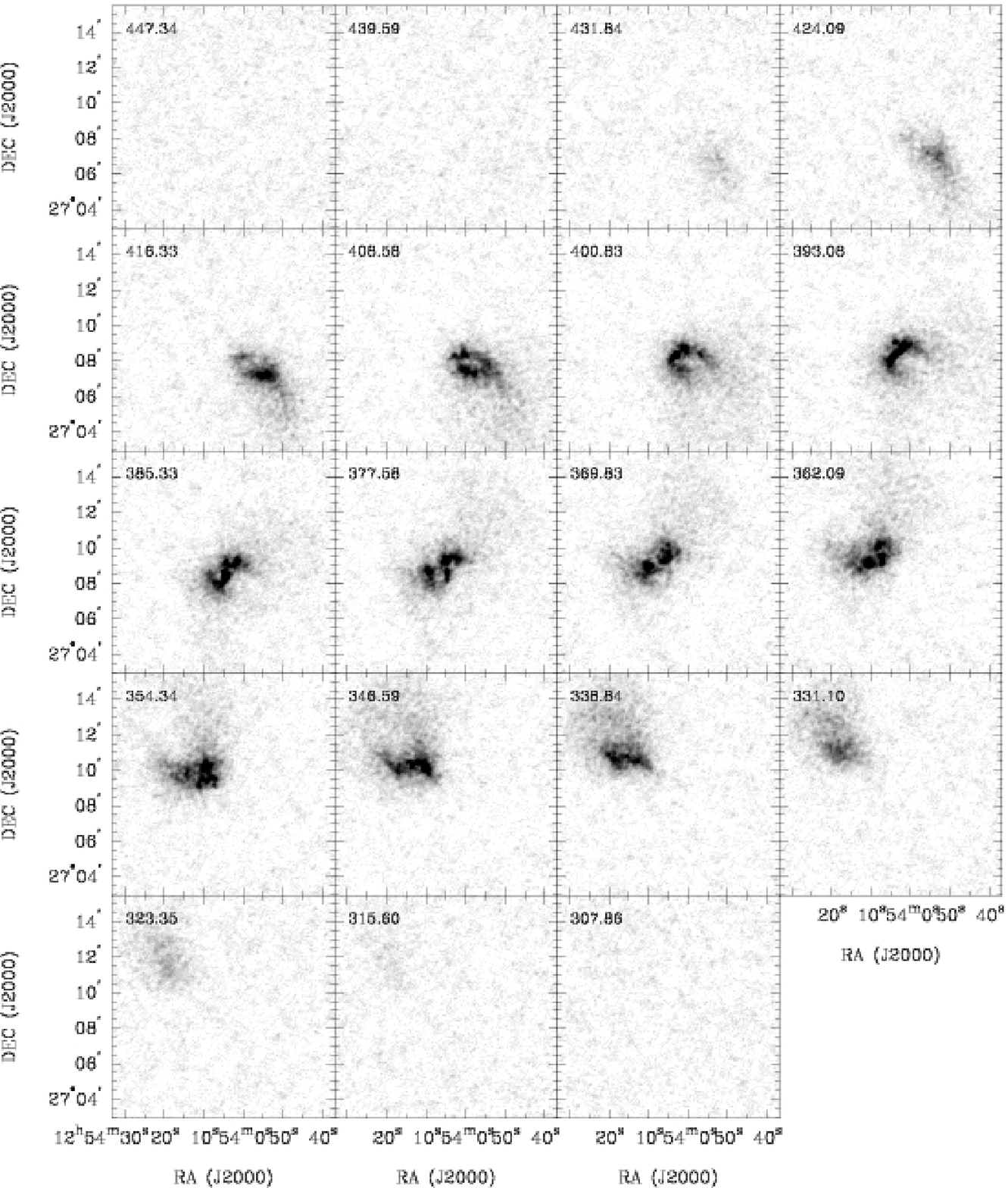}
\caption{{\bf DDO~154:} Channel maps based on the NA cube (greyscale
  range: --0.02 to 10 mJy\,beam$^{-1}$).  Every third channel is shown
  (channel width: 2.6\,km\,s$^{-1}$). The area shown in each panel is
  identical to the area shown on the next figure}
\end{figure}

\clearpage
\begin{figure}
\vspace{0cm}  \epsscale{1.1}
\plotone{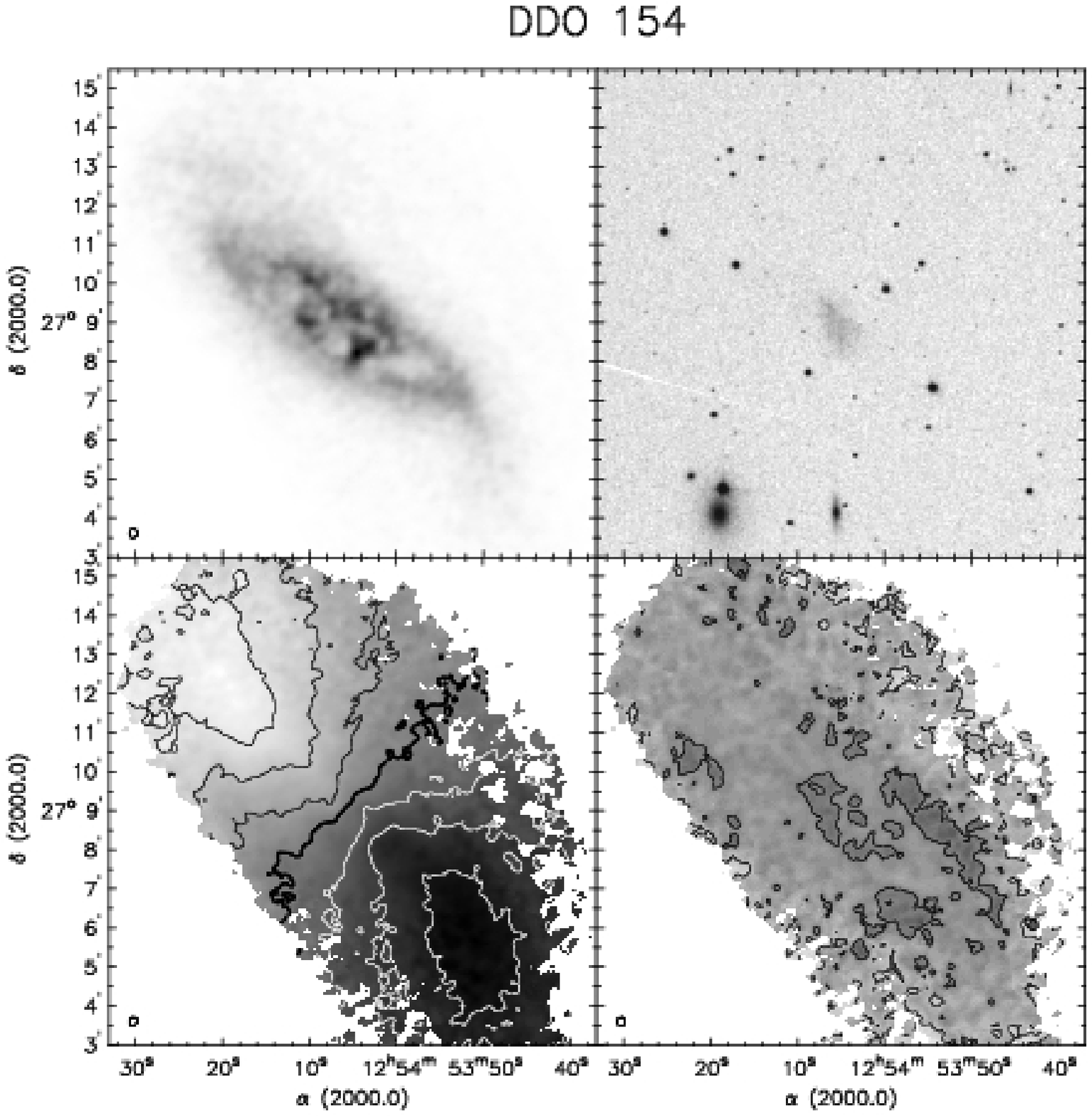}
\vspace{-2.5cm}
\caption{{\bf DDO~154}. {\em Top left:} integrated \hi\ map (moment
  0).  Greyscale range from 0--454 Jy\,km\,s$^{-1}$. {\em Top right:}
  Optical image from the digitized sky survey (DSS). {\em Bottom
    left:} Velocity field (moment 1). Black contours (lighter
  greyscale) indicate approaching emission, white contours (darker
  greyscale) receding emission. The thick black contour is the
  systemic velocity ($v_{\rm sys}$=375.5 \,km\,s$^{-1}$), the
  iso--velocity contours are spaced by
  $\Delta\,v$=12.50\,km\,s$^{-1}$.  {\em Bottom right:} Velocity dispersion map
  (moment 2). Contours are plotted at 5, 10 \,km\,s$^{-1}$.}
\end{figure}

%
% NGC\,4826
%

\clearpage
\begin{figure}
\epsscale{1.0}
\plotone{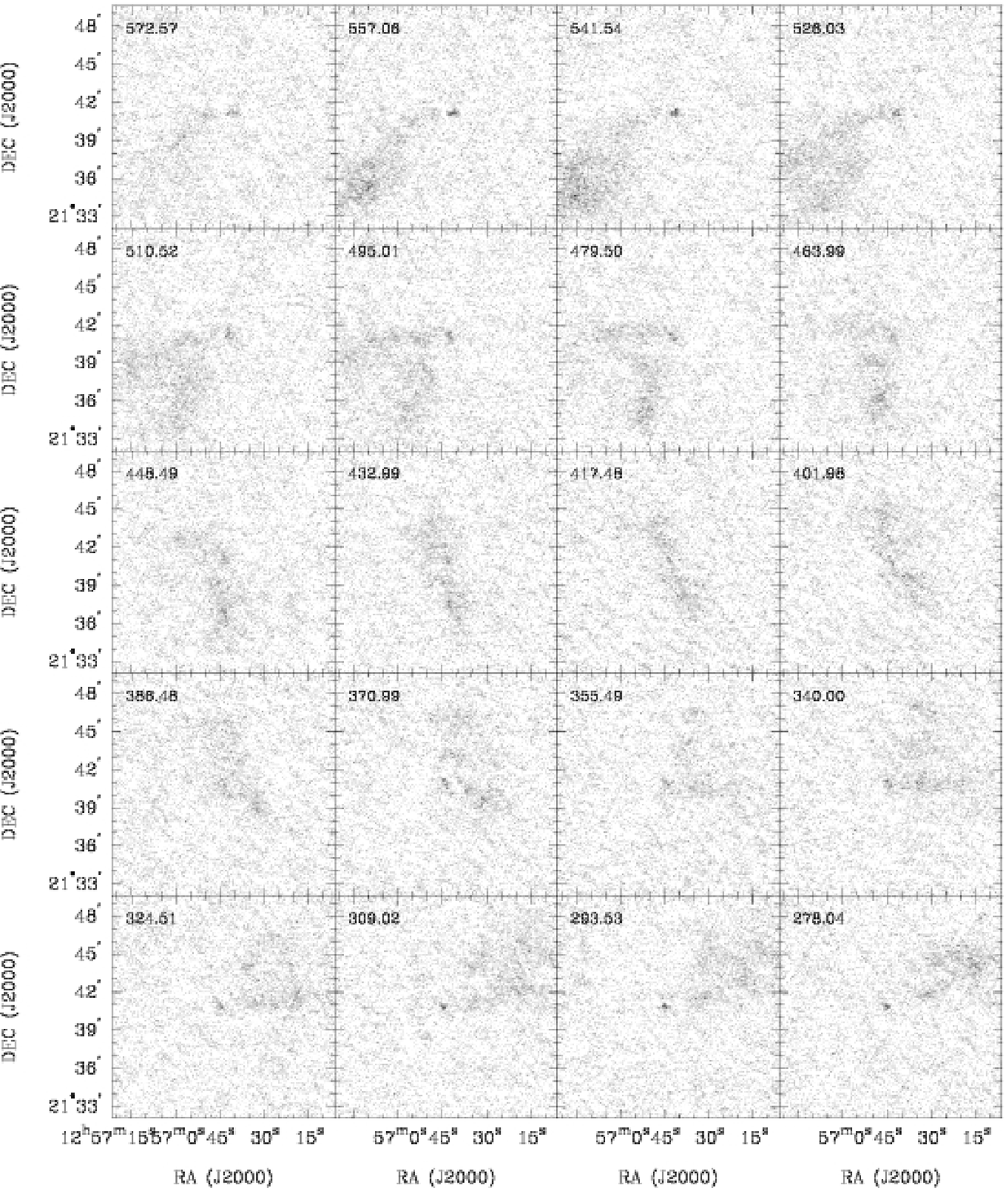}
\caption{{\bf NGC~4826:} Channel maps based on the NA cube (greyscale
  range: --0.02 to 5 mJy\,beam$^{-1}$).  Every third channel is shown
  (channel width: 5.2\,km\,s$^{-1}$). The area shown in each panel is
  identical to the area shown on the next figure}
\end{figure}

\clearpage
\begin{figure}
\vspace{0cm}  \epsscale{1.1}
\plotone{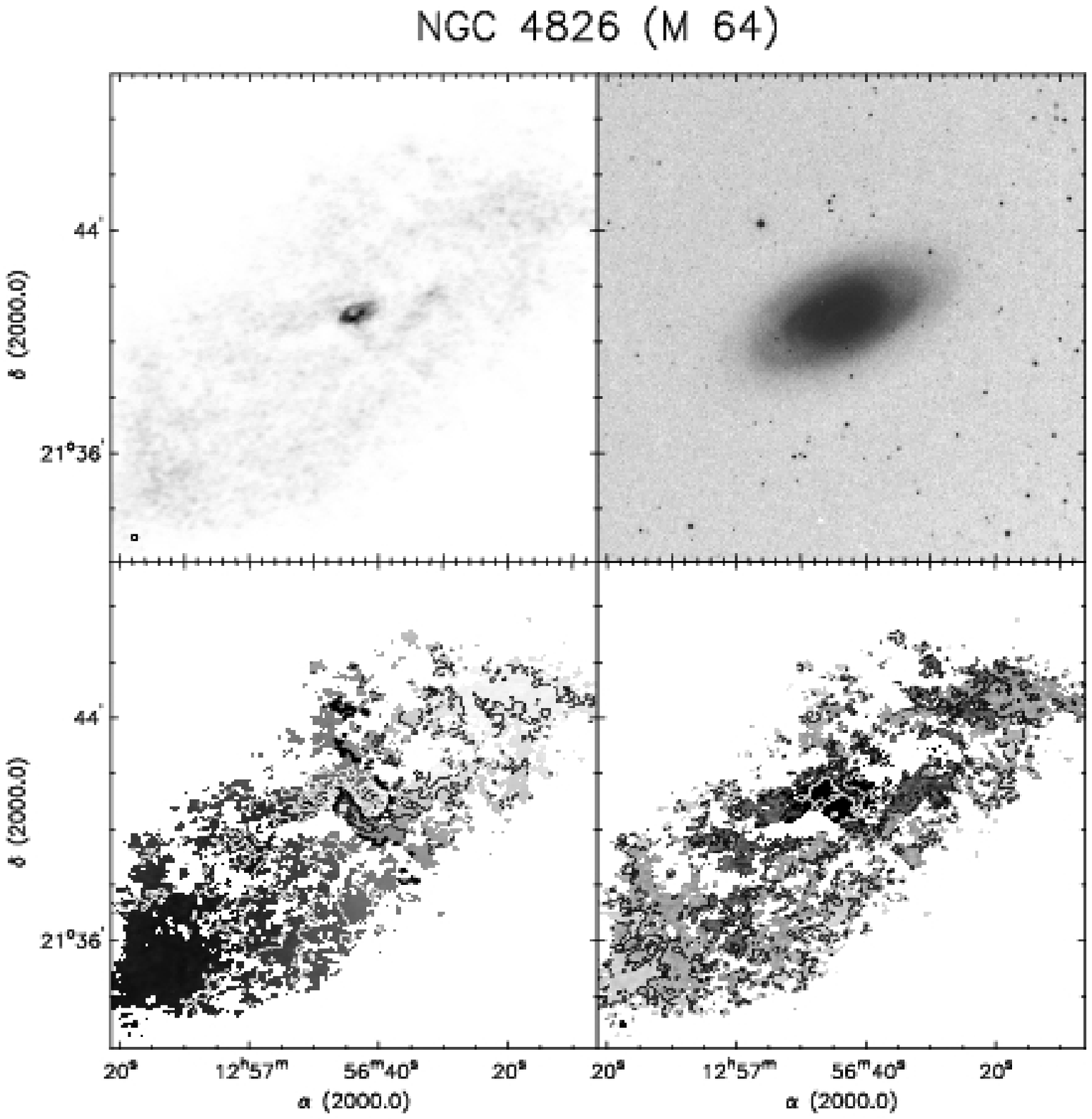}
\vspace{-2.5cm}
\caption{{\bf NGC~4826}. {\em Top left:} integrated \hi\ map (moment 0).
  Greyscale range from 0--205 Jy\,km\,s$^{-1}$. {\em Top right:}
  Optical image from the digitized sky survey (DSS). {\em Bottom
    left:} Velocity field (moment 1). Black contours (lighter
  greyscale) indicate approaching emission, white contours (darker
  greyscale) receding emission. The thick black contour is the
  systemic velocity ($v_{\rm sys}$=407.90 \,km\,s$^{-1}$), the
  iso--velocity contours are spaced by $\Delta\,v$=25.0\,km\,s$^{-1}$.
  {\em Bottom right:} Velocity dispersion map (moment 2). Contours are plotted
  at 5, 10 and 20\,km\,s$^{-1}$ (white contour: 50\,km\,s$^{-1}$).}
\end{figure}

%
% NGC\,5055
%

\clearpage
\begin{figure}
\epsscale{1.0}
\plotone{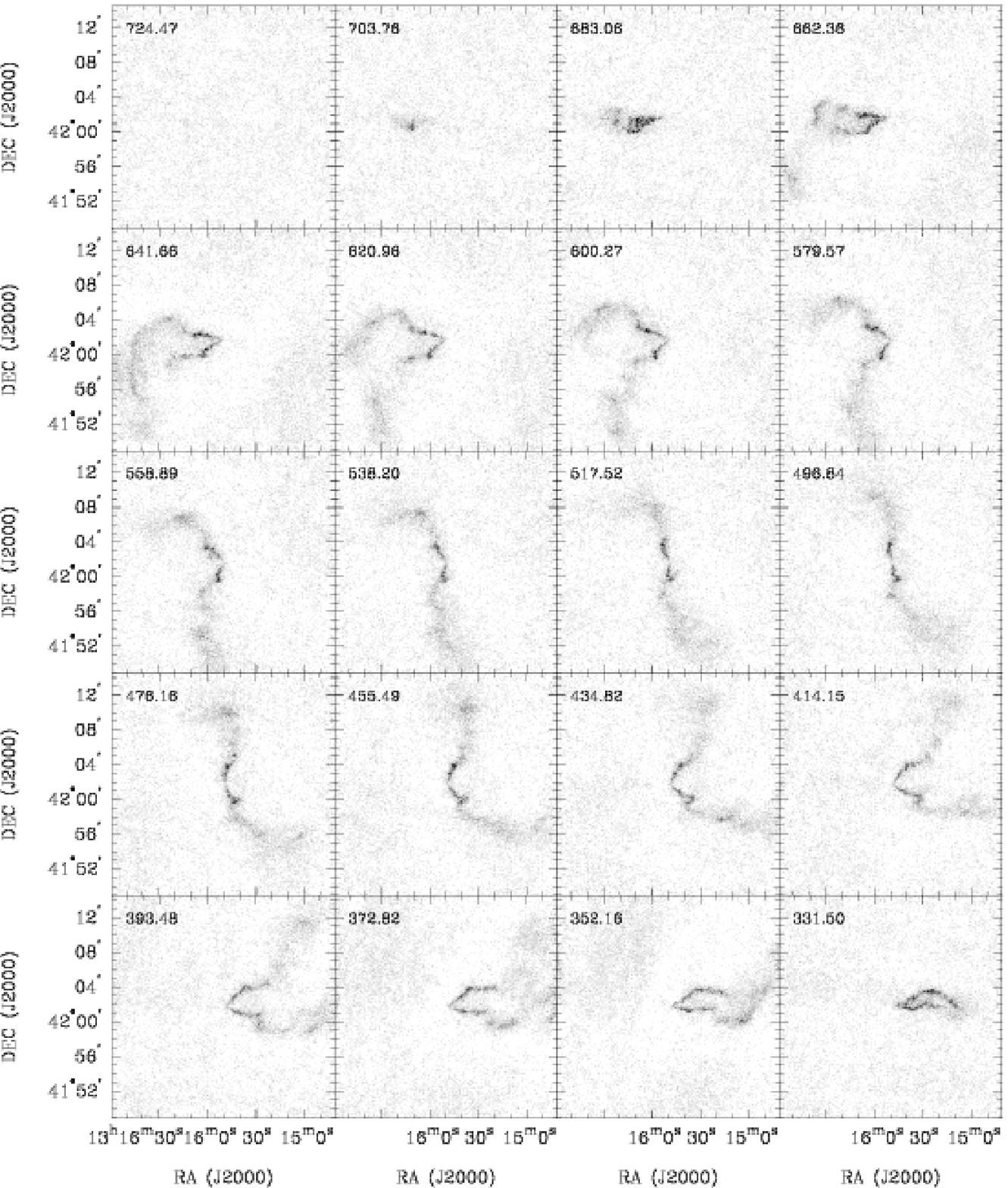}
\caption{{\bf NGC~5055:} Channel maps based on the NA cube (greyscale
  range: --0.02 to 8 mJy\,beam$^{-1}$).  Every fourth channel is shown
  (channel width: 5.2\,km\,s$^{-1}$). The area shown in each panel is
  identical to the area shown on the next figure}
\end{figure}

\clearpage
\begin{figure}
\vspace{0cm}  \epsscale{1.1}
\plotone{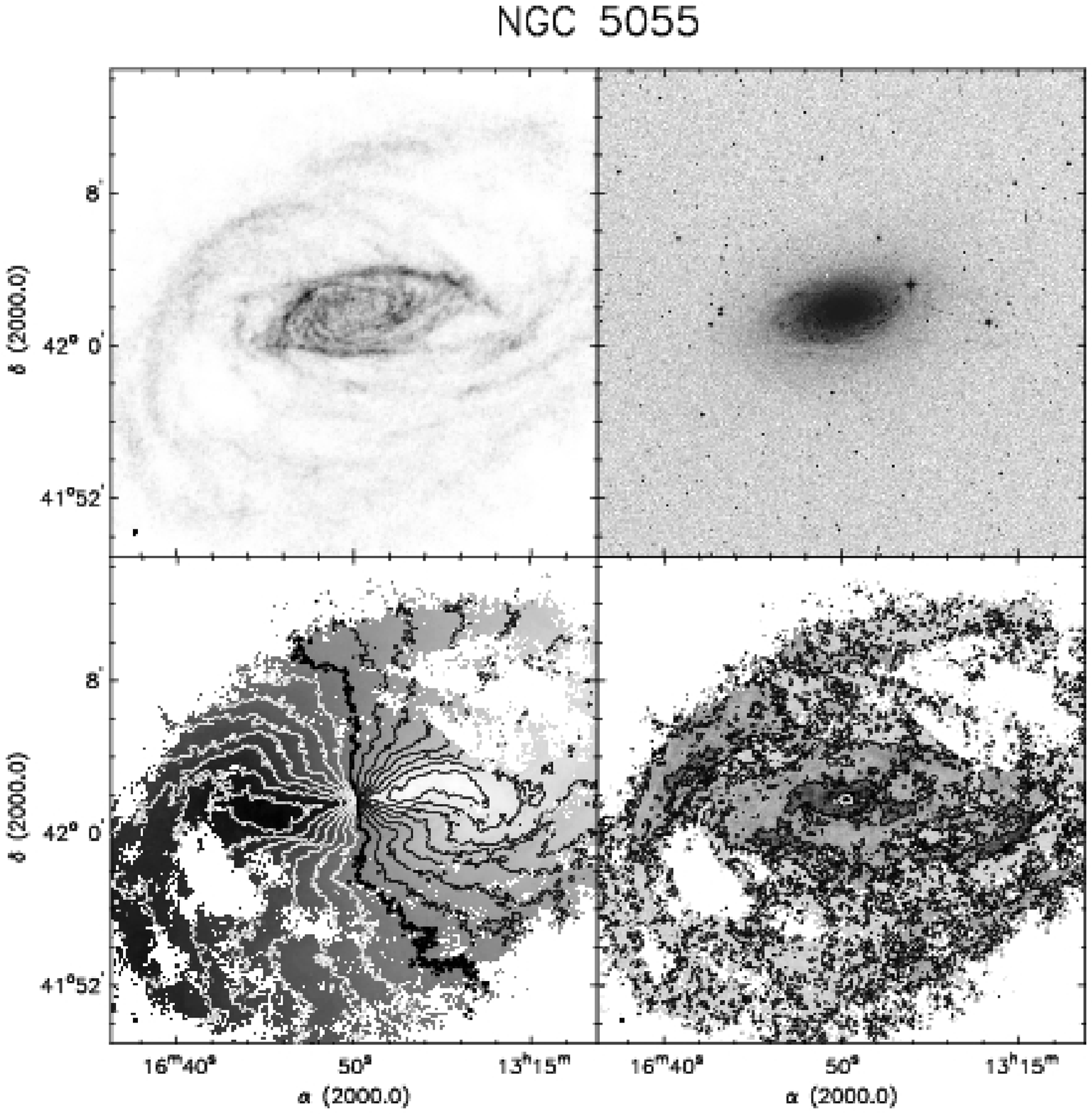}
\vspace{-2.5cm}
\caption{{\bf NGC~5055}. {\em Top left:} integrated \hi\ map (moment 0).
  Greyscale range from 0--285 Jy\,km\,s$^{-1}$. {\em Top right:}
  Optical image from the digitized sky survey (DSS). {\em Bottom
    left:} Velocity field (moment 1). Black contours (lighter
  greyscale) indicate approaching emission, white contours (darker
  greyscale) receding emission. The thick black contour is the
  systemic velocity ($v_{\rm sys}$=499.3 \,km\,s$^{-1}$), the
  iso--velocity contours are spaced by $\Delta\,v$=25\,km\,s$^{-1}$.
  {\em Bottom right:} Velocity dispersion map (moment 2). Contours are plotted
  at 5, 10 and 20\,km\,s$^{-1}$ (white contour: 50\,km\,s$^{-1}$).}
\end{figure}

%
% NGC\,5194
%

\clearpage
\begin{figure}
\epsscale{1.0}
\plotone{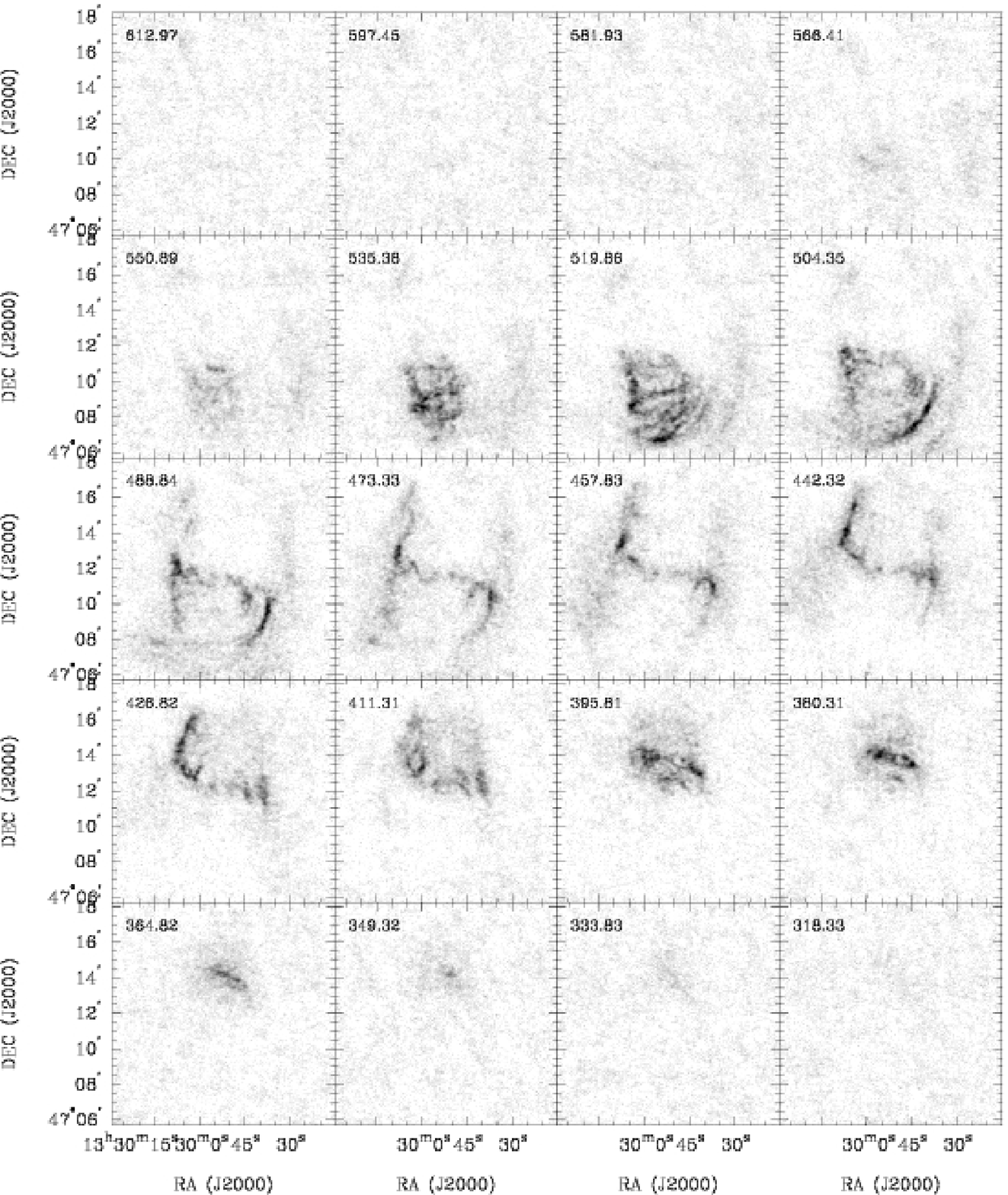}
\caption{{\bf NGC~5194:} Channel maps based on the NA cube (greyscale
  range: --0.02 to 8 mJy\,beam$^{-1}$).  Every third channel is shown
  (channel width: 5.2\,km\,s$^{-1}$). The area shown in each panel is
  identical to the area shown on the next figure}
\end{figure}

\clearpage
\begin{figure}
\vspace{0cm}  \epsscale{1.1}
\plotone{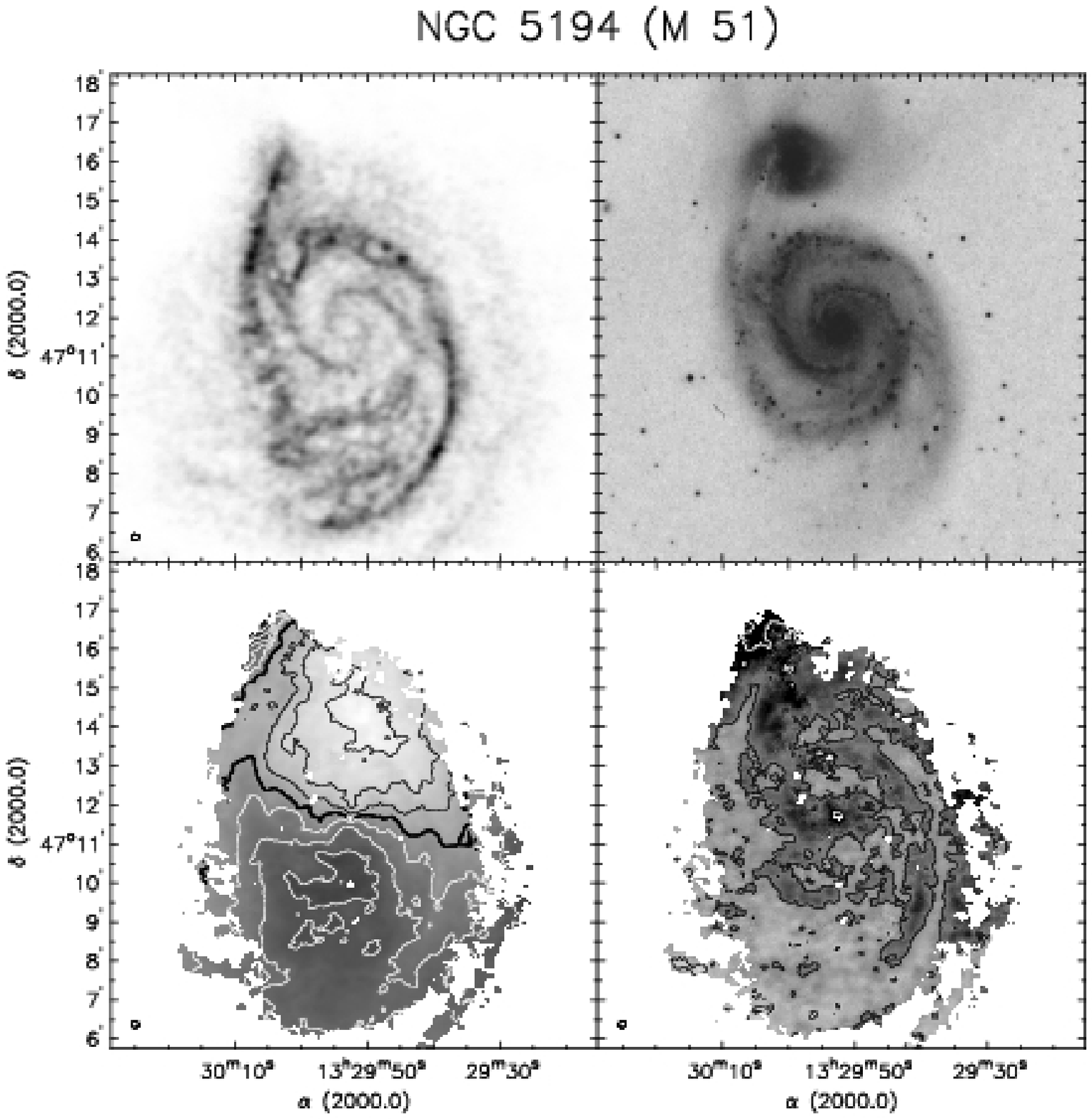}
\vspace{-2.5cm}
\caption{{\bf NGC~5194}. {\em Top left:} integrated \hi\ map (moment 0).
  Greyscale range from 0--312 Jy\,km\,s$^{-1}$. {\em Top right:}
  Optical image from the digitized sky survey (DSS). {\em Bottom
    left:} Velocity field (moment 1). Black contours (lighter
  greyscale) indicate approaching emission, white contours (darker
  greyscale) receding emission. The thick black contour is the
  systemic velocity ($v_{\rm sys}$=456.2 \,km\,s$^{-1}$), the
  iso--velocity contours are spaced by $\Delta\,v$=25\,km\,s$^{-1}$.
  {\em Bottom right:} Velocity dispersion map (moment 2). Contours are plotted
  at 5, 10 and 20\,km\,s$^{-1}$ (white contour: 50\,km\,s$^{-1}$).}
\end{figure}

%
% NGC\,5236
%

\clearpage
\begin{figure}
\epsscale{1.0}
\plotone{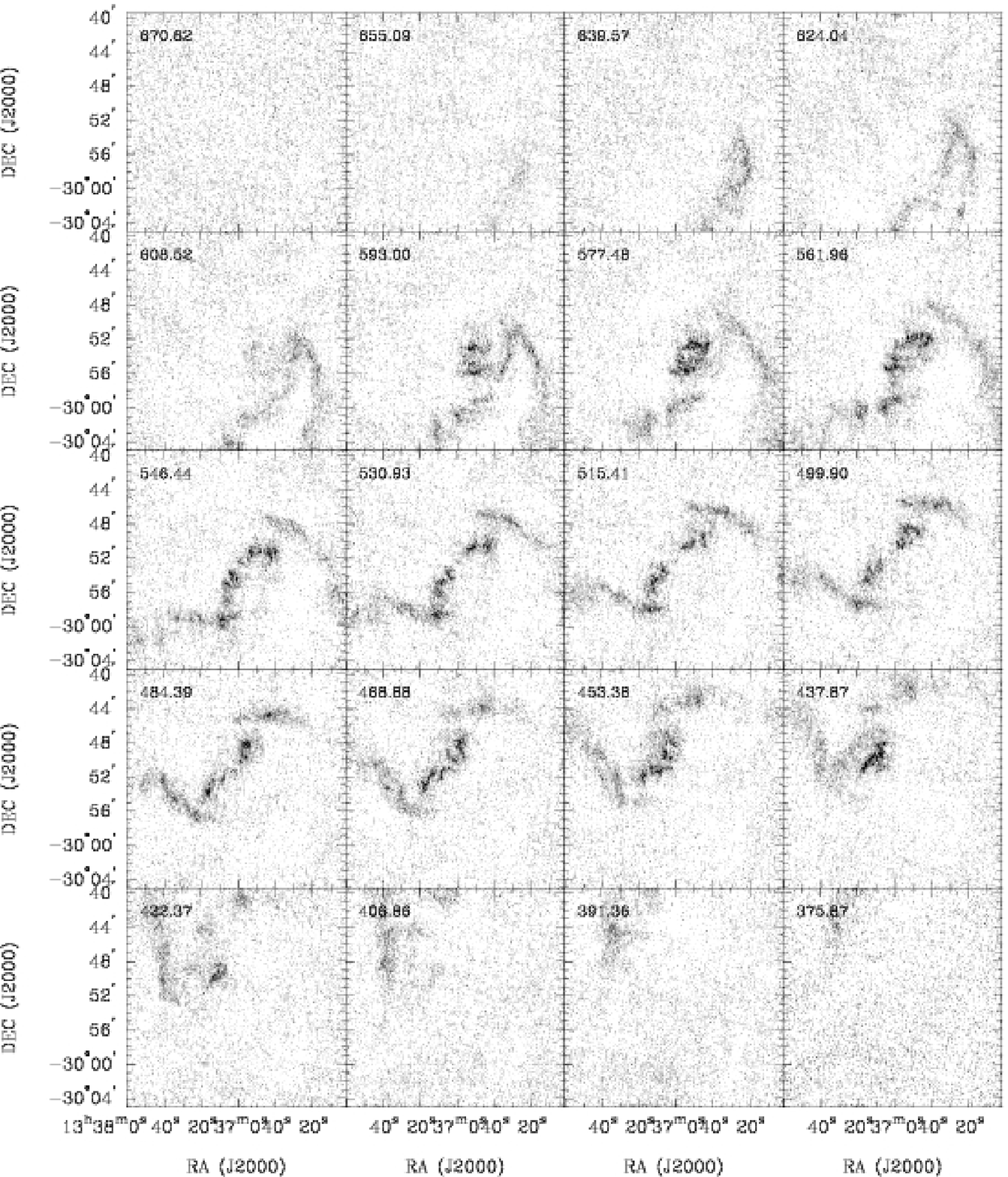}
\caption{{\bf NGC~5236:} Channel maps based on the NA cube (greyscale
  range: --0.02 to 8 mJy\,beam$^{-1}$).  Every sixth channel is shown
  (channel width: 2.6\,km\,s$^{-1}$). The area shown in each panel is
  identical to the area shown on the next figure}
\end{figure}

\clearpage
\begin{figure}
\vspace{0cm}  \epsscale{1.1}
\plotone{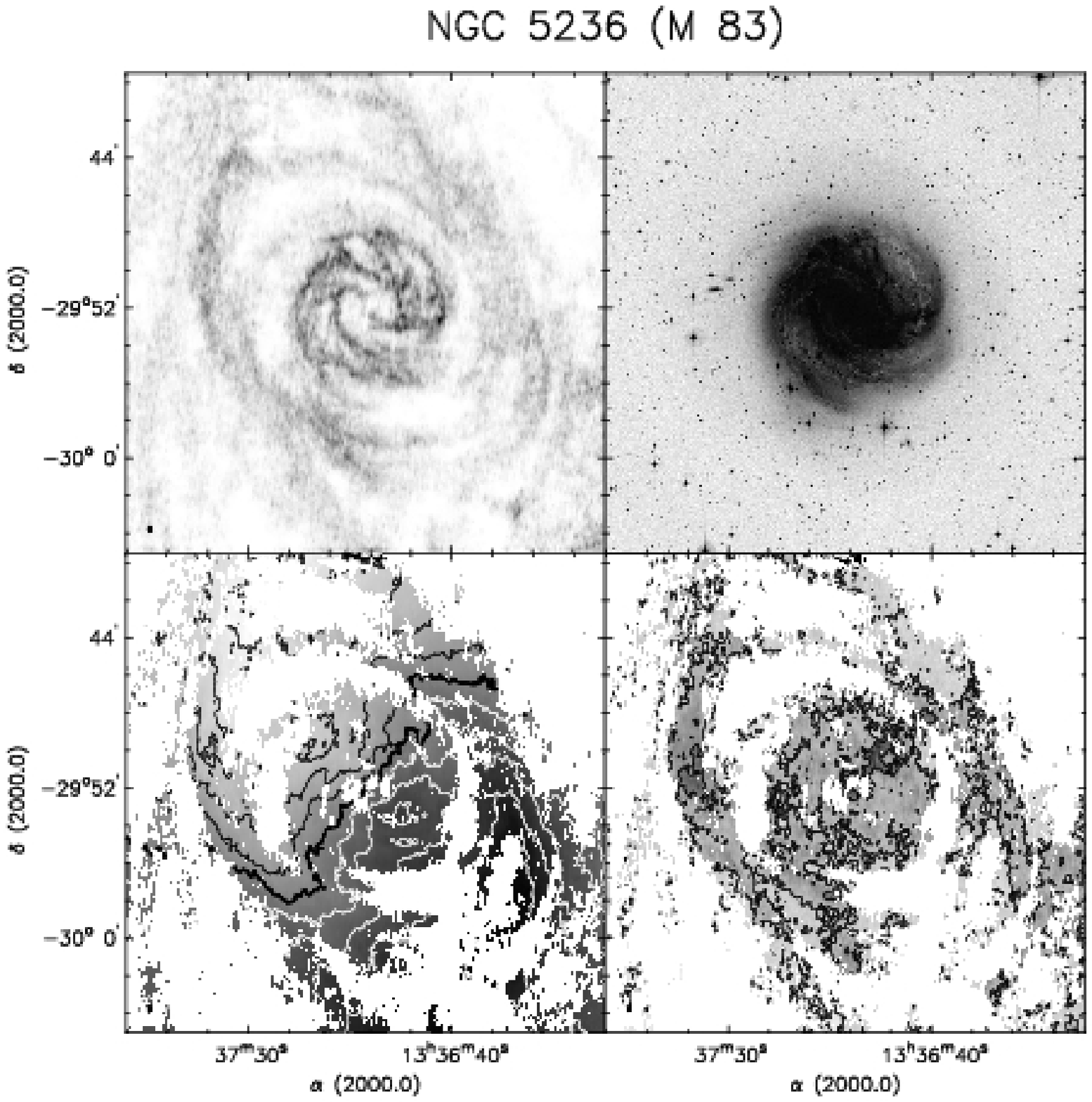}
\vspace{-2.5cm}
\caption{{\bf NGC~5236}. {\em Top left:} integrated \hi\ map (moment 0).
  Greyscale range from 0--301 Jy\,km\,s$^{-1}$. {\em Top right:}
  Optical image from the digitized sky survey (DSS). {\em Bottom
    left:} Velocity field (moment 1). Black contours (lighter
  greyscale) indicate approaching emission, white contours (darker
  greyscale) receding emission. The thick black contour is the
  systemic velocity ($v_{\rm sys}$=510.0 \,km\,s$^{-1}$), the
  iso--velocity contours are spaced by $\Delta\,v$=25\,km\,s$^{-1}$.
  {\em Bottom right:} Velocity dispersion map (moment 2). Contours are plotted
  at 5, 10 and 20\,km\,s$^{-1}$.}
\end{figure}

%
% NGC\,5457
%

\clearpage
\begin{figure}
\epsscale{1.0}
\plotone{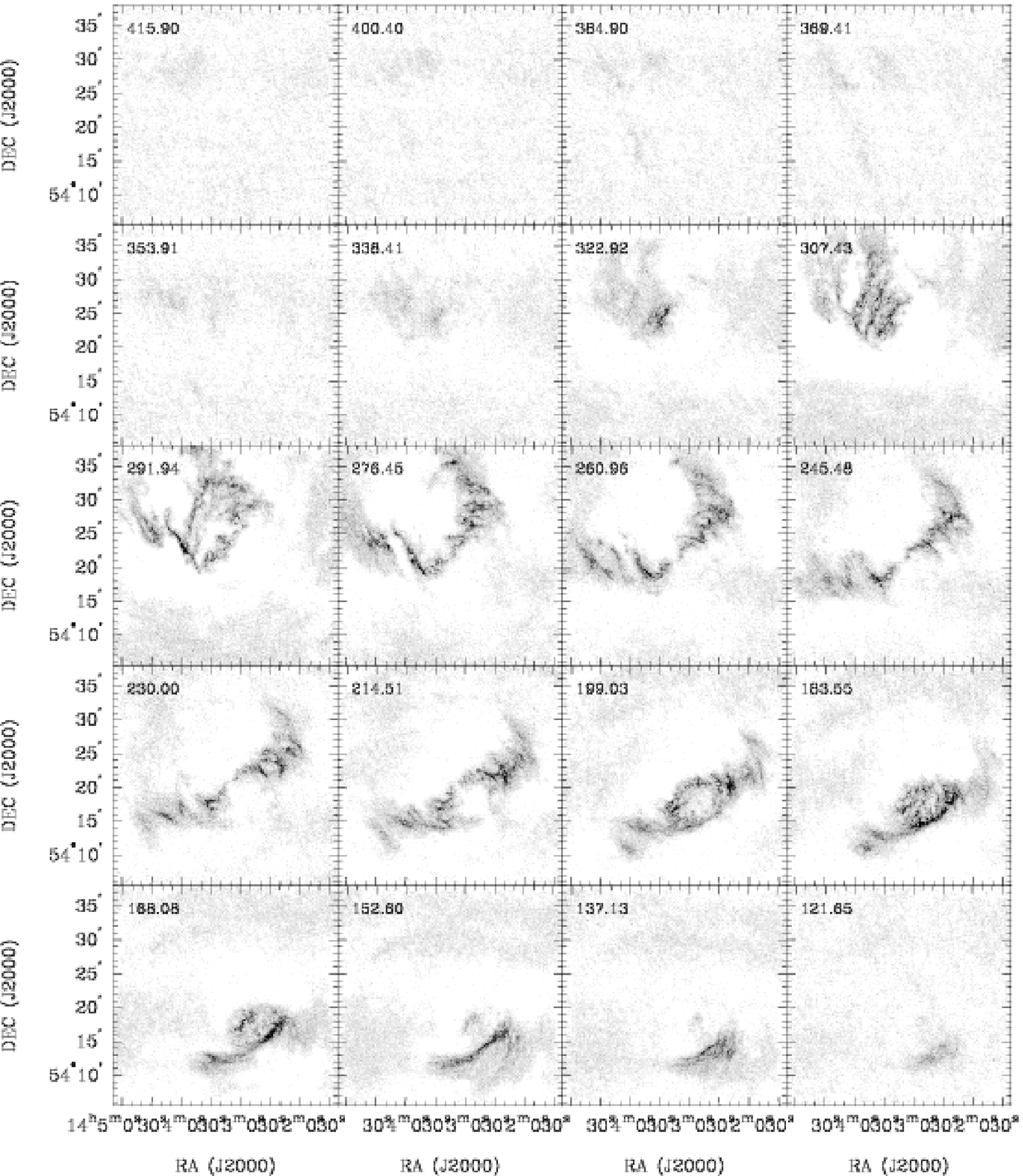}
\caption{{\bf NGC~5457:} Channel maps based on the NA cube (greyscale
  range: --0.02 to 10 mJy\,beam$^{-1}$).  Every third channel is shown
  (channel width: 5.2\,km\,s$^{-1}$). The area shown in each panel is
  identical to the area shown on the next figure}
\end{figure}

\clearpage
\begin{figure}
\vspace{0cm}  \epsscale{1.1}
\plotone{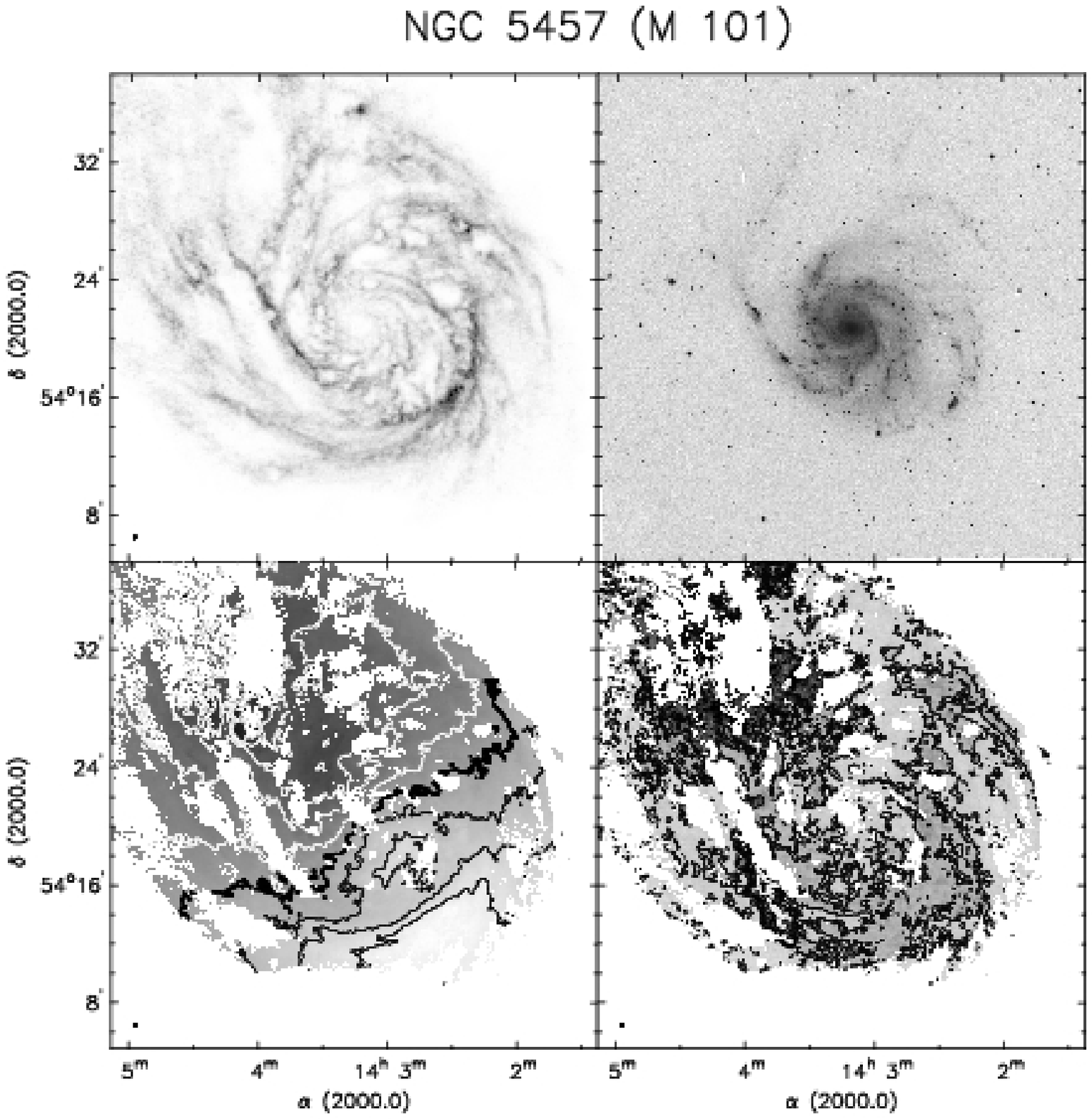}
\vspace{-2.5cm}
\caption{{\bf NGC~5457}. {\em Top left:} integrated \hi\ map (moment 0).
  Greyscale range from 0--510 Jy\,km\,s$^{-1}$. {\em Top right:}
  Optical image from the digitized sky survey (DSS). {\em Bottom
    left:} Velocity field (moment 1). Black contours (lighter
  greyscale) indicate approaching emission, white contours (darker
  greyscale) receding emission. The thick black contour is the
  systemic velocity ($v_{\rm sys}$=226.4 \,km\,s$^{-1}$), the
  iso--velocity contours are spaced bys $\Delta\,v$=25\,km\,s$^{-1}$.
  {\em Bottom right:} Velocity dispersion map (moment 2). Contours are plotted
  at 5, 10 and 20\,km\,s$^{-1}$.}
\end{figure}

%
% NGC\,6946
%

\clearpage
\begin{figure}
\epsscale{1.0}
\plotone{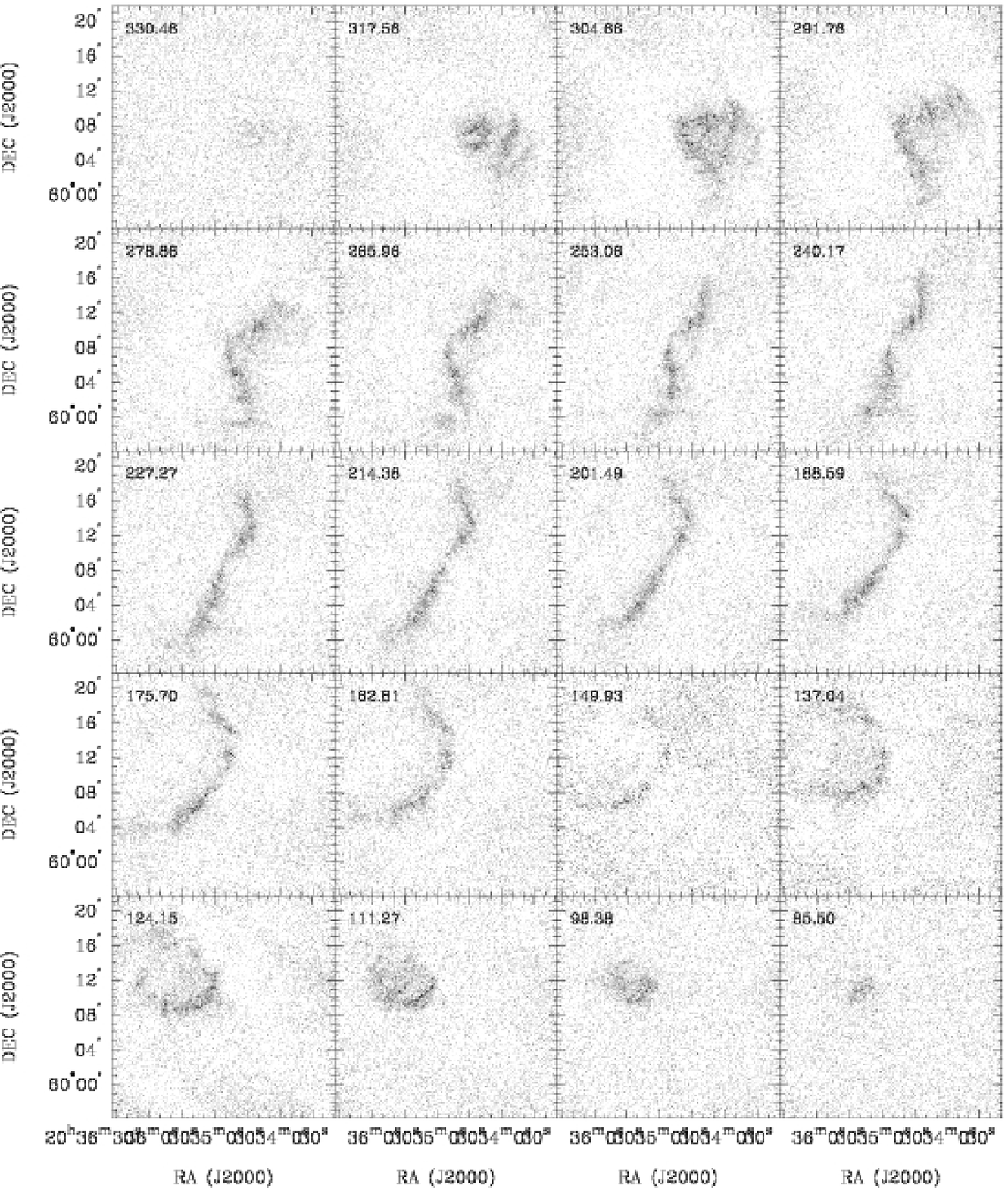}
\caption{{\bf NGC~6946:} Channel maps based on the NA cube (greyscale
  range: --0.02 to 6 mJy\,beam$^{-1}$).  Every fifth channel is shown
  (channel width: 2.6\,km\,s$^{-1}$). The area shown in each panel is
  identical to the area shown on the next figure}
\end{figure}

\clearpage
\begin{figure}
\vspace{0cm}  \epsscale{1.1}
\plotone{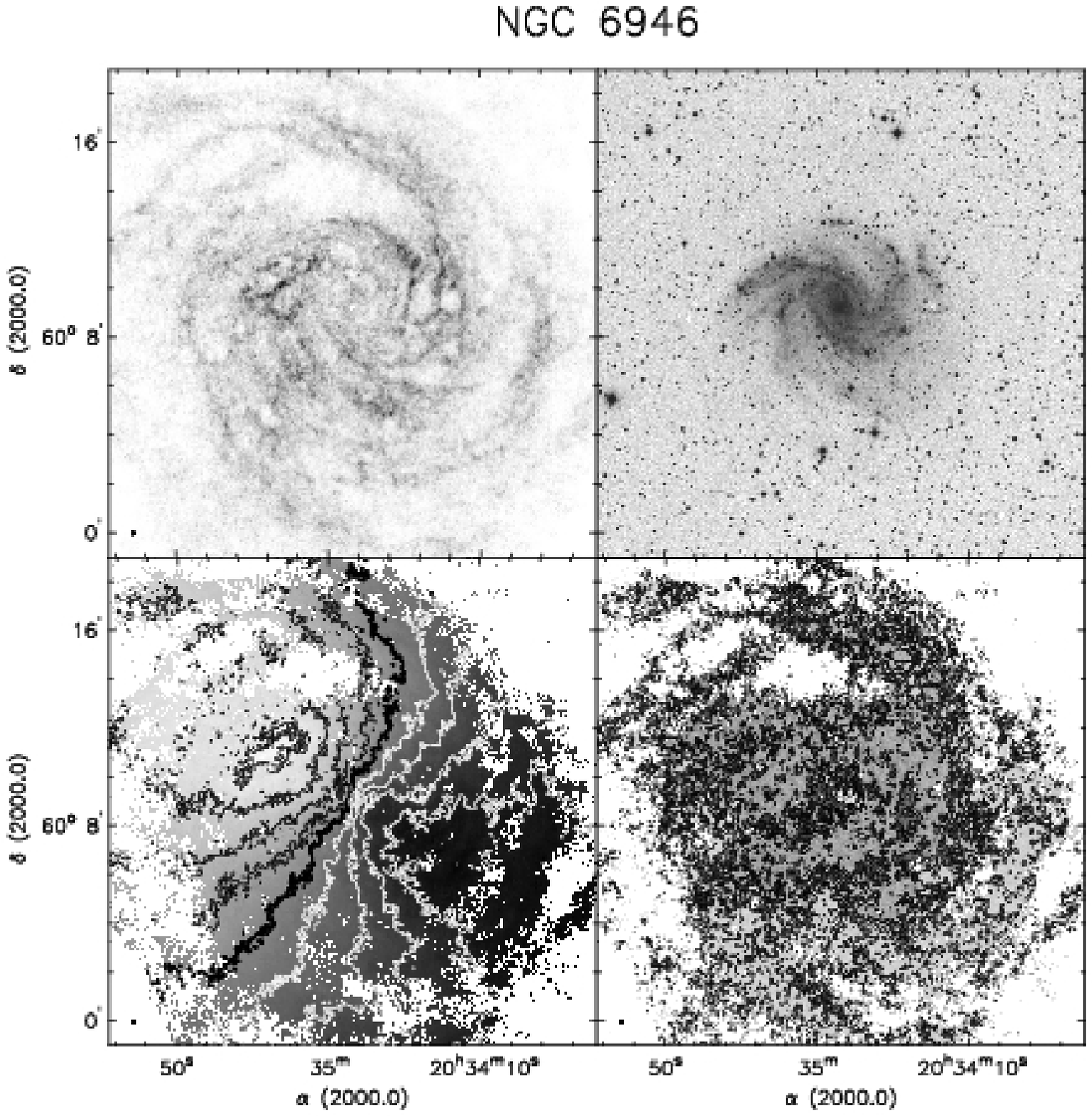}
\vspace{-2.5cm}
\caption{{\bf NGC~6946}. {\em Top left:} integrated \hi\ map (moment 0).
  Greyscale range from 0--121 Jy\,km\,s$^{-1}$. {\em Top right:}
  Optical image from the digitized sky survey (DSS). {\em Bottom
    left:} Velocity field (moment 1). Black contours (lighter
  greyscale) indicate approaching emission, white contours (darker
  greyscale) receding emission. The thick black contour is the
  systemic velocity ($v_{\rm sys}$=200.0 \,km\,s$^{-1}$), the
  iso--velocity contours are spaced by $\Delta\,v$=25\,km\,s$^{-1}$.
  {\em Bottom right:} Velocity dispersion map (moment 2). Contours are plotted
  at 5, 10 and 20\,km\,s$^{-1}$.}
\end{figure}

%
% NGC\,7331
%

\clearpage
\begin{figure}
\epsscale{1.0}
\plotone{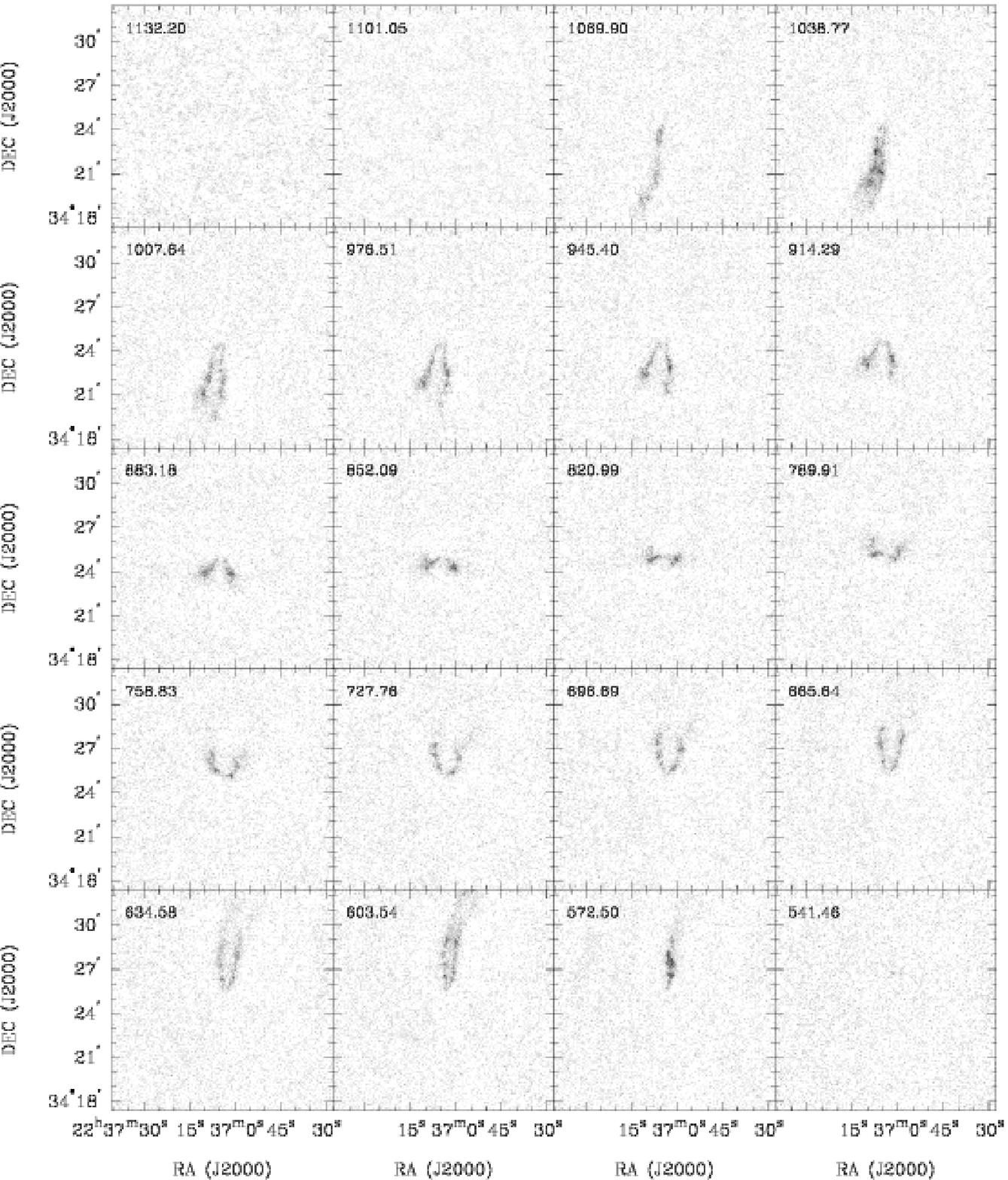}
\caption{{\bf NGC~7331:} Channel maps based on the NA cube (greyscale
  range: --0.02 to 8 mJy\,beam$^{-1}$).  Every sixth channel is shown
  (channel width: 5.2\,km\,s$^{-1}$). The area shown in each panel is
  identical to the area shown on the next figure}
\end{figure}

\clearpage
\begin{figure}
\vspace{0cm}  \epsscale{1.1}
\plotone{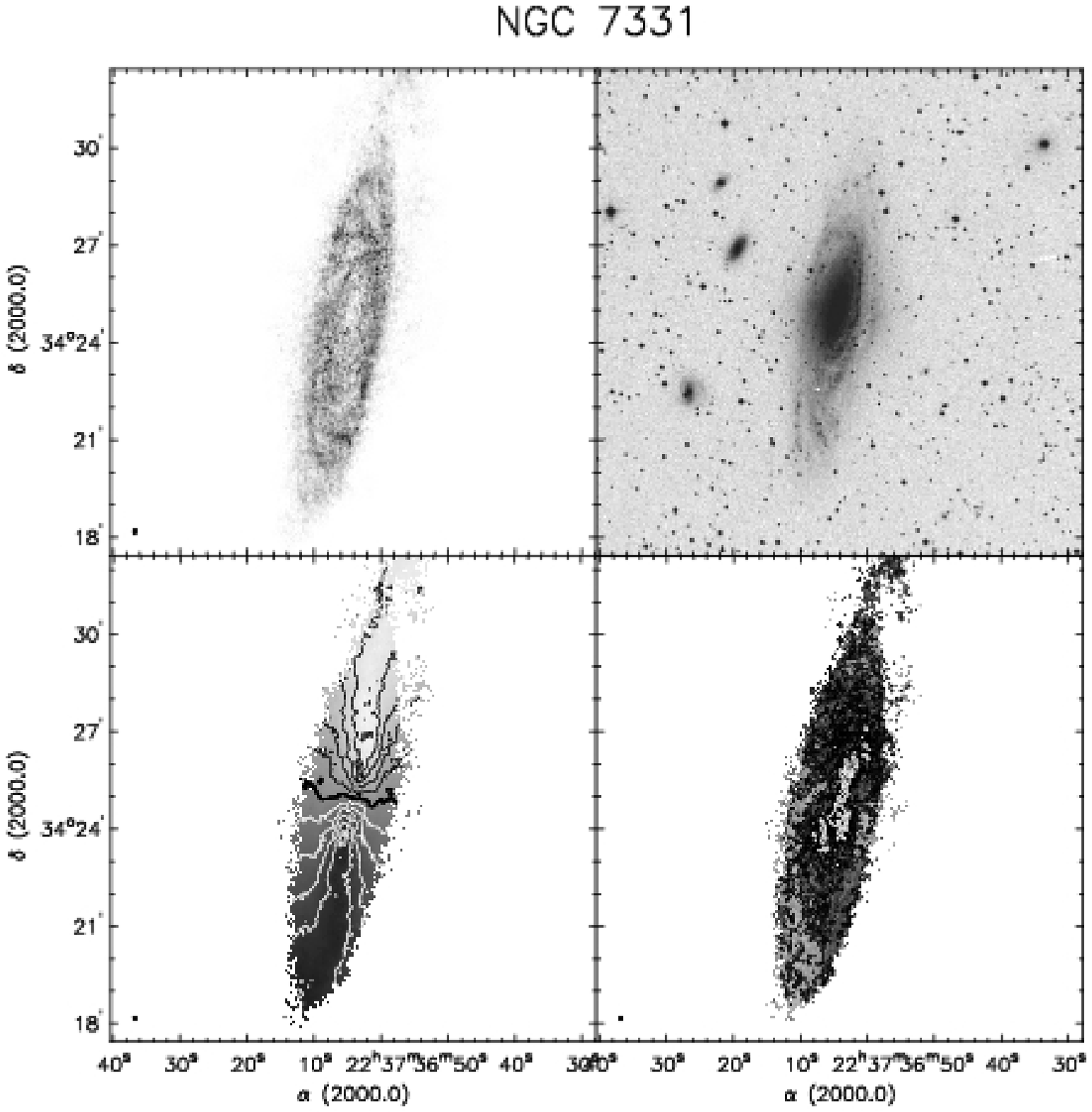}
\vspace{-2.5cm}
\caption{{\bf NGC~7331}. {\em Top left:} integrated \hi\ map (moment 0).
  Greyscale range from 0--258 Jy\,km\,s$^{-1}$. {\em Top right:}
  Optical image from the digitized sky survey (DSS). {\em Bottom
    left:} Velocity field (moment 1). Black contours (lighter
  greyscale) indicate approaching emission, white contours (darker
  greyscale) receding emission. The thick black contour is the
  systemic velocity ($v_{\rm sys}$=815.7 \,km\,s$^{-1}$), the
  iso--velocity contours are spaced by $\Delta\,v$=50\,km\,s$^{-1}$.
  {\em Bottom right:} Velocity dispersion map (moment 2). Contours are plotted
  at 5, 10 and 20\,km\,s$^{-1}$ (white contour: 50\,km\,s$^{-1}$).}
\end{figure}

%
% NGC\,7793
%

\clearpage
\begin{figure}
\epsscale{1.0}
\plotone{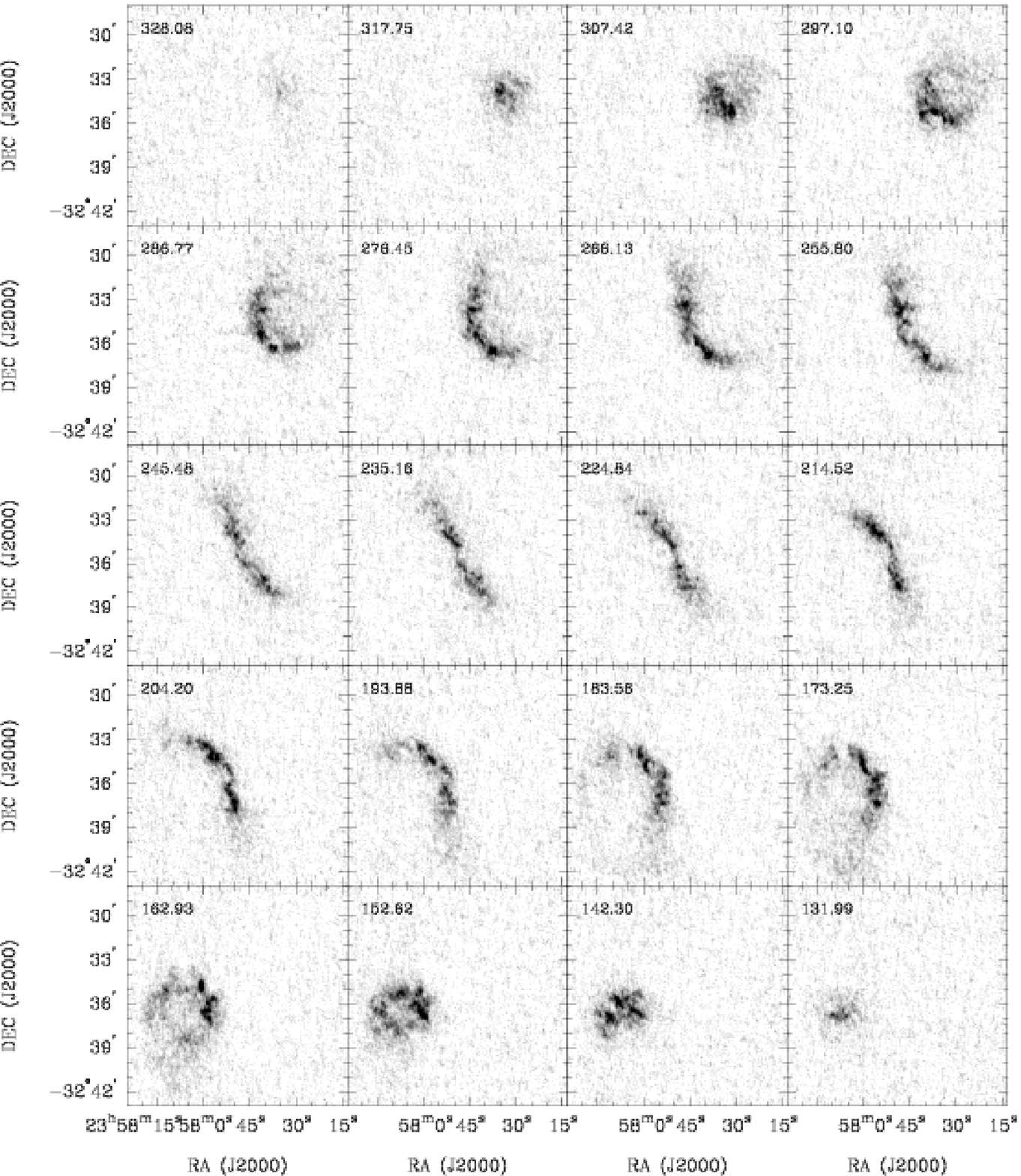}
\caption{{\bf NGC~7793:} Channel maps based on the NA cube (greyscale
  range: --0.02 to 14 mJy\,beam$^{-1}$).  Every fourth channel is
  shown (channel width: 2.6\,km\,s$^{-1}$). The area shown in each
  panel is identical to the area shown on the next figure}
\end{figure}

\clearpage
\begin{figure}
\vspace{0cm}  \epsscale{1.1}
\plotone{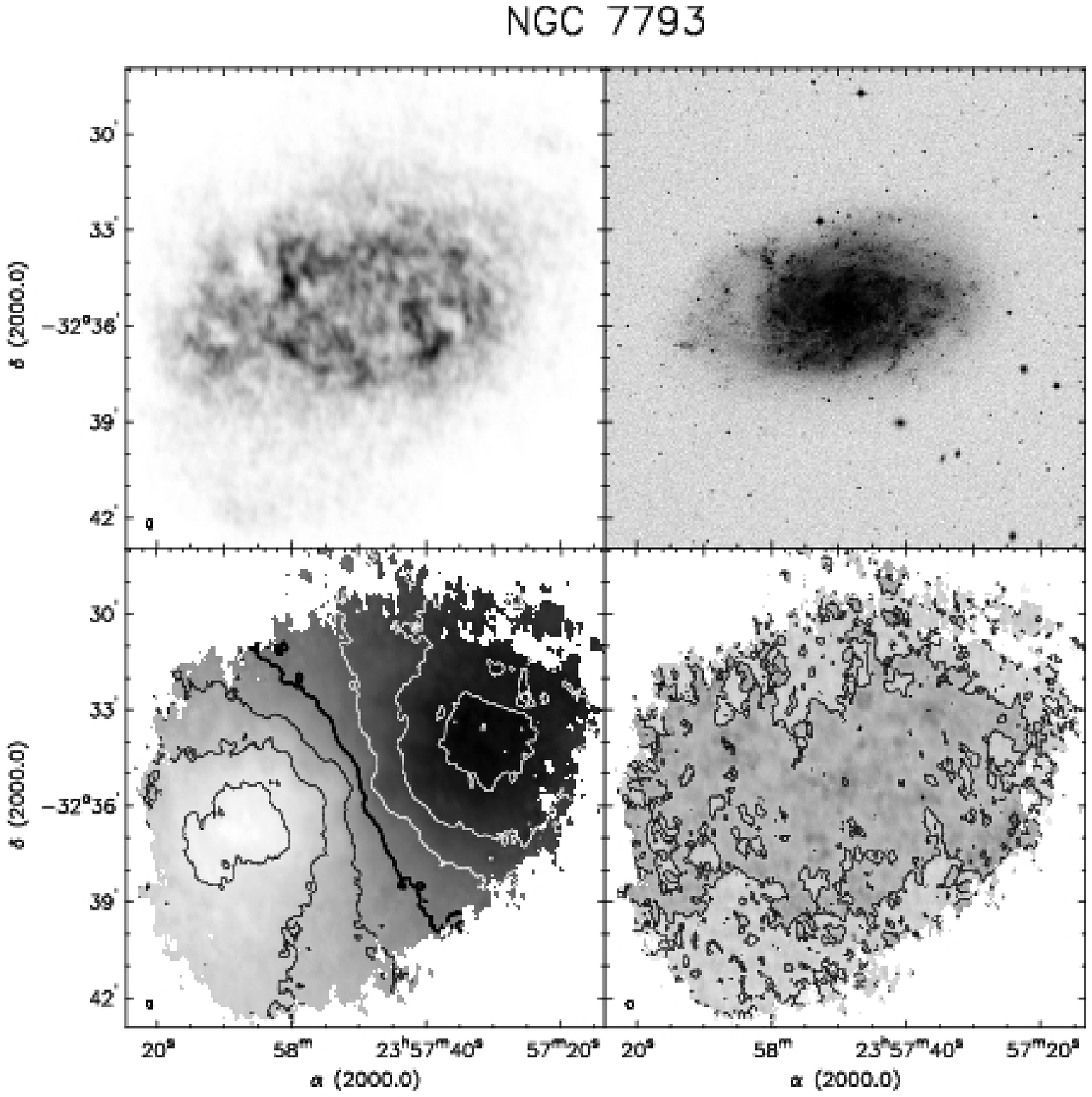}
\vspace{-2.5cm}
\caption{{\bf NGC~7793}. {\em Top left:} integrated \hi\ map (moment 0).
  Greyscale range from 0--507 Jy\,km\,s$^{-1}$. {\em Top right:}
  Optical image from the digitized sky survey (DSS). {\em Bottom
    left:} Velocity field (moment 1). Black contours (lighter
  greyscale) indicate approaching emission, white contours (darker
  greyscale) receding emission. The thick black contour is the
  systemic velocity ($v_{\rm sys}$=227.2\,km\,s$^{-1}$), the
  iso--velocity contours are spaced by $\Delta\,v$=25.0\,km\,s$^{-1}$.
  {\em Bottom right:} Velocity dispersion map (moment 2). Contours are plotted
  at 5, 10 and 20\,km\,s$^{-1}$.}
\end{figure}


\begin{thebibliography}{}


\bibitem[]{}
Baars, J. W. M., Genzel, R., Pauliny--Toth, I. I. K., \&
Witzel, A. 1977, \aap, 61, 99

\bibitem[]{}
Bagetakos, I., Brinks, E., et al., 2009, in prep.

\bibitem[]{}
Begeman et al.\, 1987, PhD Thesis, University of Groningen

\bibitem[]{} Bigiel, F., Leroy, A., Walter, F., Brinks, E., de~Blok,
  W.J.G., Madore, B., Thornley, M.D., 2008, AJ, in press

\bibitem[]{}
Bosma, 1981a, AJ, 86, 1791

\bibitem[]{}
Bosma, 1981b, AJ, 86, 1825

\bibitem[]{}
Braun, R.\ 1995, A\&AS, 114, 409 

\bibitem[]{}
Braun, R., 2007, A\&A, 461, 455

\bibitem[]{} Briggs, D. 1995, Ph.D. thesis, New Mexico Institute of
  Mining and Technology

\bibitem[]{}
Brinks, E.~\& Bajaja, E.\ 1986, A\&A, 169, 14

\bibitem[]{}
de Blok, W.~J.~G.~\& Walter, F.\ 2000, ApJL, 537, L95

\bibitem[]{} 
de Blok, W.~J.~G., McGaugh, S.~S., Bosma, A., \& Rubin,
  V.~C.\ 2001a, ApJL, 552, L23

\bibitem[]{}
de Blok, W.~J.~G., McGaugh, S.~S., \& Rubin, V.~C.\ 2001b, AJ, 122,
	2396

\bibitem[]{} de Blok, W.~J.~G., Walter, F., Brinks, E.,
        Trachternach, C., Oh, S.--H., Kennicutt, R.C., 2008, AJ, in press

\bibitem[]{}
Fisher, J.~R., \& Tully, R.~B.\ 1981, \apjs, 47, 139

\bibitem[]{}
Kamphuis, J., Sancisi, R., \& van der Hulst, T.\ 1991, \aap, 244, L29 

\bibitem[]{}
Kennicutt, R.~C., et al.\ 2003, \pasp, 115, 928 

\bibitem[]{}
Kim, S., Dopita, M. A., Staveley--Smith, L., Bessell, M. S. 1999, AJ,
	118, 2797  

\bibitem[]{}
Lee, J., 2007, PhD Thesis, University of Arizona

\bibitem[]{} Leroy, A., Walter, F., Brinks, E., Bigiel, F., de~Blok,
  W.~J.~G., Madore, B., Thornlet, M.D., 2008, AJ, in press

\bibitem[]{}
Moustakas, J., 2007, PhD Thesis, University of Arizona

\bibitem[]{}
Oh, S.--H., de~Blok, W.~J.~G., Walter, F., Brinks, E., Kennicutt, 
R.~C., 2008, AJ, in press

\bibitem[]{} 
  Ott, J., Walter, F., Brinks, E., Van Dyk, S.~D., Dirsch, B., \&
  Klein, U.\ 2001, \aj, 122, 3070

\bibitem[]{} 
  Paturel, G., Theureau, G., Bottinelli, L.,  Gouguenheim, L.,
  Coudreau-Durand, N., Hallet, N., Petit, C., 2003 A\&A 412, 57

\bibitem[]{}
Puche, D., Westpfahl, D., Brinks, E., \& Roy, J.\ 1992, AJ, 103, 1841 

\bibitem[]{} Rich, J., de~Blok, W.~J.~G., Brinks, E., Walter, F.,
  Bagetakos, I., Kennicutt, R.C., 2008, AJ, in press

\bibitem[]{}
Staveley--Smith, L., Sault, R.~J., Hatzidimitriou, D., Kesteven,
	M.~J., \& McConnell, D.\ 1997, MNRAS, 289, 225
      
\bibitem[]{} 
Swaters, R.,~A., van~Albada, T.S., van der Hulst, J.M.,
  Sancisi, R., 2002, A\&A, 390, 829

\bibitem[]{} 
  Tamburro, D., Rix, H.--W., Walter, F., Brinks, E., de~Blok,
  W.~J.~G., Mac~Low, M., 2008, AJ, in press

\bibitem[]{} 
Trachternach, C., de~Blok, W.J.G., Walter, F., Brinks, E., Kennicutt, R.~C., 
2008, AJ, in press

\bibitem[]{} 
Usero, A., Brinks, E., et al., 2009, AJ, in prep.

\bibitem[]{} 
van de Hulst, H.C.\ 1945, Nederlandsch Tijdschrift voor
  Natuurkunde, 11, 210

\bibitem[]{} 
Verheijen, M.~A.~W., \& Sancisi, R.\ 2001, \aap, 370, 765 

\bibitem[]{} 
Walter, F., \& Brinks, E.\ 1999, AJ, 118, 273

\bibitem[]{}
Walter, F., \& Brinks, E.\ 2001, AJ, 121, 3026

\bibitem[]{}
Walter, F., Weiss, A., Martin, C., \& Scoville, N.\ 2002, AJ, 123, 225

\bibitem[]{}
Westpfahl, D.~J., Coleman, P.~H., Alexander, J., \& Tongue, T.\ 1999,
	AJ, 117, 868

\bibitem[]{}
Zwaan, M. A., Briggs, F. H., Sprayberry, D., Sorar, E.,1997, ApJ 490, 173

\bibitem[]{} 
Zwaan, M., Ryan--Weber, E., Walter, F., Brinks, E., de Blok, W.~J.~G.,
  Kennicutt, R.~C., 2008, AJ, in press

\end{thebibliography}
\end{document}